\documentclass[structabstract]{aa}  
\usepackage{graphicx}
\begin{document}
   \title{Characterization of Young Stellar Clusters}

  \subtitle{}

\author{T. Santos-Silva
          \inst{1}
          \and
          J. Gregorio-Hetem\inst{1}
       }

   \offprints{T. Santos-Silva}

   \institute{ Universidade de S\~ao Paulo,
              IAG, Departamento de Astronomia, Brazil,
              \email{thaisfi@astro.iag.usp.br}           
      }

 %  \date{Received September 15, 1996; accepted March 16, 1997}

% \abstract{}{}{}{}{} 
% 5 {} token are mandatory
 
  \abstract
{}
%{\sl Aims}. 
{A  high number of embedded clusters is found in the Galaxy. Depending on the formation scenario, 
most of them can evolve to unbounded groups that are dissolved within a few tens of  Myr. A systematic study of young stellar 
clusters showing distinct characteristics provide interesting information on the evolutionary 
phases during the pre-main sequence. In order to identify and to understand these phases we
 performed a comparative study of 21 young stellar clusters.}
%{\sl Methods}. 
{Near-infrared data from 2MASS were used to determine the structural and fundamental parameters 
based on surface stellar density maps, radial density profile, and colour-magnitude
diagrams. The cluster members were selected according to the membership probability, which is based on
the statistical comparison with the cluster proper motion. Additional members were selected on basis of 
a decontamination procedure that was adopted to distinguish field-stars found in the direction of the cluster area.}
%{\sl Results}. 
{We obtained age and  mass distributions  by comparing pre-main sequence models 
with the position of cluster members in the colour-magnitude diagram.
 The mean age of our sample is $\sim$ 5 Myr, where 57\% of the objects is 
found in the 4 - 10 Myr range of age, while 43\% is $<$ 4 Myr old.
Their low  E(B-V) 
indicate that the members are not suffering high extinction ($A_V < 1$ mag), which means they are
more likely   young stellar groups than embedded clusters. 
Relations between structural and fundamental parameters
were used to verify differences and similarities that could be found among the clusters. 
The parameters of most of the objects  have the same trends or correlations.
Comparisons with other young clusters show similar relations among mass, radius and density. 
Our sample tends to have larger radius and lower volumetric density, when
compared to embedded clusters.
 These differences are compatible with the mean age of our sample, which we consider
intermediary between the embedded and the exposed phases of the stellar clusters evolution.
}
%{\sl Conclusions}.
{}


\authorrunning{Santos-Silva \& Gregorio-Hetem}
\titlerunning{Characterization of Young Stellar Clusters}

    \maketitle
%
%________________________________________________________________

%{Cluster with the same age also presents differences, like Lynga 14 with larger 
%core radius (r$_{c}$) and extinction and NGC 2659 with smaller r$_{c}$ and E(B-V), what show 
%that although both have similar ages, NGC 2659 is less embedded than Lynga 14.}

%________________________________________________________________
%%========================================Sect. 1 Introduction

\section{Introduction}

It is generally known that most stars are formed in groups or clusters. 
However, detailed studies on the initial processes 
of star formation are restricted to isolated dense cores of clouds (\cite{Shu 1987, Shu 2004}). 

The first stages of multiple star formation is usually evaluated through millimetric surveys, which are successful
probing the scenario of stellar clusters formation.
Based on 1.2mm data of the Mon OB1 region, for instance, \cite{Peretto 2005} discovered 27 proto-stellar cores
with diameters of about 0.04 pc and masses ranging from 20 to 40 M$_{\odot}$  associated to the young stellar cluster
NGC~2264. These results reveal the physical conditions for multiple massive star formation in a region that shows
a wide range of masses (\cite{Dahm 2008}) and ages (\cite{Flaccomio 1997, Rebull 2002})
indicating the occurrence of a large variety of young stellar groups.

In addition to millimetric studies,
the evolution  during the pre-main sequence (pre-MS) is more closely surveyed by using infrared data.
Particularly, near-infrared (NIR) provides information about the circumstellar  structure of the cluster members, whose 
physical conditions are directly related to the pre-MS evolutionary stage of the star. \cite{Gregorio-Hetem 2009},
for instance, have used X-ray results of sources detected in  CMa R1, combined with NIR and optical data,
to identify the young population associated to this star-forming region. They studied two young clusters
having similar  mass function, but one of them is older than the other by at least a few Myr.  A mixing of populations
seems to occur in the inter-clusters region, possibly related to sequential star formation. 

Systematic studies of young stellar clusters can directly probe several
 fundamental astrophysical problems, such as formation and evolution of open clusters, and more general problems,
 like the origin and early evolution of stars and planetary systems (\cite{Adams 2006} 2006, \cite{Adams 2010}).

\cite{Kharchenko 2005} (2005) presented angular sizes of cluster core and ``coronae" for a sample of 
520 Galactic open clusters, and \cite{Piskunov 2007} (2007)  revised these data  to obtain tidal radius
and core size 
estimated by fitting King profile to the observed density distribution.

\cite{Carpenter 2000} (2000) used 2MASS data to derive surface density maps of stellar clusters associated 
to molecular clouds, determining cluster radius and number of members based on the distributed population.

\cite{Adams 2006} (2006) explored the relations between cluster radius and stellar distribution in order to
study the dynamical evolution of young clusters. Numerical simulations were used to reproduce different
initial conditions of cluster-formation. The statistical calculations developed by \cite{Adams 2006} (2006)
were based on cluster membership and cluster radius correlations observed in the samples presented by
\cite{Carpenter 2000} (2000) and  Lada \& Lada (2003), who compiled the first extensive catalogue  of embedded 
clusters providing parameters such as mass, radius and number of members.

Lada \& Lada (2003), hereafter \cite{LL03}, verified the occurrence of a high number of embedded 
 clusters within  molecular clouds in the Galaxy. However, most of these objects probably  lose their dynamical equilibrium that would 
dissolve the group turning it into field-stars.

\cite{Pfalzner 2009} studied the evolution of a sample of 23 massive clusters younger than 20 Myr,
suggesting two distinct sequences. Depending on size and density, a bimodal distribution is found for the cluster size as a function of age.
%Moreover, in the plot of cluster density (M$_{\odot}$ pc$^{-3}$)  as a function of cluster radius (pc),
% Pfalzner (\cite{Pfalzner 2009}) verified that embedded clusters appear out of the evolutive sequence shown by the 
%sample of massive clusters.
Based on the mass-radius and density-radius dependence observed in the sample from \cite{LL03}, Pfalzner (\cite{Pfalzner 2011}) proposed a scenario of sequential formation of ``leaky" (exposed) massive 
clusters. 

On the other hand, differently of the time sequence proposed by Pfalzner (\cite{Pfalzner 2011}), the relations of cluster 
properties can be interpreted as formation condition (\cite{Adams 2006} 2006, \cite{Adams 2010}).

The structure of embedded clusters traces the physical conditions of their star-forming processes, since the origin and the evolution of 
stellar clusters are related to the distribution of dense molecular gas in their parental  cloud.
\cite{LL03} proposed two basic structural types, according to the cluster surface density.
Classical open clusters
%, like NGC~2264 (Lada, Young \& Greene 1993, Piche 1993) for instance, 
are characterised by
a high concentration in their surface distribution, whose radial profile is smooth and can be reproduced
by simple power law functions or King-like (isothermal) potential. This type of embedded cluster is considered centrally concentrated.
On the other hand, hierarchical type clusters  exhibit surface density distribution with
multiple peaks and show significant structure over a large range of spatial scale.
%, like NGC 1333 (Lada, Alves \& Lada 1996), for instance.
Although there are clear examples of both types of structures, their relative frequency is unknown.

The main goal of the present work is to characterise a large sample of 
young stellar clusters and to perform a comparative study  aiming to verify their similarities and differences,  which are related to their
evolutionary stages in the pre-MS. 

 The methods adopted by us have been commonly used 
in the characterization of open clusters like those developed by \cite{BB2009a} (2009a), for instance.
Stellar density maps are built from NIR data in order to derive parameters on the basis of radial density profile. These  parameters define the structure of the cluster,
which is related to its origin and evolution.

%A decontamination method that statistically compares the colours 
%and magnitudes of field-stars and cluster members was also an useful tool that we adopted from previous works. 

The analysis based on radial density profile is unprecedented to 86\% of our sample.
Considering the lack of systematic studies comparing pre-MS stellar clusters, 
%for a large number of clusters 
 distributed in different galactic regions, the present work aims to provide sets of
structural and fundamental parameters, determined in a uniform data analysis that may contribute to the discussion on the origin of  stellar groups. 

In Sect. 2 we describe the sample by presenting the selection criteria and the decontamination
method for  distinguishing cluster members  from field-stars. Section 3 is dedicated to determine the structural 
parameters, which are used to accurately determine distance, age and mass that are presented in  Sect. 4. 
A comparative analysis is performed in Sect. 5, while Sect. 6 summarises
the results and the main conclusions. Finally, Appendix A displays all the plots of the entire sample.

%%-----------------------------------------------  Table 1 ----------------------
\begin{table*}[ht] 
\caption{List of clusters and their structural parameters.} 
\begin{center}
{\scriptsize
\begin{tabular}{|l|c|c|c|c|c|c|c|c|c|c|}
\hline
 Cluster & $\alpha$  & $\delta$ & R & $\sigma_{bg} $ & $\sigma_{0}$ & $<n_{*}>$ & r$_{c}$ & $\delta_{c}$ & r$_{c}$/R & Class  \\ 
         & (h m s) & ($^o$ ')  &  (pc) & (pc$^{-2}$) & (pc$^{-2}$) &(pc$^{-2}$) & (pc) &  &  &  \\ \hline \hline
Collinder 205 & 09 00 31(5) & -48 59 & 3.4$\pm$1.2 & 9$\pm$3 & 51$\pm$16 & 4.8$\pm$1.9 & 0.30$\pm$0.03 & 6.6$\pm$2.6 & 0.09$\pm$0.03 & ??\\
Hogg 10 & 11 10 40(9) & -60 14 & 1.9$\pm$0.4 & 16$\pm$2 & 9$\pm$4 & 7.6$\pm$1.7 & 0.46$\pm$0.28 & 1.6$\pm$0.3 & 0.24$\pm$0.15 & H\\
Hogg 22 & 16 46 35(6) & -47 05 & 1.9$\pm$0.5 & 20$\pm$4 & 7$\pm$3 & 9.0$\pm$3.8 & 1.91$\pm$1.10 & 1.3$\pm$0.2 & 1.03$\pm$0.65 & H\\
Lynga 14 & 16 55 03(5) & -45 14 & 1.1$\pm$0.4 & 55$\pm$18 & 153$\pm$52 & 16.7$\pm$7.0 & 0.22$\pm$0.04 & 3.8$\pm$1.3 & 0.20$\pm$0.08 & CC\\
Markarian 38 & 18 15 15(5) & -19 01 & 1.6$\pm$0.5 & 18$\pm$6 & 44$\pm$15 & 6.8$\pm$2.3 & 0.13$\pm$0.03 & 3.4$\pm$1.2 & 0.08$\pm$0.03 & CC\\
NGC 2302 & 06 51 55(4) & -07 05 & 2.3$\pm$0.5 & 8$\pm$2 & 16$\pm$4 & 4.0$\pm$1.4 & 0.54$\pm$0.10 & 3.0$\pm$0.6 & 0.23$\pm$0.06 & CC\\
NGC 2362$^a$ & 07 18 42(4) & -24 28 & 1.9$\pm$0.4 & 13$\pm$3 & 62$\pm$14 & 3.3$\pm$1.0 & 0.26$\pm$0.05 & 5.8$\pm$1.7 & 0.13$\pm$0.04 & CC\\
NGC 2367 & 07 20 06(4) & -21 53 & 2.9$\pm$0.7 & 5$\pm$1 & 13$\pm$3 & 2.3$\pm$0.6 & 0.34$\pm$0.11 & 3.6$\pm$0.7 & 0.12$\pm$0.05 & ??\\
NGC 2645 & 08 39 05(6) & -46 14 & 2.1$\pm$0.5 & 13$\pm$3 & 48$\pm$12 & 7.5$\pm$2.1 & 0.22$\pm$0.04 & 4.8$\pm$1.2 & 0.10$\pm$0.03 & ??\\
NGC 2659 & 08 42 37(6) & -44 59 & 3.6$\pm$0.5 & 9$\pm$1 & 11$\pm$2 & 5.2$\pm$0.9 & 1.93$\pm$0.37 & 2.2$\pm$0.3 & 0.53$\pm$0.12 & H\\
NGC 3572 & 11 10 26(8) & -60 15 & 2.5$\pm$0.5 & 19$\pm$3 & 76$\pm$15 & 7.7$\pm$1.9 & 0.16$\pm$0.03 & 5.0$\pm$1.1 & 0.06$\pm$0.02 & H\\
NGC 3590 & 11 12 59(7) & -60 47 & 1.5$\pm$0.4 & 30$\pm$7 & 96$\pm$27 & 11.7$\pm$4.5 & 0.29$\pm$0.06 & 4.2$\pm$1.2 & 0.20$\pm$0.07 & CC\\
NGC 5606 & 14 27 48(8) & -59 38 & 2.6$\pm$0.4 & 11$\pm$2 & 25$\pm$5 & 4.8$\pm$0.9 & 0.51$\pm$0.13 & 3.2$\pm$0.6 & 0.20$\pm$0.06 & CC\\
NGC 6178 & 16 35 47(6) & -45 39 & 1.7$\pm$0.4 & 24$\pm$4 & 32$\pm$11 & 12.2$\pm$2.7 & 0.30$\pm$0.12 & 2.3$\pm$0.5 & 0.18$\pm$0.08 & H\\
NGC 6604$^b$ & 18 18 04(3) & -12 15 & 2.6$\pm$0.7 & 11$\pm$3 & 17$\pm$5 & 4.4$\pm$1.3 & 0.50$\pm$0.11 & 2.6$\pm$0.6 & 0.19$\pm$0.07 & H\\
NGC 6613 & 18 19 58(4) & -17 06 & 2.9$\pm$0.7 & 9$\pm$2 & 82$\pm$19 & 4.9$\pm$1.2 & 0.12$\pm$0.01 & 10.6$\pm$3.5 & 0.04$\pm$0.01 & CC\\
Ruprecht 79 & 09 40 59(7) & -53 51 & 3.9$\pm$0.5 & 7$\pm$1 & 4$\pm$1 & 3.6$\pm$0.5 & 2.2$\pm$0.8 & 1.5$\pm$0.2 & 0.57$\pm$0.22 & CC\\
Stock 13$^c$ & 11 13 04(8) & -58 53 & 2.3$\pm$0.3 & 15$\pm$2 & 42$\pm$14 & 3.8$\pm$0.9 & 0.11$\pm$0.05 & 3.8$\pm$1.0 & 0.05$\pm$0.02 & H\\
Stock 16 & 13 19 30(8) & -62 38 & 2.8$\pm$0.7 & 15$\pm$3 & 8$\pm$4 & 5.8$\pm$1.6 & 1.2$\pm$0.6 & 1.5$\pm$0.3 & 0.42$\pm$0.24 & H\\
Trumpler 18 & 11 11 27(9) & -60 40 & 4.6$\pm$0.8 & 9$\pm$1 & 15$\pm$4 & 2.5$\pm$0.4 & 0.38$\pm$0.10 & 2.8$\pm$0.5 & 0.08$\pm$0.03 & H\\
Trumpler 28 & 17 37 02(5) & -32 38 & 1.4$\pm$0.5 & 9$\pm$4 & 25$\pm$9 & 12.3$\pm$4.5 & 0.84$\pm$0.16 & 3.7$\pm$1.4 & 0.61$\pm$0.24 & H\\\hline

\end{tabular}
}
\label{tab1}
\end{center}
{\scriptsize
Columns description: (1) Identification; (2, 3) J$_{2000}$ coordinates;
(4)  cluster radius; (5) background density; (6) core density; (7) observed average density; (8) core radius; (9) density-contrast parameter; 
(10) ratio of core-size to cluster-radius; (11) cluster type: hierarchical (H), centrally concentrated (CC), undefined (??).
 Notes: parameters from \cite{Piskunov 2007}: (a) $R_c$=0.6$\pm$0.4 pc; (b) $R_c$= 2.4$\pm$1.4pc;
(c) $R_c$=1.0$\pm$0.8 pc. 
}
\end{table*}

%========================================Section 2

\section{Description of the sample}

%========================================Section 2.1

Several compilations of open stellar clusters are available in the literature, for example \cite{Lynga 1987}, \cite{Loktin 1994},
\cite{Mermilliod 1995}, \cite{LL03}, \cite{Kharchenko 2005} (2005),  \cite{Piskunov 2007},
WEBDA\footnote[1]{\textit{\textit{http://www.univie.ac.at/webda/}}} and DAML\footnote[2]{\textit{\textit{http://www.astro.iag.usp.br/$\sim$wilton/}}} (\cite{Dias 2002, Dias 2006}),
among others. In Sect. 2.1 we present the criteria used to select the clusters, while Sect. 2.2
describes the method used to exclude field-stars that were found in the same direction of the cluster.

\subsection{Clusters selection and observational data}

Aiming to focus our study on pre-MS objects, the DAML catalogue  was used to select young stellar clusters 
with ages in the 1 - 20 Myr range. 
Distances smaller than 2~Kpc were used as selection criterion in order to ensure  good
quality of photometric data. Despite our selection is limited to
southern objects, they are distributed in different star-forming regions enabling us to compare diverse environments.
Table 1 gives the list of 21 selected clusters.

A membership probability (P) estimated on basis of proper motion is given by DAML. For the selection of stars belonging to the clusters, 
 the values  P$>$ 50\%  were adopted to indicate the {\it possible} members, 
 hereafter denoted by P50.

In order to confirm the cluster centre coordinates available in the literature, we evaluated the distribution
of number of stars as a function of right ascension ($\alpha$) and declination ($\delta$). Gaussian profiles
were fitted to these distributions, in order to estimate their centroid.
A good agreement with the literature was found, within the  errors estimated by the fitting that are $\Delta\delta \sim$ 1 arcmin
and 0.75'$ < \Delta\alpha < $ 2.25'. Table 1 gives the error on $\alpha$ indicated in between parenthesis.

The NIR data, which provide maximum variation among colour-magnitude diagrams (CMDs) of clusters with different ages 
(\cite{BBG2004}), were extracted from 2MASS (All Sky Catalogue of Point Sources - \cite{Cutri 2003}). 
The JHK$_s$ magnitudes were  searched for all stars located within a radius of 30 arcmin from the cluster 
centre. Only good accuracy photometric data were used in our analysis, by selecting objects with AAA quality flags
that ensure the best photometric and astrometric qualities (\cite{Lee 2005}).

%========================================Section 2.2
\subsection{Field-stars decontamination}

Most of our clusters are projected against dense fields of stars in the Galactic disk, making difficult to distinguish the cluster members. 
The first step to determine the structural and fundamental parameters is to proceed the field-stars decontamination. 

This process is based on statistical analysis by comparing the stellar density of the cluster area with a reference region, near to the cluster, according to their characteristics in the CMD.  
The decontamination algorithm was employed as follows: 

\begin{itemize} 
\item the whole range of magnitudes and colours is divided into three dimensional cells  (J, J-H and J-K$_{S}$);

\item for each cell ($i$), the  stellar density  of field-stars ($\sigma _{f} $ = n$_{f}$ / a$_{f}$) is obtained by counting the number of stars (n$_{f}$) appearing in the reference region (a$_{f}$);

\item similarly,  the total stellar density ($\sigma _{t}$ = n$_{t}$ / a$_{c}$) is obtained  by counting all stars (n$_{t}$) in the cluster area (a$_{c}$); 

\item the number of  field-stars  ($n_{f}$), appearing in the cluster area, that have colours and magnitude similar to those estimated in the reference region  is calculate by:

\begin{equation}
 n_{f}=\frac{\sigma_{f}}{\sigma{t}}  \times n_{t}
\label{eq.2.2}
\end{equation}

\item Finally, the number of field-stars in the cluster area is randomly subtracted in 
each cell,  remaining  $N =\Sigma _i \ (n_{t_i} - n_{f_i}$) members of the cluster. 
\end{itemize}

In order to minimize errors introduced by the parameters choice and  uncertainties on 2MASS data, 
the decontamination algorithm is applied to several grids by adopting three different positions in the CMD. 
By defining  a cell size $ \Delta J $ = 0.5 mag, the algorithm uses a grid starting at $J_{i}$ 
and two  other grids  starting at $J_{i} \pm \frac {1} {3} \Delta J $. This dithering is also made for both colours, 
$\Delta$(J-H) = $\Delta$(J- K$_{S}$) = 0.1 mag, providing 27 different results 
for each star. By this way, we obtain the probability that a given object should be considered a field-star. All {\it possible}  members (P50) have 0\% of probability of being a field-star.

We considered as {\it candidate} members (denoted by P?) the stars that were not removed 
by the field-star decontamination.  Both, P50 and P? members were used,  
but having different weight, in the estimation of distance and age, presented in Sect. 4.2.

The adopted decontamination procedure  has been used in studies of several types of clusters: 
objects showing low contrast relative to the field-stars (\cite{BBonatto 2005}, embedded clusters (\cite{BSB2006}), young clusters (\cite{Bonatto 2006b}), among others.

%%%------------------------------------ Fig. 1 Densit maps & RDP
\begin{figure*}[ht]
\includegraphics[width=7.0cm]{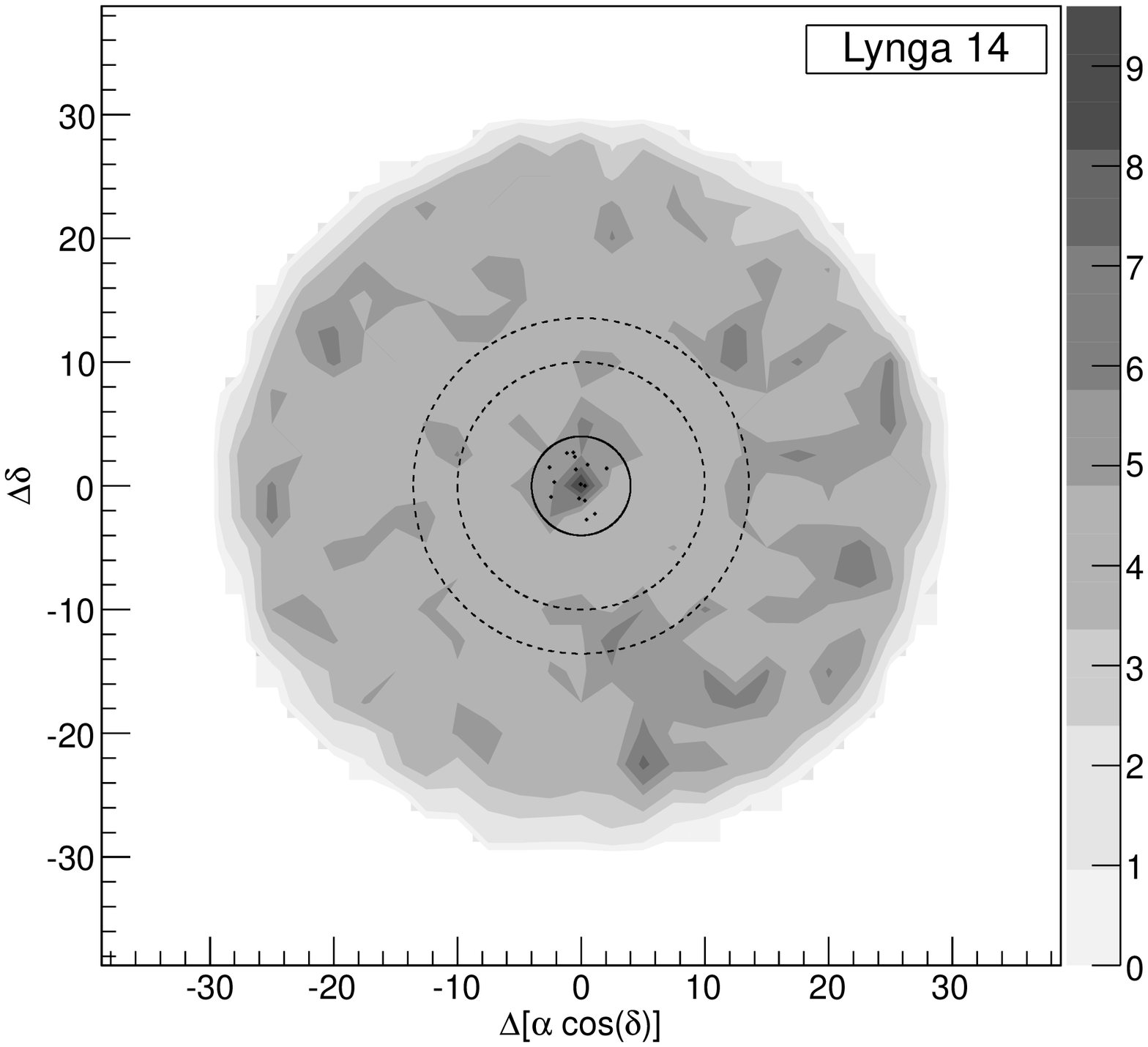}
\includegraphics[width=7.0cm]{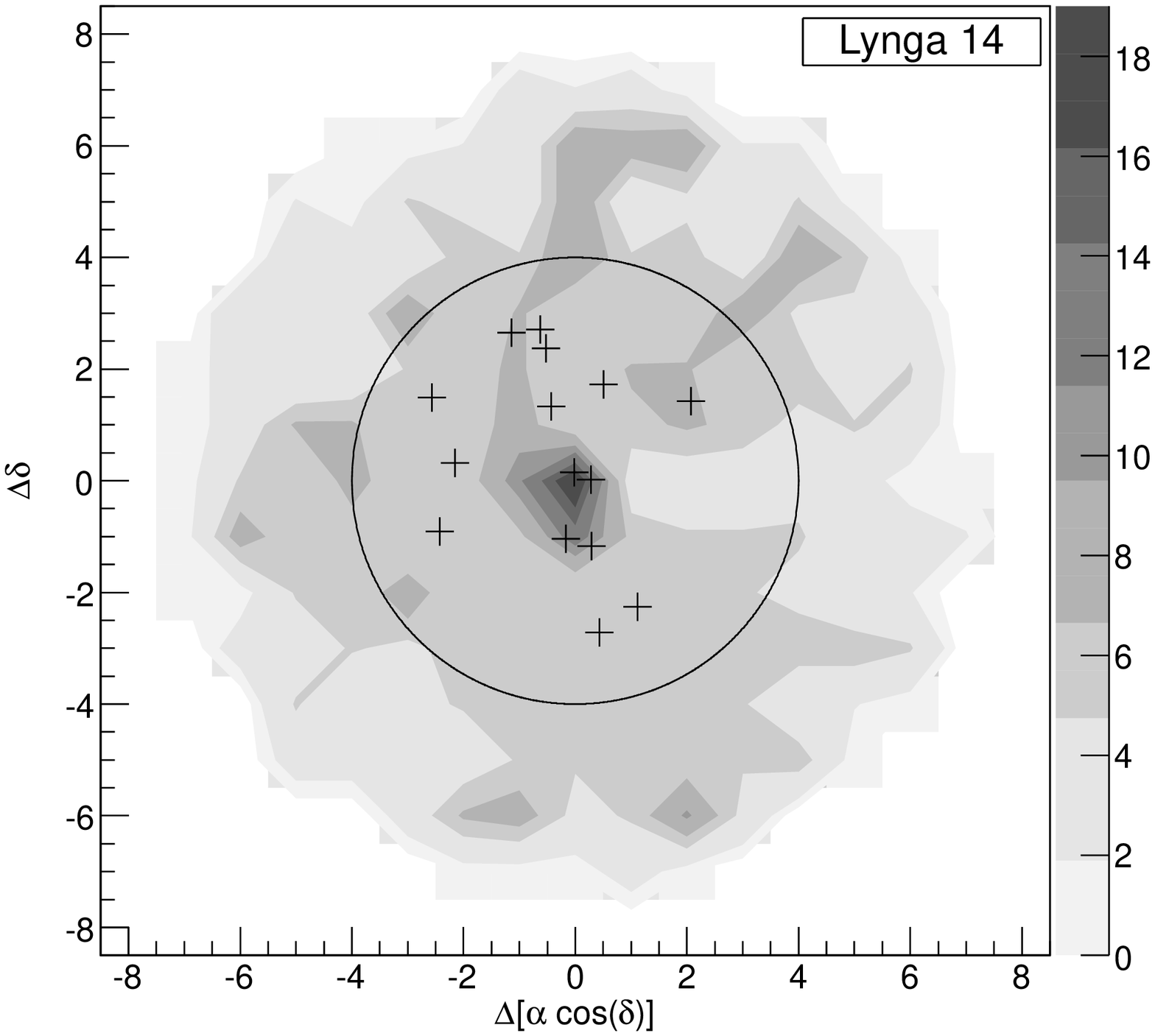}
\includegraphics[width=4.2cm]{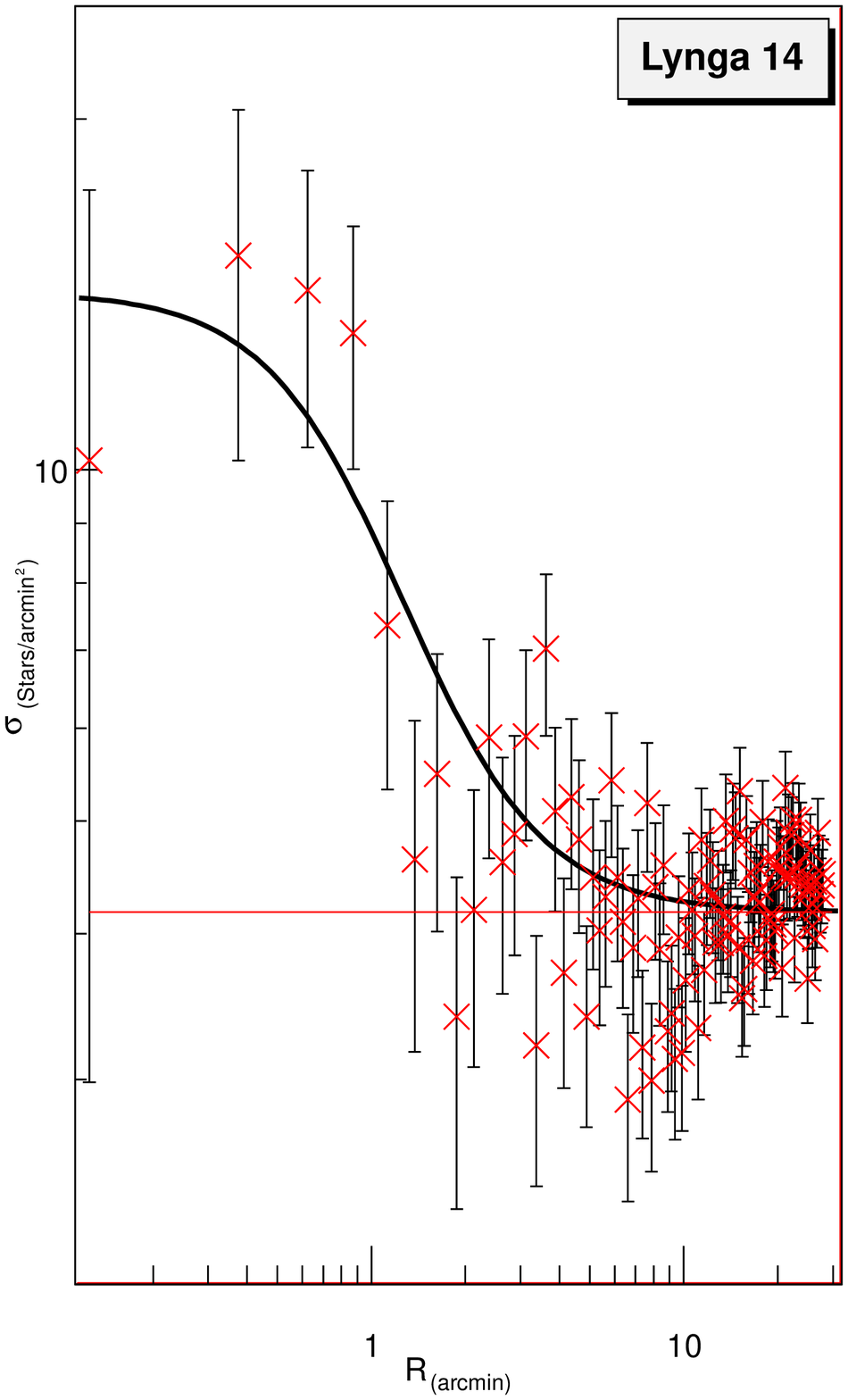}
\begin{center}
%\setcaptionmargin{1cm}
\caption{{\it Left}: Stellar surface-density map ($\sigma$ (stars/arcmin$^{2}$)) obtained for the region of 30 arcmin around
Lynga~14. The comparison field-stars area is indicated by dashed lines, while the full line indicates the cluster area.
{\it Center}: a zoom of the $\sigma$ map indicating by crosses the position of objects with membership probability P$>$70\%.
{\it Right}: The distribution of the stellar density as a function of radius. The best fitting of observed radial 
density profile, indicated by the full line, was obtained by using the model from \cite{King 1962}. A red line indicates 
the background density ($\sigma_{bg}$).} 
\label{rdp}
\end{center}
\end{figure*}
%%%------------------------------------ end of Fig. 1
%========================================Section 3
\section{Structural parameters}

The structural parameters were determined on basis of stellar surface density, derived from stellar
 surface-density maps and radial density profile that are detailed in the Sects. 3.1 and 3.2. 
 
 %Similar analyses are used in previous studies of young open clusters such as NGC~6611 (Bonatto, Santos
 % Jr. \& Bica 2006), and Pismis~5, vdB~80, NGC~1931 and BDSB~96 (Bonatto \& Bica 2009). 

The first step, in the calculation of the stellar surface density, is to enhance the contrast 
between surface distribution of cluster members and field-stars by using a colour-magnitude filter.  

Pre-MS isochrones were adopted to establish the colour-magnitude filter limiting a CMD  region that 
should contain only cluster members. 
%This filter ensures a refined field-stars decontamination.
Once the observed magnitudes were unreddened, using the visual extinction given in the literature, 
we disregarded 
all the objects lying out of the range defined by the colour-magnitude
filter.  This filter is successful to accentuate the 
structures and to reduce the fluctuations caused by the presence of field-stars (\cite{BBonatto 2005}).

%========================================Sect. 3.1
\subsection{Stellar surface-density maps}

We obtained spatial distribution maps of the stellar surface density ($\sigma$) given by the number of stars per arcmin${^2}$
 for the clusters and their surroundings. Appendix A presents the results for the whole sample, while
Fig. 1 shows the map derived for the cluster Lynga~14, as illustration.
The left panel shows the entire studied area, whose 
 surface density was calculated in cells of $ | \Delta (\alpha cos (\delta_{center})) | = | \Delta \delta | $ = 2.5 arcmin$^{2}$,
 where $\Delta \alpha$ and $\Delta \delta$ are the steps on right ascension and declination, respectively.
 The dashed annulus corresponds to the reference field-stars region and the central circle indicates the 
cluster area. 
%The position of the P70 members is indicated by crosses. 

A zoom view (Fig. 1 central panel) displays a more detailed structure that was obtained by using smaller cells (1.0 arcmin$^{2}$) for the star 
counting process. 
Following the classification proposed by  \cite{LL03}, the surface density maps
were visually inspected to characterise the clusters.  Last column of Table 1 indicates if the type is Hierarchical (H) or 
Centrally Concentrated (CC), according to 
the definition discussed in Sect. 1.  About 38\% of the sample (8/21) shows a single major peak 
of density being classified by CC. Half of the sample (10/21) has multiple peaks and were considered 
of H type, while three objects presenting filamentary density distributions, have undefined type (marked as ``??" in Table 1).

%========================================Sect. 3.2
\subsection{Radial density profile}

Aiming to quantify the stellar distribution of the clusters, we evaluated their radial density 
profile (RDP) by using concentric rings to calculate the surface density.  

Some of the structural parameters were obtained by fitting
a theoretical RDP to the observed data. The adopted function is 
similar to the empirical model from \cite{King 1962}, given by:

\begin{equation}
\sigma(r)=\sigma_{bg} + \frac{\sigma_{0}}{1+(r/r_{c})^{2}} 
\label{eq.3.2.1}
\end{equation}

\noindent{where $\sigma(r)$ is the  stellar surface density (stars/arcmin${^2}$), {\it r} is the
radius (arcmin) of each concentric annulus used in the star counting, $\sigma_{0}$ and  $r_{c}$ are respectively the density and the radius of the cluster core.
The average density measured in the reference region ($\sigma_{bg}$) was calculate separately, in
order to diminish the number of free parameters in the RDP fitting.

Figure 1 (right panel) shows an example of observed radial distribution of surface density 
 and the  best fitting of the King's profile, which was obtained
by adopting the $ \chi^{2}$ method based on the parameters $\sigma_{0}$ and $r_{c}$.

Aiming to determine the cluster radius (R) more accurately, we verified the point where the cluster stellar
density reaches the background density. Figure 1 displays a  red line corresponding to 
$\sigma = \sigma_{bg}$ that was used to find R. 

A quantitative estimation on how compact is the cluster can be obtained from the density-contrast parameter
proposed by \cite{BB2009a} (2009a):

\begin{equation}
\delta_{c} = 1 + \frac{\sigma_{0}}{\sigma_{bg}}
\label{eq.3.2.2}
\end{equation}

\noindent{where  compact clusters have $7 \ \leq \ \delta_{c} \ \leq 23$. According this
this criterion, only NGC~6613 can be considered  compact  ($\delta_c \sim$ 10).}

Two other parameters are usual in comparative analysis: the average density ($<n_{*}>$), 
calculated by dividing the total number of
 observed members by the cluster area, and the ratio of core-size to cluster-radius (r$_c$/R). 
We could expect small values of r$_c$/R for the younger clusters,
since their members would not have had enough time to disperse away from the centre.  

Table 1 gives the structural parameters and the uncertainties derived from the RDP fitting. 

Three of our objects are found in the catalogue of Galactic open clusters presented by 
\cite{Piskunov 2007}.  Based on proper
 movement, they identified bright stars( V $<$ 14 mag) and used the  radial profile of the region 
 containing  possible members (P=14-61\%) to define the ``coronae" radius of the cluster, while the
  concentration of the probable members (P$>$61\%) defines the core radius (\cite{Kharchenko 2005} 2005). 
   Therefore, their sample is 5 to 10 times less numerous than our sample that includes low-mass 
   stars detected by 2MASS.  
   Besides, we focused on the main  stellar distribution of the cluster (radius $<6'$), 
   while they studied larger areas (radius $\sim 15'$), seeking  for the tidal radius ($R_t$).

These different definitions of cluster radii, imply in systematically smaller values when comparing our 
results with the structural parameters listed by them. However, 
these results cannot be directly compared because we are not dealing with the same 
kind of stellar groups. 

In fact,  the tidal radius obtained
by \cite{Piskunov 2007} (2007)  for NGC~6604 ($R_t$=8.8pc);  NGC~2362 (6.2pc); and Stock 13 (7pc) is more
compatible with the tidal radius that we estimated for these clusters: $r_t$= 7.7$\pm$2.3;  8.1$\pm$2.4; and 7.0$\pm$2.1, respectively, by adopting the expression used by  Saurin, Bica \& Bonatto (\cite{SBB2012}):

\begin{equation}
r_{t} =  (\frac{M}{M_{gal}})^{1/3} d_{GC}
\label{}
\end{equation}

\noindent{where M is the cluster mass, M$_{gal}$ is the Galactic mass and d$_{GC}$ is the 
Galactocentric distance, given by:}

\begin{equation}
M_{gal} =  \frac{V^{2}_{GC} d_{GC}}{G}
\label{}
\end{equation}

\noindent{where V$_{GC}$ = 254$\pm$16 km/s is the circular rotation velocity
of the Sun at R$_{GC}$ = 8.4$\pm$0.6 kpc (\cite{Reid 2009}).}

On the other hand, the R $\propto$ N$^{0.5}$ dependence discussed in Sect. 5.1 
distinguishes the sample studied by  \cite{Piskunov 2007} (2007)  from other young clusters 
presenting size and membership relations similar to our sample.
It seems to be more consistent comparing the cluster radius that we obtained 
%for NGC~6604 (R= 2.6$\pm$0.7 pc); NGC~2362(1.9$\pm$0.4 pc); and Stock~13 (2.3$\pm$0.3 pc),
 with the core radius obtained by \cite{Piskunov 2007}, given as notes
in Table 1.  

%Our measurements tend to be larger for NGC~2362 and Stock~13 because the
%number of members that we considered is higher than the Piskunov et al. (2007) sample, 
%by a factor of 4.4 and 6.5, respectively.

%%-----------------------------------------------Sect. 4
\section{Revisiting the fundamental parameters}

Adopting the same procedure described in Sect. 2.2, the field-stars decontamination was refined by
using the accurate estimative of cluster radius. The size of the reference region was also re-evaluated
according to the cluster area, allowing us to more closely define the sample of {\it candidate} members.

In Sect. 4.1 we adopt the extinction available in the literature to correct the observed colours,
which were fitted to the MS intrinsic colours. An iterative fitting process was adopted to accurately determine
E(B-V). 

Despite the distance and age of our clusters are available in the literature, these
parameters were checked by us on the light of the well determined structural parameters, as
described in Sect. 4.2.

%%-----------------------------------------------Sect. 4.1
\subsection{Colour excess}

The (J-H) and (H-K$_{s}$) colours were used  to estimate the extinction, by evaluating the position
of the cluster members compared to the  MS. 
In Fig. 2 (left panel), we plot the intrinsic colours of MS 
and giant stars given by \cite{BBrett 1988}, as well as the corresponding reddening vectors given by \cite{RLebofsky 1985}.
For comparison, the Zero Age Main-Sequence (ZAMS) from \cite{Siess 2000} (2000) is also plotted, which has quite the same
distribution of the MS, mainly for massive stars. 

Adopting the normal extinction law $A_{V} = 3.09 E(B-V) $ from \cite{SMathis 1979} and the relation 
$ \frac {A_{\lambda}} {A_{V}} $ from \cite{Cardelli 1989}, the observed colours of the cluster members
were unreddened and fitted to the MS intrinsic colours.  
This fitting is based on  massive stars, mainly P50 members.
Table 2 gives the E(B-V) that provides the best fitting, which is in good agreement with those available in 
the literature, within the 
estimated errors. Only a few objects had E(B-V) incompatible with the literature. In these cases, the procedure described in Sect. 3 for field-stars decontamination
by using colour-magnitude filter  was reapplied with the extinction derived by us, and 
the structural parameters were determined for this refined sample of cluster members. 

The (J-H)$_o$ and (H-K$_{s}$)$_o$  colours were also used to reveal the
stars with K-band excess. Young clusters are expected to have a large number of members showing high E(H-K$_s$)
(\cite{Lada 1996}). These stars appear in the right side of the MS reddening vector in the colour-colour diagram (Fig. 2).
The fraction f$_K$ was calculated by dividing the number of stars having large (H-K$_{s}$)$_o$
by the total number of cluster members.

%%%------------------------------------Fig. 2 Diagrams
\begin{figure}[]
\begin{center}
\includegraphics[width=1.0 \columnwidth,angle=0]{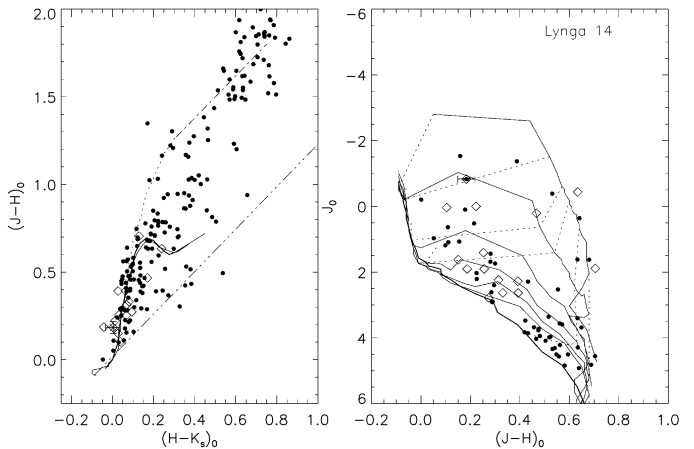}
%\includegraphics[width=0.45 \columnwidth,angle=0]{diag_ly14e.eps}
%\setcaptionmargin{1cm}
\caption{{\it Left}: Colour-colour diagram for Lynga~ 14. The MS and the ZAMS are indicated by full lines, while the locus of giant
stars is represented by a dotted line. Reddening vectors from \cite{RLebofsky 1985} are shown by  dot-dashed lines. 
{\it Right}: Colour-magnitude diagram showing the isochrones and evolutionary pre-MS tracks from \cite{Siess 2000} (2000). 
Cluster members are indicated by open diamonds (P50) 
and dots (P?).}
\label{cmd}
\end{center}
\end{figure}
%%%------------------------------------ end of Fig. 2

%%-----------------------------------------------Sect. 4.2

\subsection{Evaluation of distance and age}

The unredenned magnitude and colour J$_o$ $\times$ (J-H)$_o$ of the cluster members were compared to
pre-MS models (\cite{Siess 2000} 2000) and also MS Padova models (\cite{Girard 2002}) that were required to fit
the colours of massive stars.

Figure 2 (right panel) shows the CMD  obtained for Lynga~14, for illustration.
The distances were confirmed by searching for $(m-M)_{Jo}$ that best fits the position of massive
stars. The error bars were estimated from the minimum and the 
maximum distance module that provide good MS fittings. Within the uncertainties, the distances estimated from
the MS fitting are in good agreement  with the literature (differences are lower than 30\%), excepting for Trumpler 18. 
Our results are presented in Table 2 and were used to
convert into parsec the angular measurements of the structural parameters. 

%The large dispersion of the cluster members in the CMD motivated us  to adopt a mean value for the age.
%In this case, the error bar represents the whole range of ages suggested for the cluster members. 

The  number of stars as a function of age was obtained by counting the objects in between 
different pairs of isochrones in the CMD. We use bins of 5Myr, excepting the two first that 
correspond to  0.2 - 1 Myr and 1 - 5 Myr ranges. For MS objects (above the 7 M$_{\odot}$ track) 
we adopted the age estimated by fitting the data with Padova model. 
Figure 3 (left panel) shows the histogram of age distribution as a function of fractional number 
of members,  where {\it possible} members (P50) and {\it candidate} members (P?) are displayed 
separately. Figures A3 to A7 show the plots used to evaluate age and mass for the whole sample. 
 
The determination of age on basis of these histograms 
may not be obvious. Some of the clusters have members separated in two different ranges, 
making difficult the choice between the two options.
The first choice of age is based on the most prominent peak formed by the P50 members.
If the second peak has a similar number of members, we adopt the smaller value
of age that is listed in Table 2, while the second option is given as notes in the same table.

 Due to the large uncertainties on the age estimation, our conclusions are based
on mean values. Considering  that 12/21 objects of the sample (57\%) are in the range
of 4 - 10 Myr, and 9/21 (43\%) are younger than 4 Myr, we suggest for our clusters
a mean age of $\sim$ 5 Myr.

%%%------------------------------------Fig. 3 Age histogram and Mass Function
\begin{figure}[]
\begin{center}
\includegraphics[width=0.60\columnwidth, angle=0]{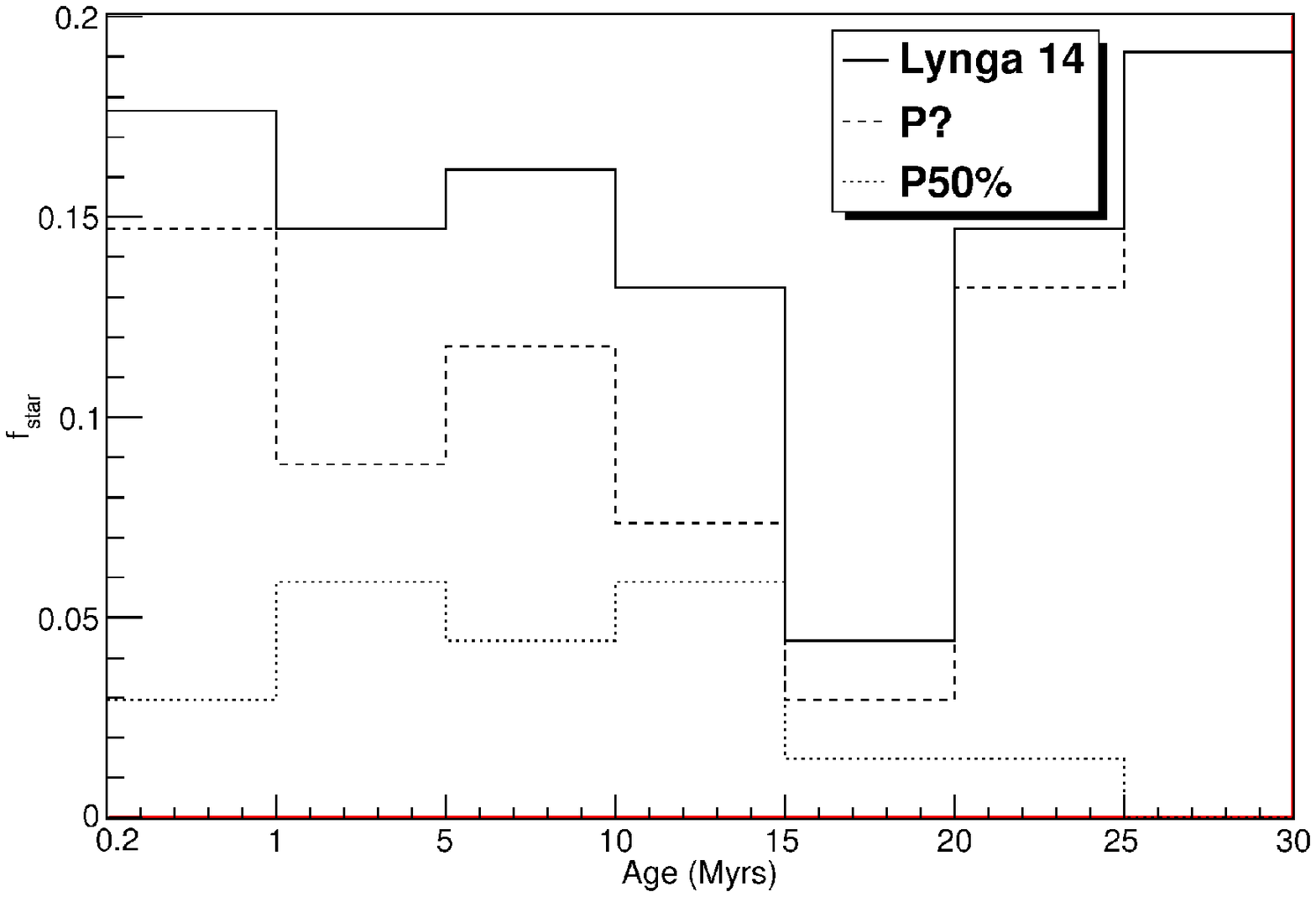}
\includegraphics[width=0.38\columnwidth, angle=0]{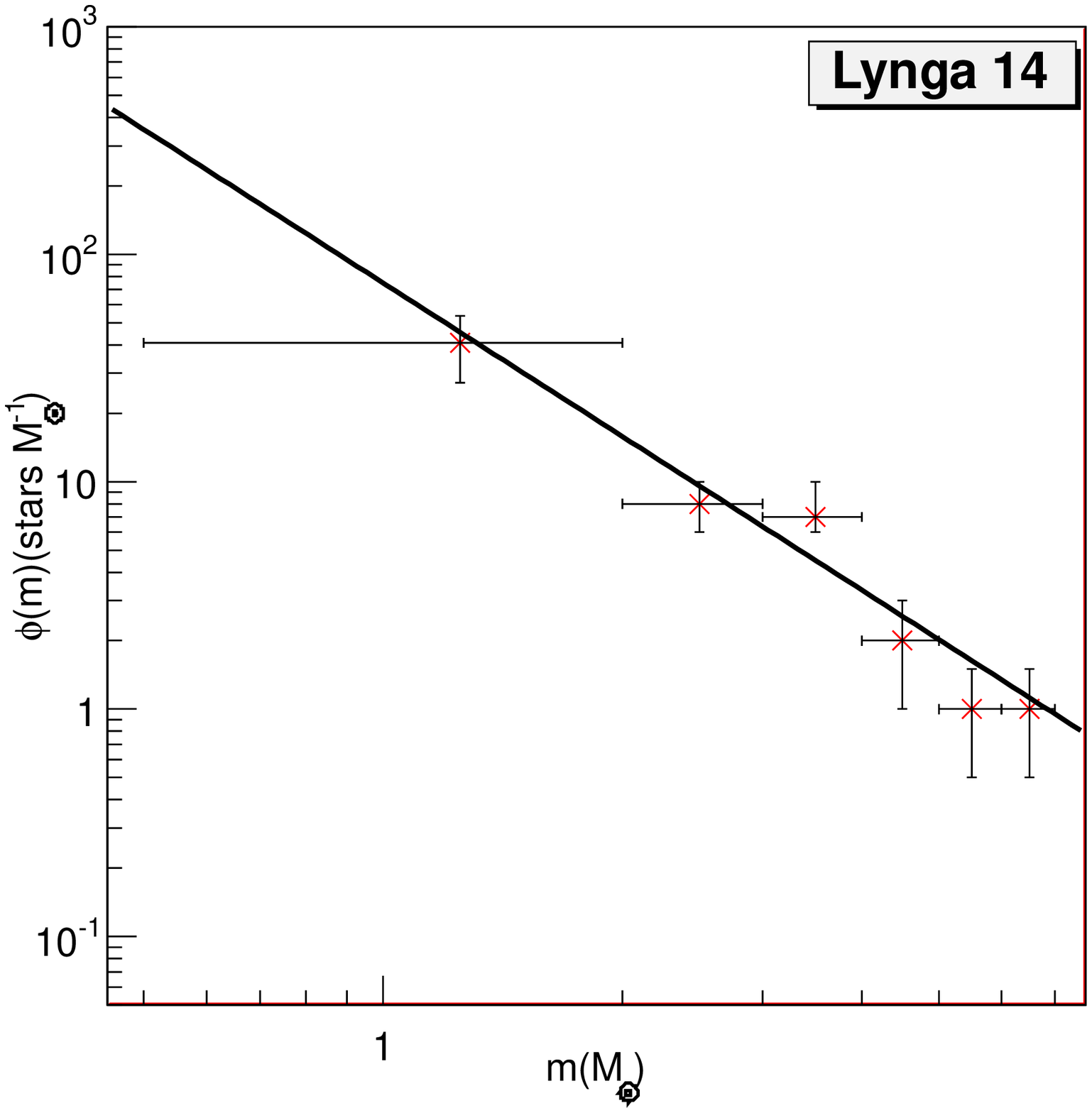}
%\setcaptionmargin{1cm}
\caption{{\it Left}: Age distribution of Lynga~14 members (thick line) showing the contribution of P50 (dotted line) and
P? (dashed line) objects.
{\it Right}: Observed mass distribution indicated by crosses with error bars. The thick line represents the mass function $\phi$(m)
fitting.} 
\label{histo}
\end{center}
\end{figure} 
%%%------------------------------------ end of Fig. 3

%%-----------------------------------------------  Table 2 ----------------------
\begin{table*}[ht] 
\caption{Fundamental parameters.} 
\begin{center}
{\scriptsize
\begin{tabular}{|l|c|c|c|c|c|c|c|c|c|c|}
\hline
Cluster & Age  & E(B-V)  & d & (m-M) & f$_{k}$ & M$_{obs}$ &  N$_{obs}$ & M$_{T}$ &  N$_{T}$ & $\chi$\\ 
 & Myrs & mag & (pc) & mag &  (\%) & M$_{\odot}$ & (stars)  &  M$_{\odot}$ & (stars)  & \\ \hline \hline

Collinder 205 & 5.0$^{+5.0}_{-4.5}$ & 0.56$\pm$0.07 & 1800$^{+500}_{-500}$ & 11.3$^{+0.5}_{-0.7}$ & 14$^{+17}_{-10}$ & 390 $\pm$ 79 & 174 $\pm$ 38 & 876 $\pm$ 262 & 2617 $\pm$ 701 & 1.17$\pm$0.33\\ 

Hogg 10 & 3.0$^{+2.0}_{-2.0}$ & 0.23$\pm$0.05 & 2200$^{+700}_{-400}$ & 11.7$^{+0.6}_{-0.4}$ & 22$^{+18}_{-15}$ & 218 $\pm$ 43 & 88 $\pm$ 10 & 318 $\pm$ 81 & 465 $\pm$ 149 & 0.59$\pm$0.17\\ 

Hogg 22$^a$ & 3.0$^{+2.0}_{-2.0}$ & 0.43$\pm$0.07 & 1700$^{+550}_{-250}$ & 11.2$^{+0.6}_{-0.3}$ & 28$^{+19}_{-20}$ & 285 $\pm$ 47 & 97 $\pm$ 33 & 376 $\pm$ 99 & 429 $\pm$ 220 & 0.54$\pm$0.25\\ 

Lynga 14$^b$ & 3.0$^{+2.0}_{-2.5}$ & 0.50$\pm$0.10 & 950$^{+200}_{-200}$ & 9.9$^{+0.4}_{-0.5}$ & 6$^{+16}_{-3}$ & 127 $\pm$ 30 & 68 $\pm$ 16 & 374 $\pm$ 111 & 1374 $\pm$ 430 & 1.24$\pm$0.21\\ 

Markarian 38 & 5.5$^{+4.5}_{-4.5}$ & 0.10$\pm$0.07 & 1500$^{+200}_{-250}$ & 10.9$^{+0.3}_{-0.4}$ & 21$^{+11}_{-12}$ & 137 $\pm$ 28 & 57 $\pm$ 3 & 190 $\pm$ 57 & 251 $\pm$ 107 & 0.52$\pm$0.27\\ 

NGC 2302$^c$ &  2.5$^{+2.5}_{-2.0}$ & 0.19$\pm$0.05 & 1700$^{+300}_{-400}$ & 11.2$^{+0.3}_{-0.6}$ & 26$^{+19}_{-11}$ & 168 $\pm$ 32 & 70 $\pm$ 19 & 238 $\pm$ 57 & 345 $\pm$ 111 & 0.60$\pm$0.19\\ 

NGC 2362 & 4.0$^{+6.0}_{-3.0}$ & 0.06$\pm$0.05 & 1480$^{+400}_{-200}$ & 10.9$^{+0.5}_{-0.3}$ & 31$^{+17}_{-13}$ & 332 $\pm$ 55 & 124 $\pm$ 25 & 500 $\pm$ 126 & 927 $\pm$ 331 & 0.92$\pm$0.25\\ 

NGC 2367 & 3.0$^{+2.5}_{-2.5}$ & 0.15$\pm$0.05 & 2200$^{+400}_{-200}$ & 11.7$^{+0.4}_{-0.2}$ & 20$^{+12}_{-15}$ & 155 $\pm$ 29 & 60 $\pm$ 6 & 267 $\pm$ 64 & 586 $\pm$ 167 & 1.03$\pm$0.21\\ 

NGC 2645$^d$ & 5.0$^{+5.0}_{-4.0}$ & 0.28$\pm$0.08 & 1800$^{+800}_{-200}$ & 11.3$^{+0.8}_{-0.3}$ & 27$^{+22}_{-18}$ & 238 $\pm$ 49 & 104 $\pm$ 14 & 498 $\pm$ 102 & 1366 $\pm$ 265 & 1.10$\pm$0.20\\ 

NGC 2659 & 5.0$^{+10}_{-4.5}$ & 0.25$\pm$0.05 & 2000$^{+500}_{-200}$ & 11.5$^{+0.5}_{-0.2}$ & 28$^{+21}_{-22}$ & 495 $\pm$ 103 & 215 $\pm$ 27 & 857 $\pm$ 237 & 1801 $\pm$ 608 & 0.89$\pm$0.22\\ 

NGC 3572 & 3.0$^{+7.0}_{-2.5}$ & 0.15$\pm$0.07 & 1900$^{+400}_{-200}$ & 11.4$^{+0.4}_{-0.2}$ & 28$^{+15}_{-16}$ & 371 $\pm$ 72 & 149 $\pm$ 22 & 645 $\pm$ 175 & 1422 $\pm$ 485 & 0.97$\pm$0.26\\ 

NGC 3590 & 2.5$^{+1.5}_{-2.5}$ & 0.35$\pm$0.07 & 1680$^{+500}_{-300}$ & 11.1$^{+0.6}_{-0.4}$ & 34$^{+23}_{-27}$ & 175 $\pm$ 36 & 79 $\pm$ 20 & 328 $\pm$ 90 & 768 $\pm$ 264 & 0.95$\pm$0.27\\ 

NGC 5606$^e$ & 2.5$^{+2.5}_{-2.3}$ & 0.32$\pm$0.05 & 2200$^{+300}_{-200}$ & 11.7$^{+0.3}_{-0.2}$ & 16$^{+14}_{-9}$ & 262 $\pm$ 49 & 98 $\pm$ 11 & 417 $\pm$ 81 & 768 $\pm$ 147 & 0.86$\pm$0.13\\ 

NGC 6178$^f$ & 5.0$^{+2.5}_{-2.5}$ & 0.15$\pm$0.08 & 1430$^{+150}_{-150}$ & 10.7$^{+0.3}_{-0.2}$ & 6$^{+9}_{-5}$ & 220 $\pm$ 52 & 106 $\pm$ 7 & 379 $\pm$ 284 & 1430 $\pm$ 1345 & 1.43$\pm$1.38\\ 

NGC 6604 & 6.0$^{+4.0}_{-5.0}$ & 0.67$\pm$0.07 & 1600$^{+400}_{-300}$ & 11.0$^{+0.5}_{-0.4}$ & 18$^{+16}_{-11}$ & 279 $\pm$ 52 & 90 $\pm$ 7 & 326 $\pm$ 95 & 245 $\pm$ 141 & 0.26$\pm$0.21\\ 

NGC 6613 & 5.0$^{+5.0}_{-4.0}$ & 0.45$\pm$0.07 & 1550$^{+300}_{-250}$ & 11.0$^{+0.4}_{-0.4}$ & 26$^{+15}_{-18}$ & 345 $\pm$ 73 & 133 $\pm$ 9 & 557 $\pm$ 113 & 1087 $\pm$ 185 & 0.93$\pm$0.13\\ 

Ruprecht 79 & 5.0$^{+10}_{-4.5}$ & 0.28$\pm$0.05 & 2700$^{+400}_{-300}$ & 12.2$^{+0.3}_{-0.3}$ & 40$^{+20}_{-25}$ & 459 $\pm$ 86 & 174 $\pm$ 8 & 677 $\pm$ 119 & 1156 $\pm$ 154 & 0.95$\pm$0.14\\ 

Stock 13 & 4.0$^{+6.0}_{-3.0}$ & 0.07$\pm$0.05 & 2000$^{+500}_{-350}$ & 11.5$^{+0.5}_{-0.4}$ & 32$^{+18}_{-15}$ & 179 $\pm$ 31 & 65 $\pm$ 11 & 308 $\pm$ 80 & 686 $\pm$ 236 & 1.02$\pm$0.26\\ 

Stock 16 & 7.0$^{+8.0}_{-6.5}$ & 0.28$\pm$0.05 & 2000$^{+500}_{-350}$ & 11.5$^{+0.5}_{-0.4}$ & 30$^{+19}_{-23}$ & 254 $\pm$ 66 & 139 $\pm$ 13 & 1043 $\pm$ 239 & 4676 $\pm$ 995 & 1.50$\pm$0.18\\ 

Trumpler 18 & 5.0$^{+5.0}_{-4.0}$ & 0.10$\pm$0.07 & 2850$^{+400}_{-350}$ & 12.3$^{+0.3}_{-0.4}$ & 26$^{+15}_{-14}$ & 512 $\pm$ 85 & 164 $\pm$ 8 & 572 $\pm$ 214 & 334 $\pm$ 543 & 0.00$\pm$0.17\\ 

Trumpler 28 & 2.0$^{+3.0}_{-1.5}$ & 0.65$\pm$0.05 & 1050$^{+250}_{-200}$ & 10.1$^{+0.5}_{-0.5}$ & 10$^{+25}_{-6}$ & 178 $\pm$ 35 & 73 $\pm$ 7 & 233 $\pm$ 58 & 268 $\pm$ 86 & 0.42$\pm$0.16\\ \hline

\end{tabular}
}
\label{tab1}
\end{center}
{\scriptsize
Columns description: (1) Identification; (2) age; (3)colour excess;  (4) Distance;
(5) distance modulus; (6) fraction of stars showing K excess;
(7, 8) observed mass and number of members; (9, 10) total mass and number of members;
(11) mass function slope.\\
Notes: Clusters showing a second peak on age distribution: (a) 12.5$^{+2.5}_{-2.5}$; (b) 12.0$^{+3.0}_{-7.0}$; (c) 12.5$^{+2.5}_{-2.5}$; (d) 12.0$^{+3.0}_{-7.0}$; (e) 11.5$^{+3.5}_{-6.5}$; 
(f) 12.5$^{+2.5}_{-2.5}$.
}

\end{table*}

%%-----------------------------------------------Sect. 4.3
\subsection{Mass Function}

Similar to the age estimation, the counting in between the CMD tracks was adopted to determine the
distribution of masses. Nine evolutionary tracks ranging from 0.1M$_{\odot}$ to 7 M$_{\odot}$ were used 
to estimate the mass of pre-MS stars. For the MS objects we adopted the  Padova model that best fits 
the observed colours, as described in Sect. 4.2.
By this way,  the sum of  MS
and pre-MS stars corresponds to the number of observed objects, which is given in Table~2 along with 
 the respective observed mass. 

Instead of using a histogram to display the mass distribution,  
we calculate the mass function given by \cite{Kroupa 2001}:

\begin{equation}
\phi(m) \propto m^{-(1+\chi)}
\label{eq.3.2.2}
\end{equation}

% following Bonatto and Bica (2009a)

Figure 3 (right panel) displays the observed mass distribution based on the sum of both MS and pre-MS
 stars. By fitting the observed distribution, we obtained the slope of the mass function, represented
  by $\chi$ (see last column of Table 2).

About half of the sample has a mass distribution flatter  than  the initial cluster mass function 
(ICMF) verified in a large variety of clusters (e.g. \cite{Elmegreen 2006, Oey 2011}). 
The ICMF slope ($\chi \sim$ 1.0) is slightly shallower than Salperter's IMF ($\chi \sim$ 1.35). 

The IMF suggested by \cite{Kroupa 2001} assumes slopes $\chi = 0.3\pm0.5$ for M $<$ 0.5M$_{\odot}$ and 
 Salperter's IMF for M $>$ 1 M$_{\odot}$. We used the  the observed mass and number of members
 to estimate the average stellar mass
 ($\bar{m}$)
 and verified that most of them have $\bar{m} >$ 1 M$_{\odot}$, probably due to a lack
 of low mass stars in our sample. This is related to the 2MASS detection limit that constrains 
 the presence of faint sources in our sample. 
 A sample incompleteness gives for our clusters  large $\bar{m}$, when compared to
 embedded clusters. Bonatto \& Bica (\cite{BBica2010, BBica2011}), for instance, uses
 $\bar{m}$ = 0.6 M$_{\odot}$, which is lower than our results by a factor of 2, at least. 

 Based on the IMF fitting of the observed distribution of mass ($\chi$ slope), we corrected
the incompleteness of our sample by synthetically deriving 
the number of faint stars that should be considered in the real membership.
Table 2 gives the total number of members
N$_T$ = N$_{obs}$ + N$_{imf}$, where N$_{obs}$ is related to the observed objects and N$_{imf}$  
is the number of lacking faint stars, 
estimated by integrating the IMF in the range of low-mass (below
the limit of detection). The corresponding total mass is also given in Table 2, along with the 
observed mass. 
The correction on the sample completeness is needed to improve the estimation 
of the cluster parameters that are used in the comparison with other samples.

%%%%----------------------------------------New Fig 04 - RxN, MxN, MxR
\begin{figure*}[]
\begin{center}
%\vspace{0.3cm}
%\includegraphics[bb= 40 33 410 288, width=7.0cm,angle=0]{fig_pf1a.eps}
%\includegraphics[bb= 33 33 374 337, width=6.8cm,angle=0]{fig_pf2.eps}
\includegraphics[width=1.0\columnwidth, angle=0]{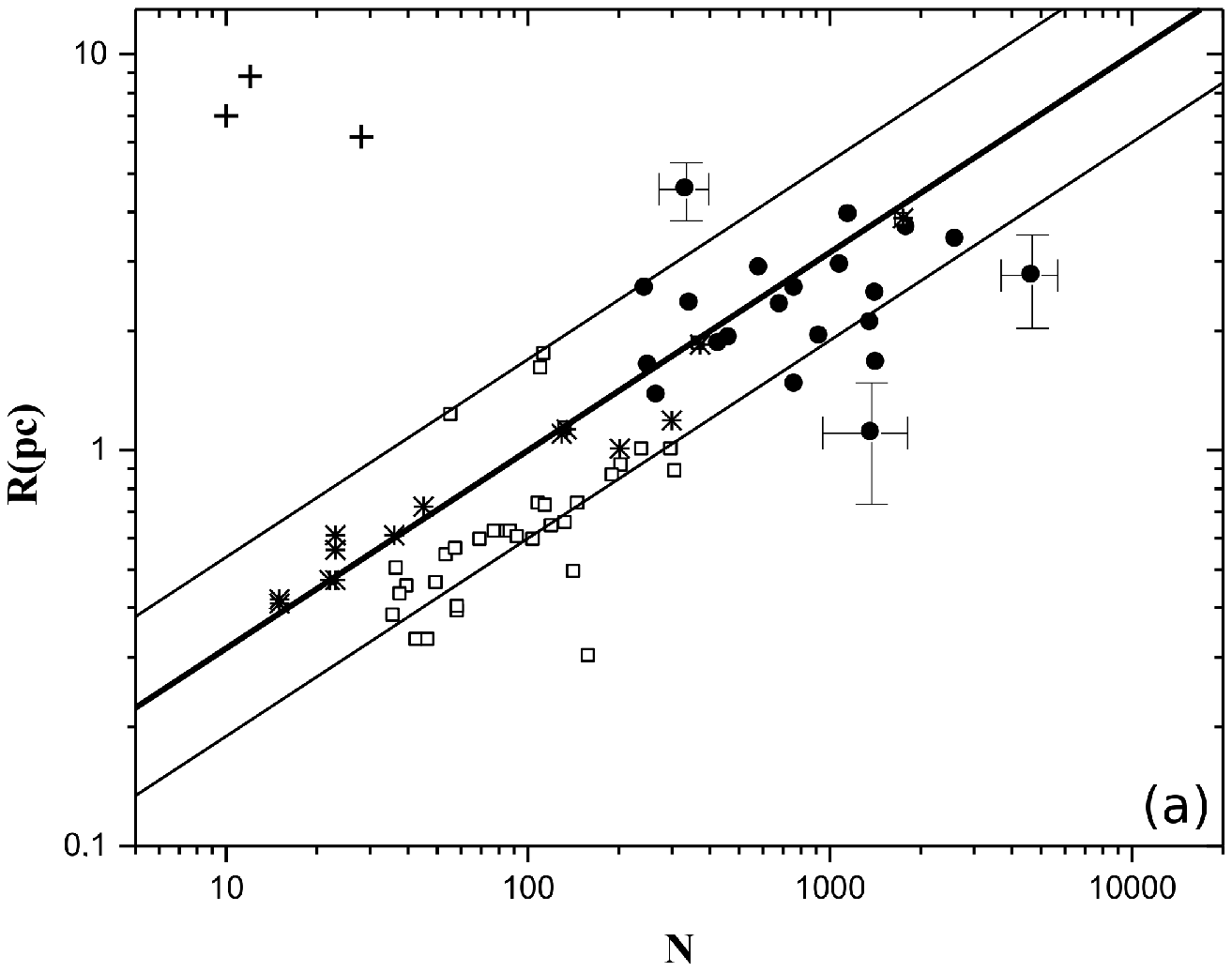}
\includegraphics[width=1.0\columnwidth, angle=0]{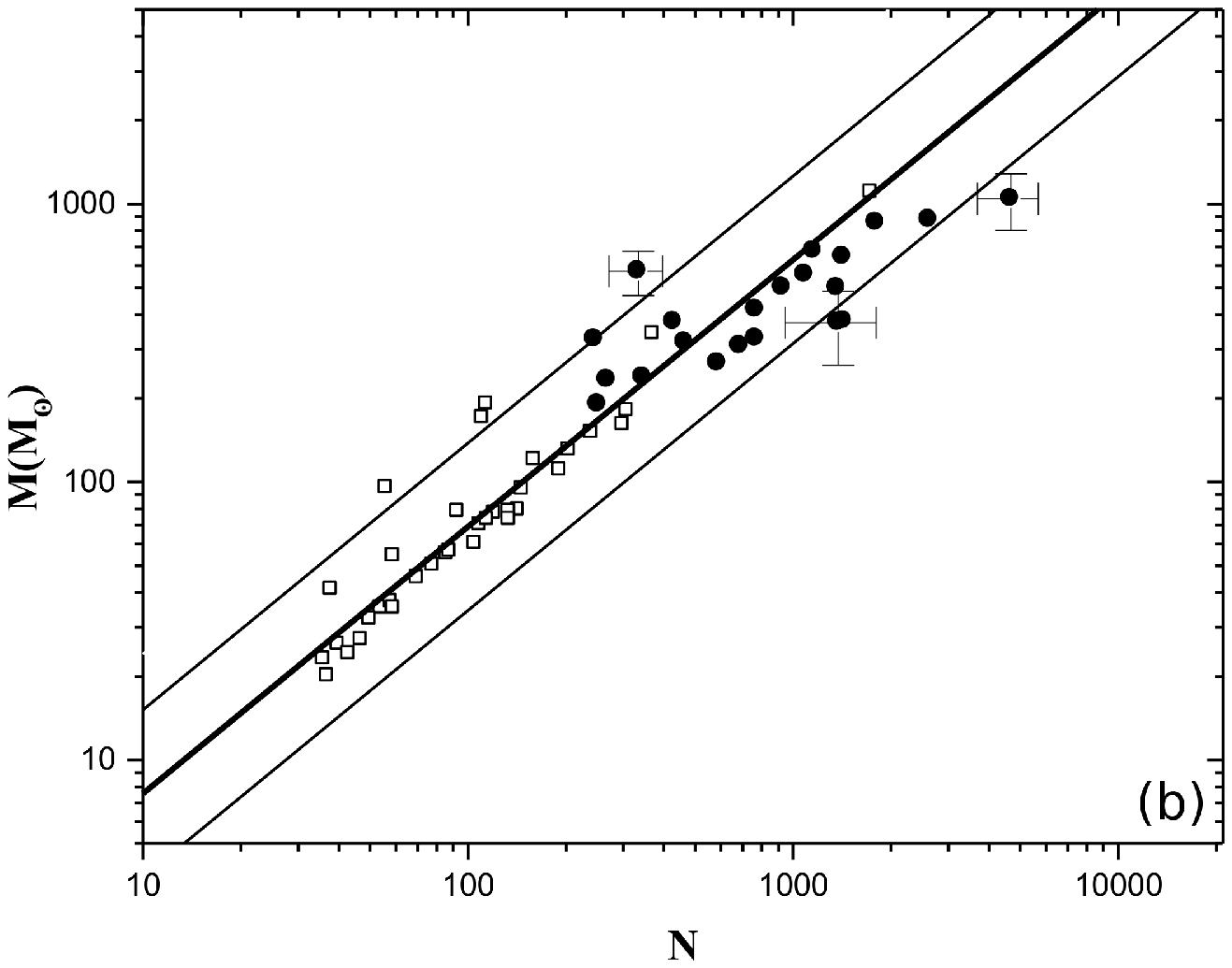}
\includegraphics[width=1.0\columnwidth, angle=0]{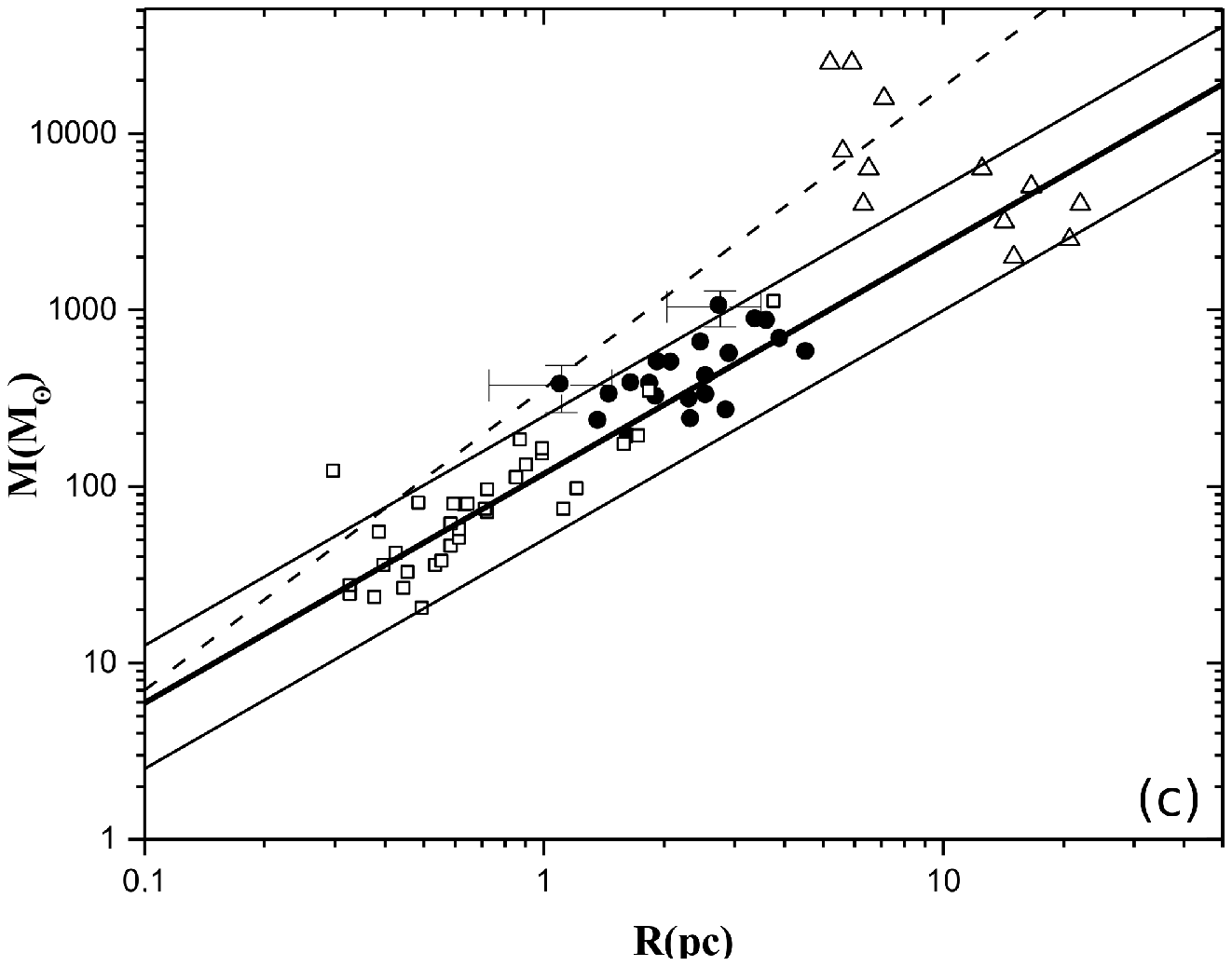}
\includegraphics[width=1.0\columnwidth, angle=0]{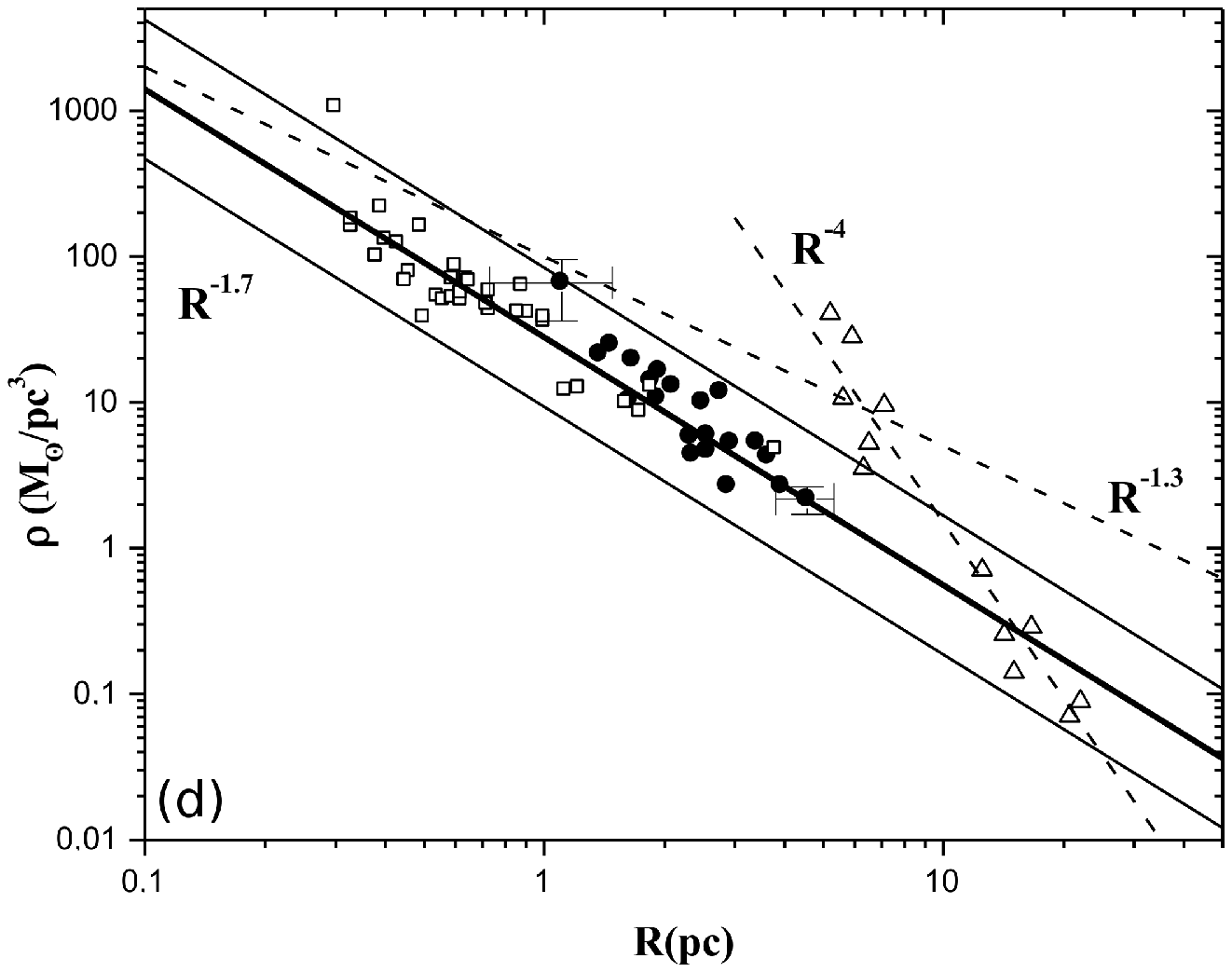}
%\setcaptionmargin{1cm}
{\scriptsize
\caption{Comparing our objects (filled circles) with  embedded clusters 
(open squares) studied by \cite{LL03} and other samples. Representative error bars are
used in a few data points, for illustration.
{\bf (a)}:  Radius of the cluster as a function of number of members. Our sample, as well as the 
embedded clusters  follow the same dependence $R \propto N^{0.5}$ (thick line) found for the stellar
clusters (asterisks) studied by \cite{Carpenter 2000} (2000). Thin lines indicate
the limits suggested by  \cite{Adams 2006} (2006). The results obtained by \cite{Piskunov 2007} (2007)  for three
 objects of our sample (crosses) are used to illustrate differences on the cluster size definition.
{\bf (b)}:  Mass of the cluster as a function of number of members showing the same dependence 
$M \propto N^{1}$ for our sample and embedded clusters (thick line), with limits scaled by a factor 
$\sim$ 2 (thin lines).
{\bf (c)}: Mass-radius dependence showing a mean distribution of M $\sim$ 118 R$^{1.3}$ (tick line)
that spreads by a factor 2 (thin lines). 
The parameters of massive (``leak") exposed clusters (triangles) are also plotted, as well as the dependence 
$M =359  R^{1.7}$ suggested by Pfalzner (\cite{Pfalzner 2011}) for the embedded clusters (dashed line).
{\bf (d)}: Volumetric density as a function of radius, showing the dependence $\rho$ = 28 R$^{-1.7}$ 
presented by our sample and \cite{LL03} data (thick line), limited by thin lines that are scaled by a factor 2.
Dashed lines represent the relations proposed by Pfalzner (\cite{Pfalzner 2011}).
}
} 
\label{fig4}
\end{center}
\end{figure*} 
%%%------------------------------------ end of Fig. 4

%%----------------------------------------Section 5
\section{Comparative Analysis}

We investigate possible  correlations among the cluster characteristics 
by comparing the structural and fundamental  parameters, respectively from Table 1 and Table 2.

First, the correlations among mass, radius and number of members are compared to similar 
relations obtained by \cite{Adams 2006} (2006) and Pfalzner (\cite{Pfalzner 2011}) using cluster properties from other works.
 In Sect. 5.4 the structural parameters  (obtained on basis of surface density) are compared among 
each other, and Sect. 5.5 presents the relations with age.

%%----------------------------------------Sect. 5.1
\subsection{Radius}

%A trend of cluster radius (R) increasing  with both, total number of 
%members (N) - excepting NGC~2367 and Stock~13 - and total mass (M).

Figure 4a shows the cluster radius distribution as a function of number of members 
(the same as fig.2 from \cite{Adams 2006} 2006). Our sample is compared with 14 stellar clusters listed by
\cite{Carpenter 2000} (2000), and 34 embedded clusters from \cite{LL03}, which have available size 
(radius in pc). 

In order to illustrate the different cluster size definitions
discussed in Sect. 3.2, we also plot the
cluster radius ($R_t$) and  the respective number of members ($N_1$) that were
obtained by \cite{Piskunov 2007}
for three of our clusters (NGC~2362, NGC~6604, and Stock~13). 
It can be noted that the relation $R_t \times N_1$ (crosses in Fig. 4a) is  
incompatible with the distribution of the other samples, possibly due to the fact that
\cite{Kharchenko 2005} (2005) and cite{Piskunov 2007} study different 
stellar groups that encompass our clusters (see Sect. 3.2).
  
 We verified that 52\% of our sample follows the
dependence R =0.1N$^{0.5}$ that is the same   
relation proposed by \cite{Adams 2006} (2006)
for the  clusters listed by \cite{Carpenter 2000} (2000).
%$R_c(N)=1.7 (N/300)^{1/2}$ 
Two of our clusters, NGC~6604 and Trumpler~18, as well as three clusters 
from \cite{LL03} (NGC~2282, Gem~1, and Gem~4) 
also follow this trend, but scaled up by a factor 1.7, which is the superior limit 
suggested by \cite{Adams 2006} (2006)  for the distribution
of cluster size as a function of N. 

Coinciding with most of the \cite{LL03} clusters, 33\% of our sample are distributed along 
the lower curve in Fig. 4a that is  similar to the \cite{Carpenter 2000} (2000) sample,
but scaled down by a factor 1.7.

Lynga~14 appears bellow this correlation, suggesting that its membership is 
larger than the expected number of members,
when compared to other clusters having same size. This characteristic is also noted
for some clusters from \cite{LL03}, in particular S~106.
%Note that even neglecting the number of faint stars (estimated from IMF),
%Lynga~14 remains out of the correlation.

We also evaluate for our sample the commonly used half-mass radius $r_{1/2}$, 
which encloses half of the total mass of the cluster. Considering that we do not determine 
individual mass for each member of the clusters, their observed radial mass distribution 
cannot be established.  In order to have an approximate estimation of $r_{1/2}$,  
we adopted the integrated mass distribution M(r) given by \cite{Adams 2006} (2006):

\begin{equation}
\frac{M(r/r_o)}{M_{tot}} = \bigg[\frac{(r/r_o)^a}{\big(1 + (r/r_o)\big)^a}\bigg]^p 
\end{equation}

\noindent where $r_o$ is the scale length that we assume to be the radius of the cluster. 
The validity of using Eq.7 for our sample was checked for Lynga~14, Collinder~205 and Hogg~10, 
for which we could estimate the observed M(r) and were used as test cases.   
The ``virial" model, used by \cite{Adams 2006} (2006) in the simulations for N=100 members, has coefficients
(a=3 and p $\sim$ 0.41) that  seem to well reproduce the radial profile of the checked clusters. 
We verified that these simulations provide a ratio of $r_{1/2}/r_o$ of about 60-63\%. 

By adopting the relation $r_{1/2}$ = 0.615 $r_o$, we obtain the range of 0.69 - 2.5 pc for the half-mass
radius estimated in our sample. We compared this result to the study that \cite{Adams 2006} (2006)  developed
for NGC~1333, which model is most like the N=100 simulations 
with ``cold" starting states (a=2 and p $\sim$ 0.55) and gives $r_{1/2}$  = 0.117 to 0.238 for 
$r_o$=0.3 to 0.4 pc. The relation  $r_{1/2}/r_o$ is about 40-60\% in this case, which
is probably due to the fact that our clusters are more evolved than the embedded phase.

%%----------------------------------------Sect. 5.2

\subsection{Mass}

Figure 4b shows the relations between mass and number of members.
 The \cite{LL03} sample has a dependence M = a N$^{1.0}$, where a = 0.6 $^{+0.6}_{-0.3}$,
which means that they are
distributed between the lines scaled (up and down) by a factor 2.
 Our objects also follow this dependence.

It is interesting to note that  the clusters NGC~2282, Gem~1, and Gem~4,
from \cite{LL03}, are found above the superior limit of the distribution.
The same occurs for our clusters Trumpler~18 and NGC~6604,
which showed a similar trend in the R $\times$ N plot (Fig. 4a).

The reason for a deviation of the expected relation is that
these objects are different of most of the clusters. They have a smaller number 
of low-mass members, which is confirmed by their flat IMF ($\chi <$ 0.4).
In the opposite sense, NGC~6178 and Stock~16 are found below the lower limit
curve, being compatible with their steep IMF ($\chi >$ 1.4).

In order to discuss if these differences could be interpreted as formation condition, as proposed by 
\cite{Adams 2006} (2006), or differently if they are related to a time sequence, as suggested by 
Pfalzner (\cite{Pfalzner 2011}), we compare the relations between mass and radius.
 The thick line in Fig. 4c shows that our sample has the same dependence  
obtained by fitting the \cite{LL03} data: M = 118 R$^{1.3}$. 

Figure 4c also displays the data of the ``leaky" massive clusters studied by Pfalzner (\cite{Pfalzner 2011}), 
who proposed that these objects would be the ending of a time sequence. However, 
the mass-radius dependence M $\sim$ 359 R$^{1.7}$ (dashed line), which Pfalzner (\cite{Pfalzner 2011}) used to illustrate 
the suggested time sequence, is steeper than the distribution of \cite{LL03} clusters. 

Instead of confirming a sequence
that ends in the exposed massive clusters, our results show that clusters from Pfalzner (\cite{Pfalzner 2011})
having radius $>$ 10 pc follow the same M {\it vs.} R relation presented by \cite{LL03} data and also our
sample. On the other hand, the massive clusters with R $<$ 10 pc are found above the upper line
in Fig. 4c, similar to Lynga~14 and S~106 (\cite{LL03}), which differences we interpret to be more 
likely due to different formation conditions. 

%%----------------------------------------Sect. 5.3
\subsection{Volumetric density}

By assuming a spherical distribution, we estimated the volumetric density, defined by 
$\rho = 3 M / (4 \pi R^3)$,  which is plotted in Fig. 4d as a function of radius.
For both samples: our clusters and those from \cite{LL03},  we found a  similar relation 
 $\rho = 28 R^{-1.7}$. This is quite similar to the results obtained by \cite{Camargo 2010} that find 
$\rho \propto R^{-1.92}$. 

 Pfalzner (\cite{Pfalzner 2011}) used a different dependence $\rho \sim 100 R^{-1.3}$ 
to represent the \cite{LL03} data, that were compared to  the relation  
$\rho \propto  R^{-4}$ obtained for the massive exposed clusters (dashed lines in Fig. 4c). However, 
the distribution of our clusters, as well as \cite{LL03} data, does not agree with the mass {\it vs.} density relation
proposed by Pfalzner (\cite{Pfalzner 2009, Pfalzner 2011}). As suggested in Sect. 5.2,
the large massive clusters seem to follow the same trend of our clusters, while those with R $<$ 10 pc
are scaled up by a factor larger than 2.

%%----------------------------------------Sect. 5.4
\subsection{Surface density}

Were analysed the structural parameters looking for correlations among
the clusters themselves, and to verify a possible relation between the cluster and 
its environment. Hatched areas in Fig. 5 illustrate the  trends that were found
for most of the clusters.

Excepting for Trumpler~28, the observed average density of the cluster ($<n_{*}>$) 
increases with the background density ($\sigma_{bg}$), as shown in Fig. 5a. This
indicates that dense clusters are found in dense background fields.

Considering that N increases with R (see Fig. 4a) it is expected an anti-correlation
of $<n_{*}>$ {\it vs.} R, which is indeed noted in Fig. 5b. By consequence, $\sigma_{bg}$  
also appears anti-correlated with R.

Core density ($\sigma_0$) is another structural
parameter that also diminishes when cluster radius increases, but
a considerable dispersion is seen in Fig. 5c, where almost six clusters are 
out of the observed trend.
 
In spite of the fact that R is not expected to be related to the cluster distance 
(D),  Fig. 5d shows a trend of R increasing with D. 
This is not surprising, since the similar angular sizes of our objects
lead to large linear size (given in parsec) for more distant clusters. 

 We plot in Fig. 5e the anti-correlation between core radius (r$_c$) and
$\sigma_0$ that is observed for most of the clusters, excepting Lynga~14 and
NGC~3590. The core parameters also showed some trends when compared with
other structural parameters, like the correlation between $\sigma_0$ and $\delta_c$ 
(density contrast); and r$_c$/R (ratio of core-size to cluster radius)
decreasing with the raise of $\sigma_0$.

The evaluation of structural parameters is unprecedented for most of our objects,
but our results  are comparable with those available in the literature for other clusters. 
We verified that the ranges of values obtained by us are compatible with several kind of young 
clusters like: NGC~6611, an embedded  and dynamically evolved cluster (\cite{BSB2006}); NGC~4755, a post-embedded cluster (\cite{Bonatto 2006b}); the low-mass open clusters  NGC~1931, vdB~80, BDSB~96 and  Pismis~5 (\cite{BB2009a}, 2009a);  NGC~2239, a possible ordinary open cluster (\cite{BB2009b}); and the dissolving clusters  NGC~6823 (\cite{BBD2008}), Collinder~197, vdB~92 (Bonatto \& Bica, \cite{BBica2010}) and Trumpler~37 (Saurin, Bica \& Bonatto, \cite{SBB2012}).
 
%The ranges covered by the structural parameters, discussed in this section, are in agreement with those found in the literature,
%for instance NGC~1931 and  Pismis~5 studied by Bonatto \& Bica (2009). In the next section we develop the comparison among 
%fundamental parameters.

%%%------------------------------------ New Fig 5 (including Figs 6 and 7)
\begin{figure*}[]
\begin{center}
\includegraphics[width=4.5cm,angle=0]{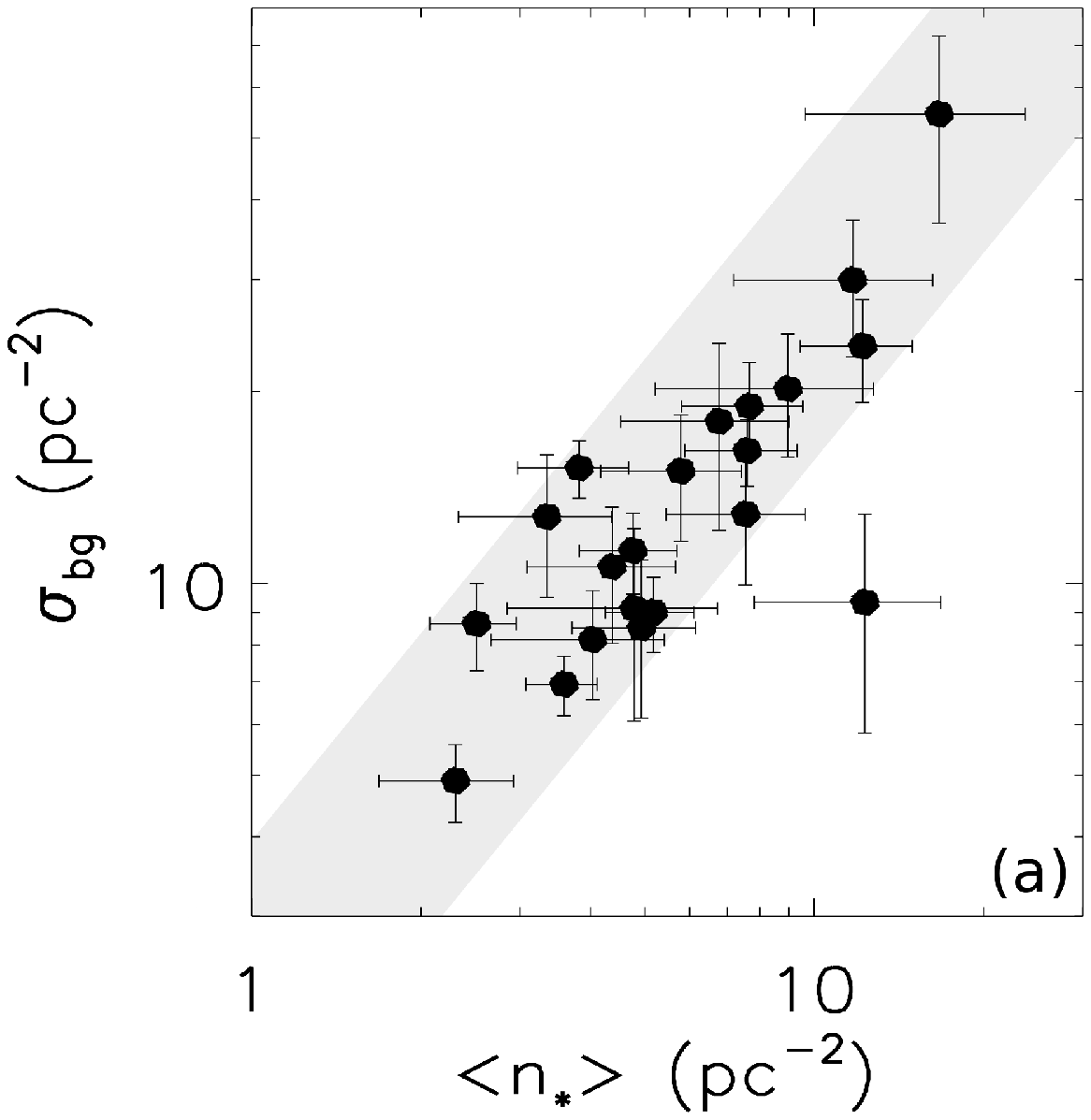}
\includegraphics[width=4.5cm,angle=0]{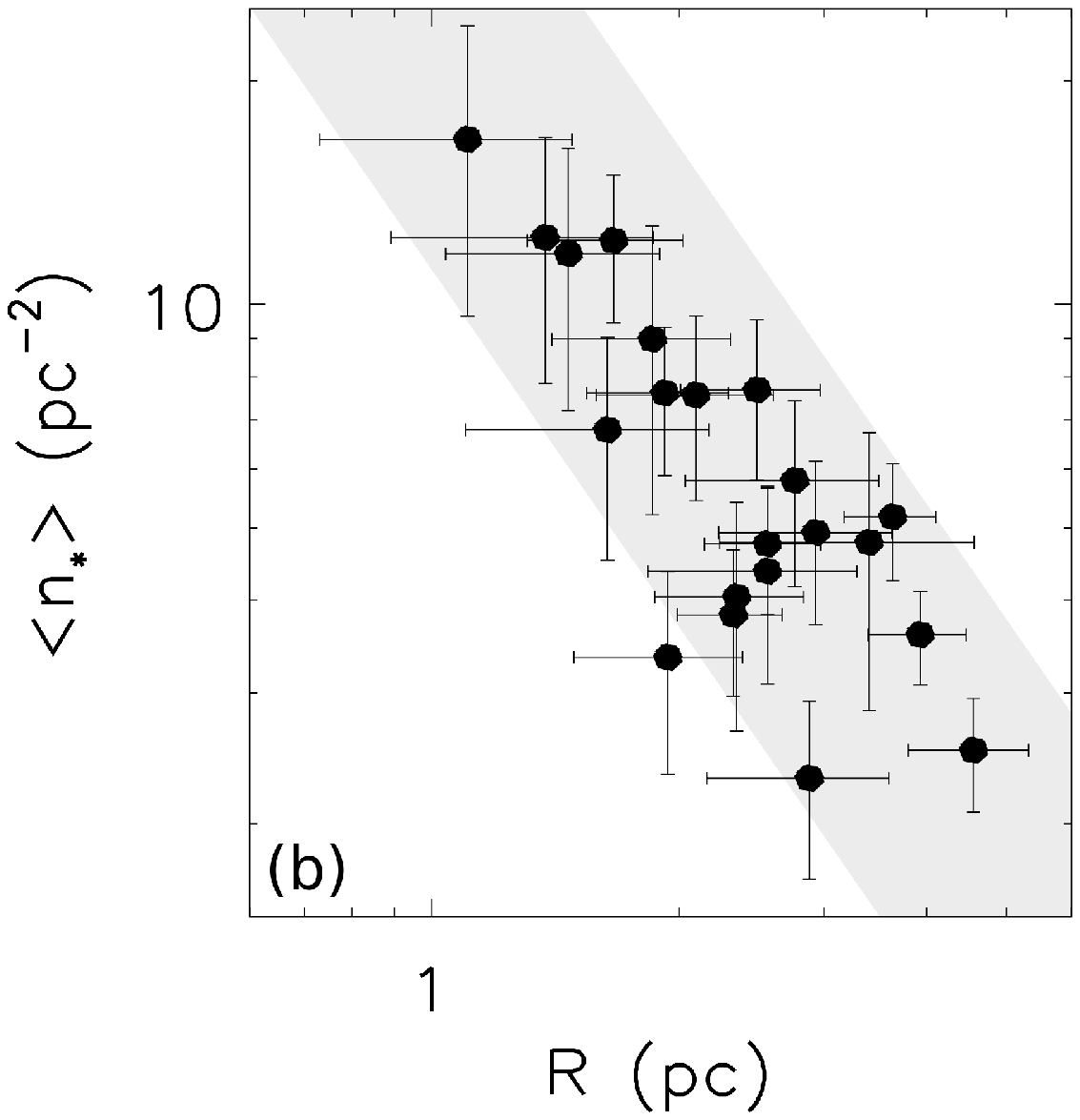}
\includegraphics[width=4.5cm,angle=0]{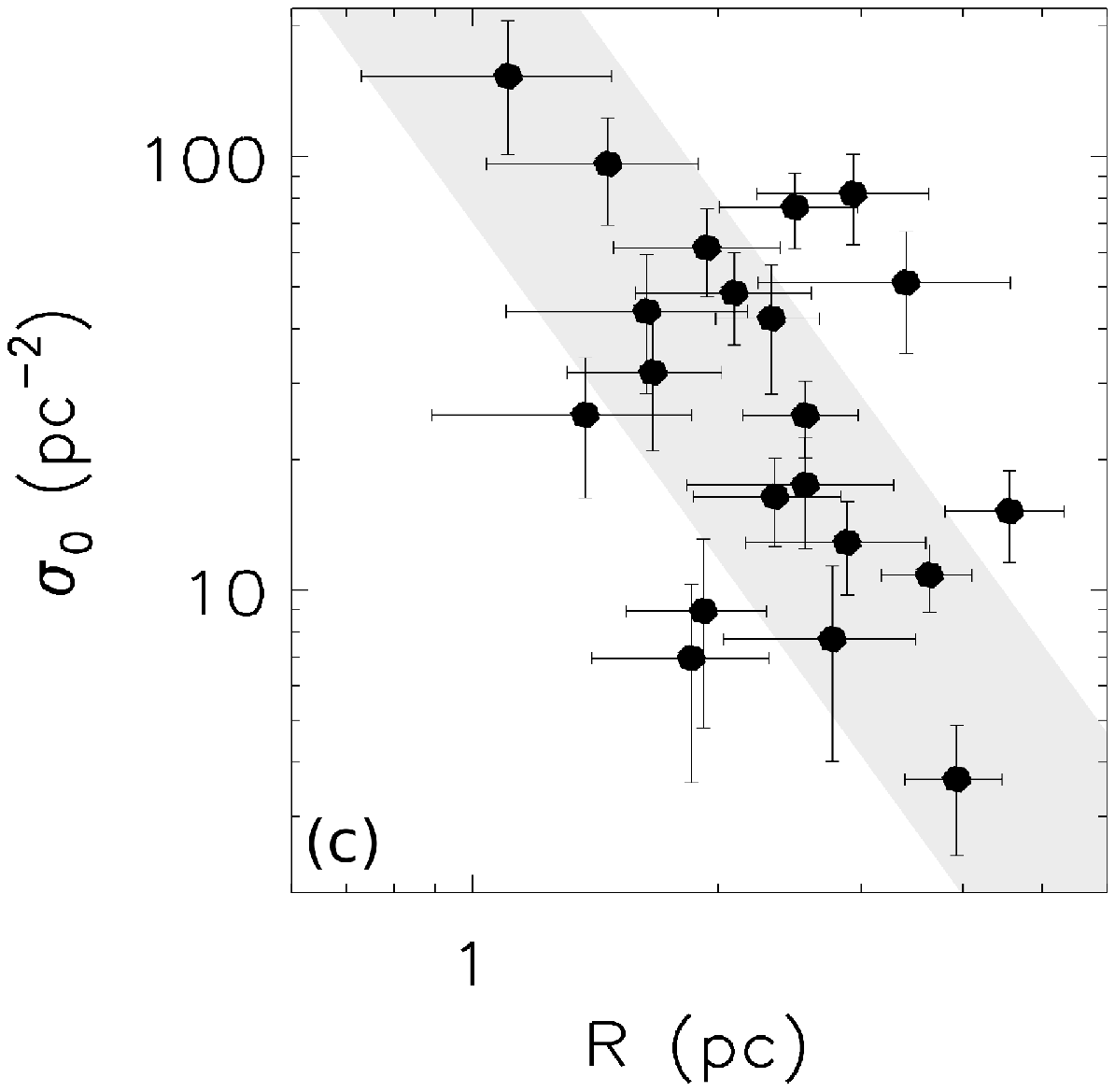}
\includegraphics[width=4.5cm,angle=0]{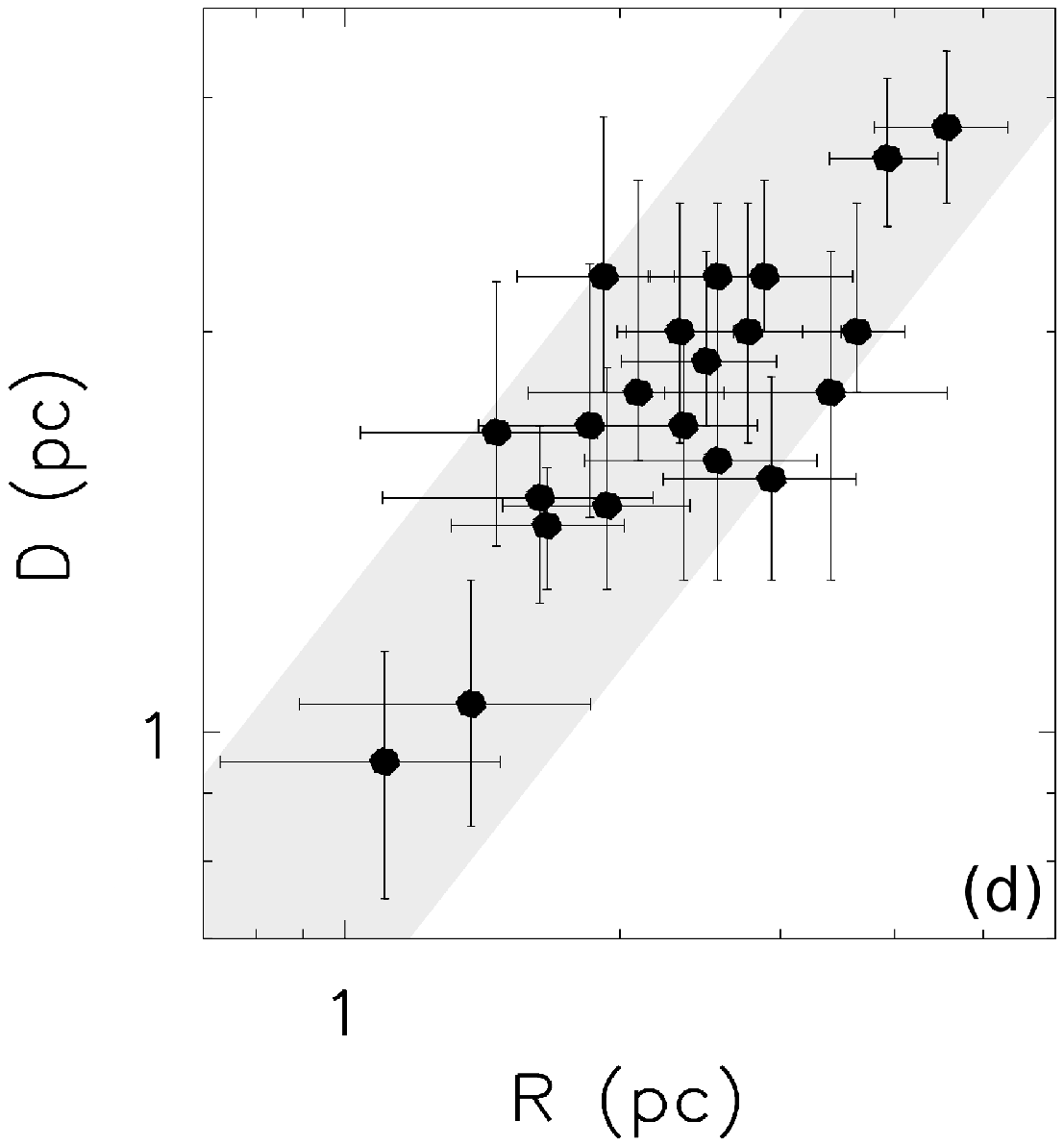}
\includegraphics[width=4.5cm,angle=0]{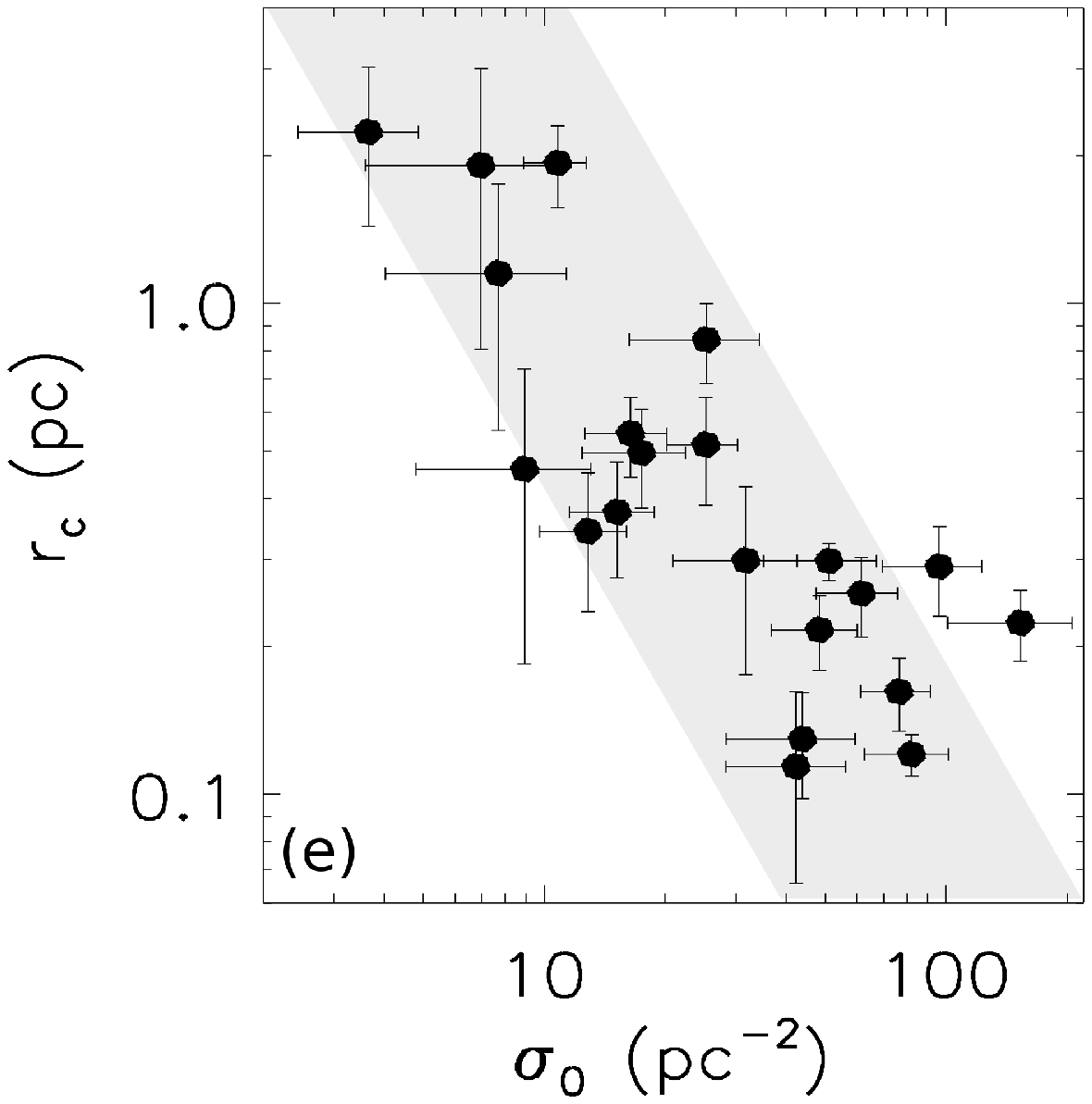}
\includegraphics[width=4.5cm,angle=0]{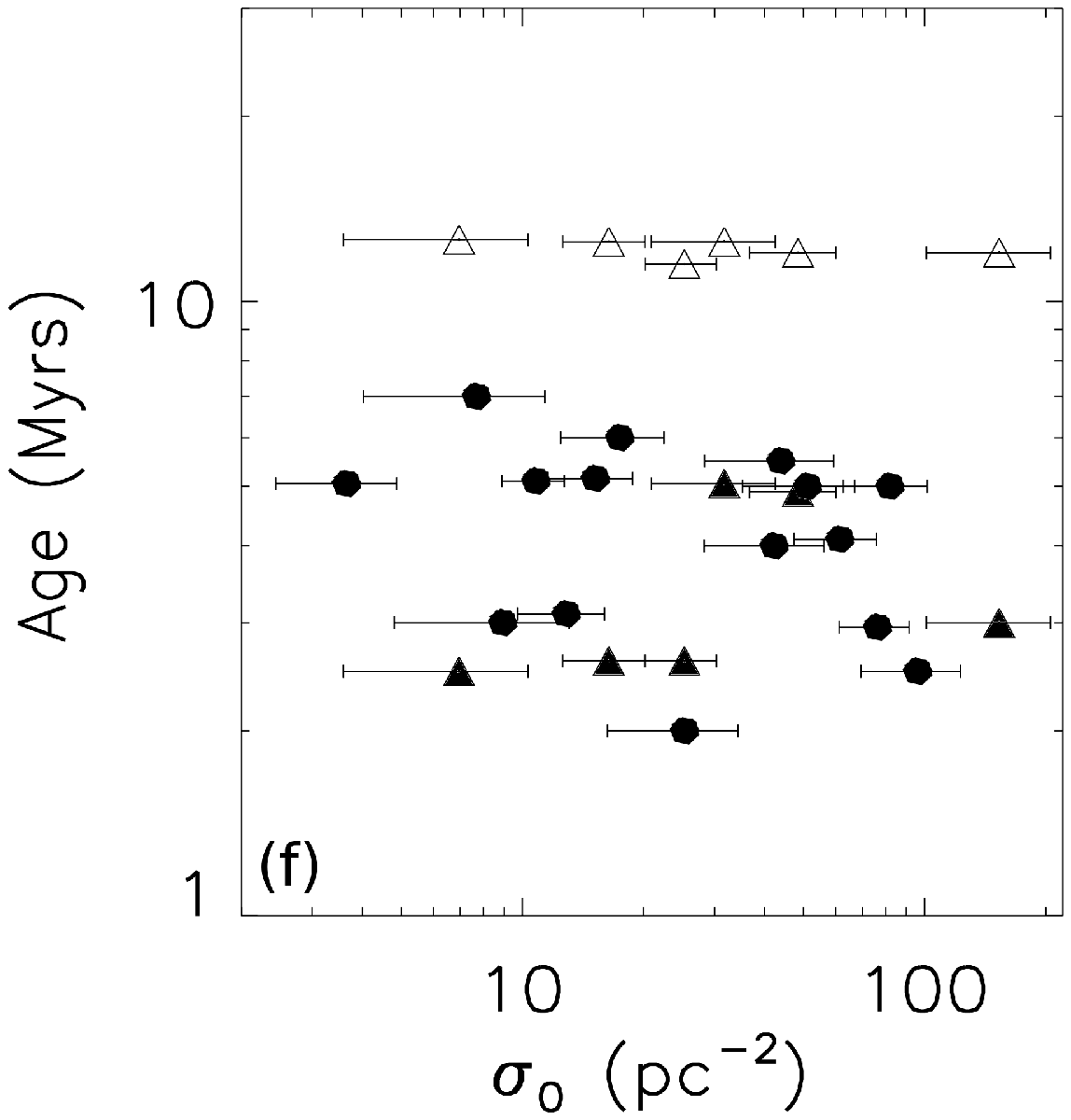}
\includegraphics[width=4.5cm,angle=0]{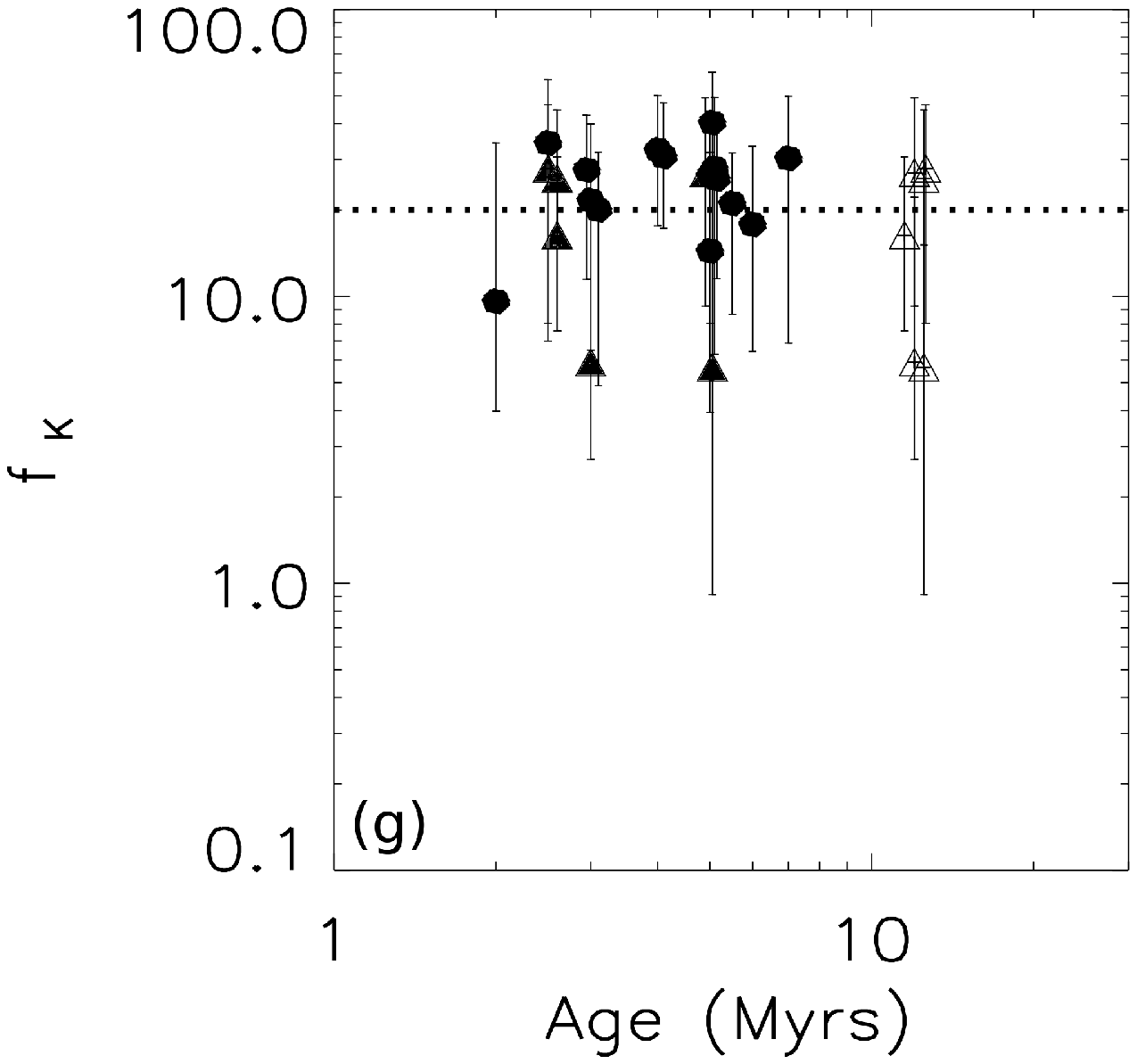}
\includegraphics[width=4.5cm,angle=0]{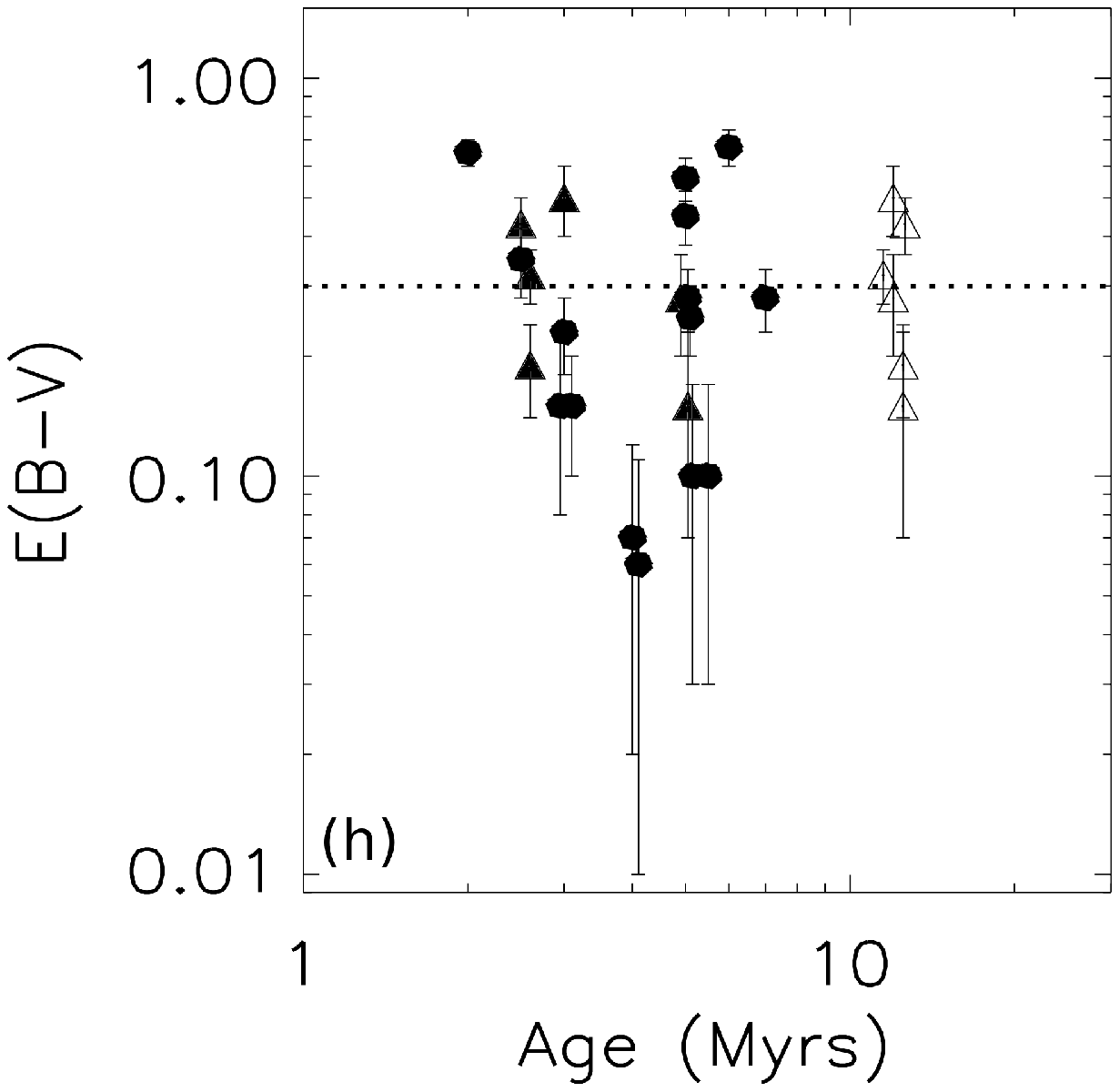}
%\includegraphics[width=4.3cm,angle=0]{fig05e.eps}
%\includegraphics[width=4.3cm,angle=0]{fig05c.eps}
%\includegraphics[width=4.3cm,angle=0]{fig05f.eps}
%\setcaptionmargin{1cm}
{\scriptsize
\caption{
Comparing structural parameters: 
(5a) background density {\it vs.} average density.
Cluster radius compared to (5b) average density; 
(5c) core density; and (5d) distance.
(5e) core radius {\it vs.} core density.
Age correlations with: (5f) core density; (5g) the fraction of cluster 
members showing excess in the K-band; and (5h) E(B-V).  Dotted lines
indicate respectively A$_V \sim$ 1 mag and the limit of f$_K = $ 20\%.
Triangles are used to indicate objects with two options of age (filled symbols correspond to the
first choice, listed in Table 2).
}} 
\label{stru1}
\end{center}
\end{figure*} 
%%%------------------------------------ end of Fig. 5

%%----------------------------------------Sect. 5.5

\subsection{Age}

 It would be expected that  $\sigma_0$ should} decrease with age once the members 
of older clusters could have had time to disperse, diminishing its surface 
stellar density.  However no clear 
trend is observed in Fig. 5f, which displays different symbols to indicate
objects that have two options of age, but only one of them was adopted. 

%In addition to the anti-correlation between age and $\sigma_0$ described above, 
%Fig. 7b shows a trend of r$_c$/R increasing with age, excepting for Trumpler~28. 

The fraction of stars having large E(H-K), f$_K$, is expected to be related to age because of
the colour excess in the K-band  traces circumstellar matter, which is indicative of youth,
 as mentioned in Sect. 4.1 (\cite{Lada 1996}).
Figure 5g shows the distribution of f$_K$ as a function of age, with a dotted line indicating
the f$_K$ = 20\% limit, above which objects younger than 5 Myr are supposed to be found. 
Most of our objects have f$_K >$ 20\%, but three of the youngest clusters are below this limit, 
showing no correlation of age with f$_K$.

 Probably this lack of correlation is due to the fact that some
of our clusters have a mixing of populations, which means: part of the members is $<$ 4 Myr, while
others are 4 - 10 Myr. 

On the other hand, the lack of correlation is also found for 
Stock~16 that has f$_K$ = 30\% and mean age of 7.0$^{+8.0}_{-6.5}$ Myr. 
An explantion is a possible field-stars contamination in the number of
objects with K-band excess in the colour-colour diagram, mainly those 
having 0.1 $<$ (J-H)$_o$ $<$ 0.3 mag. In the
colour-magnitude diagram, these stars are counted in the range of ages 
older than 20 Myr,  but should not be considered. The large error bars
on both,  age and f$_K$ estimation, give us only a qualitative analysis
of the relation between these parameters.

Age is also expected to be related with other parameters such as colour excess. Since  E(B-V) is
indicative of visual extinction, it can be used to infer how embedded the cluster is, and by consequence 
to verify the youth of the cluster. In fact, Fig. 5h shows three of the youngest objects  ($<$ 5 Myr)
appearing above the line representing E(B-V) = 0.3 mag, which means A$_V > $ 1 mag. However, five other
young clusters are below this limit, which indicates that they are not deeply embedded.
For this reason, we concluded that there is no correlation between age and E(B-V) for our sample.

%%----------------------------------------Section 6
\section{Summary of the results and conclusions}

We determined the structural parameters as a function of superficial density and
radial distribution profile of a sample of 21 clusters, selected on basis of their 
youth and intermediate distance.
A statistical procedure using colour-magnitude criteria provided a double-checked decontamination 
of field-stars. The remaining stars were considered members and their observed
colours were used to more accurately determine the visual extinction affecting the cluster. 
The unreddened colours were compared with theoretical isochrones
in the CMD aiming to confirm distance and age. 
The same was done to determine individual masses, based on the cluster members position 
compared to  evolutionary tracks.

In principle, centrally concentrated clusters should be the youngest ones  since they would not
had enough time to disperse. In fact, most  of our clusters
show this characteristic. However, the constrained range of ages in our sample impeaches us to
be more conclusive about differences or similarities on the evolution of the studied clusters.

We conclude that all the 21 studied clusters are very similar, probably due to the selection criteria choosing 
restricted ranges of size, distance and age. By consequence, there is no large variation on number of members, 
radius and mass of the clusters. On the other hand, the galactic distribution of the objects causes differences
among the clusters environments.

When compared with other young clusters (\cite{LL03}, \cite{Carpenter 2000} 2000),   our sample follows the same trends,
but has larger radius and lower volumetric density. This means a less concentrated distribution of members that may be related to the expected spatial dispersion, when the cluster gets older.

The distinction between star clusters and associations has been discussed in several works. 
 Following the definitions presented by \cite{LL03}, our clusters are classified as  
stellar groups because they have more than 35 physically related stars and their mass density 
exceeds 1 M$_{\odot}$/pc$^{-3}$. Since our objects are optically visible, they cannot be considered 
embedded clusters. Even considering the large error bars on the age estimation,
the mean age of the clusters ($\sim$ 5 Myr) is  a clear indication that
our sample is formed by young stellar associations. 

 We also checked for our sample the relation between age and crossing 
time ($ \tau_{cr}$), following \cite{GPZ2011}, for instance.
They proposed a criterion in which bound systems (open star clusters) would have age/$ \tau_{cr}> 1 $ 
 and unbound systems would have age/$\tau_{cr} < 1$.
 For this test, we adopted the crossing 
time defined by:
 $\tau_{cr} =\frac{2R}{\sigma_{V}}$,
where R is the cluster radius and the stellar velocity dispersion is given by:
$\sigma_{V} = \sqrt{\frac{GM}{R}}$, according to Saurin, Bica \& Bonatto (\cite{SBB2012}).

We verified that  none of our clusters
has stars with age that exceeds the crossing time.  Therefore, they probably
would evolve like stellar associations. 
If the error bars on the age/$ \tau_{cr} $
calculations are considered, only Lynga~14 could evolve as an open cluster.
This result is in agreement with the suggestion by 
\cite{LL03} and Pfalzner (\cite{Pfalzner 2009}), for instance, that few star clusters are expected to be bound.
However, it must be kept in mind that these are qualitative conclusions, due to the 
large uncertainties on age estimation.

%We conclude that our sample fulfils the caveat between embedded and exposed clusters. The mass-radius
%and mass-density dependences found for our objects are consistent with those shown by the
%\cite{LL03} sample, but do not confirm the time sequence proposed by Pfalzner (\cite{Pfalzner 2011}).
 
 Our sample fulfils the caveat between the samples of embedded and exposed clusters studied by Pfalzner (\cite{Pfalzner 2011}). However,  our results do not confirm the mass-radius or density-radius
dependences suggested by her. 

 In fact, we verified that massive clusters from Pfalzner (\cite{Pfalzner 2011}) are
distributed in two groups. Those with large sizes, lower masses, and 
intermediate ages (4 - 10 Myr) follow the same trends shown by embedded clusters (\cite{LL03}), 
as well as our sample. The younger massive clusters ($<$ 4 Myr)  studied by Pfalzner (\cite{Pfalzner 2011})
have lower sizes and higher masses, appearing out of the correlations shown in Fig. 4, as well as
Lynga~14 (this work) and S~106 (\cite{LL03}). 
As proposed by \cite{Adams 2006} (2006), the differences among these clusters could be interpreted as formation condition.

 An interesting perspective of the present work is to increase the studied sample by
including other clusters having larger radius (4 - 10 pc), which could complet the gap
between our clusters and the sample of massive clusters.

%%==================================================

\begin{acknowledgements}

TSS thanks financial support from CNPq (Proc. No. 142851/2010-8).
JGH thanks partial support from CAPES/COFECUB (Proc. No. 712/11).
%This work has made use of the SIMBAD, VizieR, and
%Aladin databases operated at CDS, Strasbourg, France.
This publication makes use of data products from the Two Micron All Sky Survey, which is a joint project of the 
University of Massachusetts and the Infrared Processing and Analysis Center/California Institute of Technology, 
funded by the National Aeronautics and Space Administration and the National Science Foundation.

\end{acknowledgements}

\begin{appendix} %First online appendix --------------------------A

\section{Plots of the entire sample}

All plots used in the analysis of structural and fundamental parameters are displayed in this 
Appendix. Figures A1 and A2 show the stellar surface-density maps and the distribution of the 
stellar density as a function of radius (same as Fig. 1).

Figures A3 to A7 present  colour-colour and colour-magnitude diagrams in the left panel 
(same as Fig. 2). The centre and right panels show respectively the histogram of age 
and the mass distribution (same as Fig. 3).

%%%%----------------------Fig. A1
\begin{figure*}[]
\begin{center}
\includegraphics[width=3.4cm]{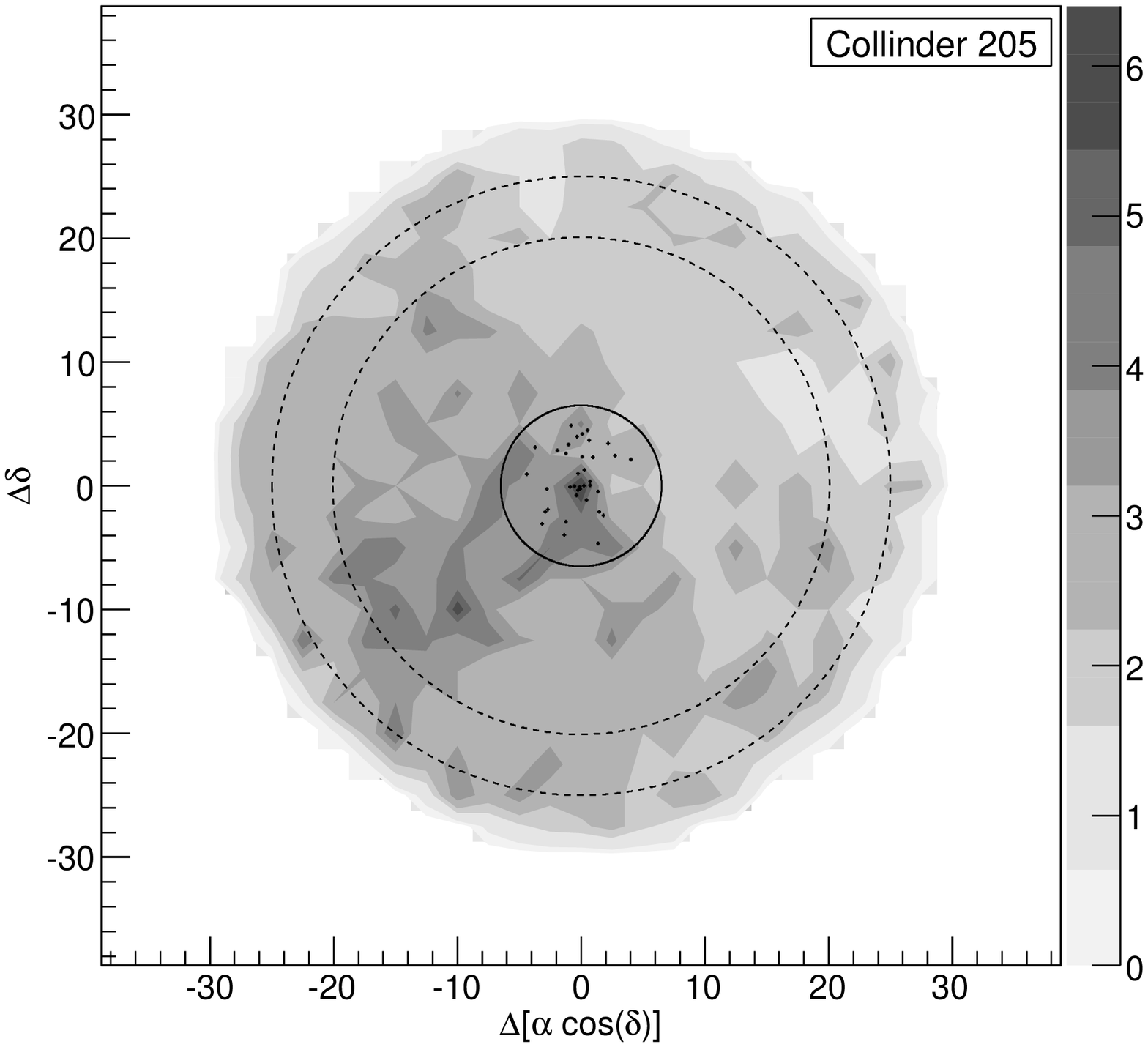}
\includegraphics[width=3.4cm]{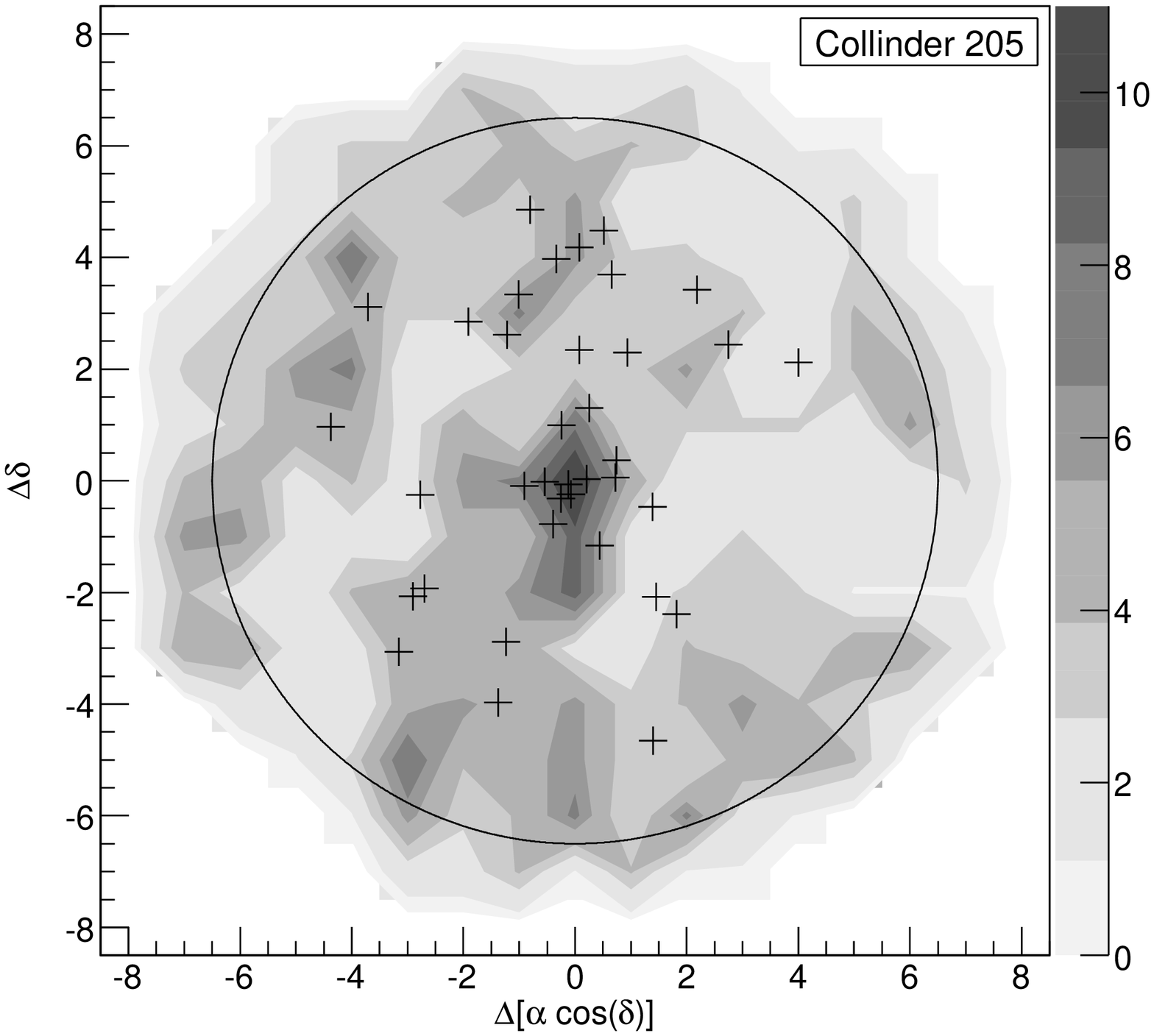}
\includegraphics[width=2.05cm]{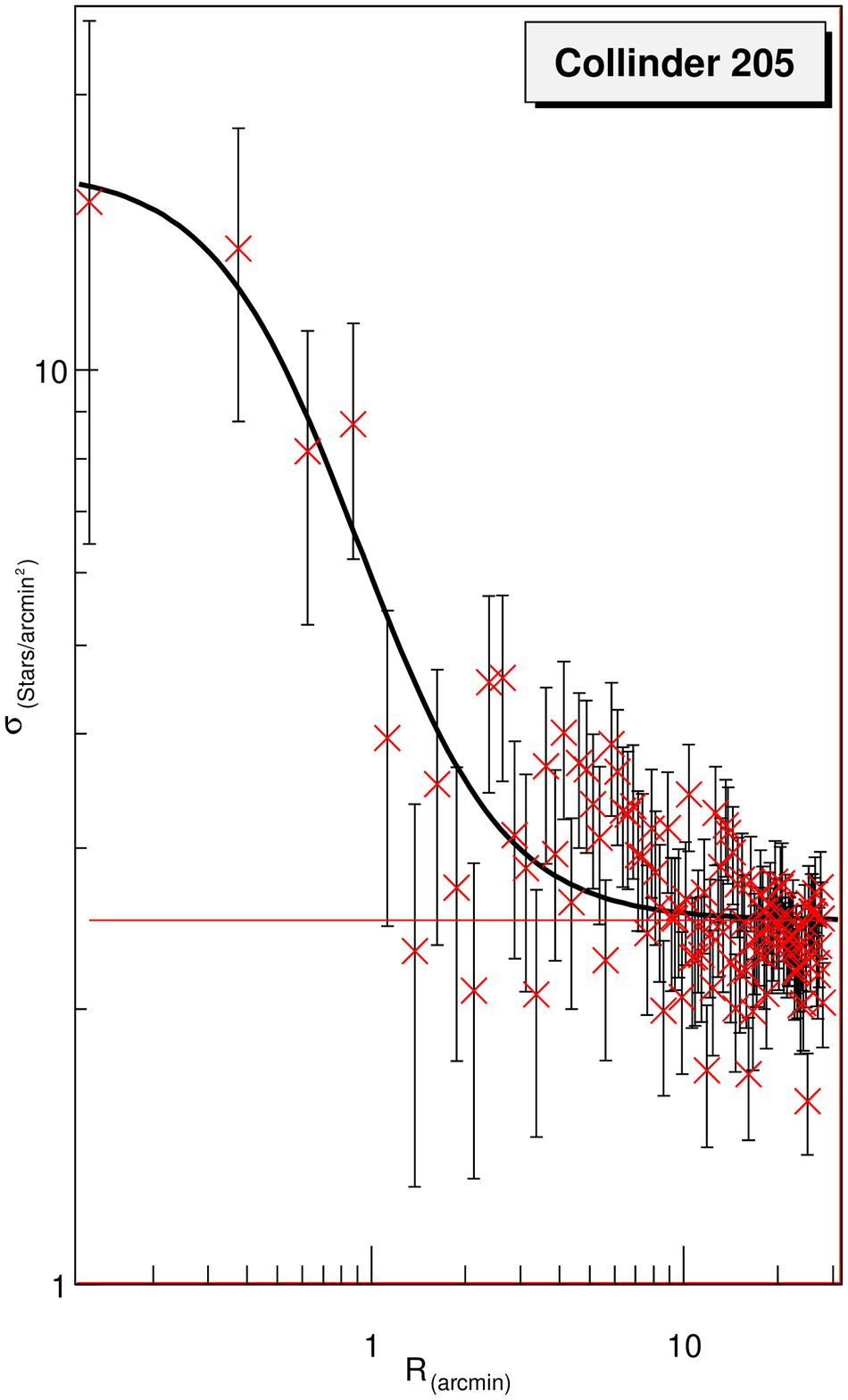}
\includegraphics[width=3.4cm]{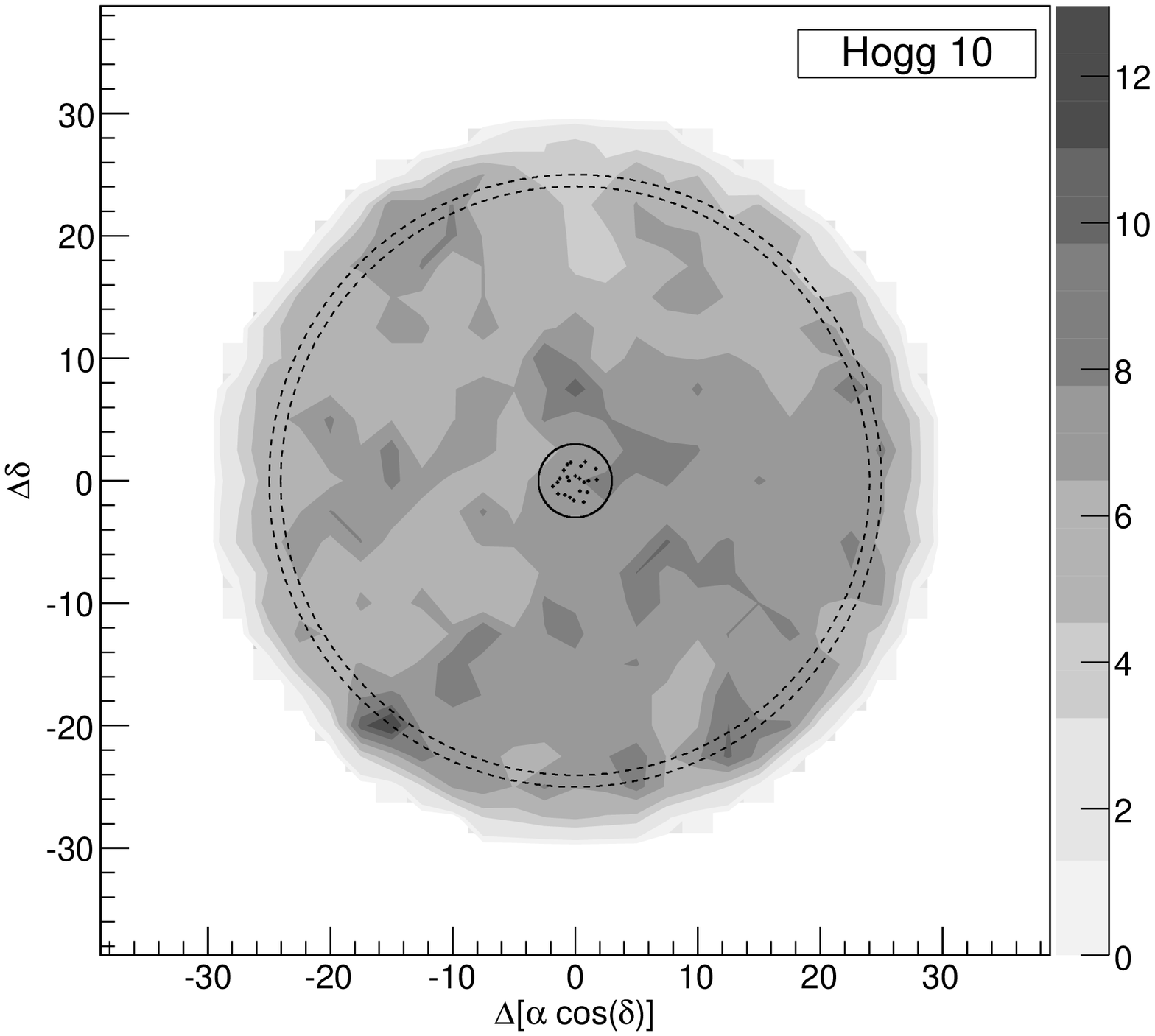}
\includegraphics[width=3.4cm]{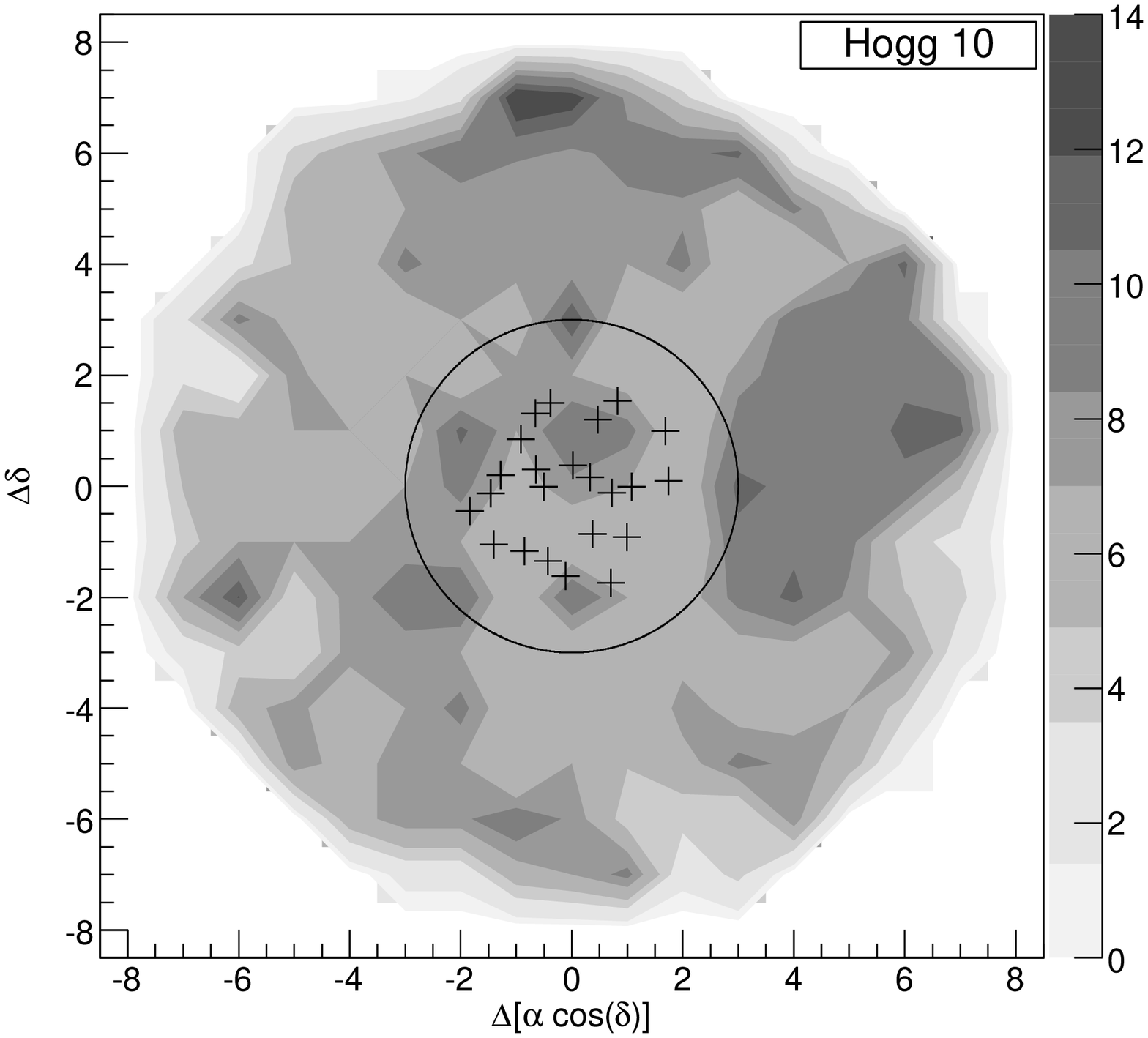}
\includegraphics[width=2.05cm]{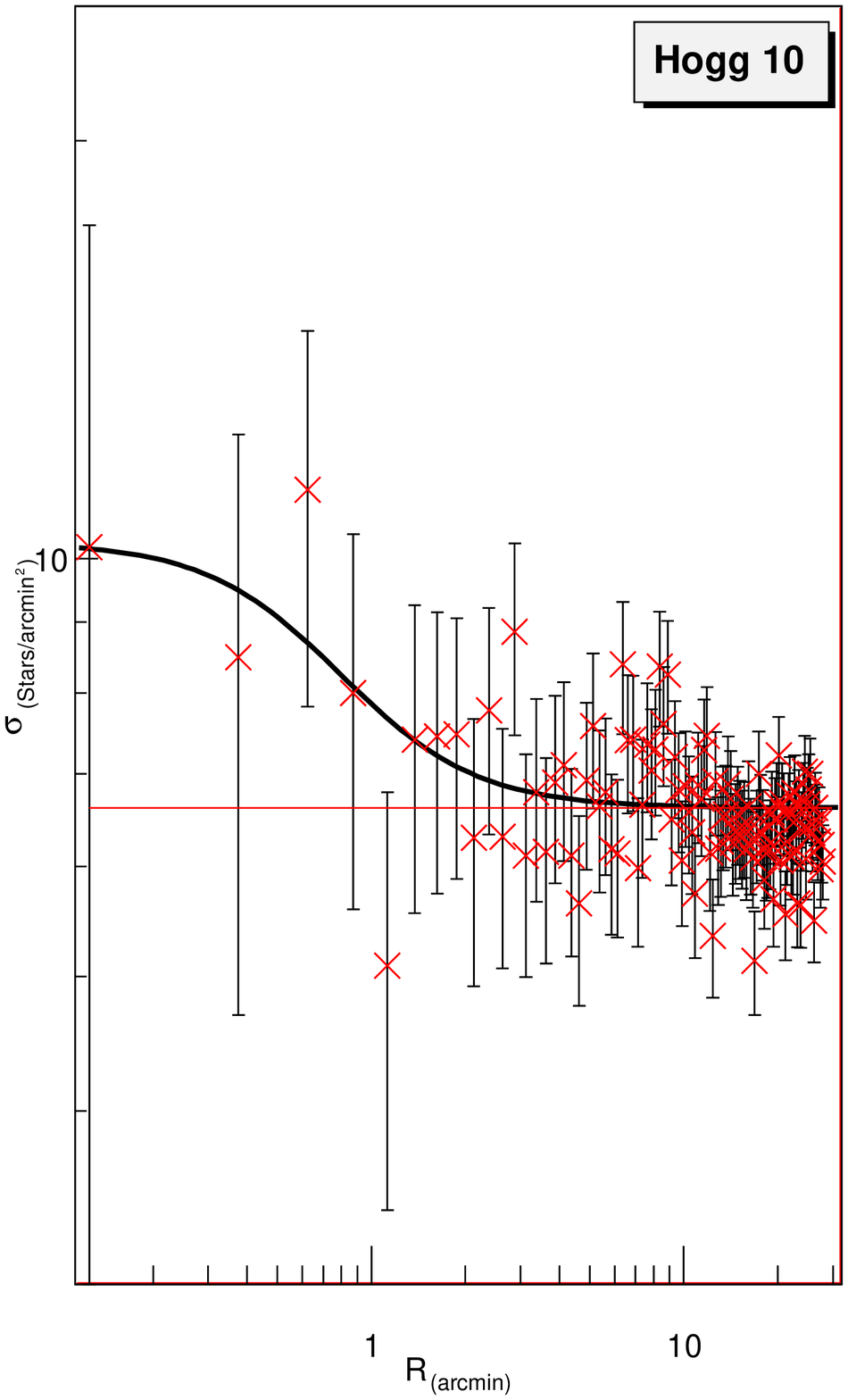}
\includegraphics[width=3.4cm]{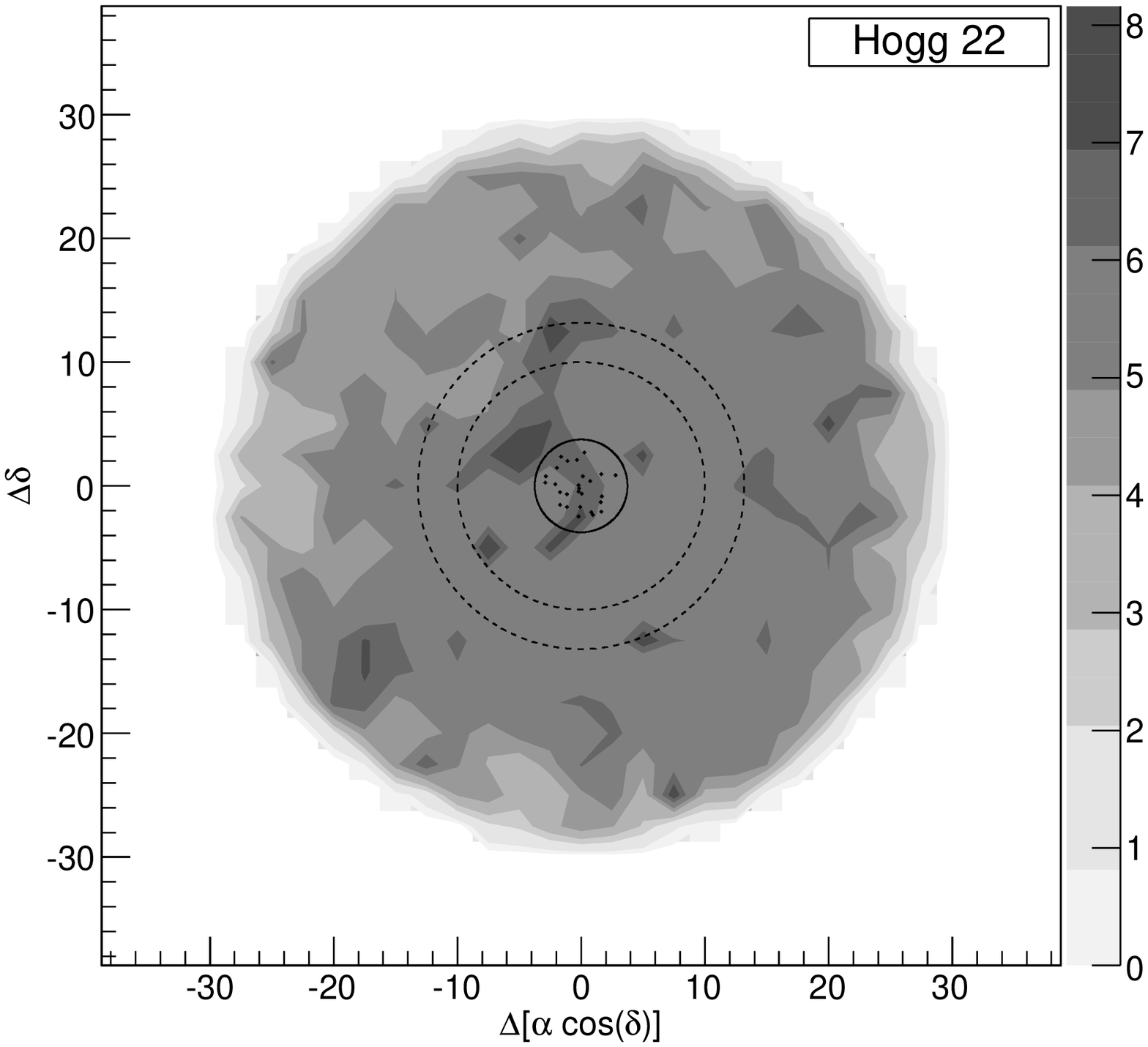}
\includegraphics[width=3.4cm]{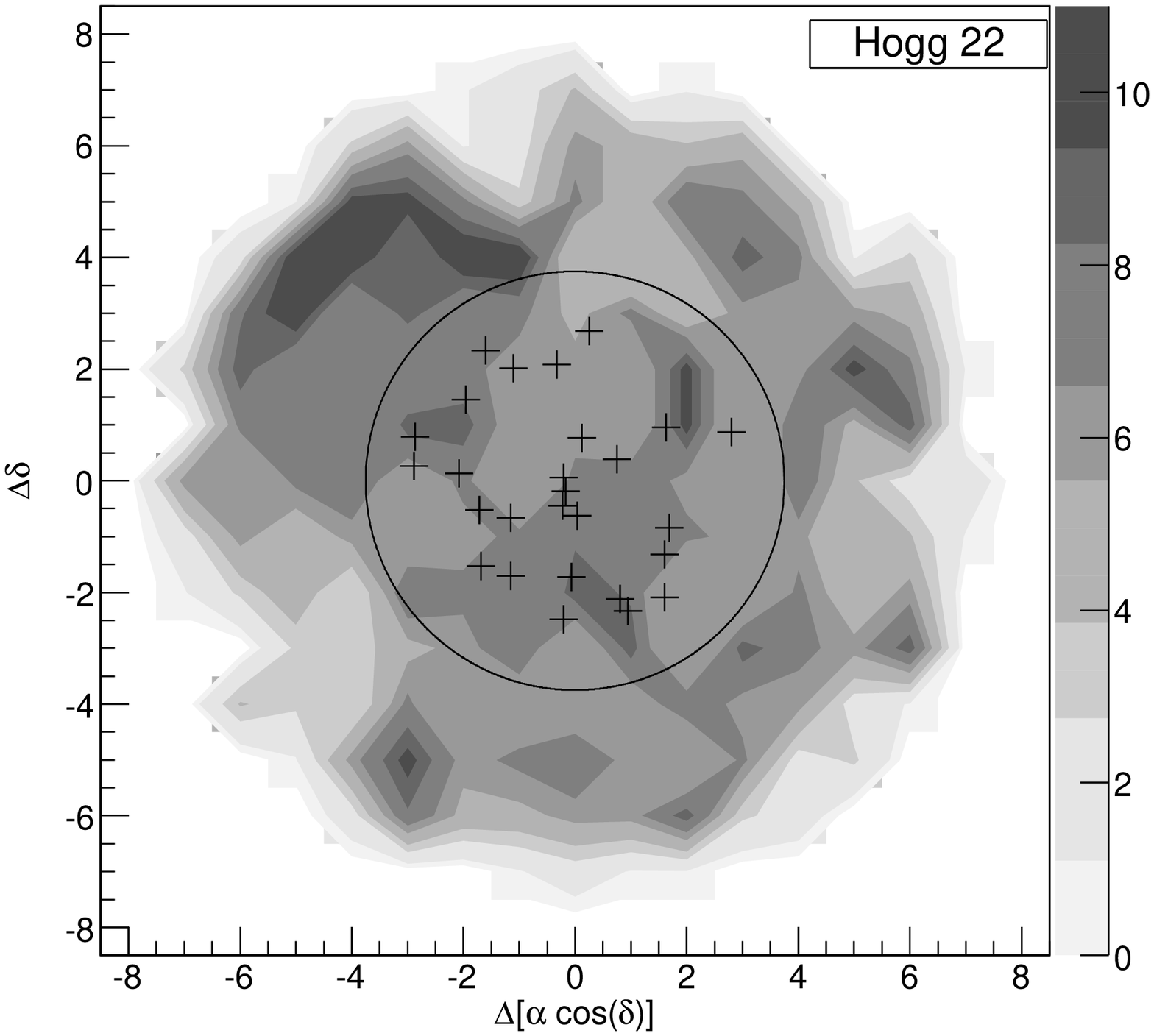}
\includegraphics[width=2.05cm]{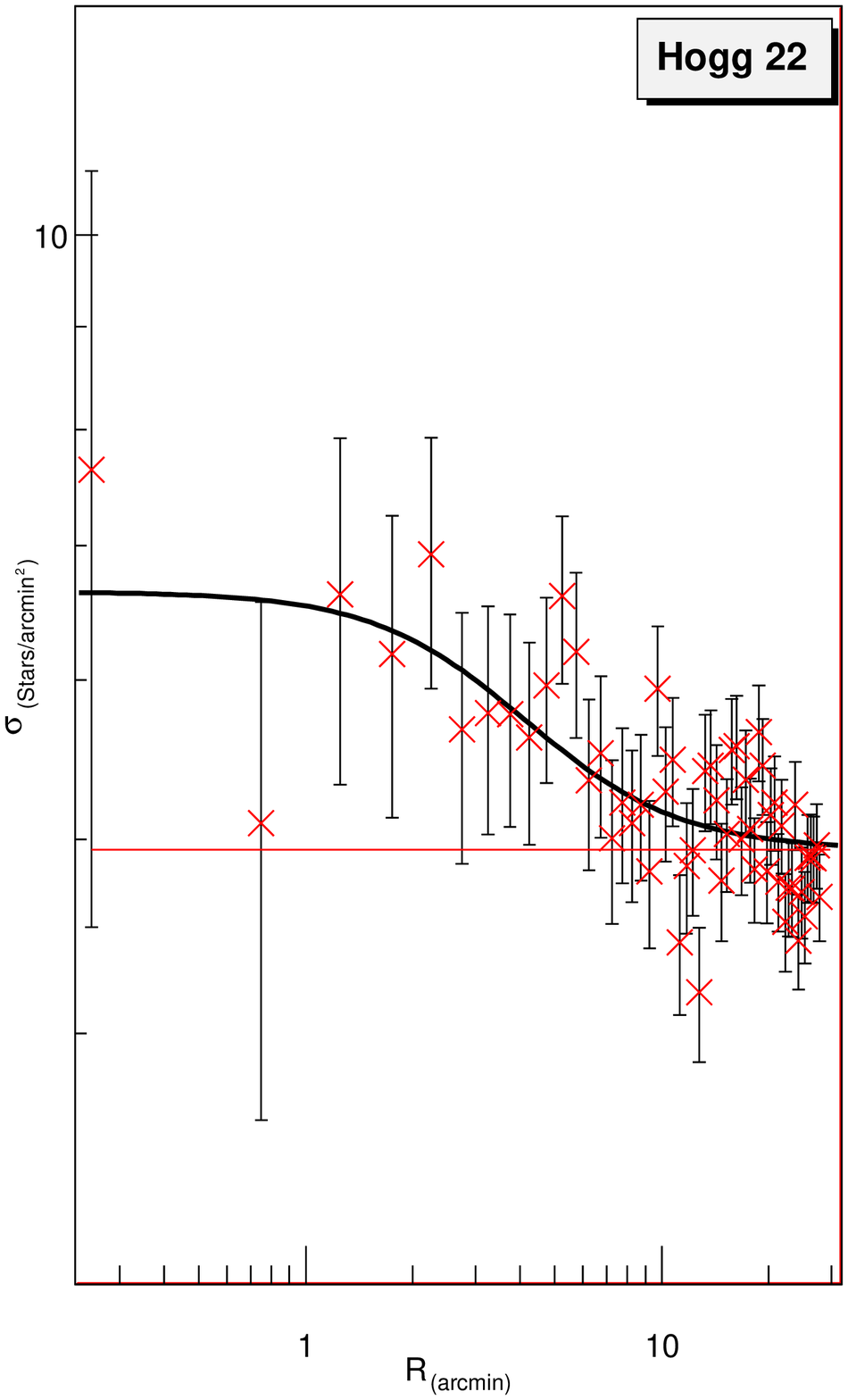}
\includegraphics[width=3.4cm]{ly14_sup_t.eps}
\includegraphics[width=3.4cm]{ly14_sup_5.eps}
\includegraphics[width=2.05cm]{king_ly14.eps}
\includegraphics[width=3.4cm]{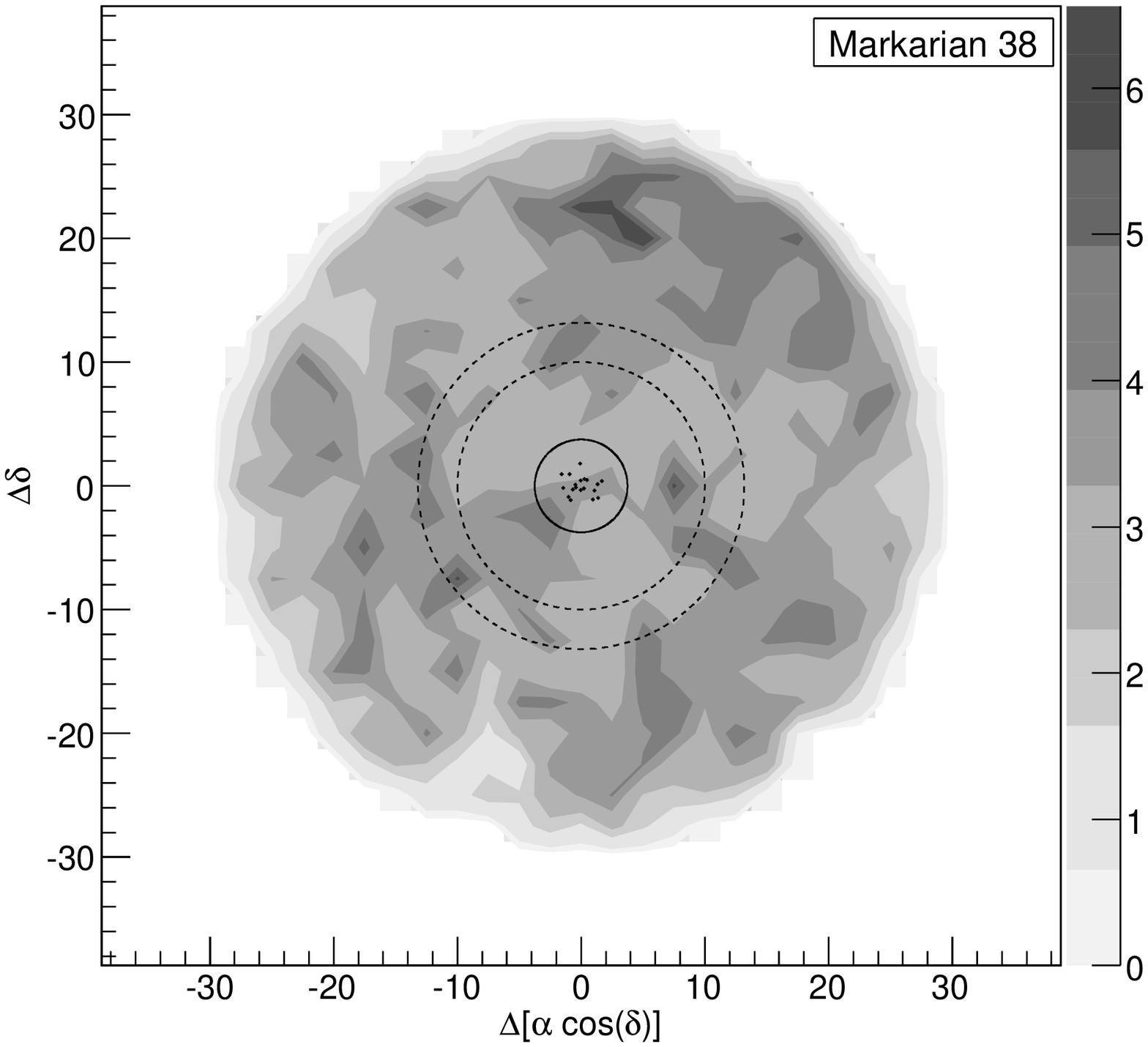}
\includegraphics[width=3.4cm]{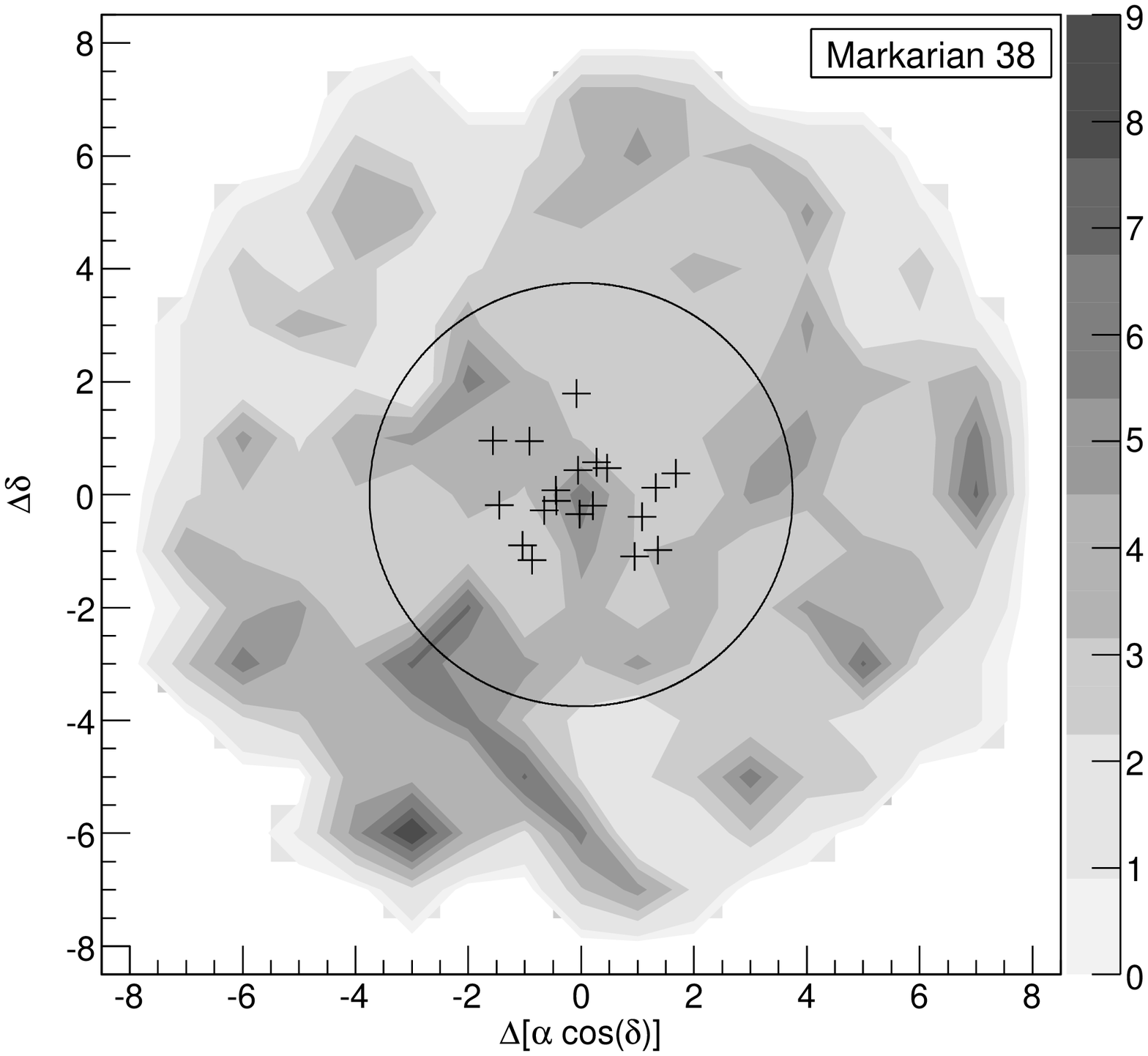}
\includegraphics[width=2.05cm]{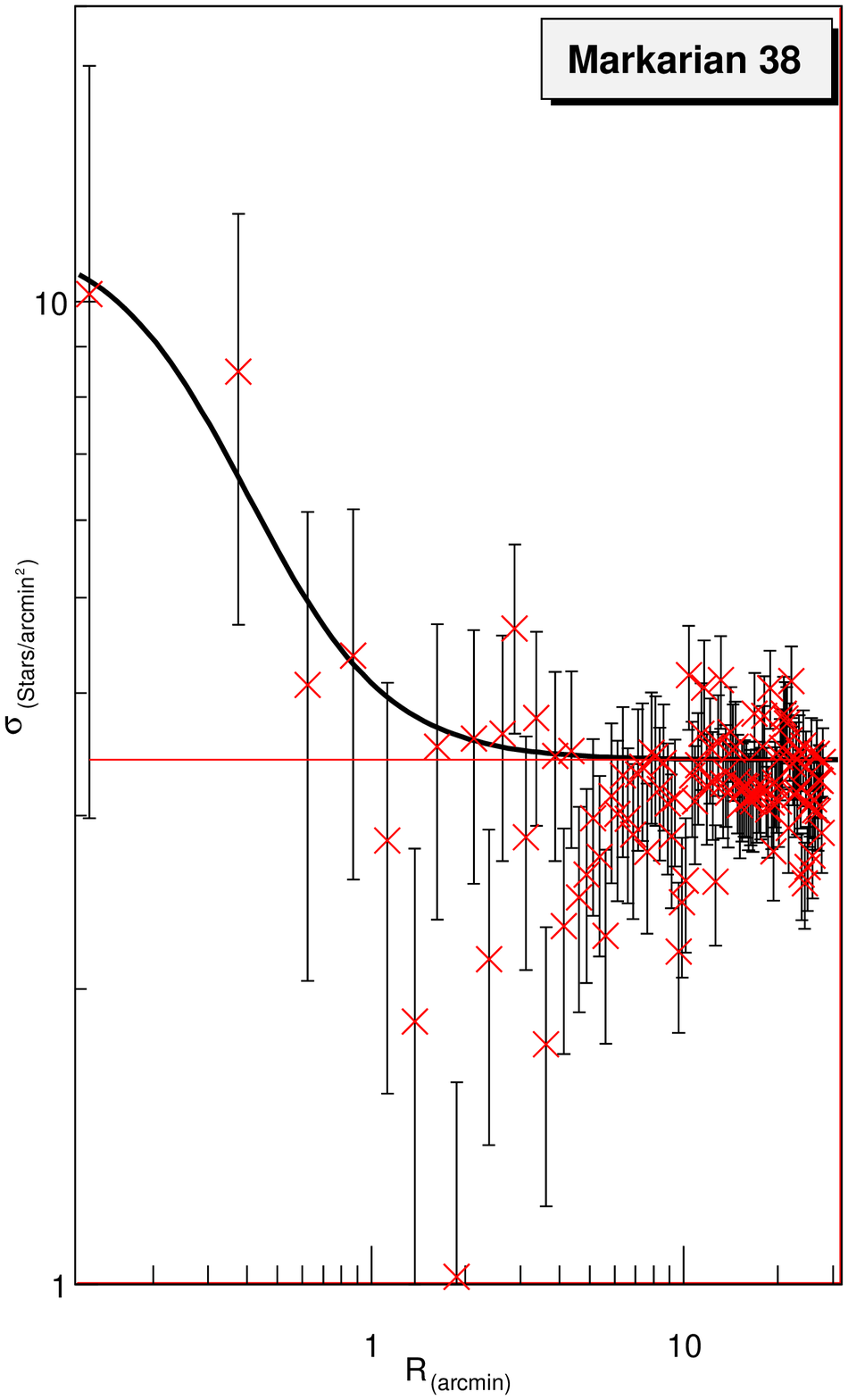}
\includegraphics[width=3.4cm]{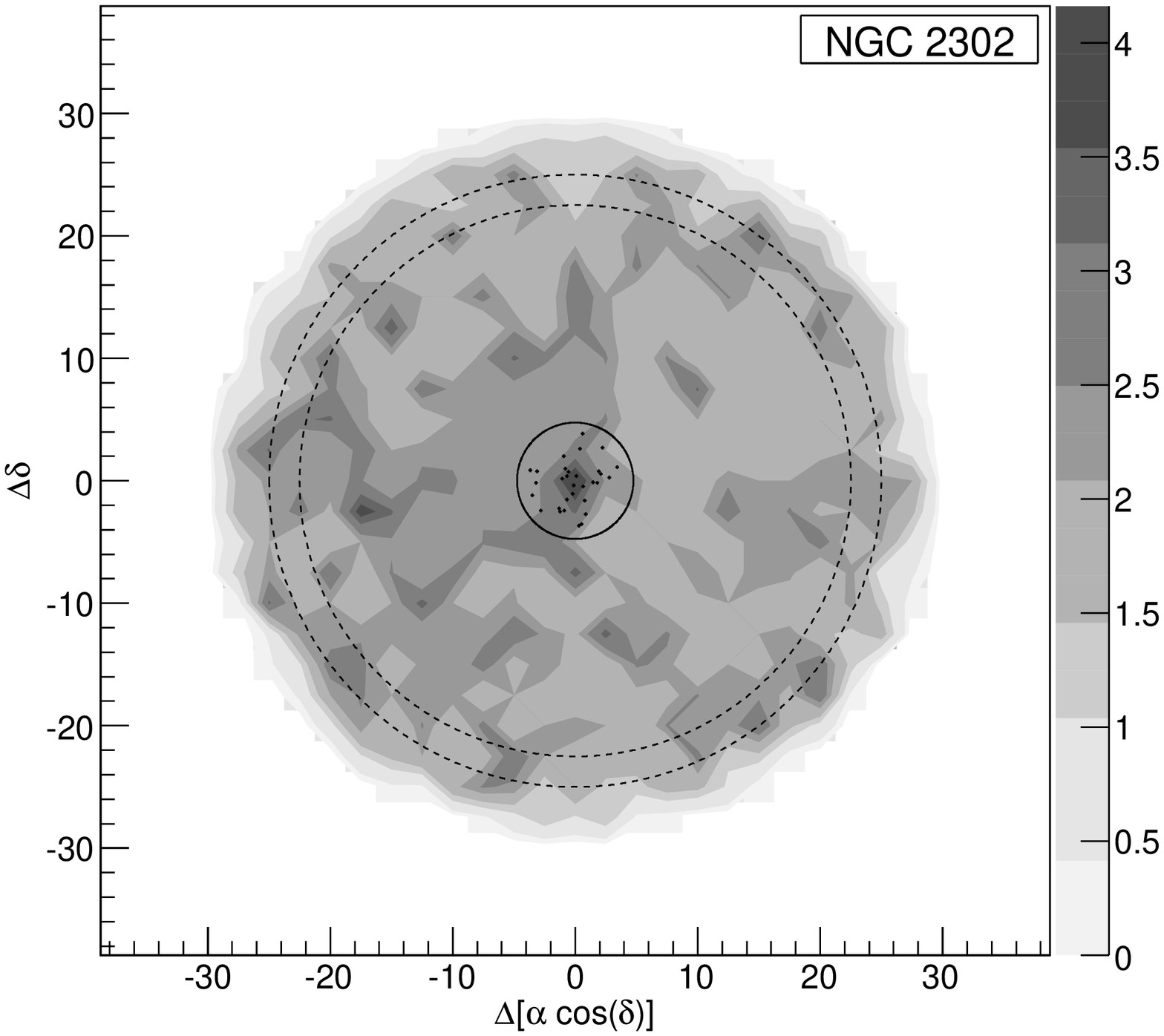}
\includegraphics[width=3.4cm]{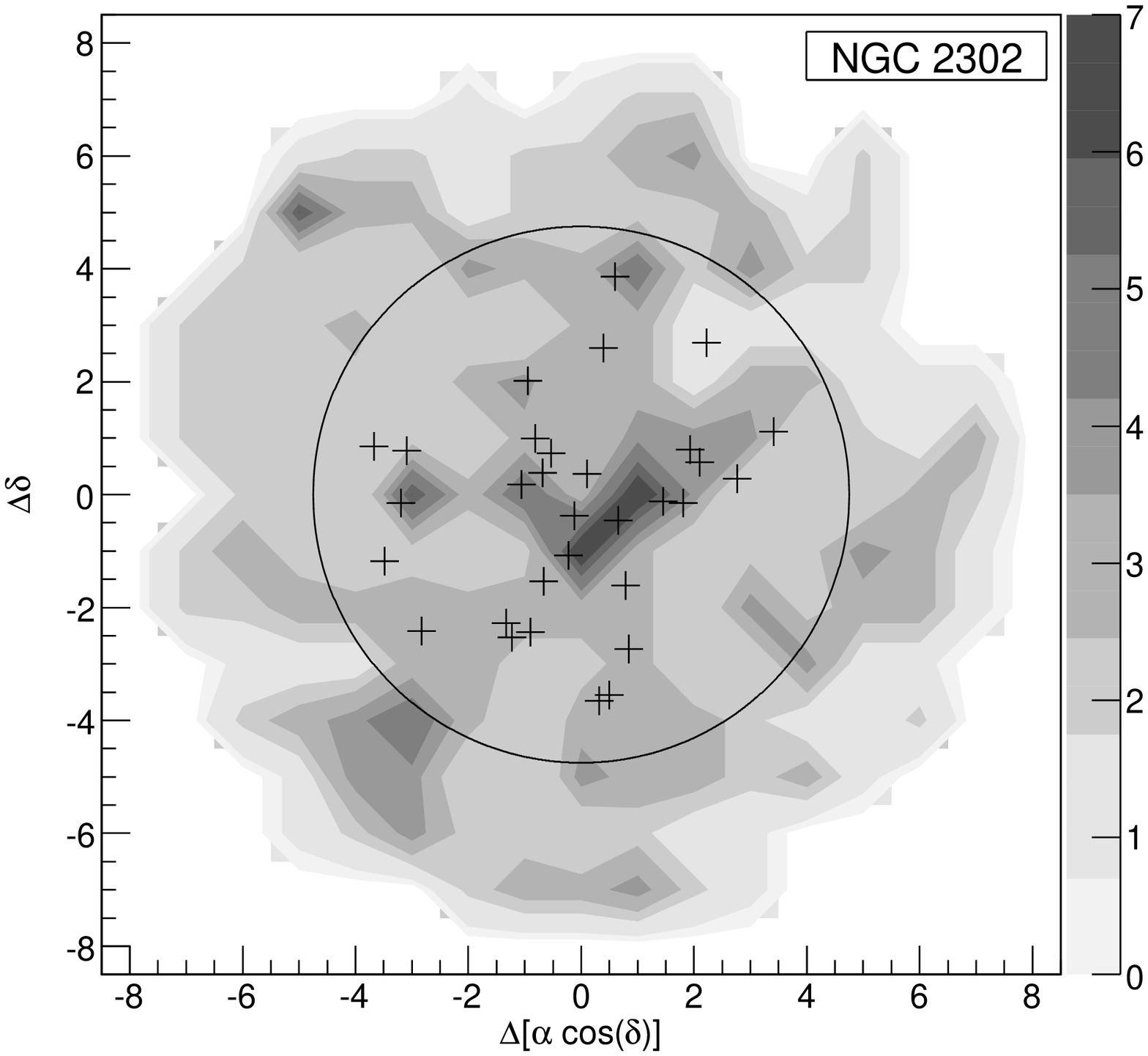}
\includegraphics[width=2.05cm]{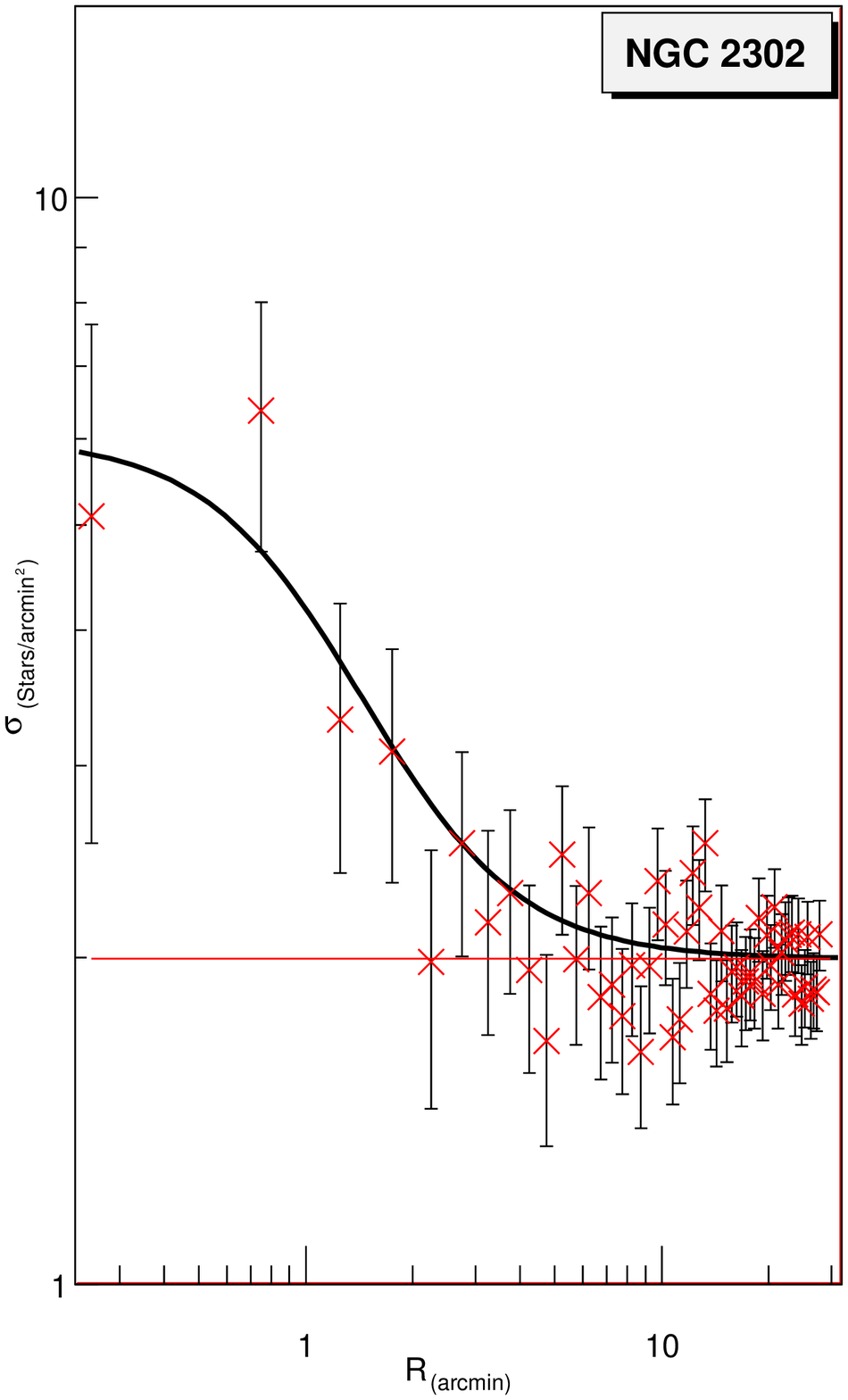}
\includegraphics[width=3.4cm]{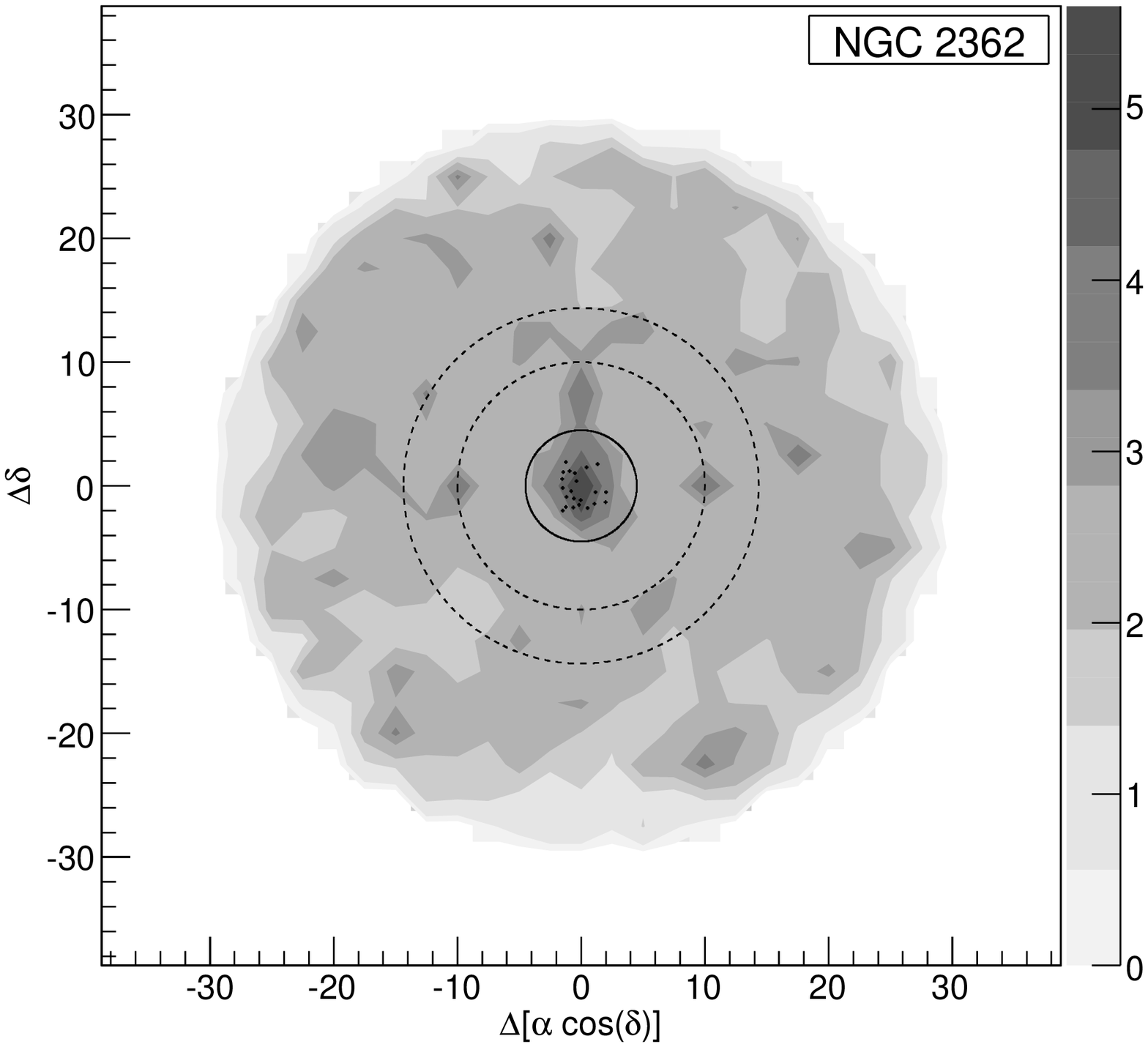}
\includegraphics[width=3.4cm]{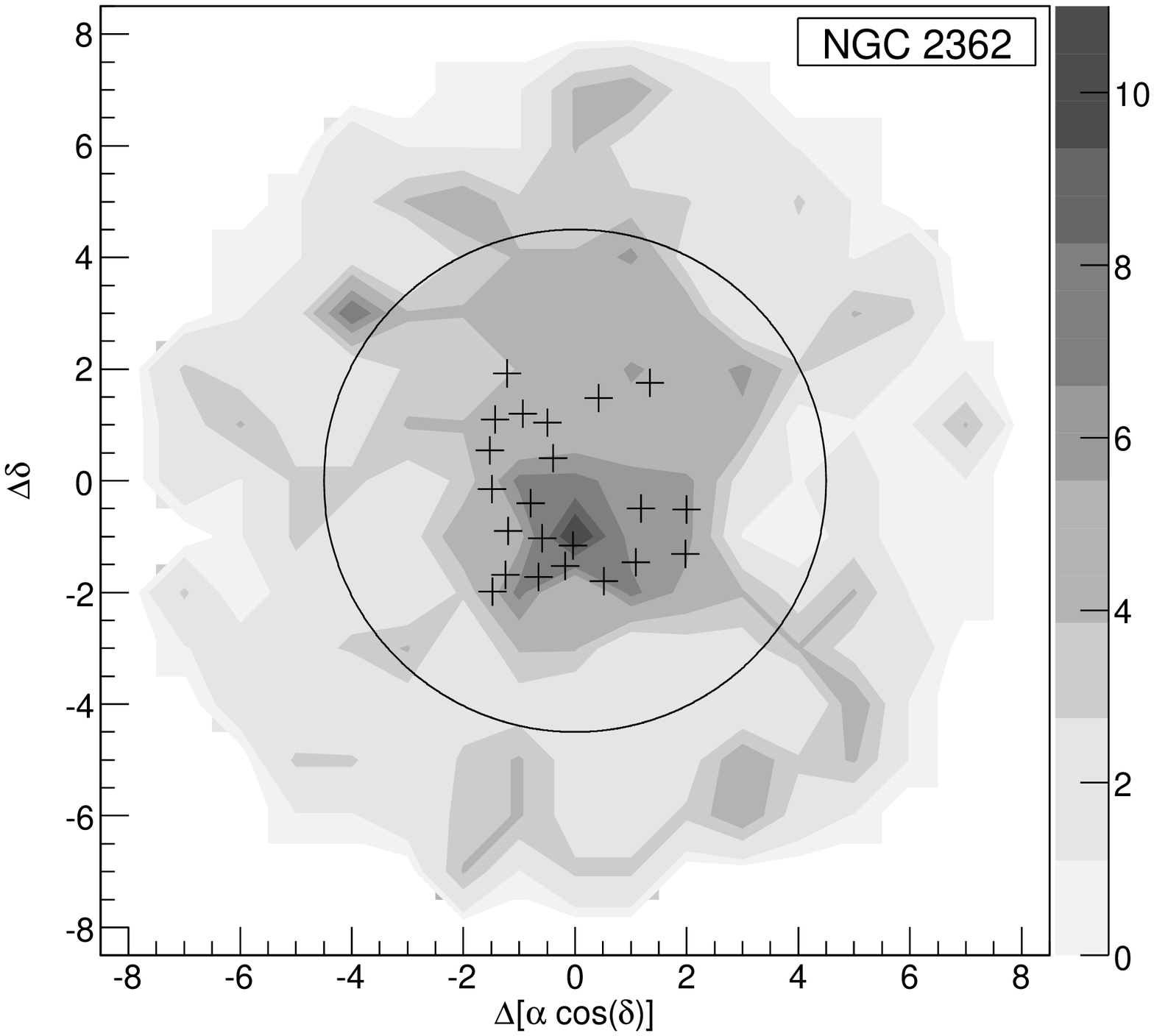}
\includegraphics[width=2.05cm]{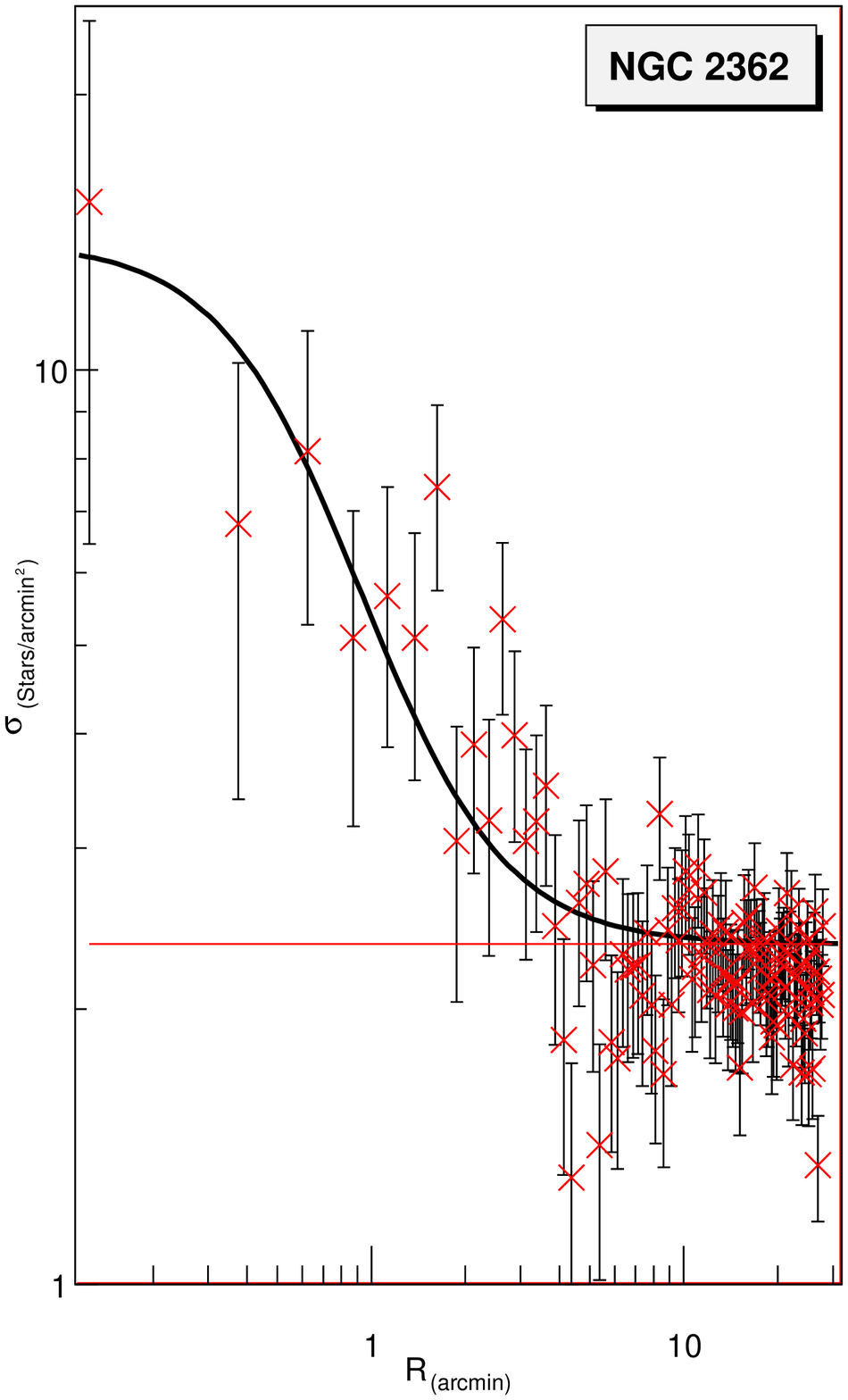}
\includegraphics[width=3.4cm]{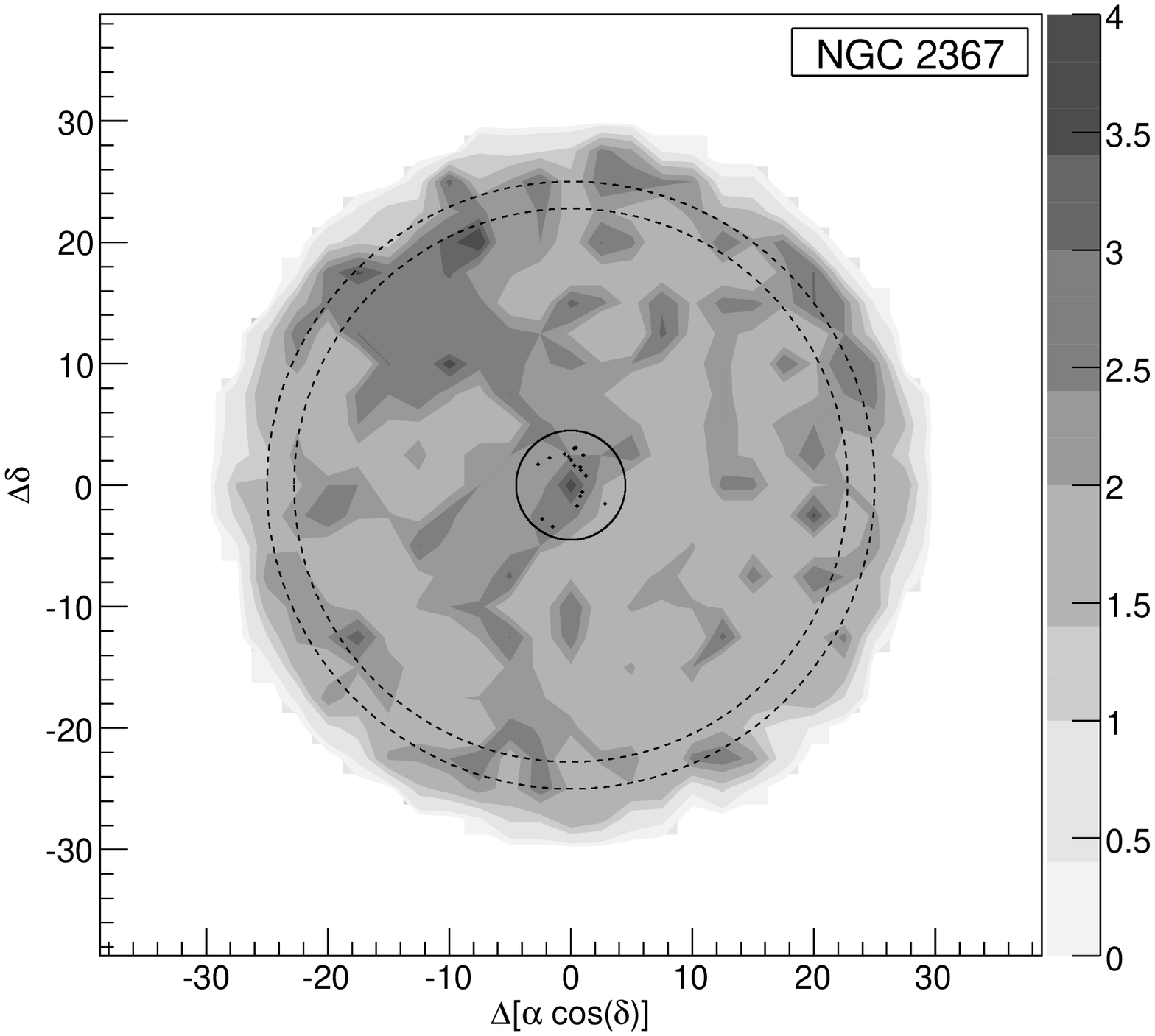}
\includegraphics[width=3.4cm]{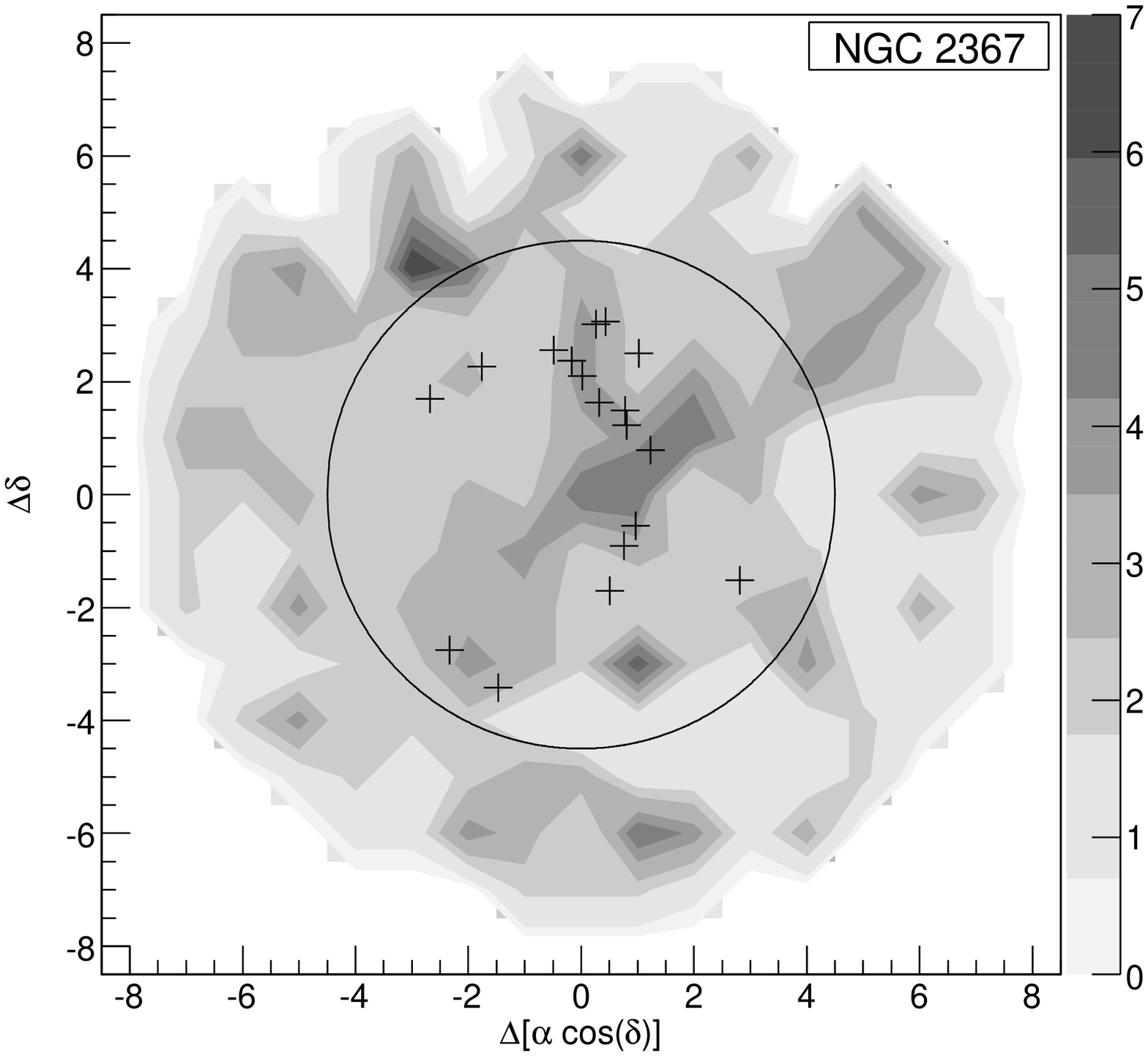}
\includegraphics[width=2.05cm]{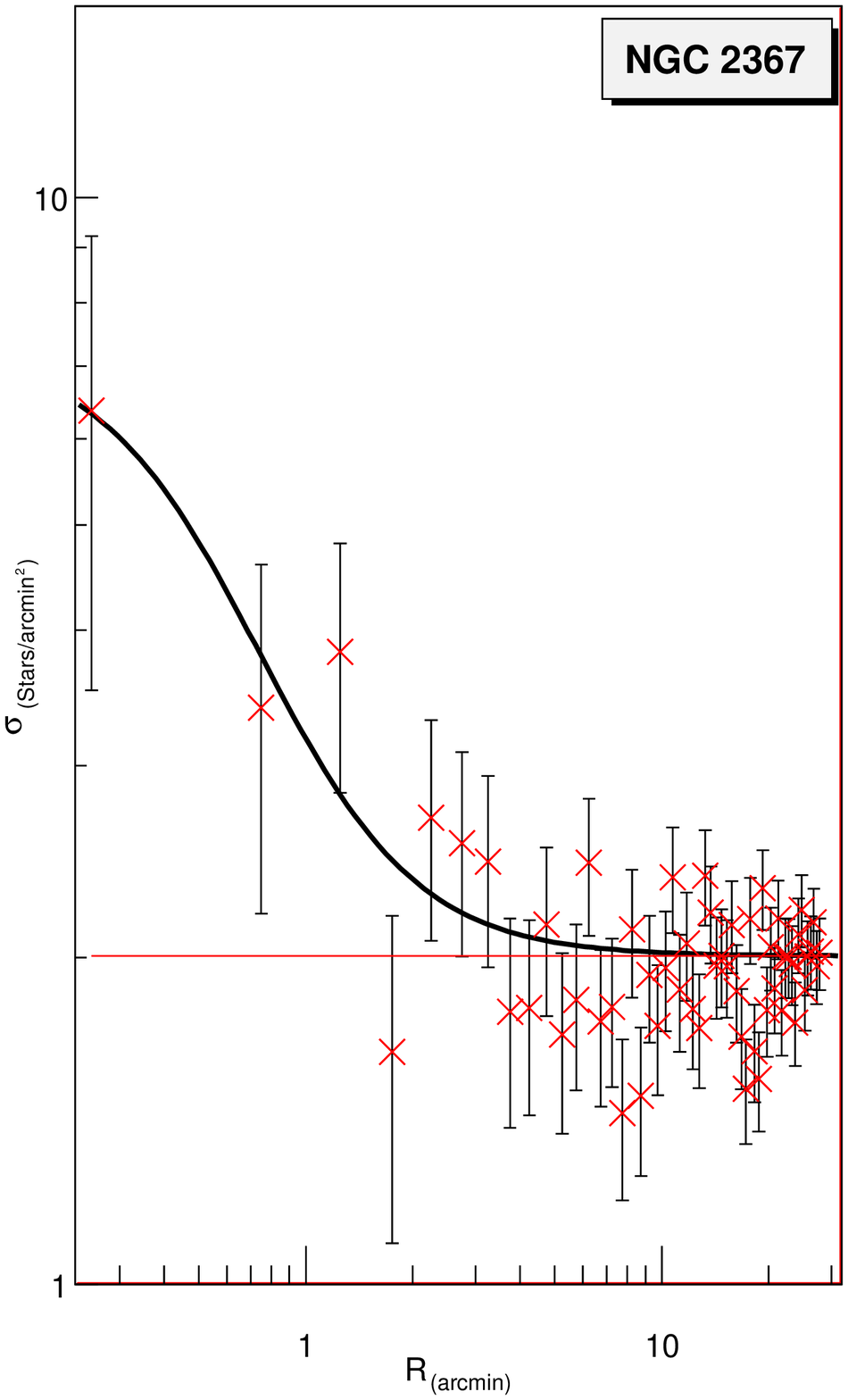}
\includegraphics[width=3.4cm]{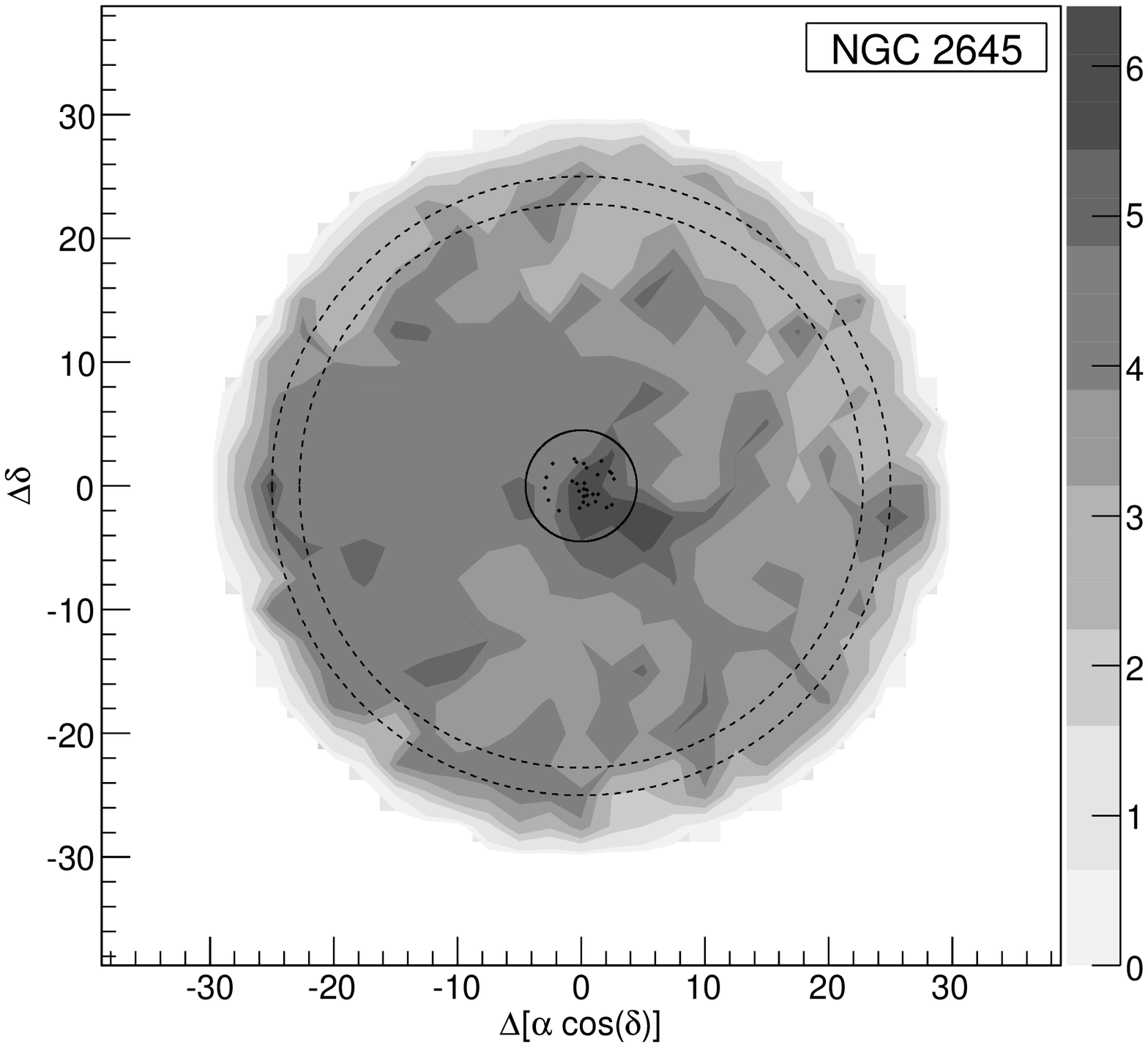}
\includegraphics[width=3.4cm]{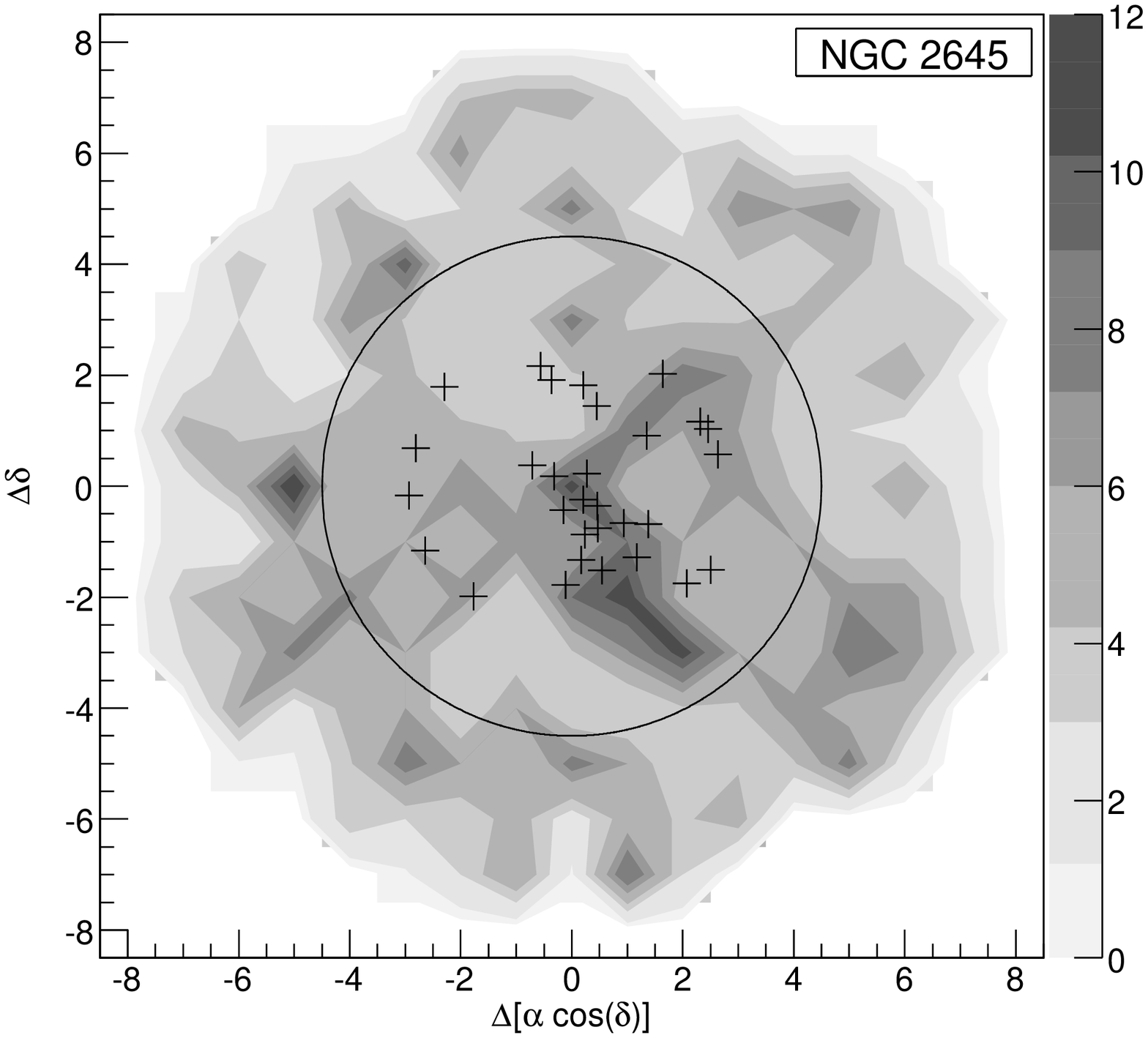}
\includegraphics[width=2.05cm]{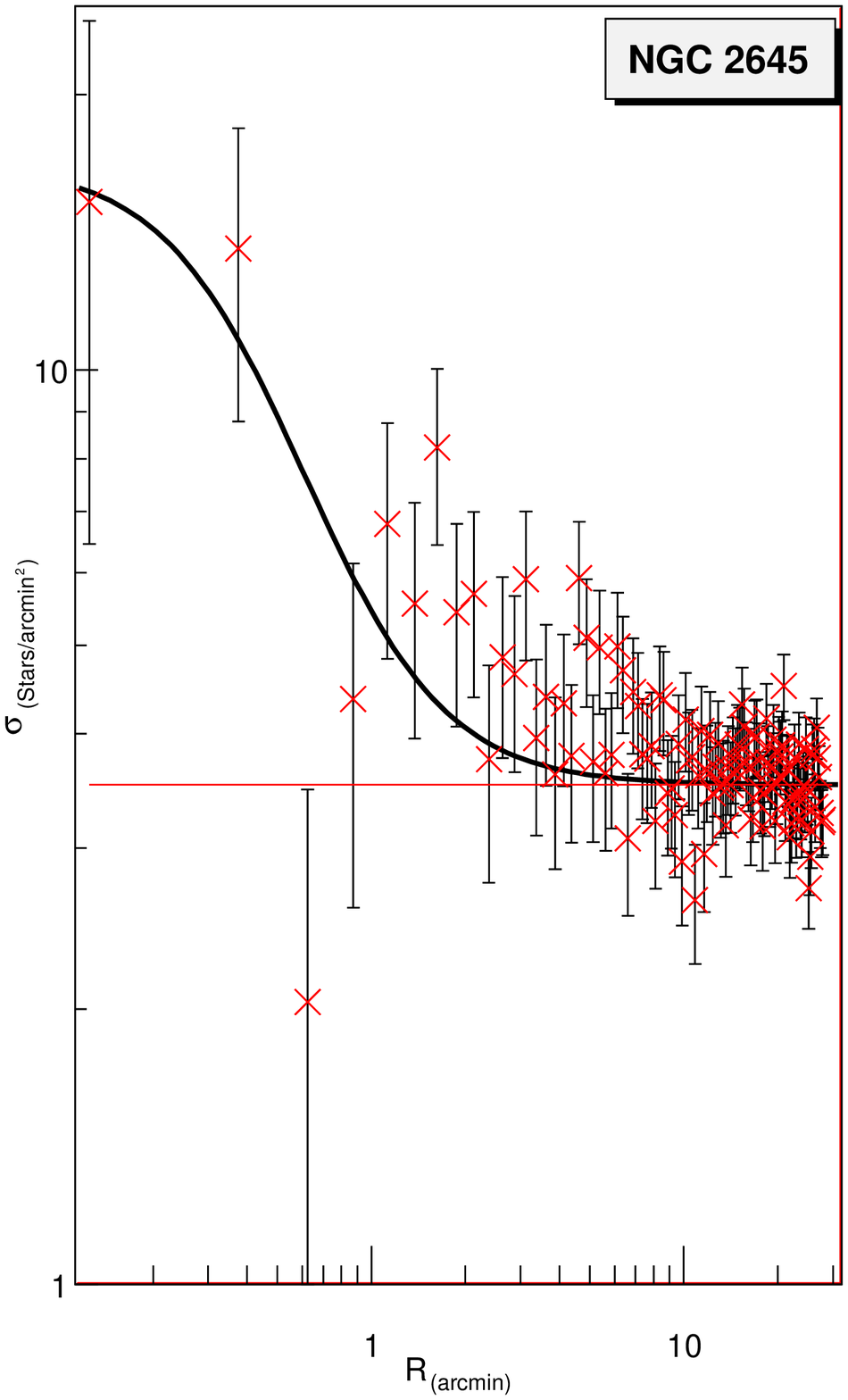}
\includegraphics[width=3.4cm]{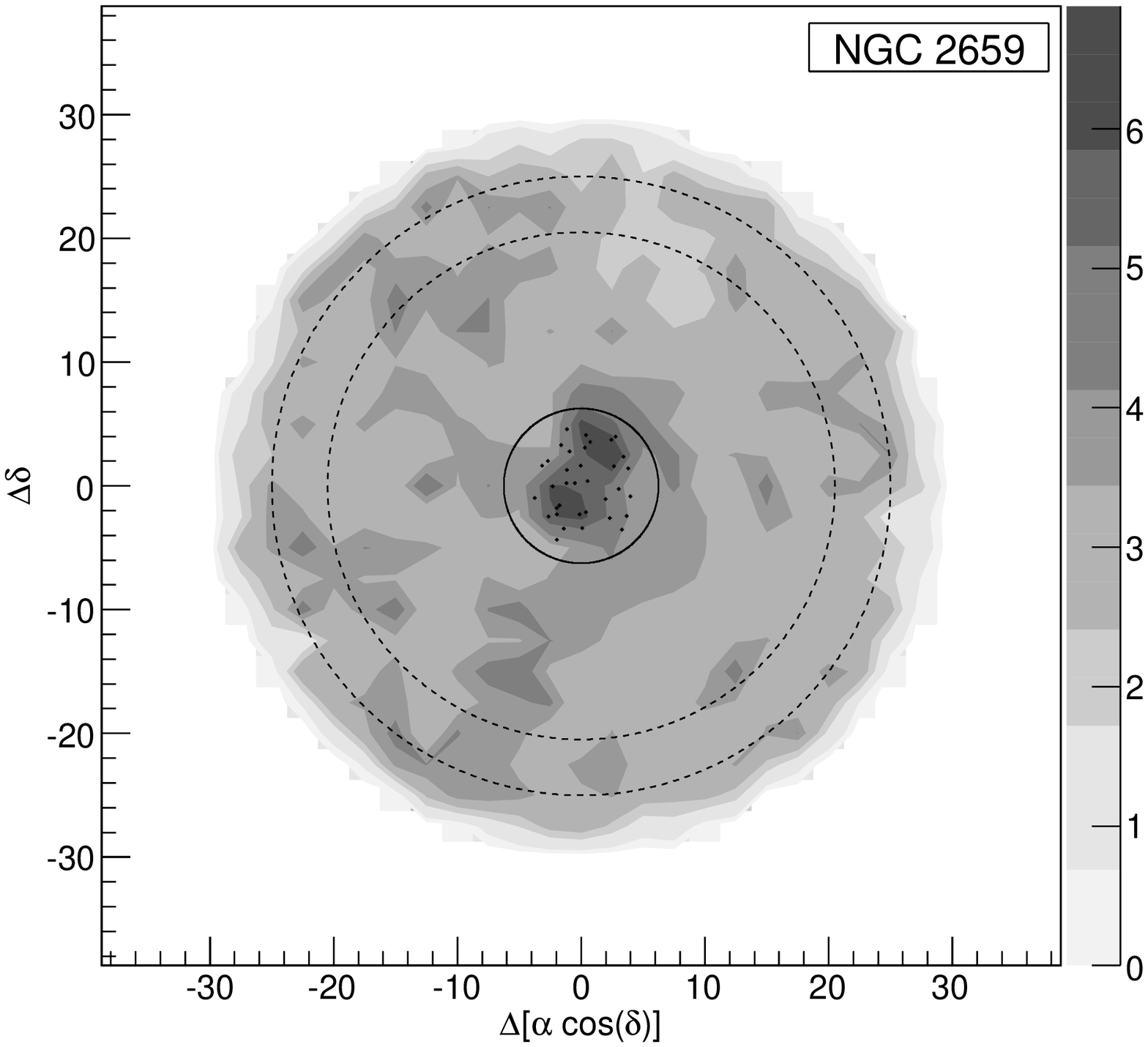}
\includegraphics[width=3.4cm]{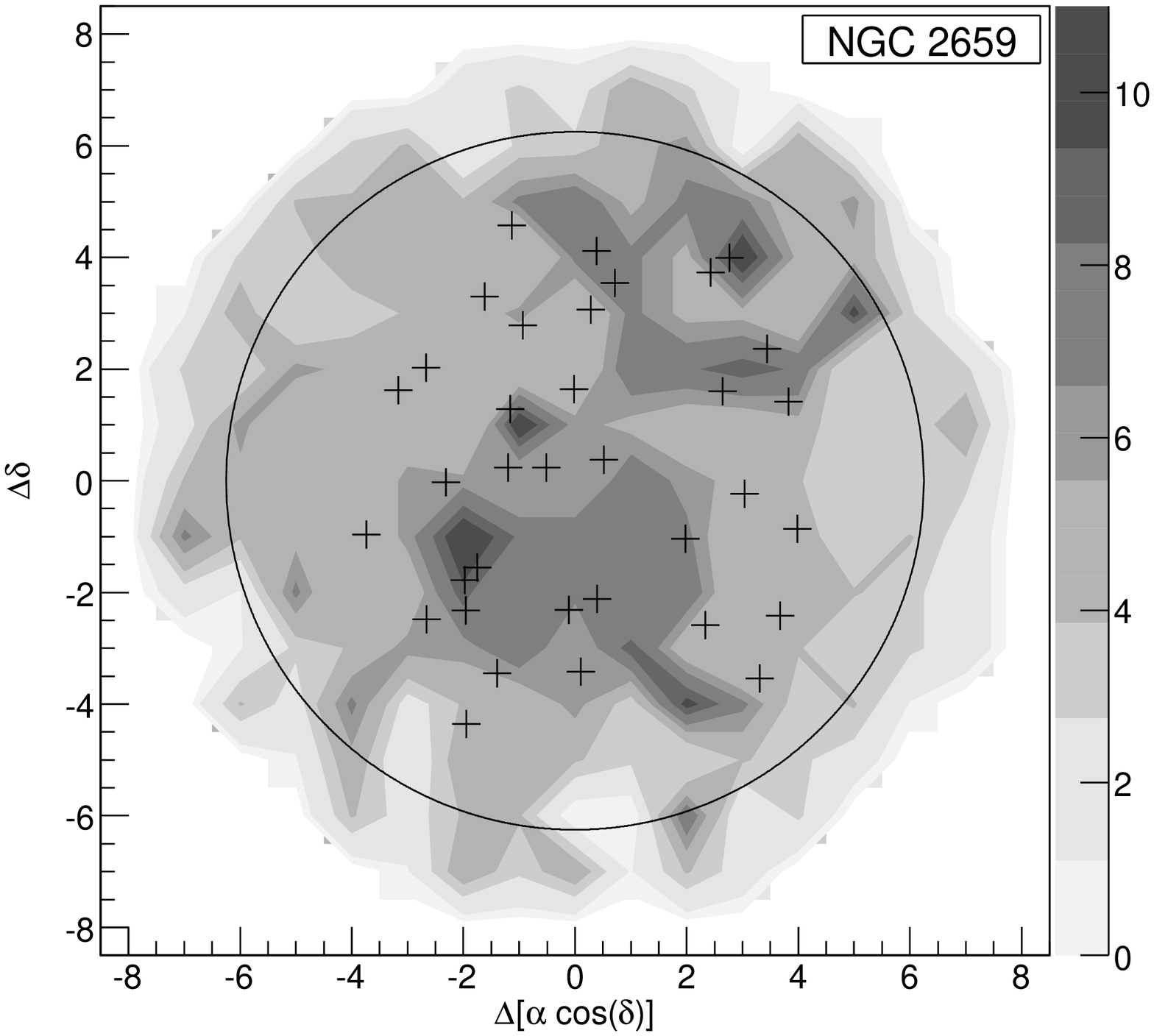}
\includegraphics[width=2.05cm]{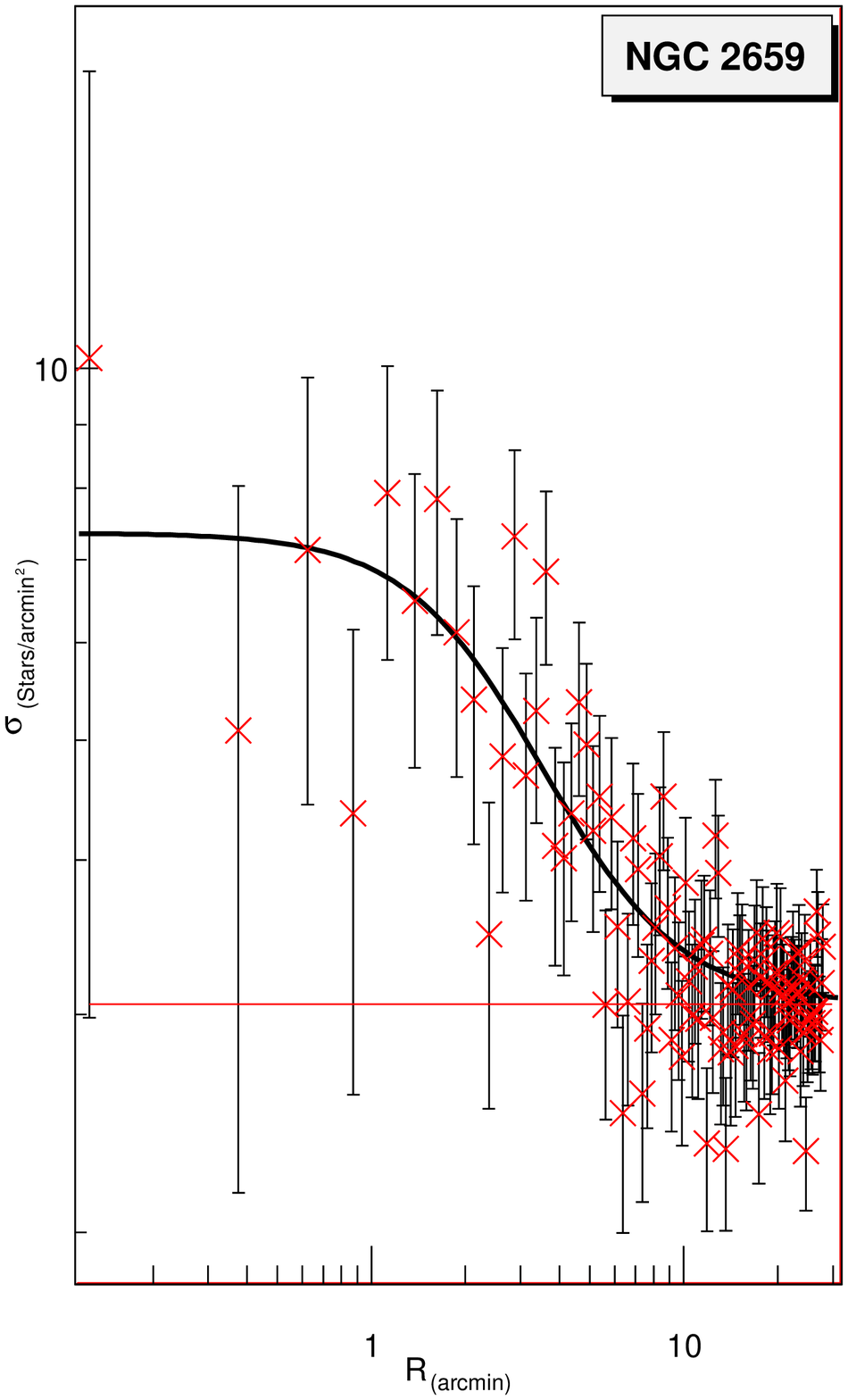}
\includegraphics[width=3.4cm]{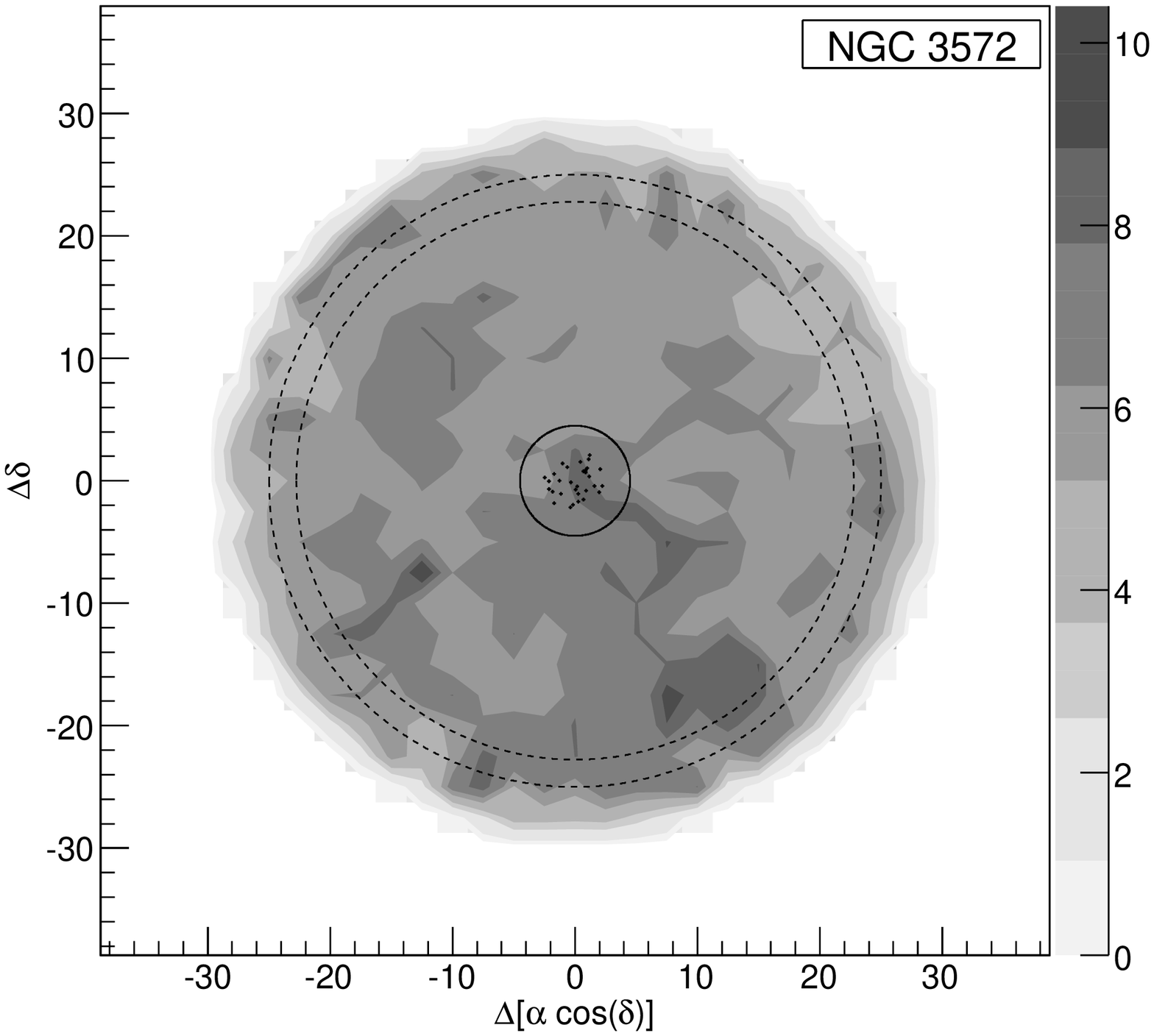}
\includegraphics[width=3.4cm]{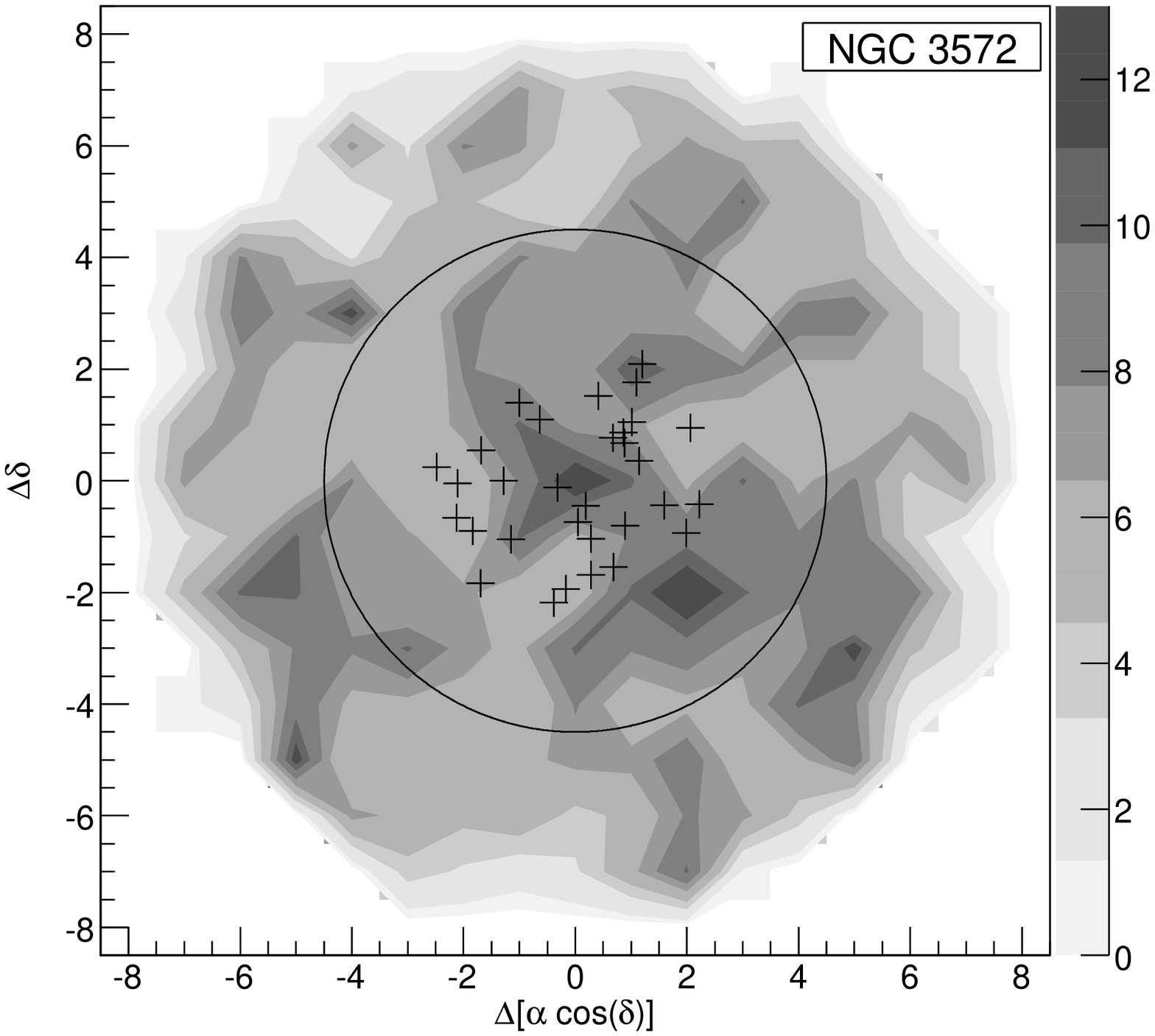}
\includegraphics[width=2.05cm]{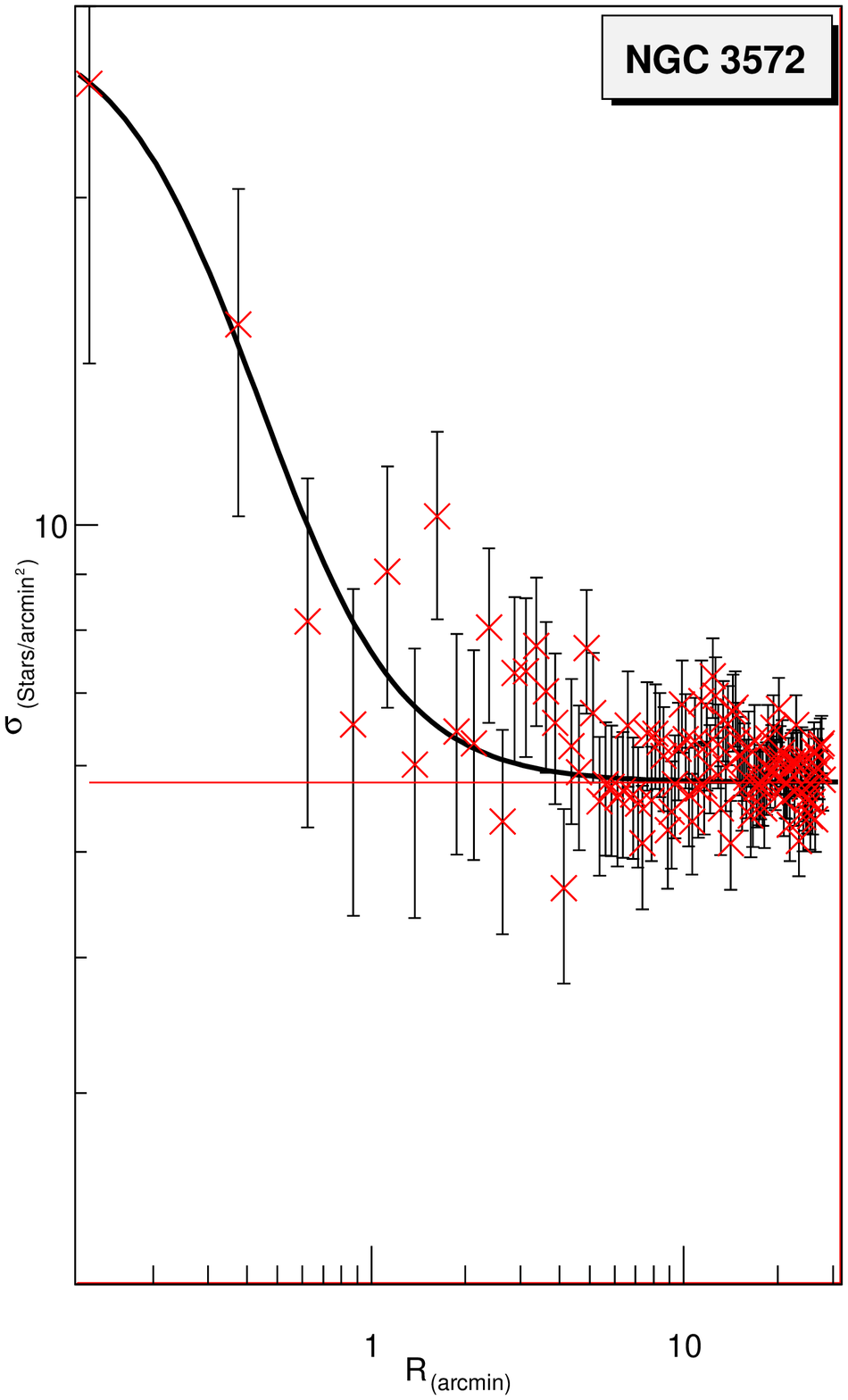}
\includegraphics[width=3.4cm]{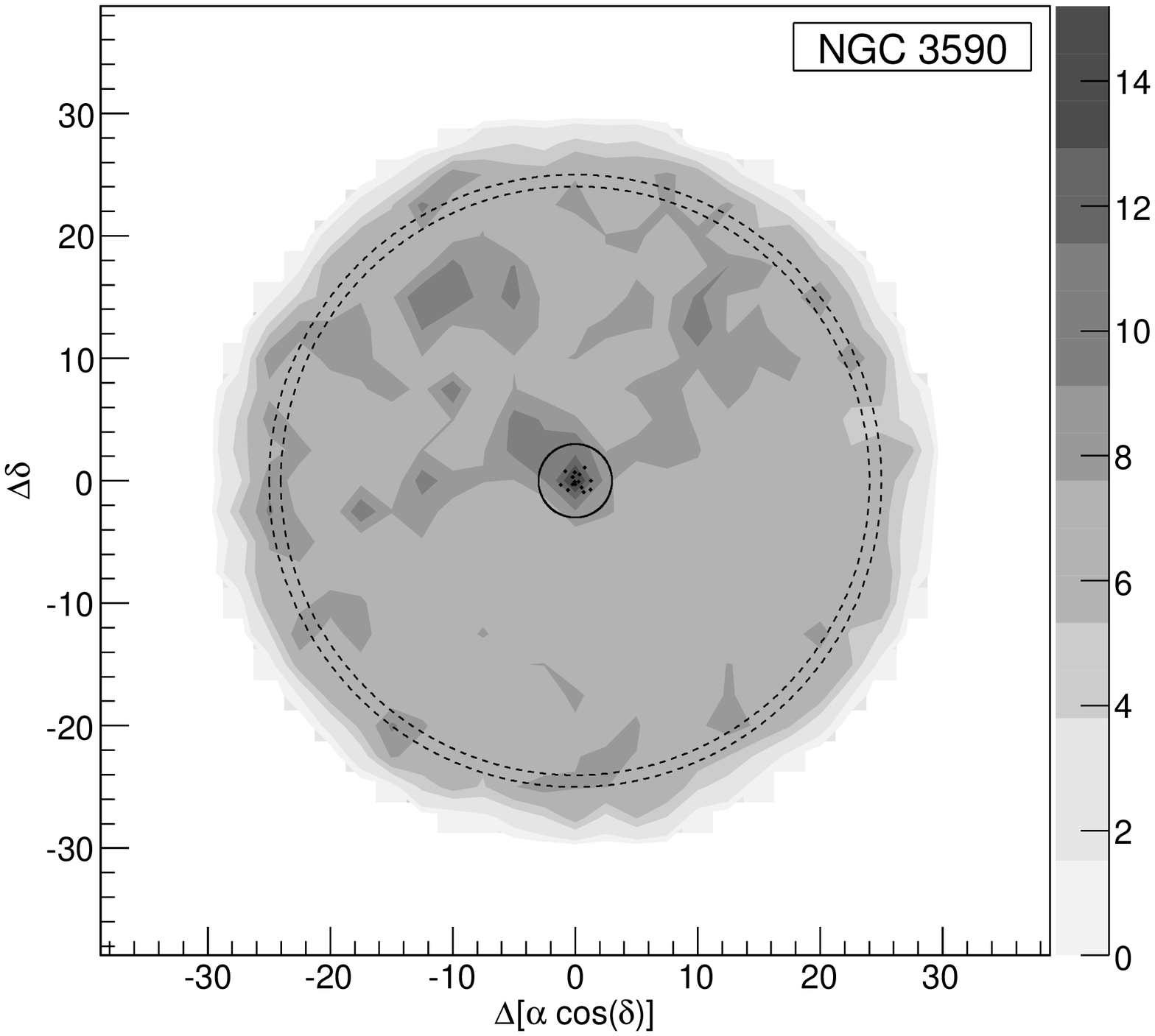}
\includegraphics[width=3.4cm]{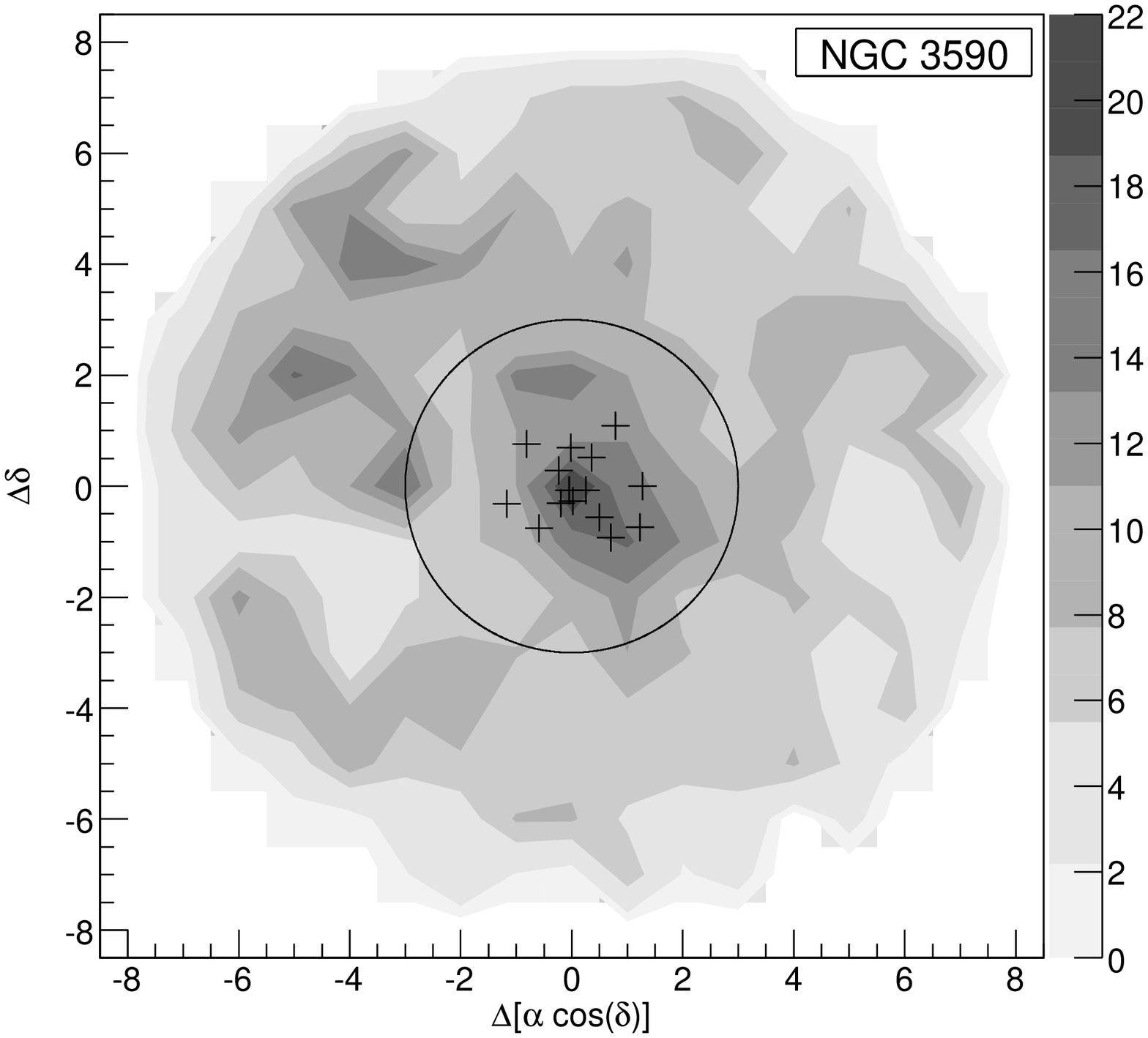}
\includegraphics[width=2.05cm]{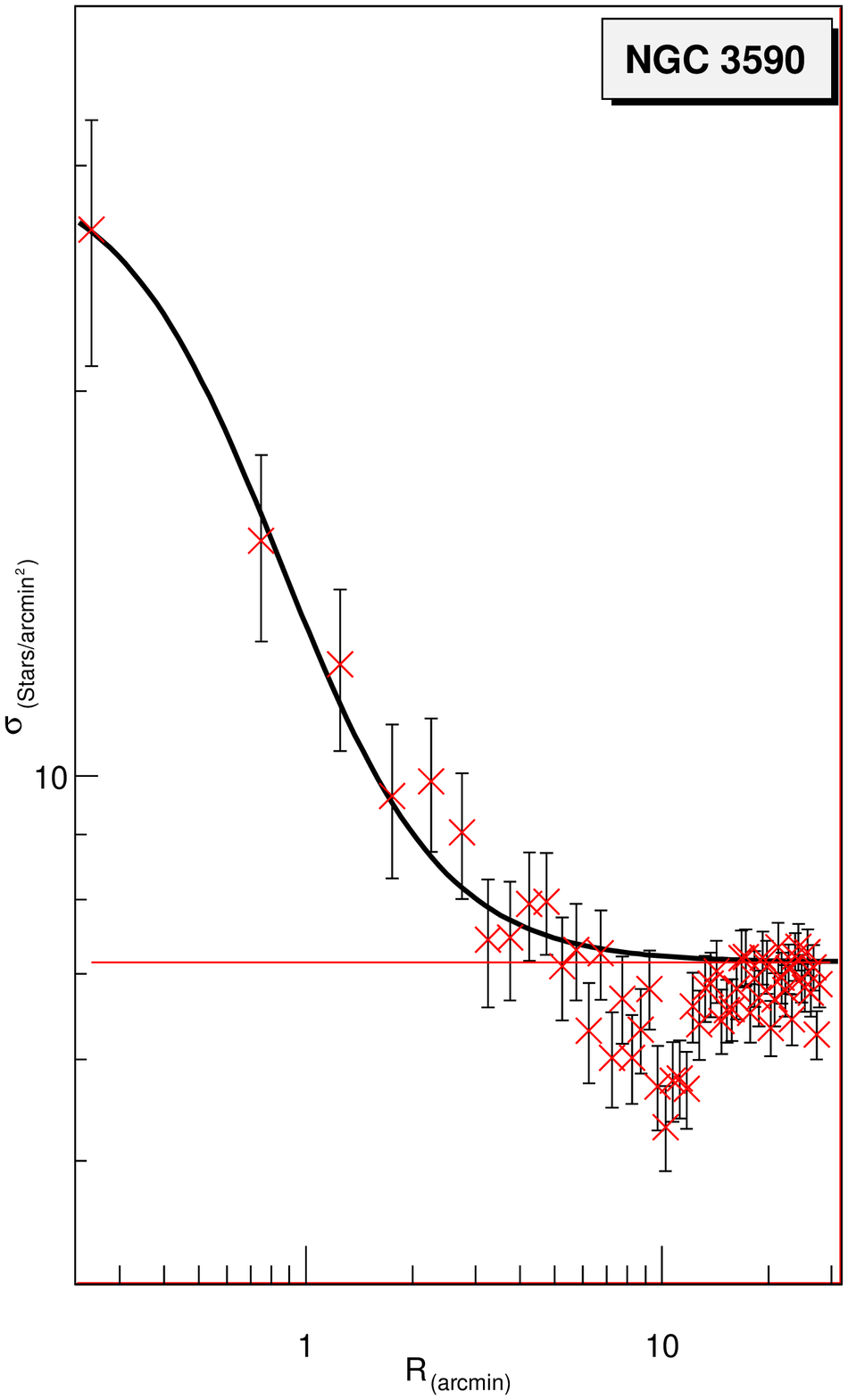}
\includegraphics[width=3.4cm]{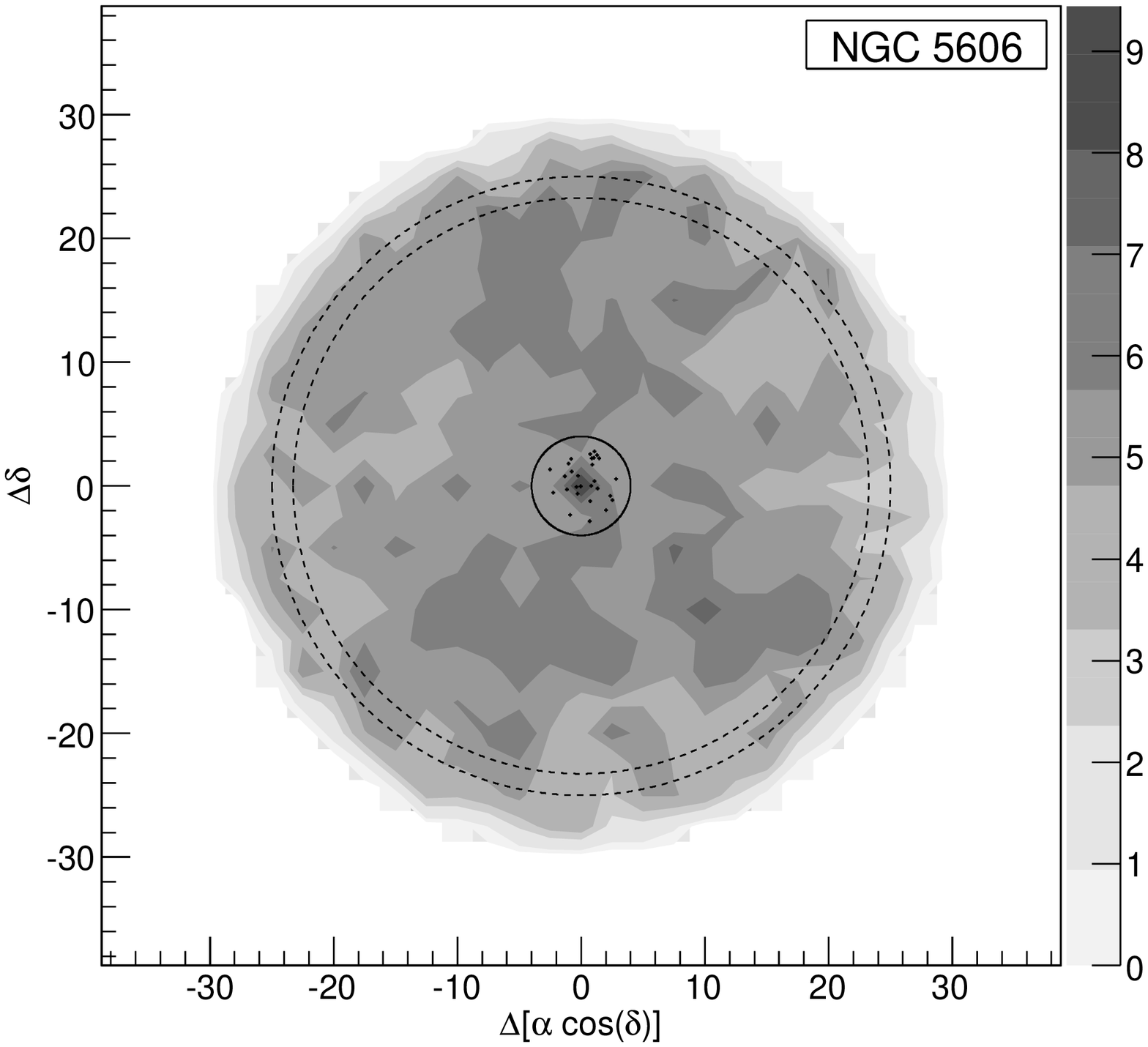}
\includegraphics[width=3.4cm]{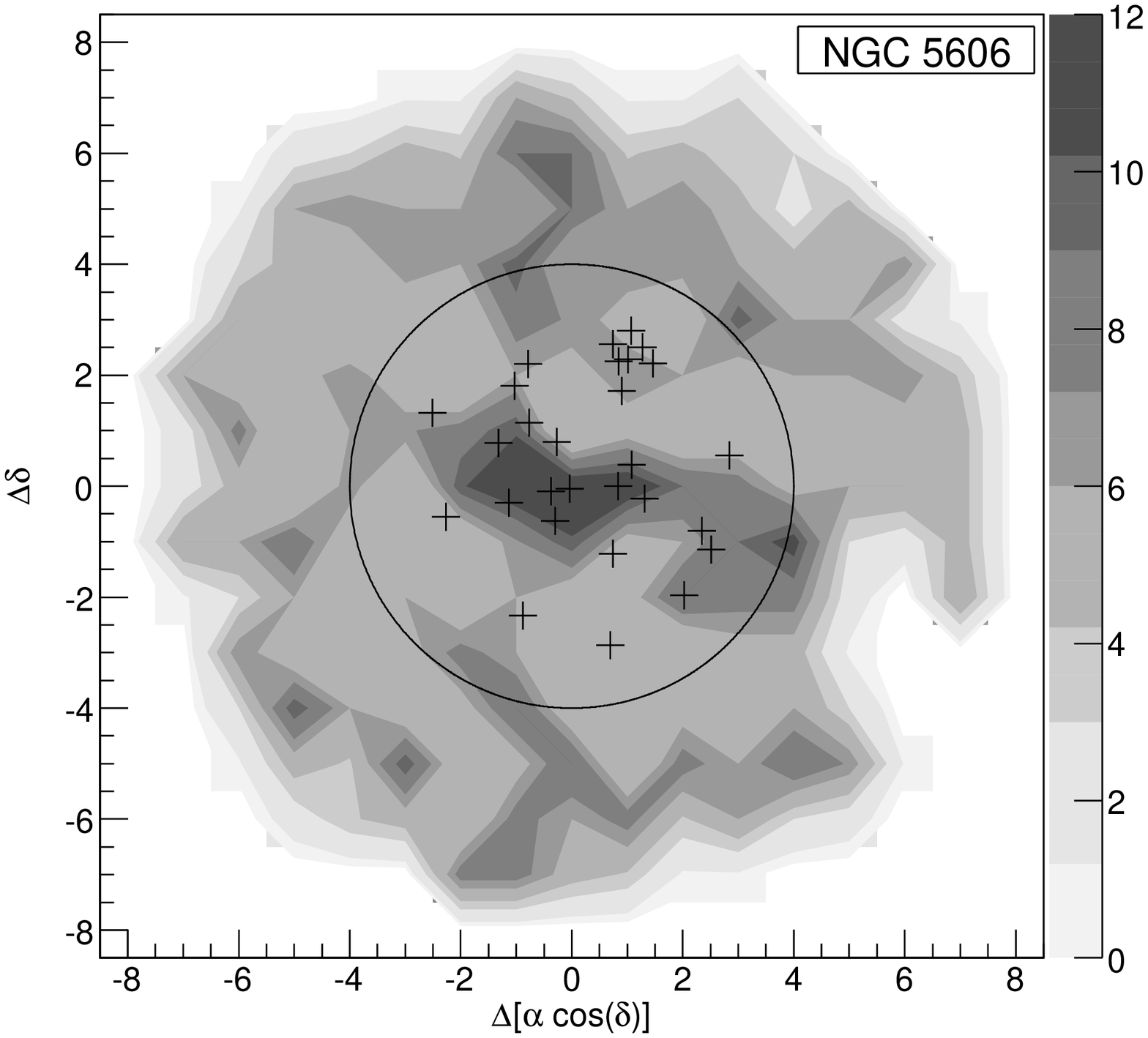}
\includegraphics[width=2.05cm]{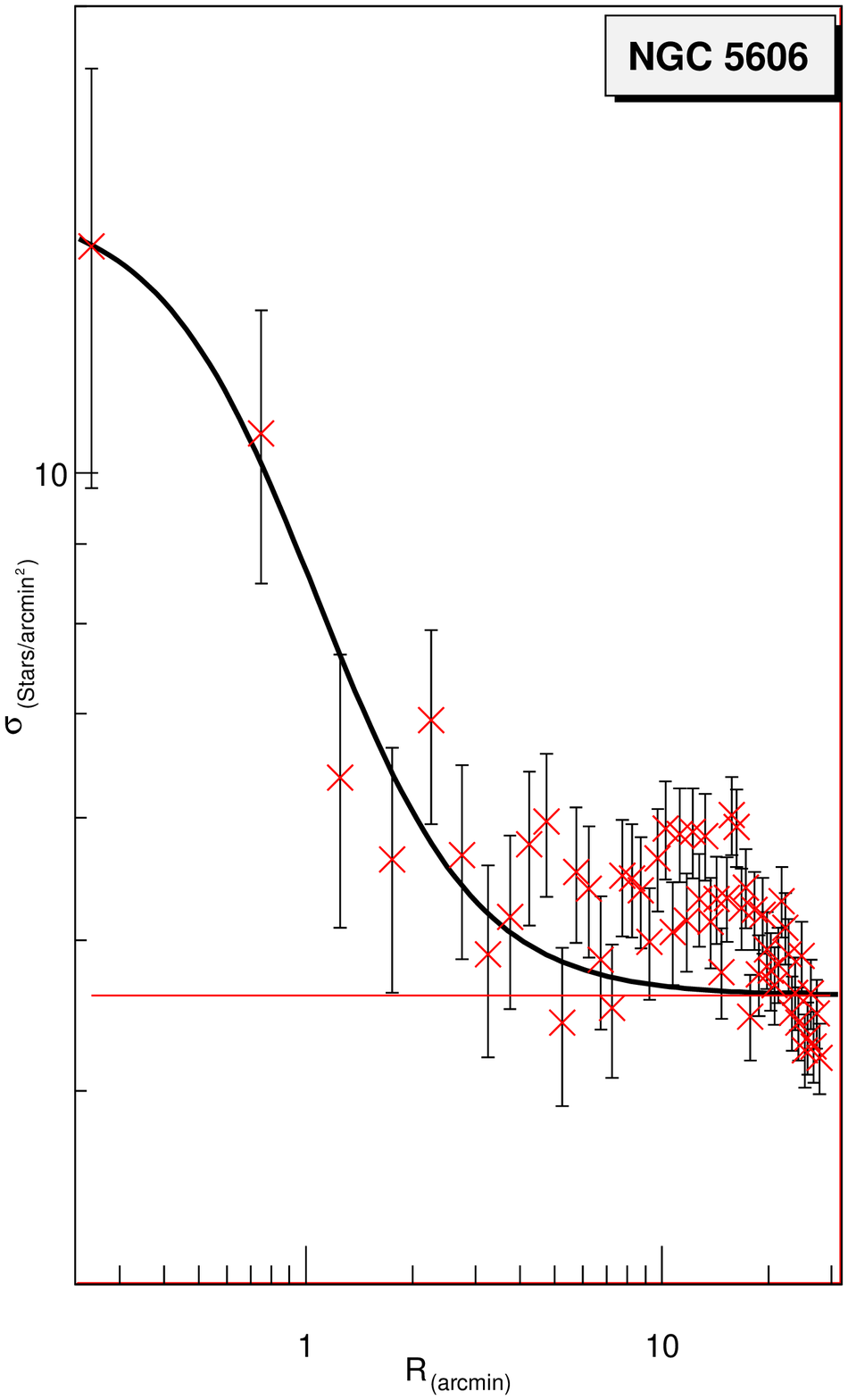}
\includegraphics[width=3.4cm]{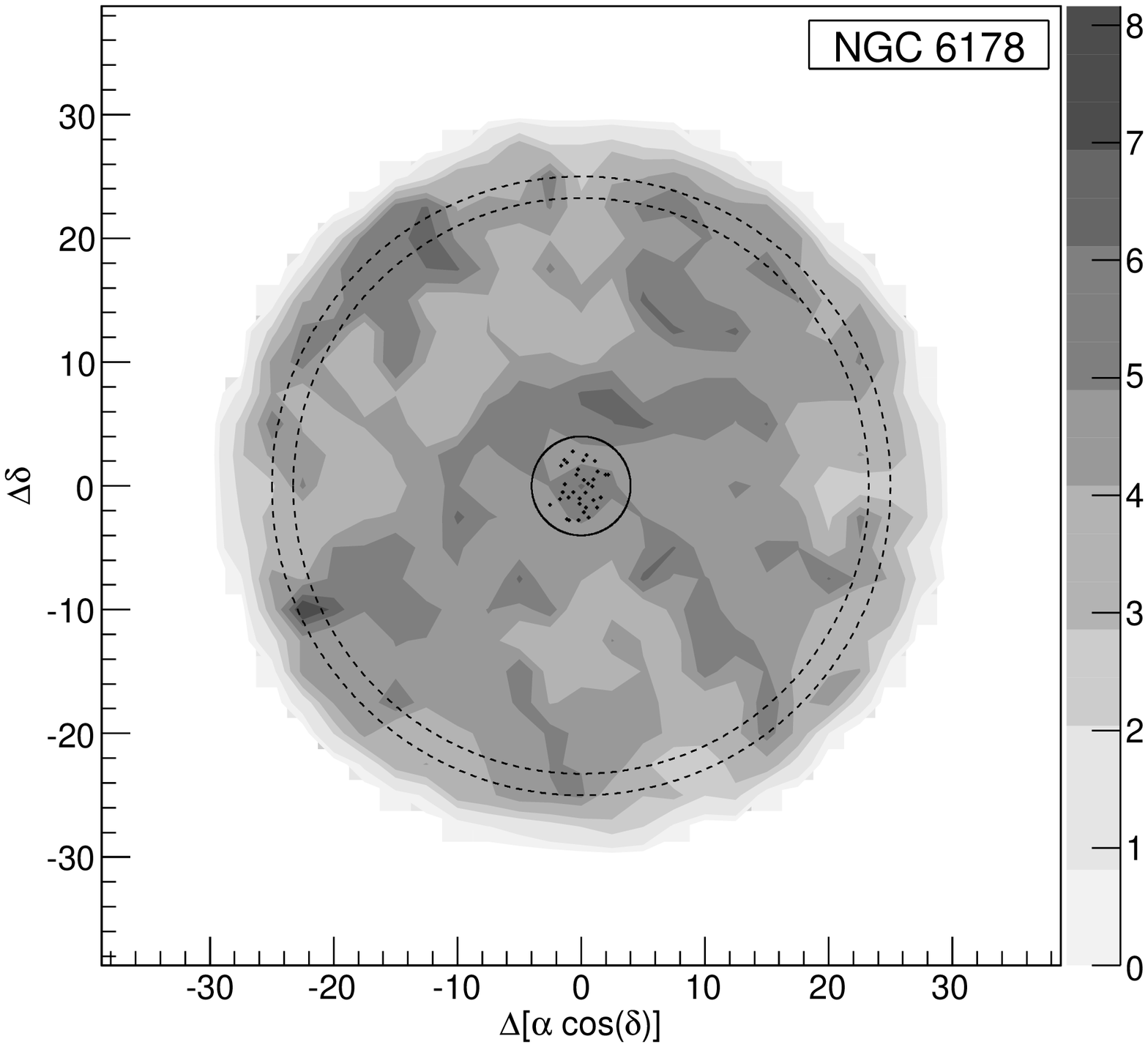}
\includegraphics[width=3.4cm]{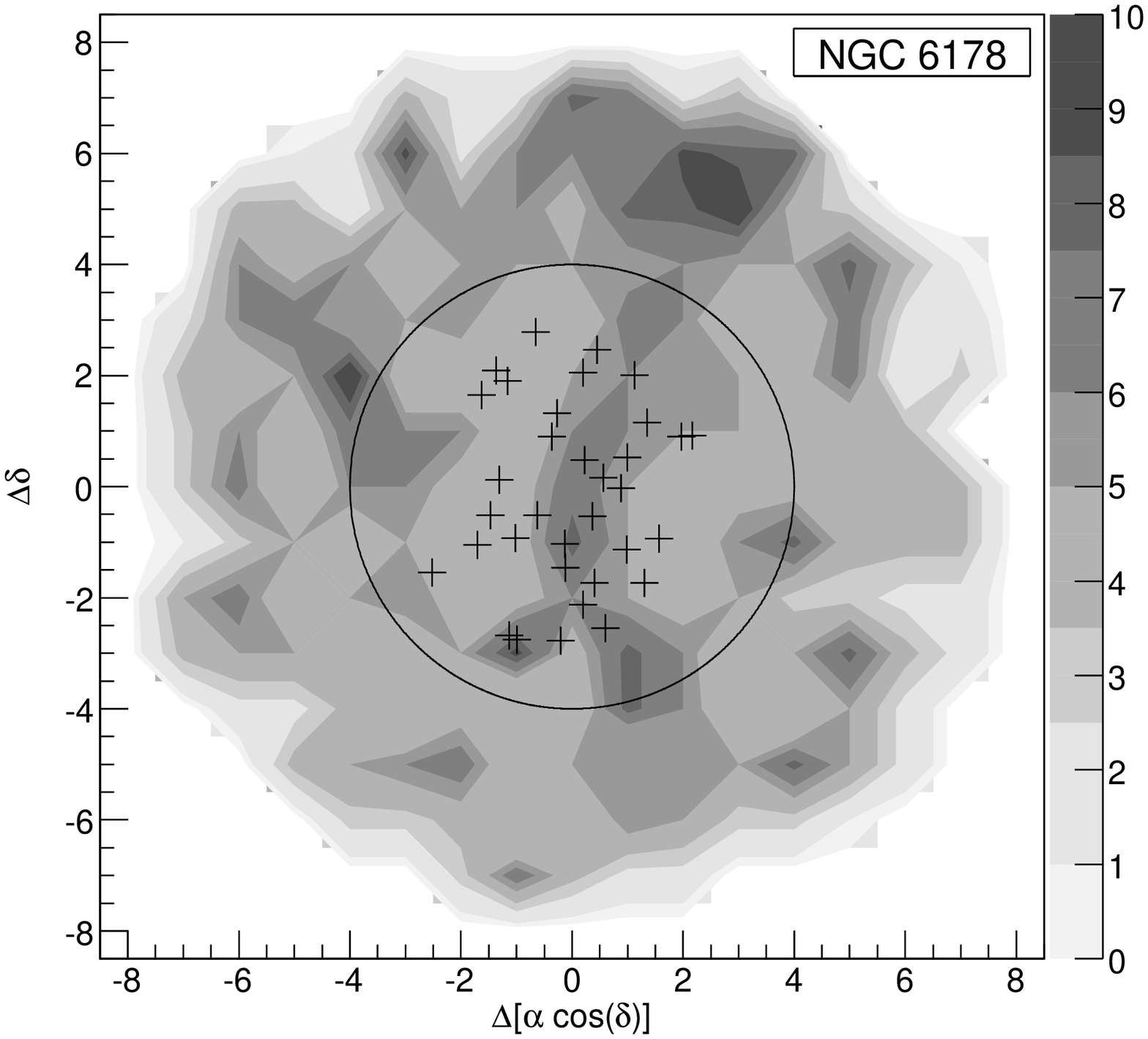}
\includegraphics[width=2.05cm]{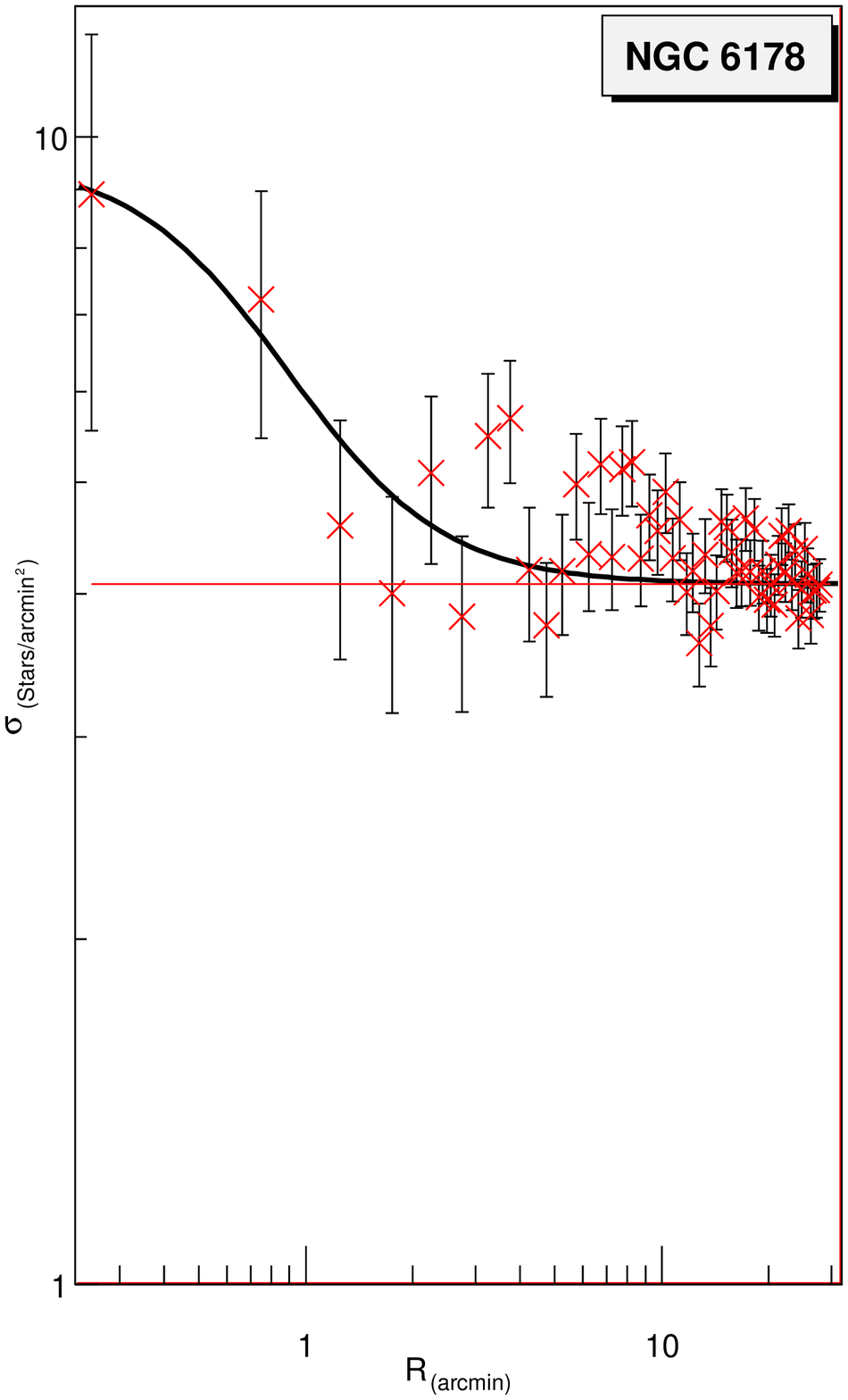}

\caption{{\it Left}: Stellar surface-density map ($\sigma$ (stars/arcmin$^{2}$)) obtained for the region of 30 arcmin around
the clusters. The comparison field-stars area is indicated by dashed lines, while the full line indicates the cluster area.
{\it Centre}: a zoom of the $\sigma$ map indicating by crosses the position of objects with membership probability P$>$70\%.
{\it Right}: The distribution of the stellar density as a function of radius. The best fitting of observed radial 
density profile, indicated by the full line, was obtained by using the model from King (1962). A dashed line indicates 
the background density ($\sigma_{bg}$).}  

\label{hmass}
\end{center}
\end{figure*}

%%%%----------------------Fig. A2
\begin{figure*}[]
\begin{center}

\includegraphics[width=3.4cm]{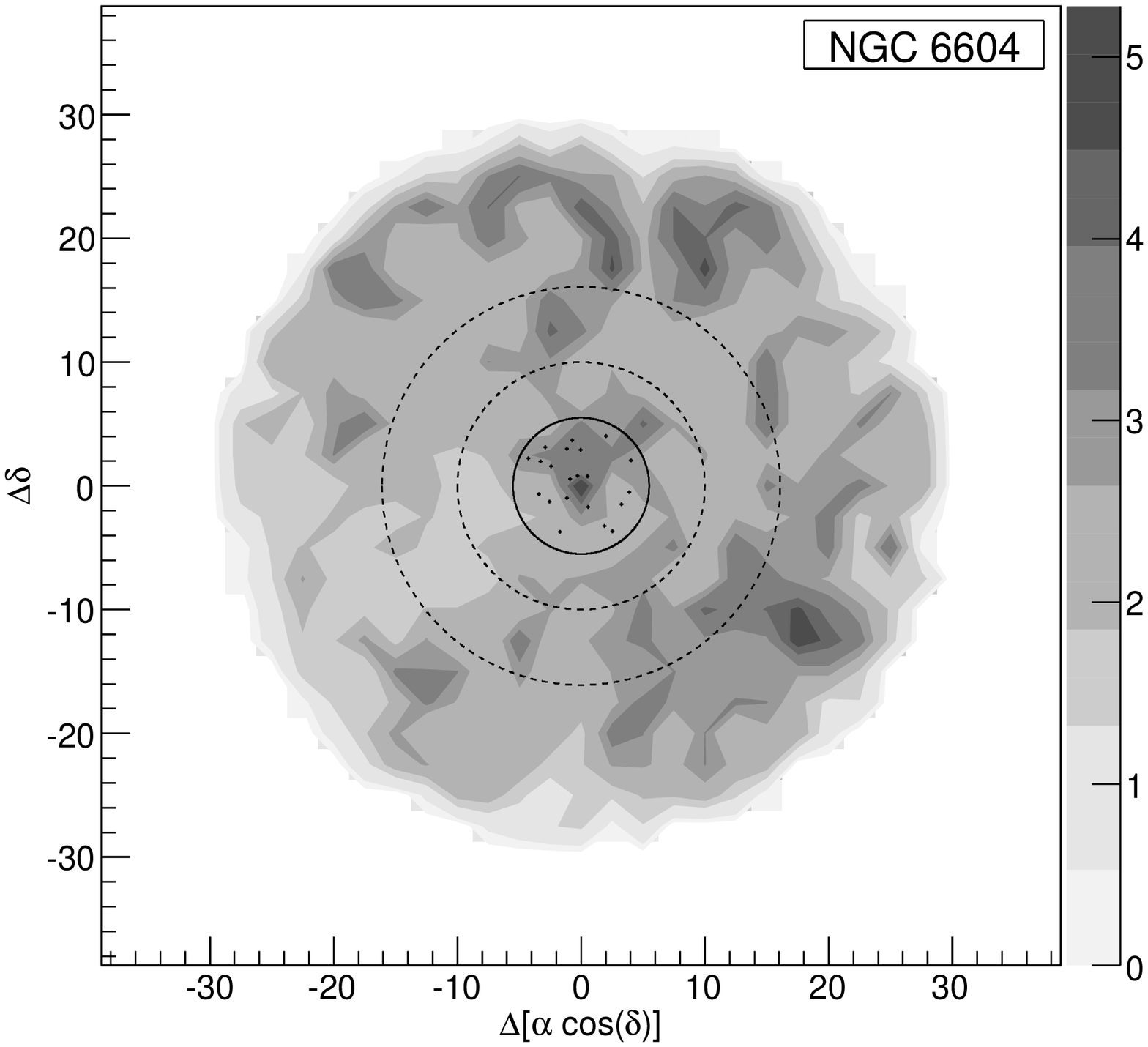}
\includegraphics[width=3.4cm]{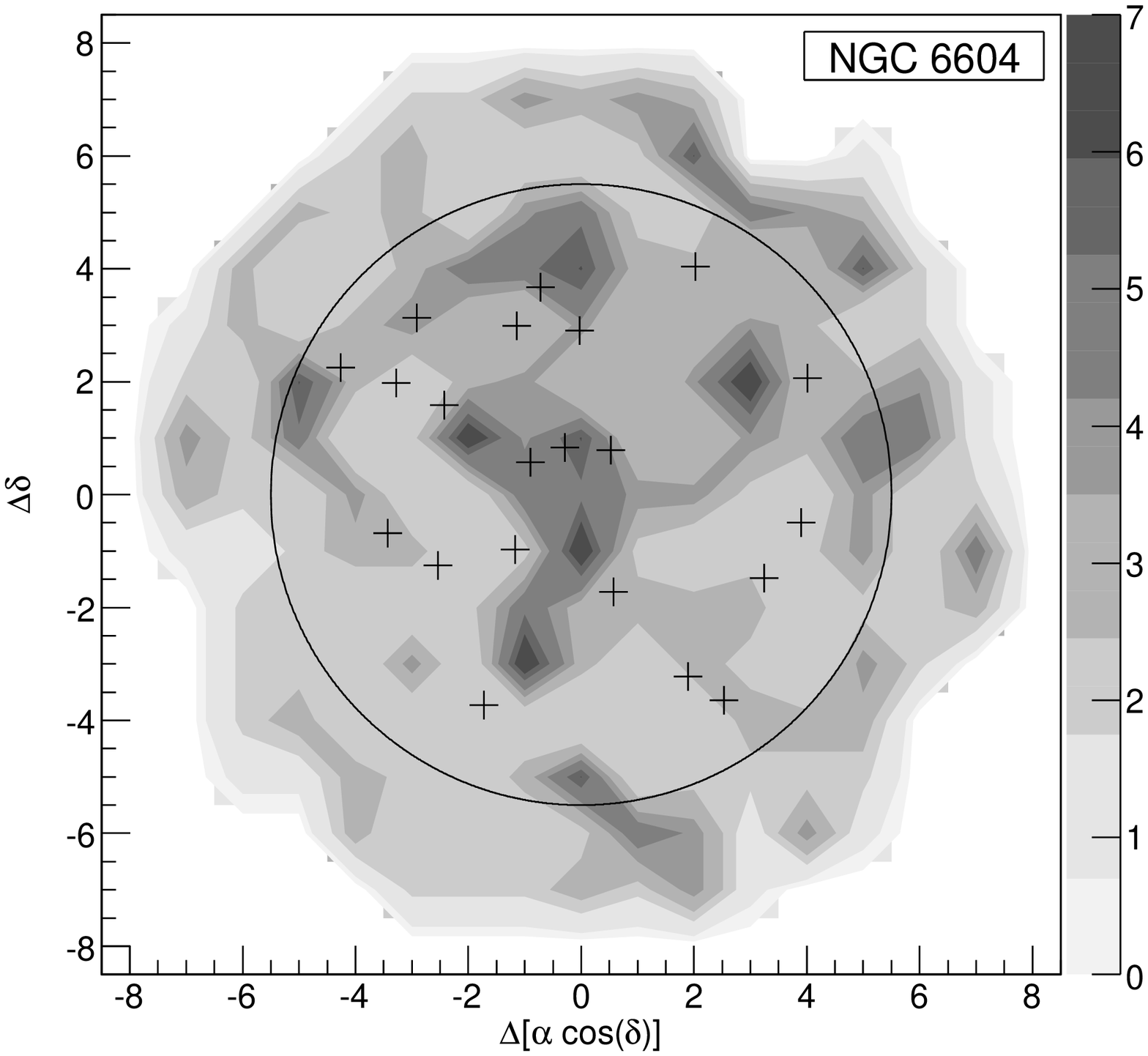}
\includegraphics[width=2.05cm]{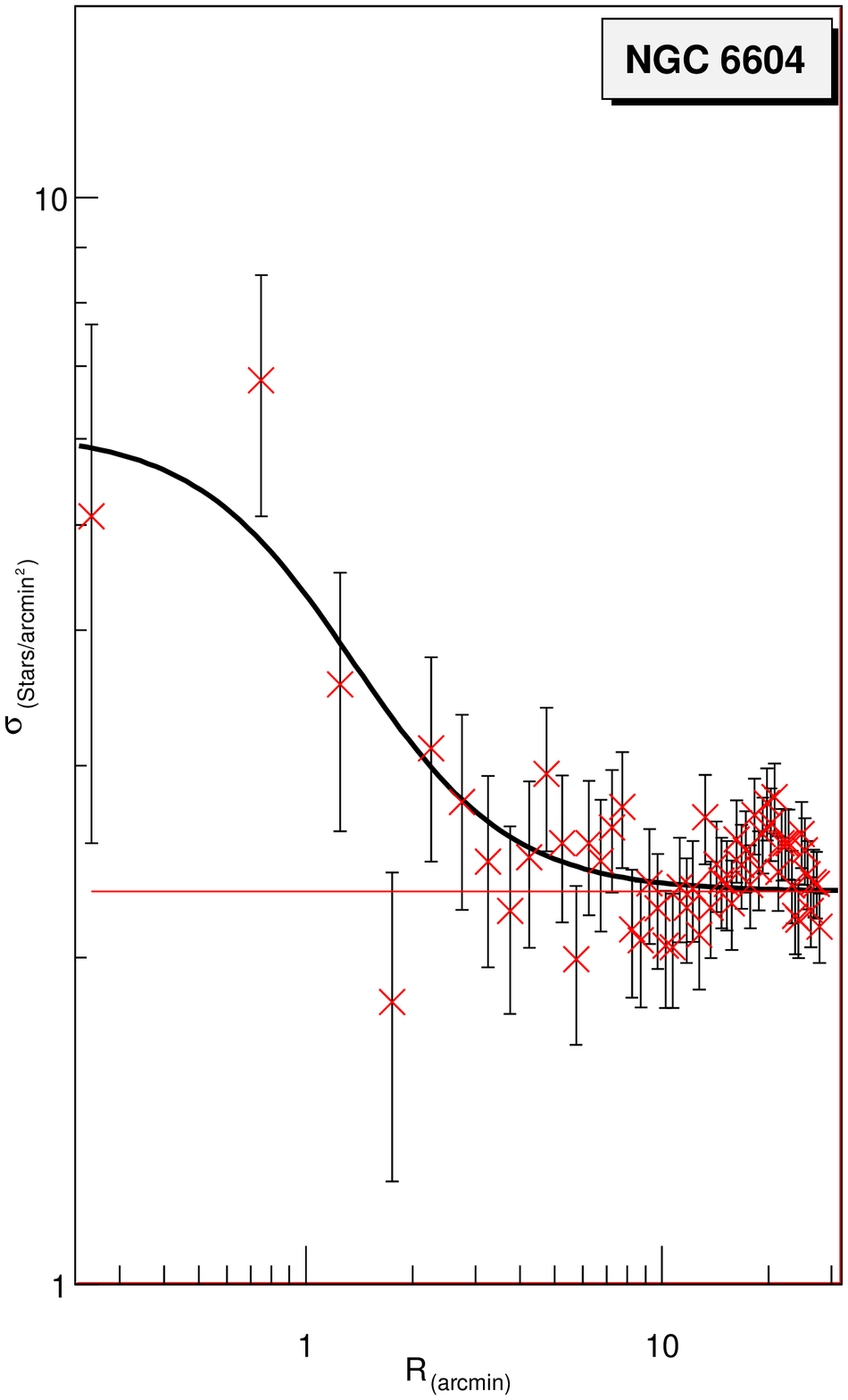}
\includegraphics[width=3.4cm]{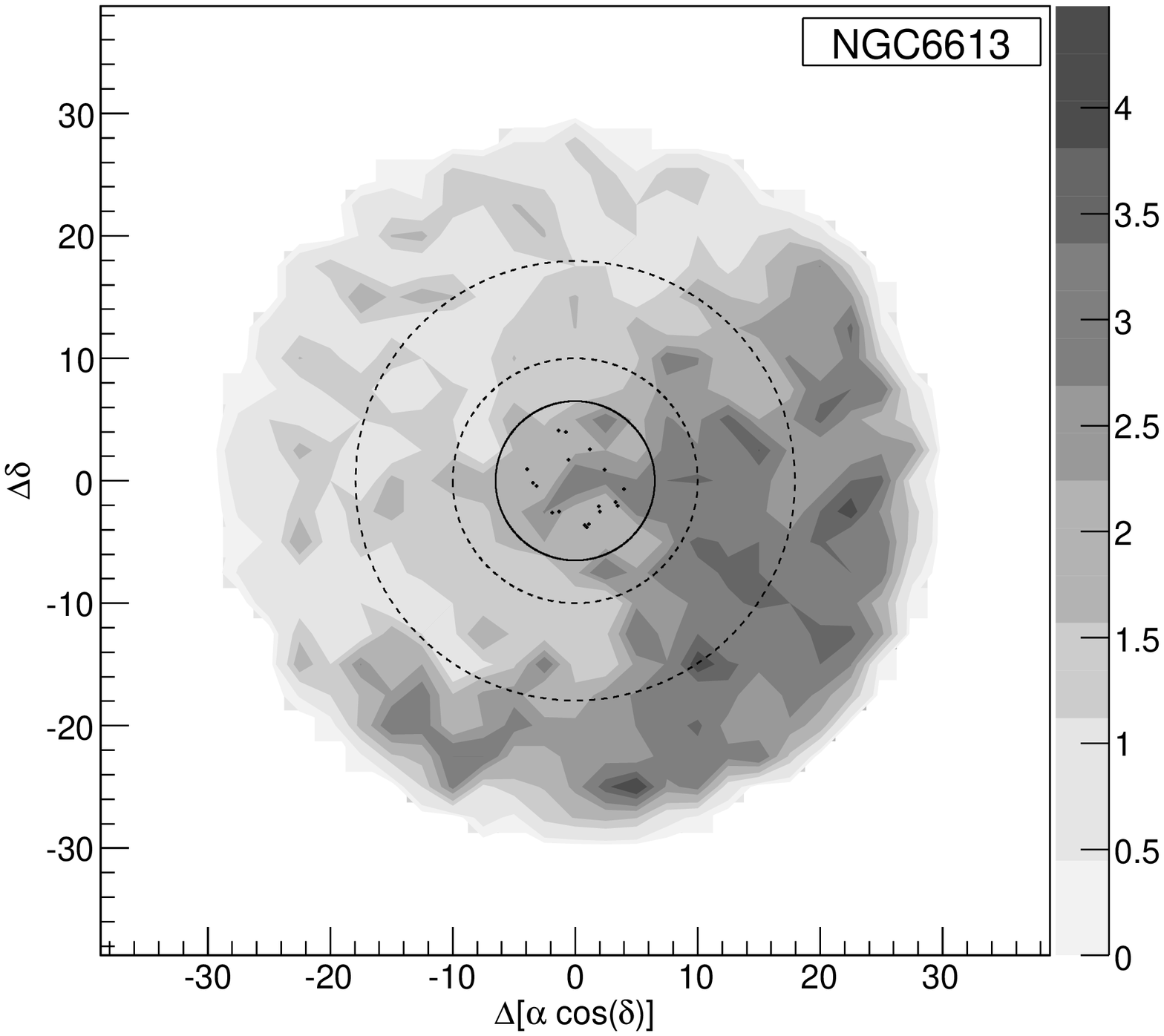}
\includegraphics[width=3.4cm]{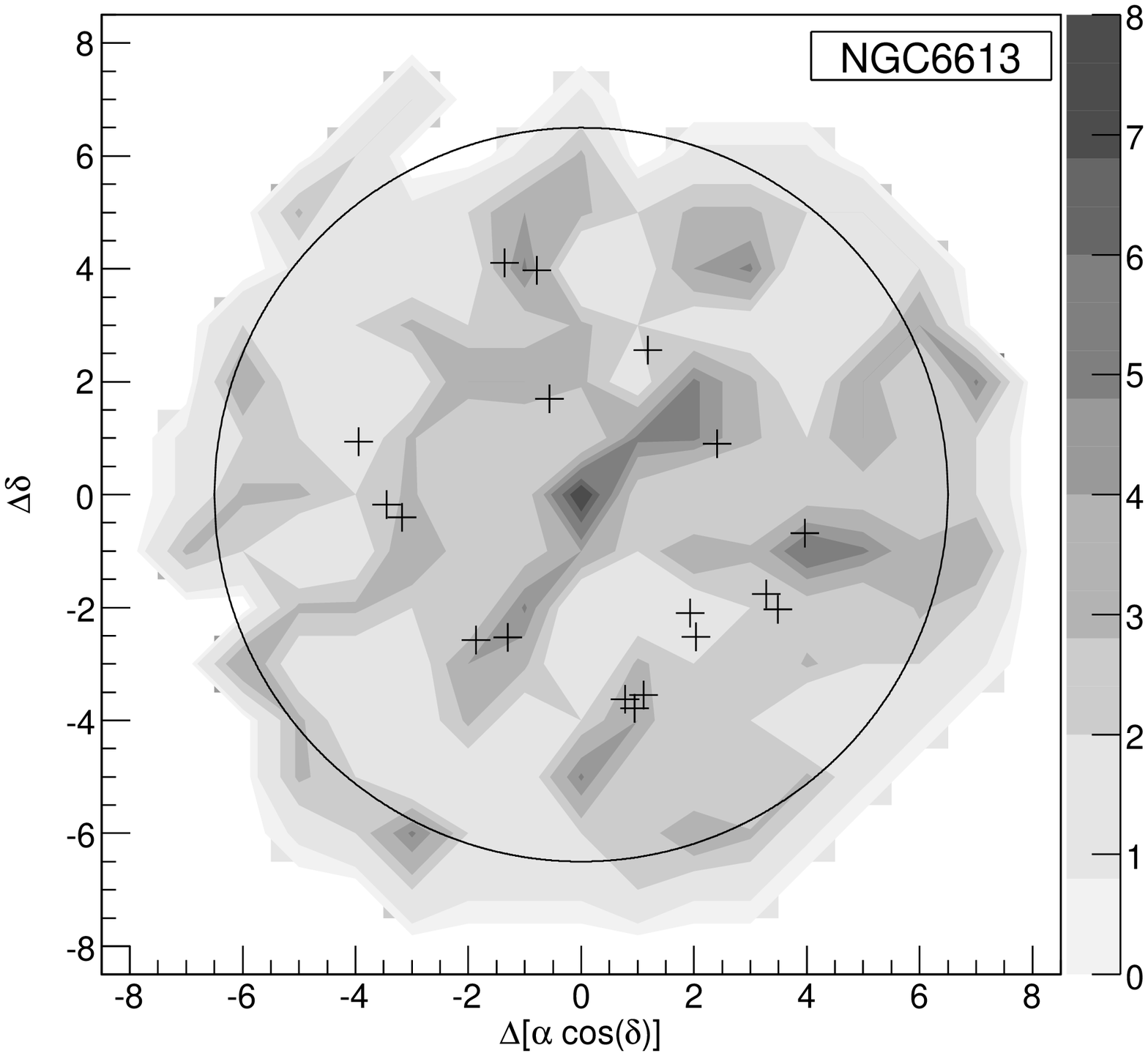}
\includegraphics[width=2.05cm]{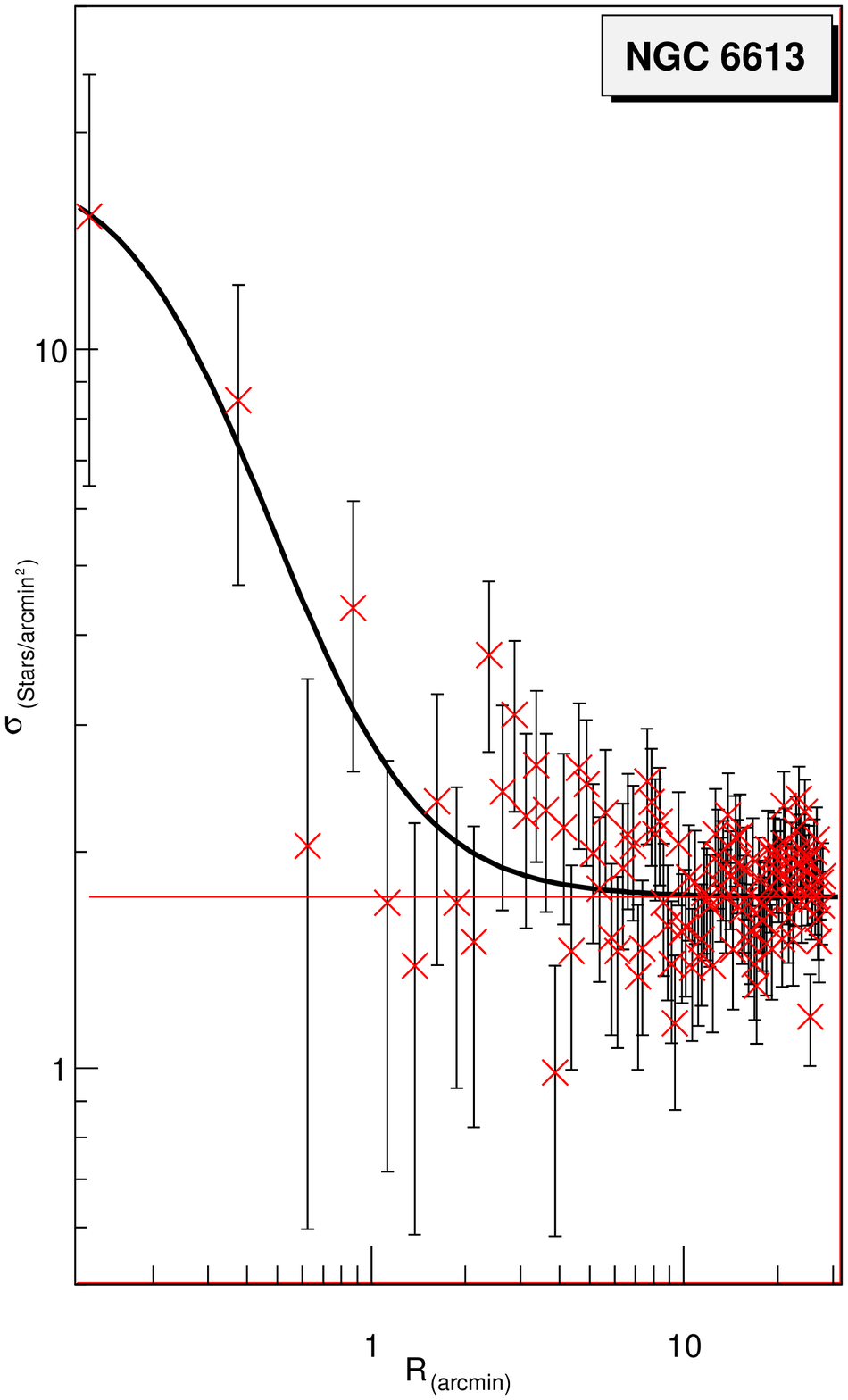}
\includegraphics[width=3.4cm]{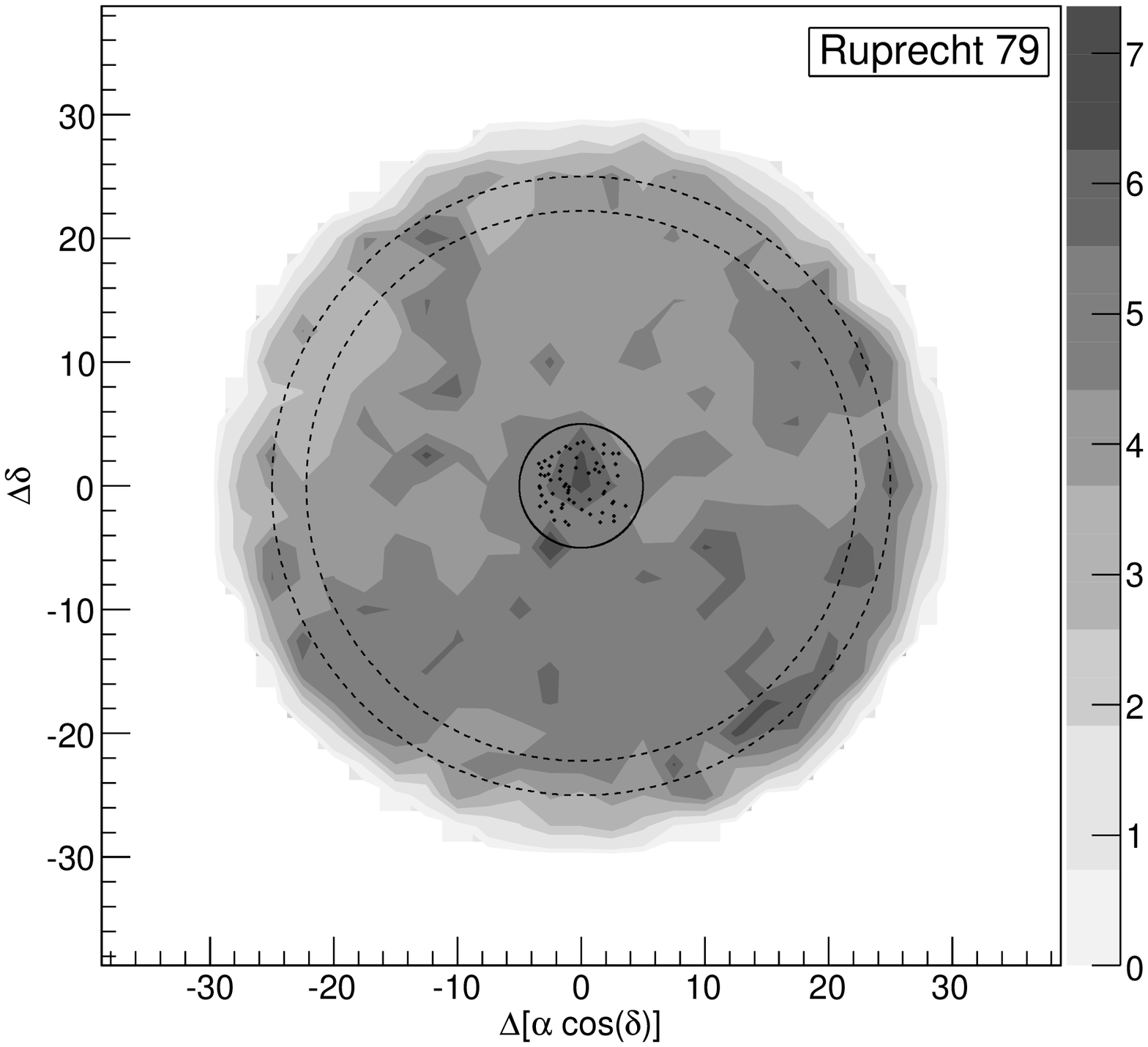}
\includegraphics[width=3.4cm]{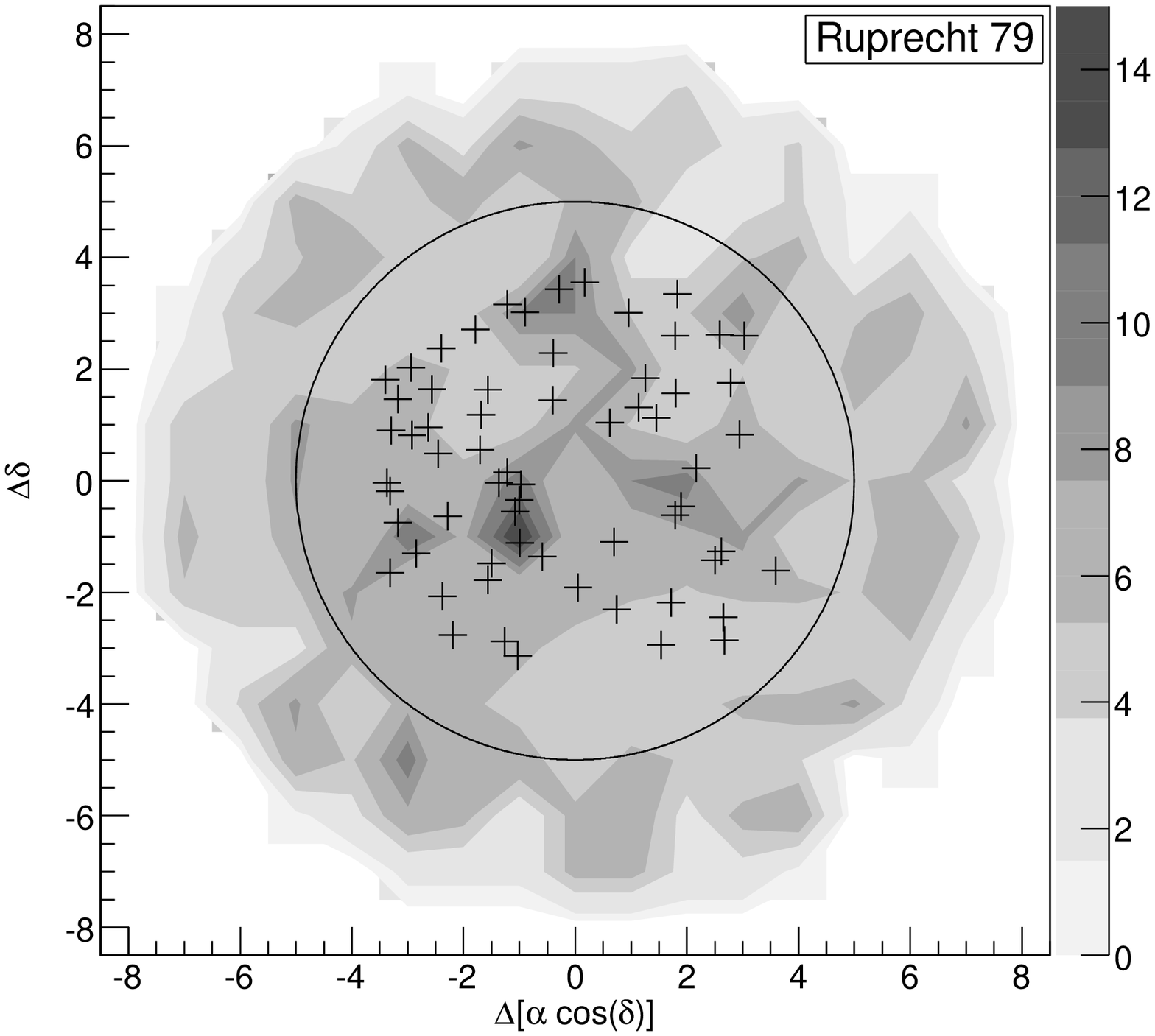}
\includegraphics[width=2.05cm]{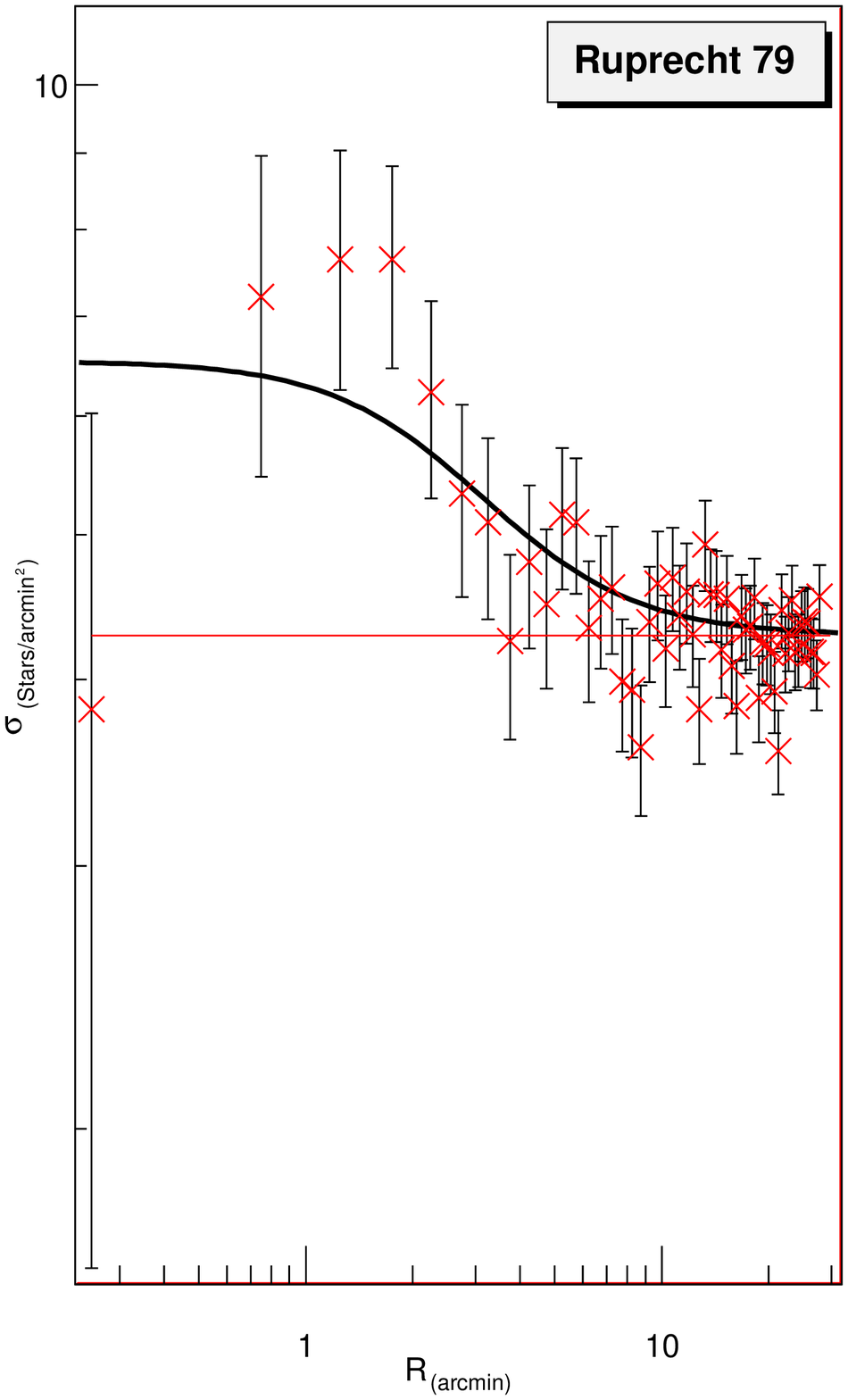}
\includegraphics[width=3.4cm]{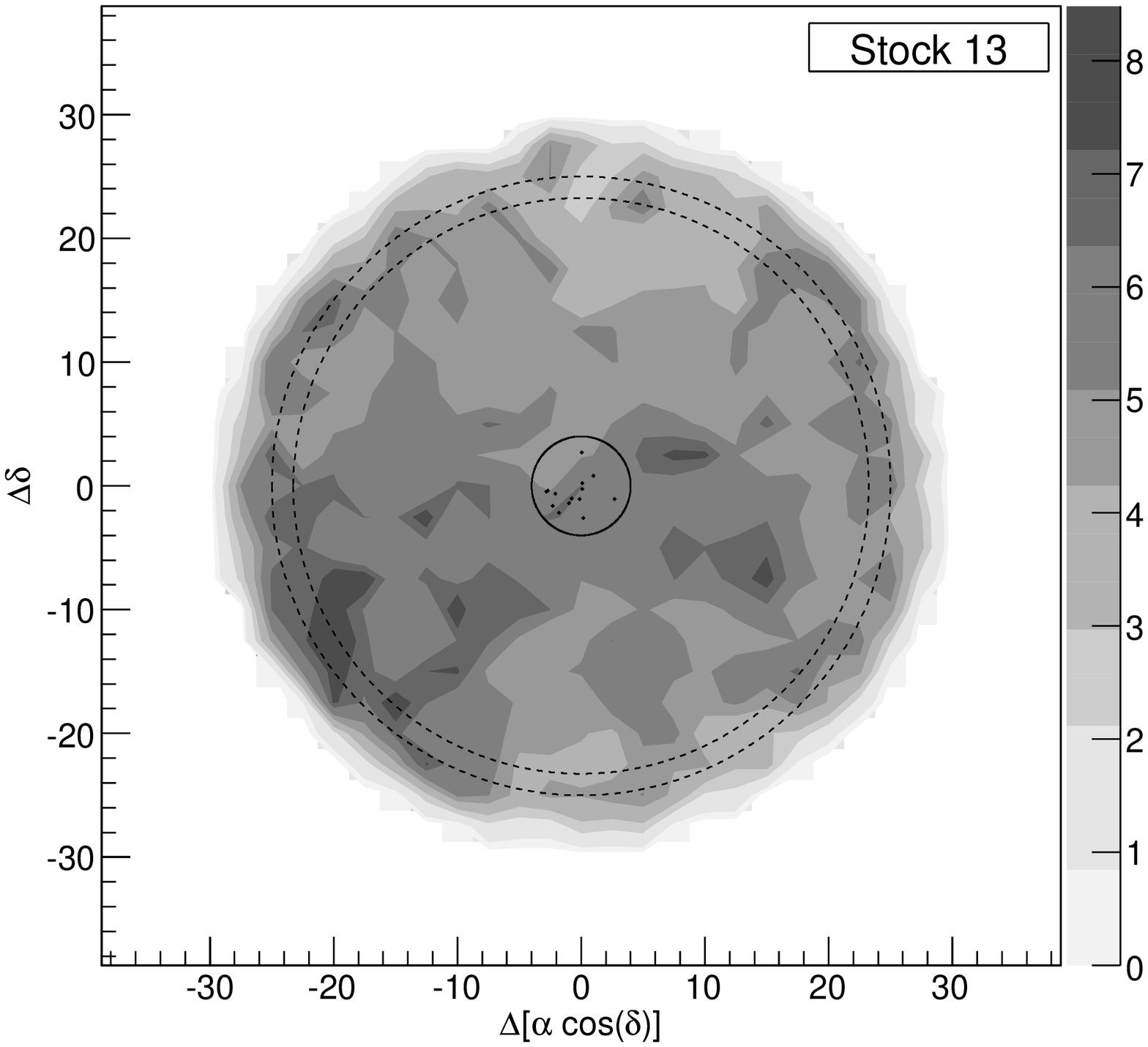}
\includegraphics[width=3.4cm]{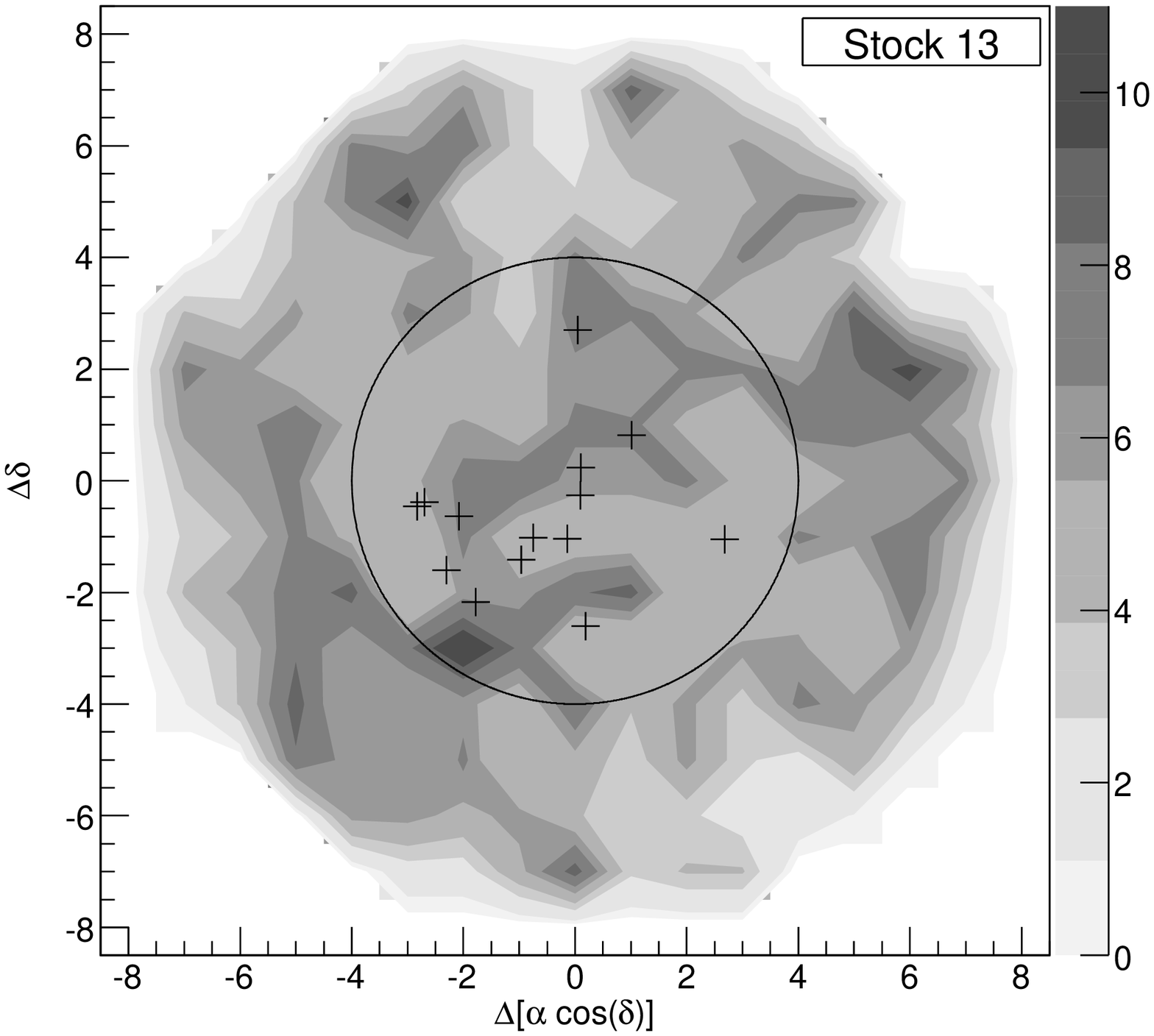}
\includegraphics[width=2.05cm]{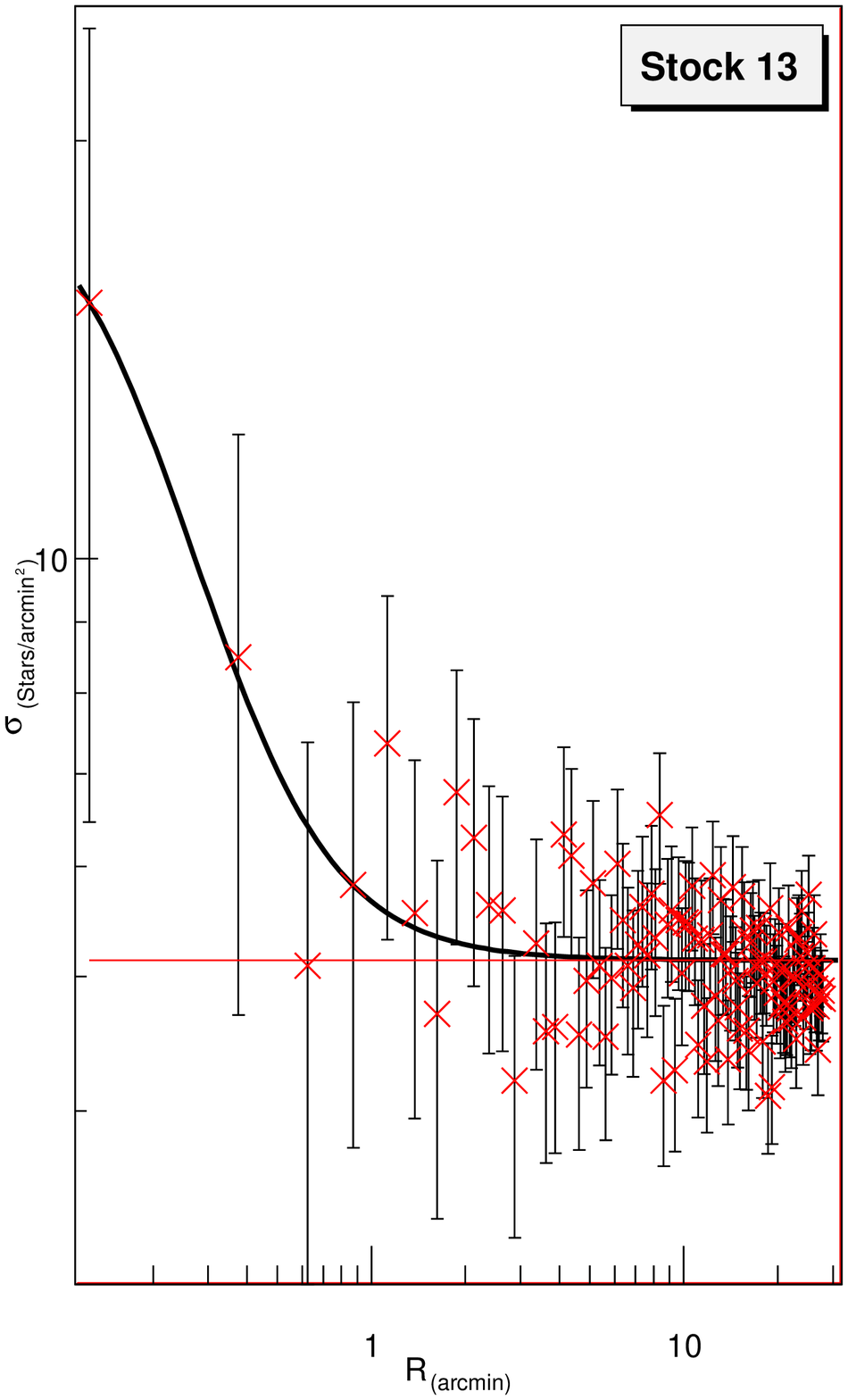}
\includegraphics[width=3.4cm]{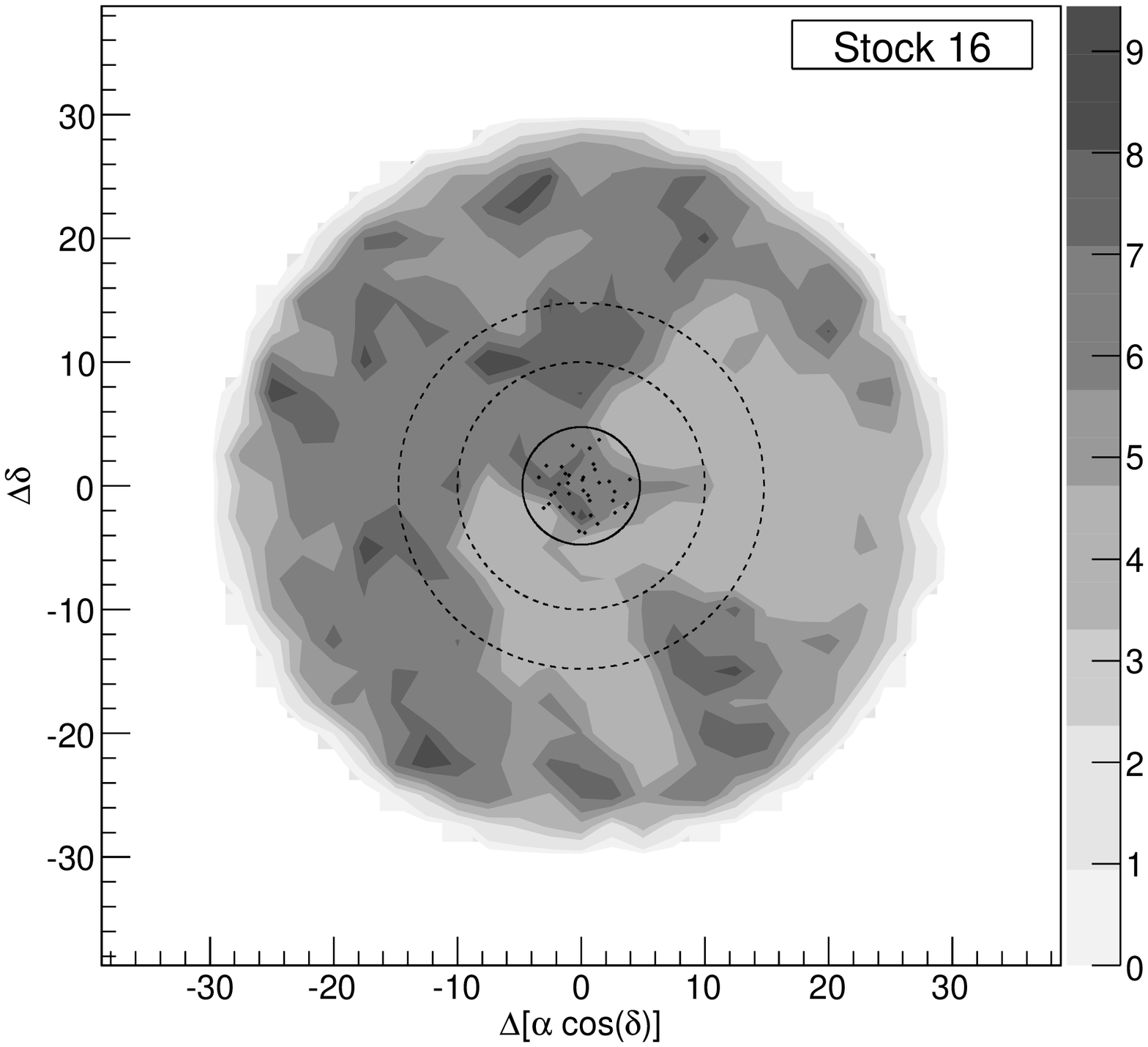}
\includegraphics[width=3.4cm]{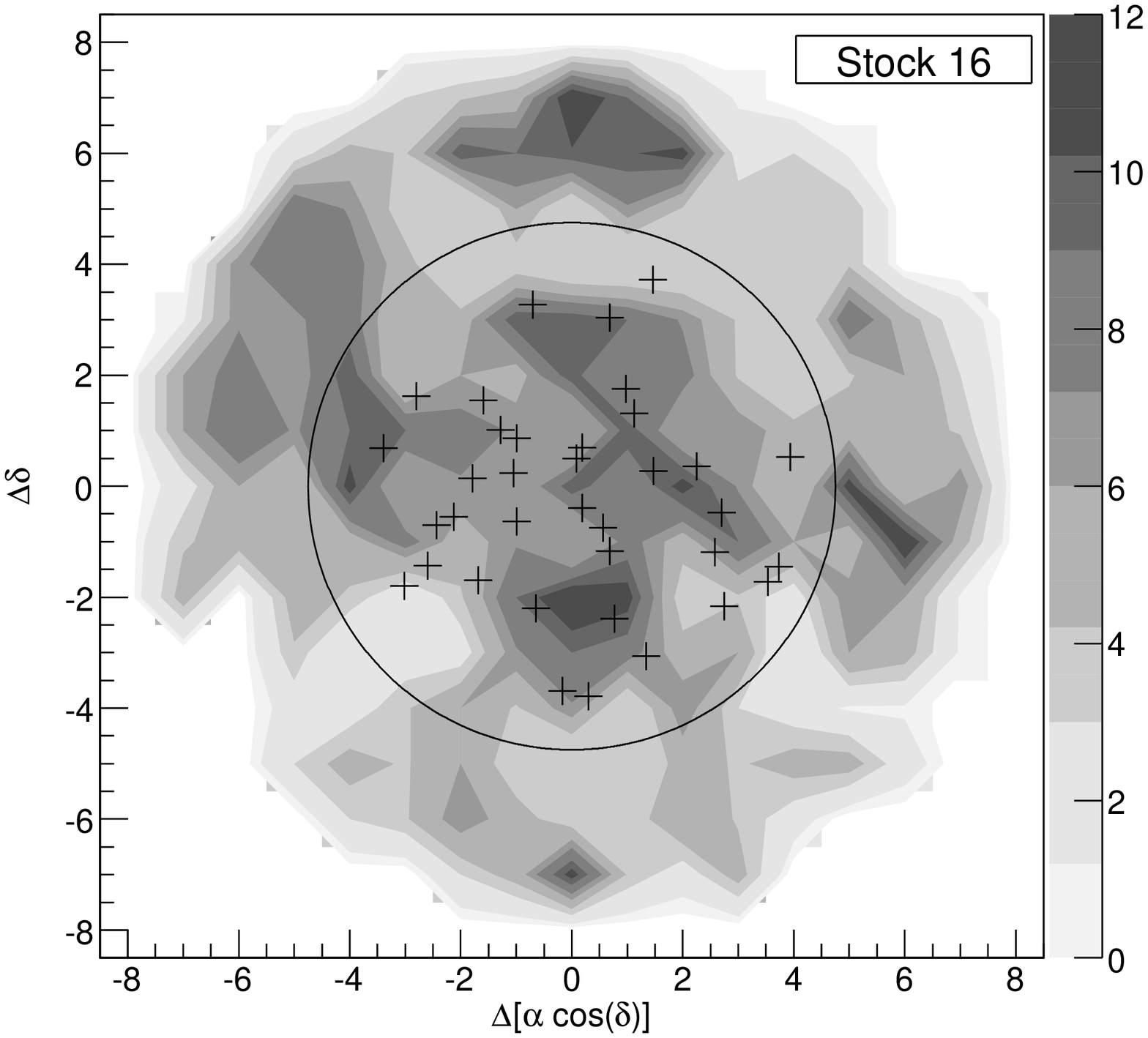}
\includegraphics[width=2.05cm]{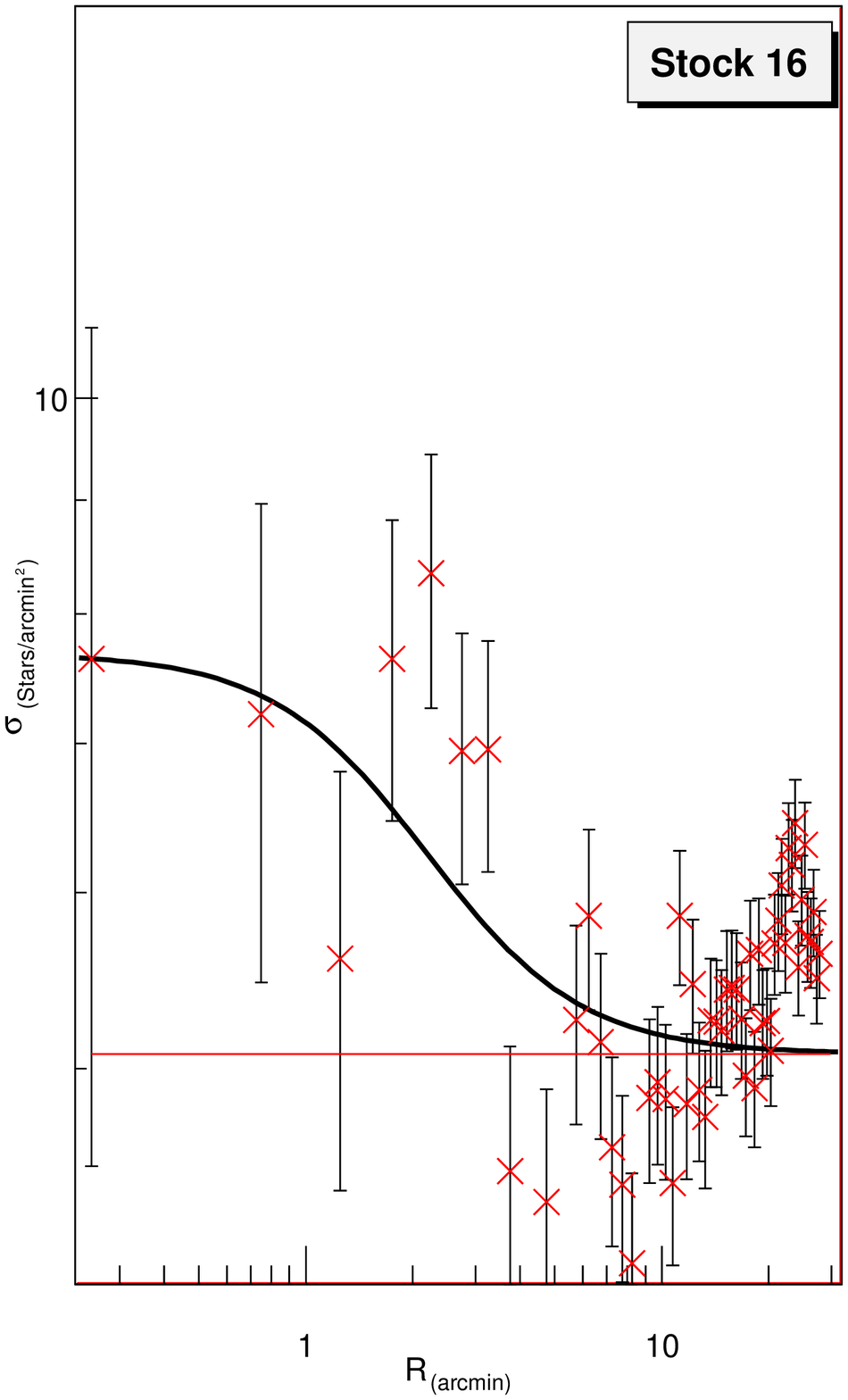}
\includegraphics[width=3.4cm]{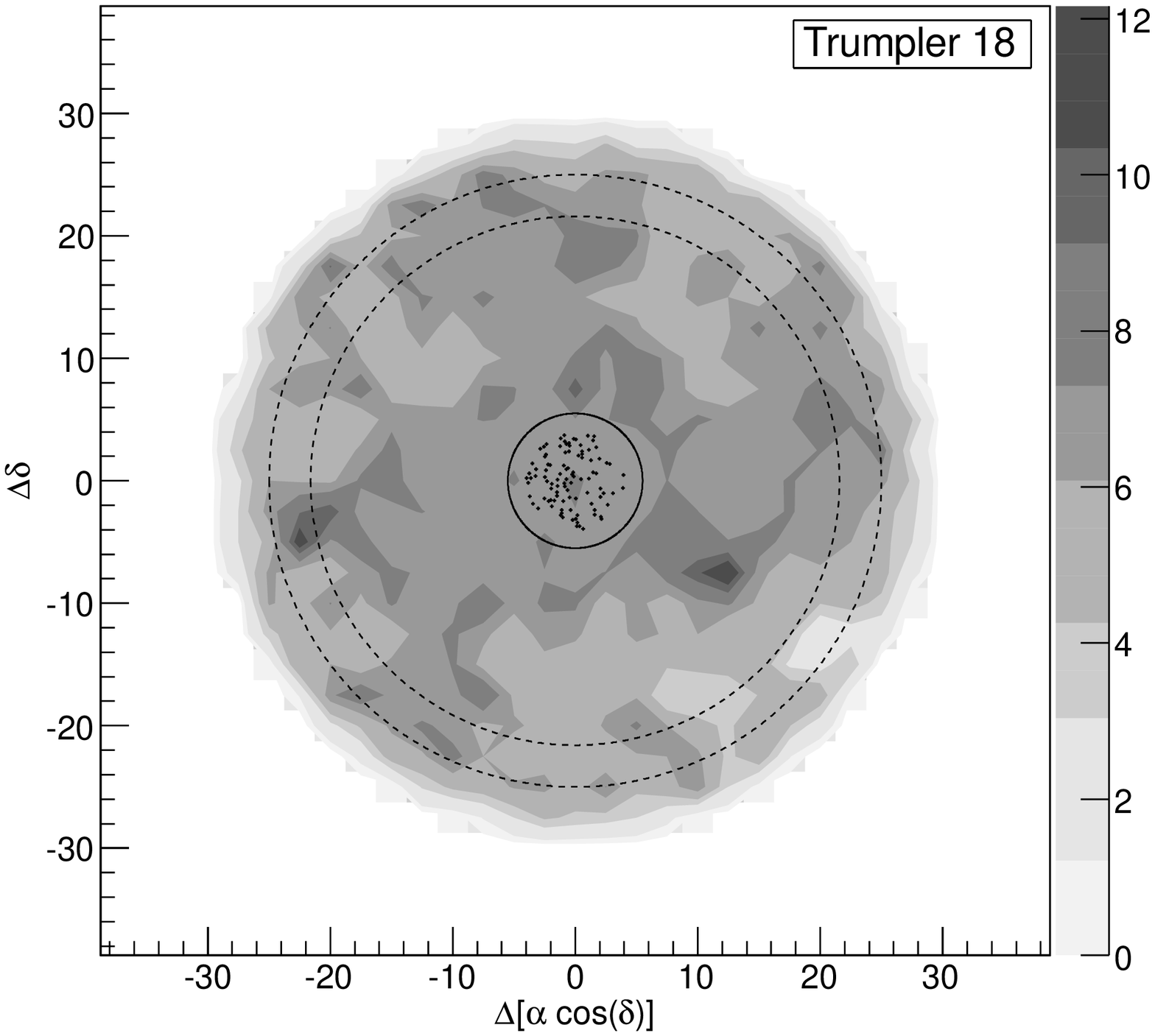}
\includegraphics[width=3.4cm]{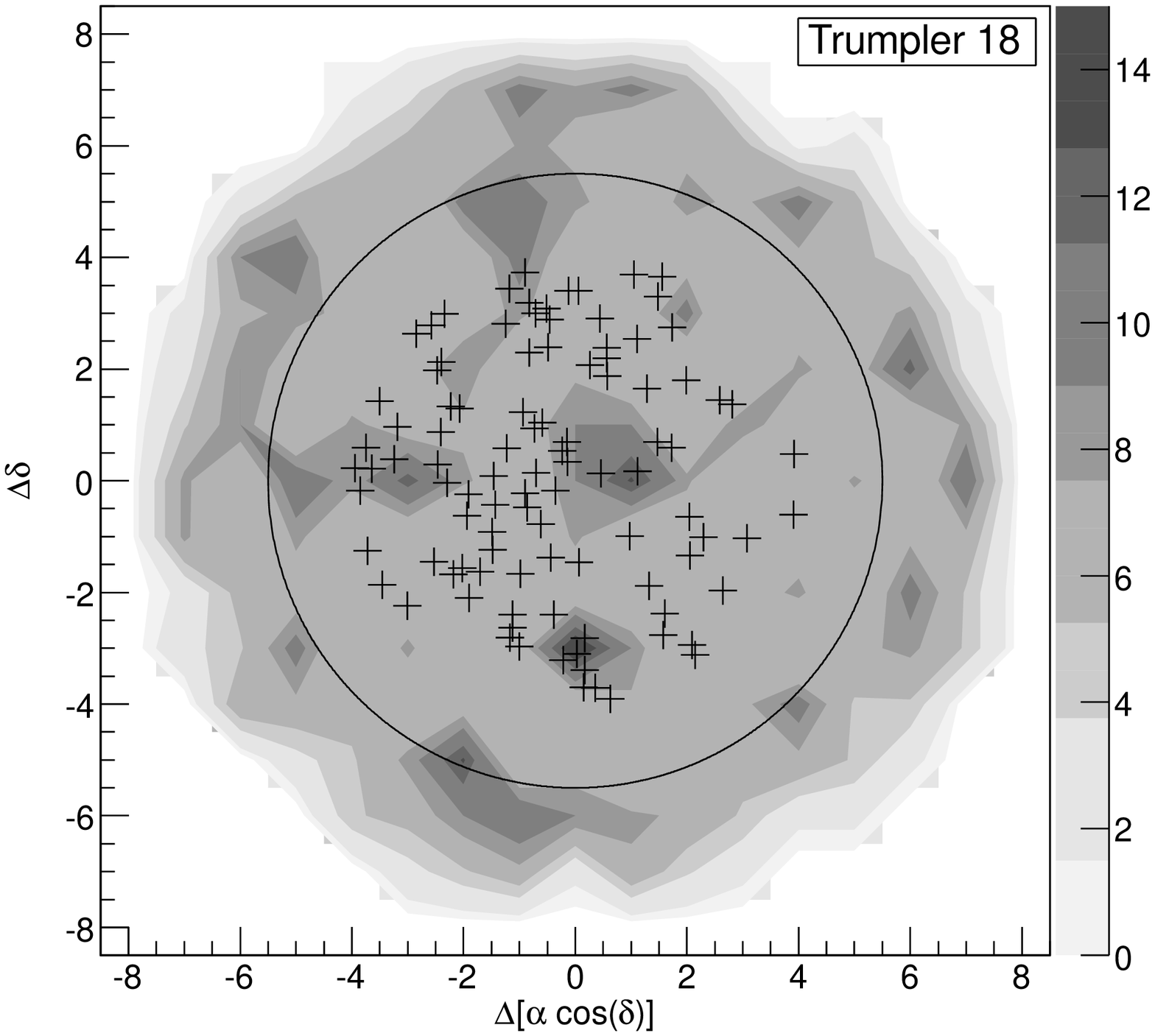}
\includegraphics[width=2.05cm]{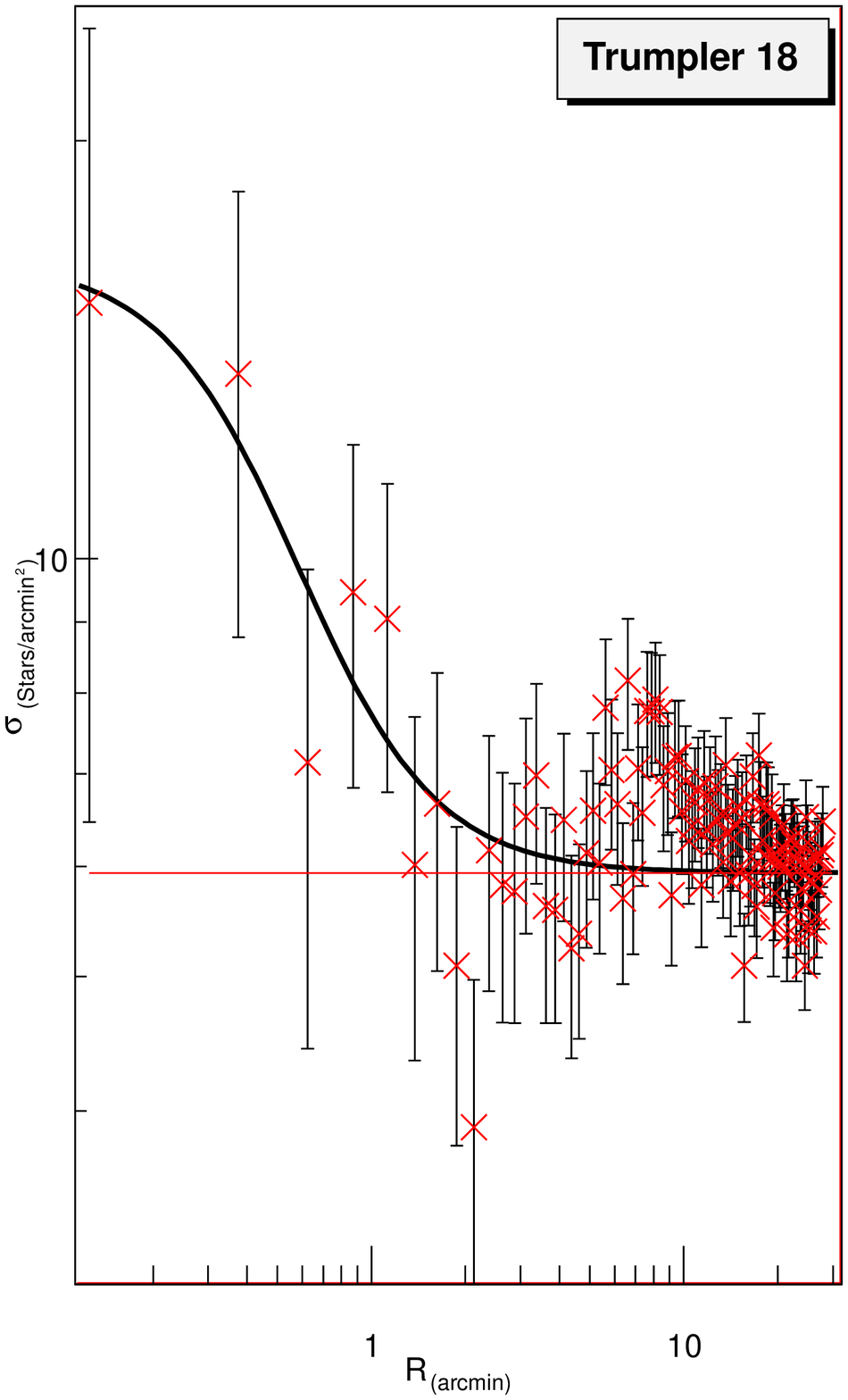}
\includegraphics[width=3.4cm]{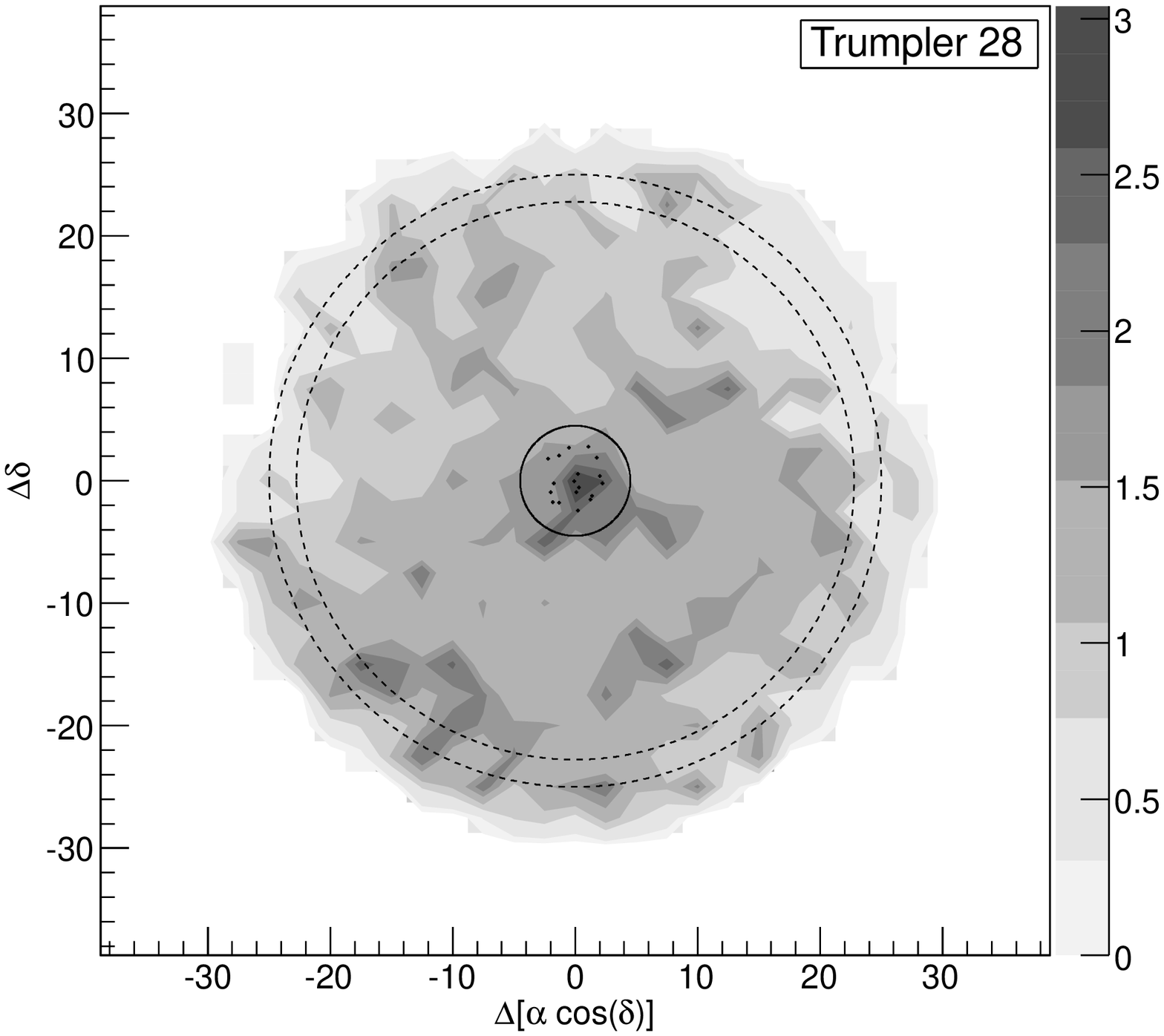}
\includegraphics[width=3.4cm]{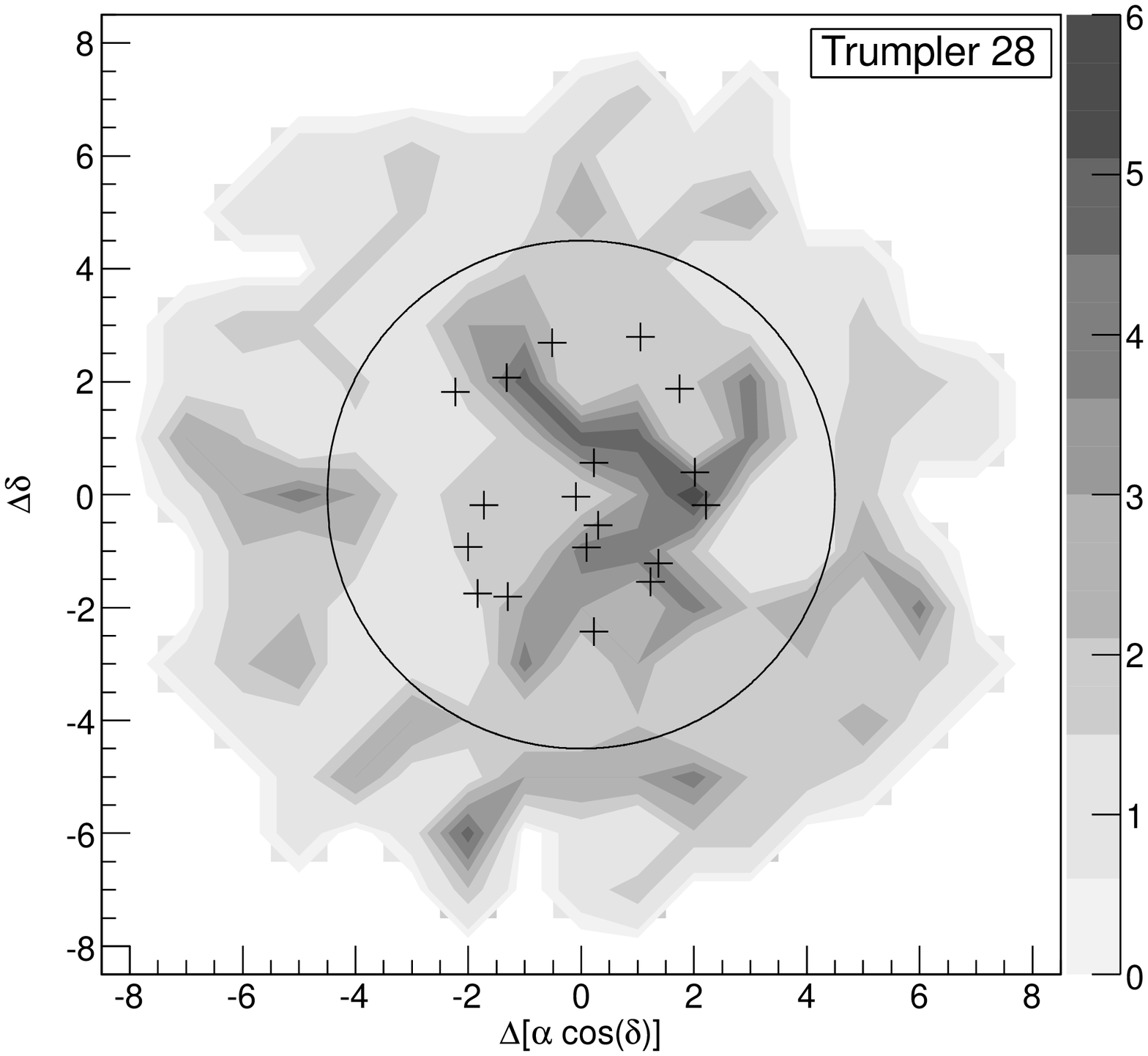}
\includegraphics[width=2.05cm]{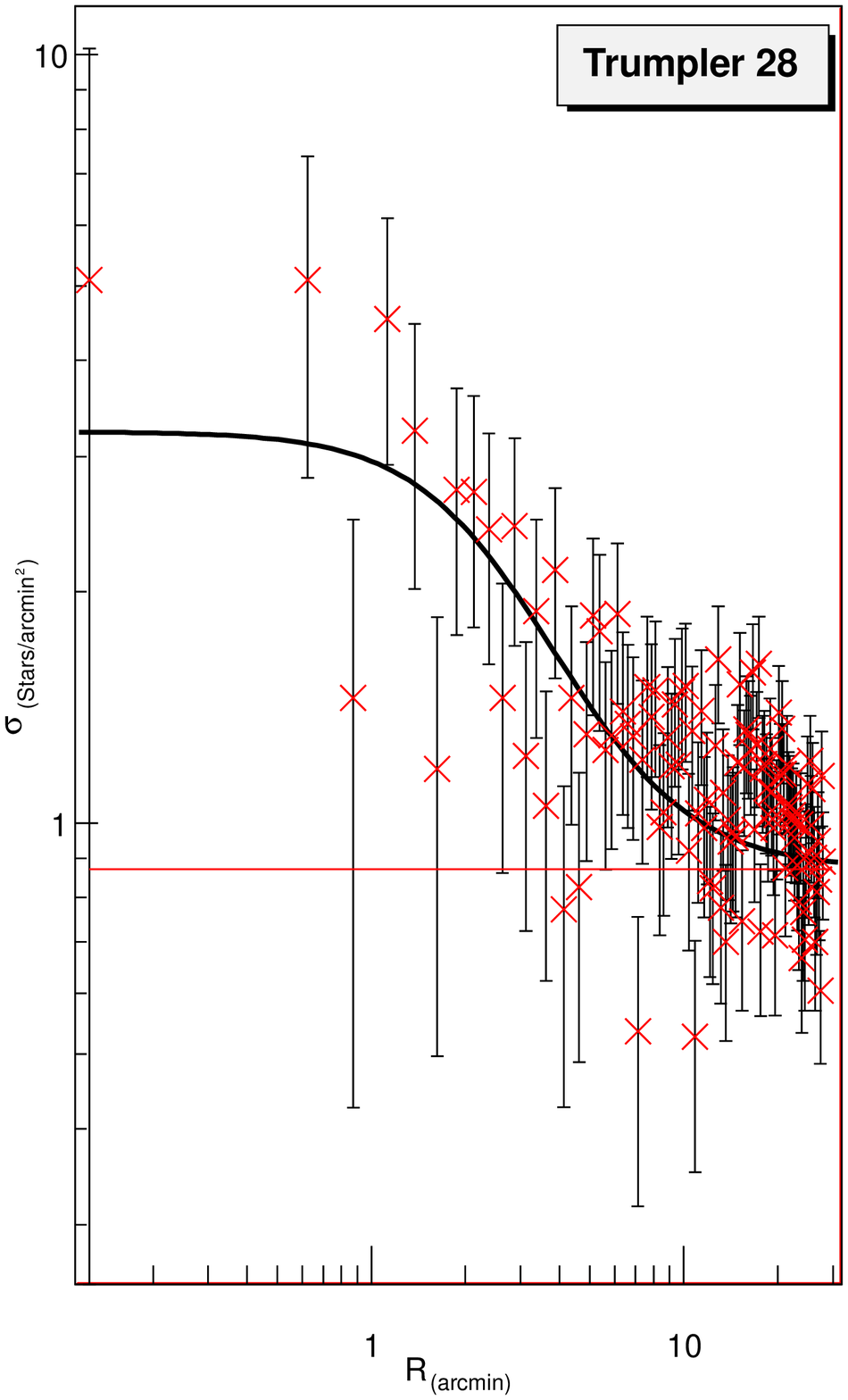}
\caption{Same as Fig. A1}  
\label{hmass}
\end{center}
\end{figure*}

%%%------------------------------------  Fig. A3

\begin{figure*}[]
\begin{center}

\includegraphics[width=6.3cm]{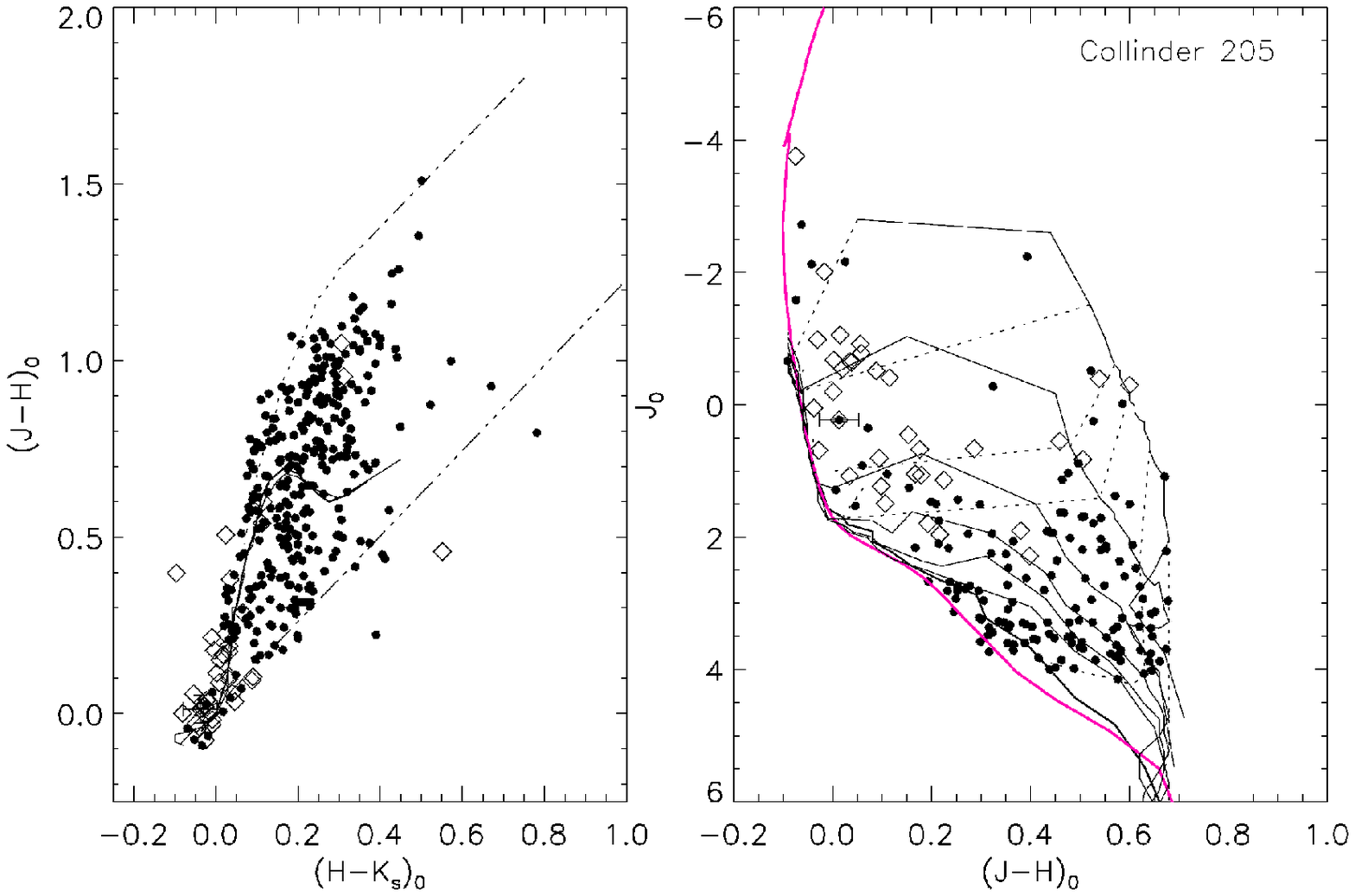}
\includegraphics[width=6.8cm]{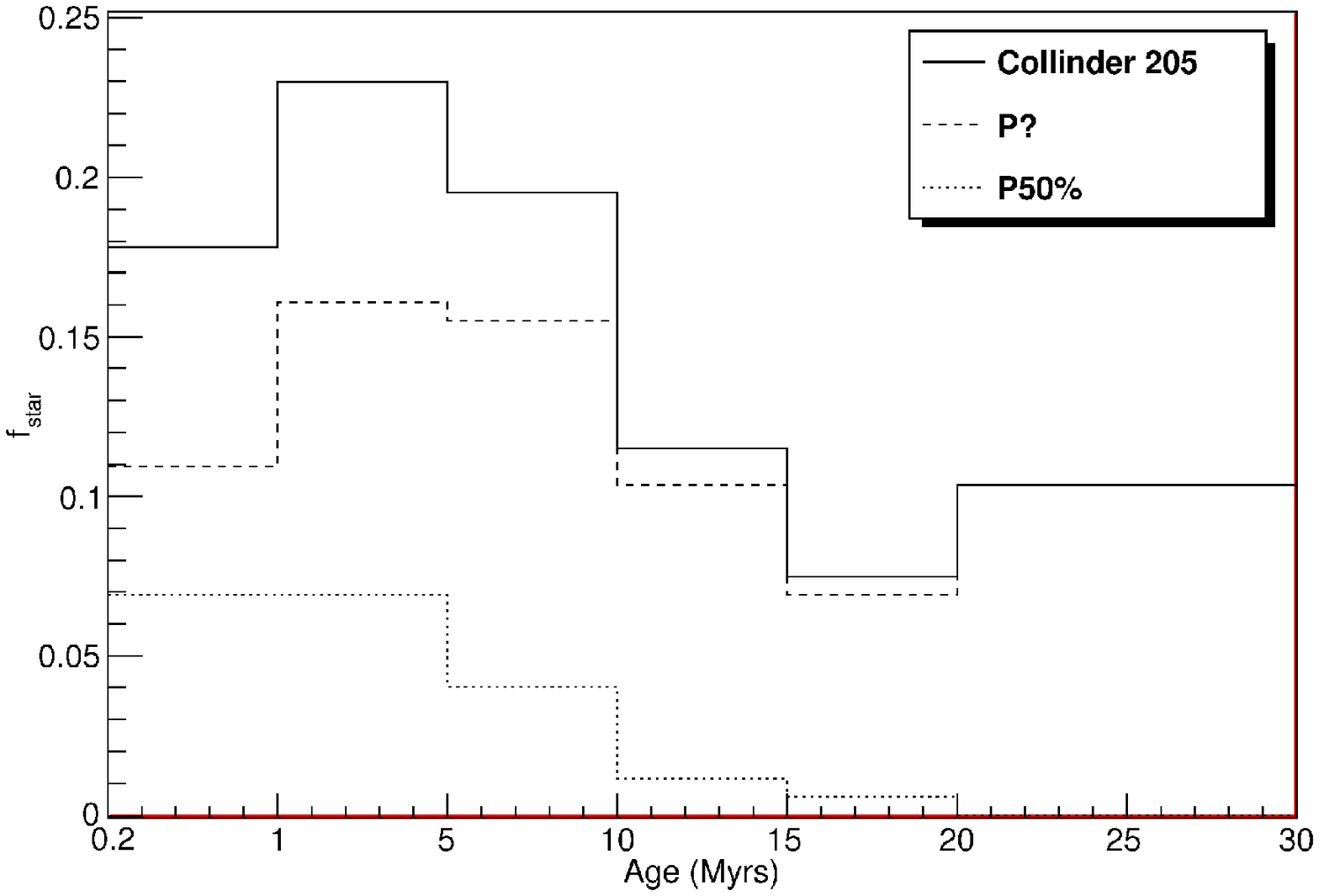}
\includegraphics[width=4.6cm]{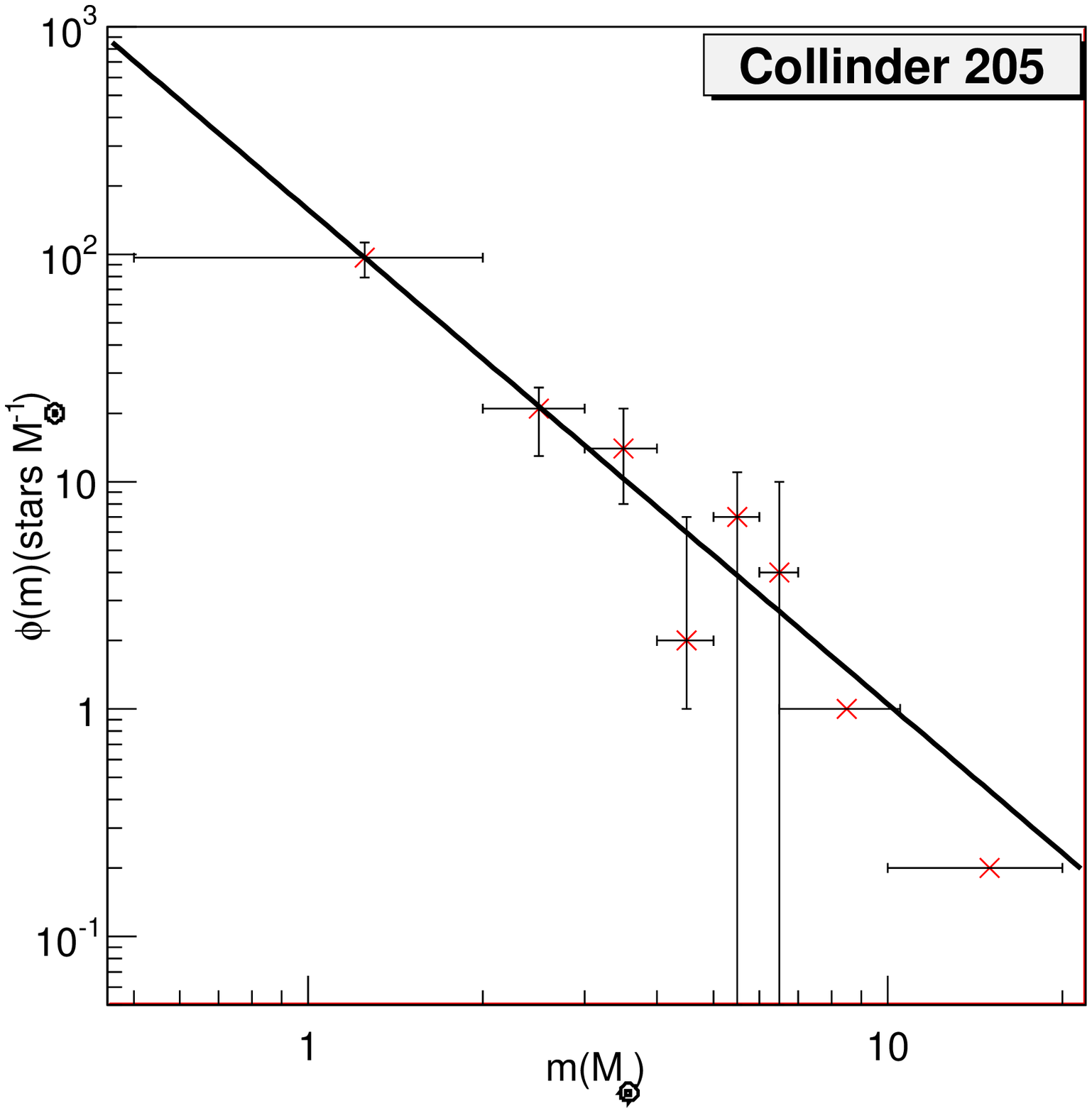}

\caption{{\it Left panels}: Colour-colour and colour-magnitude diagrams. The MS and the ZAMS are indicated by full lines, while the locus of giant
stars is represented by a dotted line. Reddening vectors from Rieke \& Lebofsky (1985) are shown by  dot-dashed lines. The isochrones and evolutionary pre-MS tracks from Siess et al. (2000). 
Cluster members are indicated by open diamonds (P50) 
and dots (P?).
{\it Centre}: Age distribution of clusters members (thick line) showing the contribution of P50 (dotted line) and
P? (dashed line) objects.
{\it Right}: Observed mass distribution indicated by crosses with error bars. The thick line represents the mass function $\phi$(m) fitting.) }  
\label{hmass}
\end{center}
\end{figure*}
\begin{figure*}[]
\begin{center}

%%%%----------------------Fig. A4
\includegraphics[width=6.3cm]{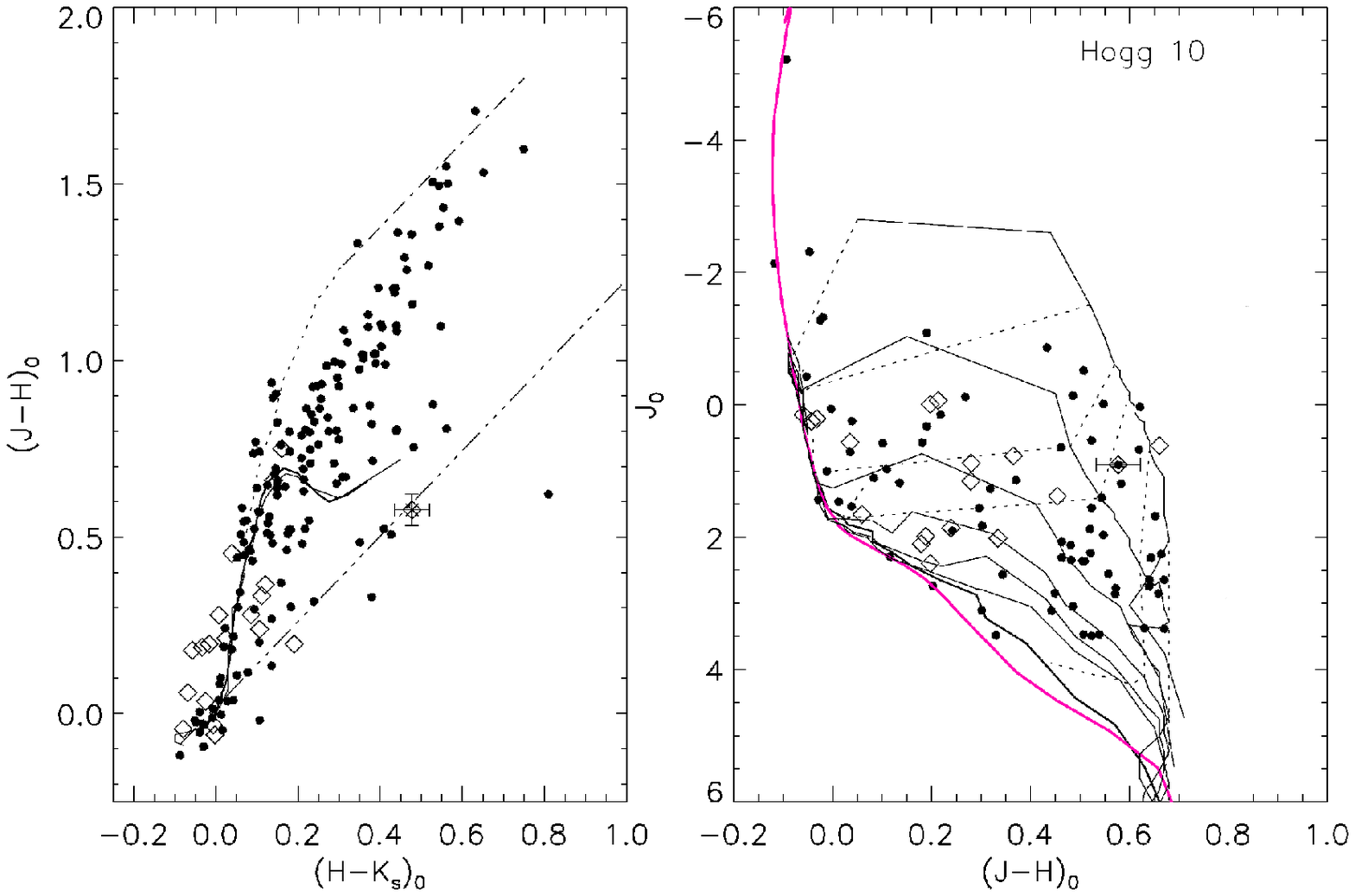}
\includegraphics[width=6.8cm]{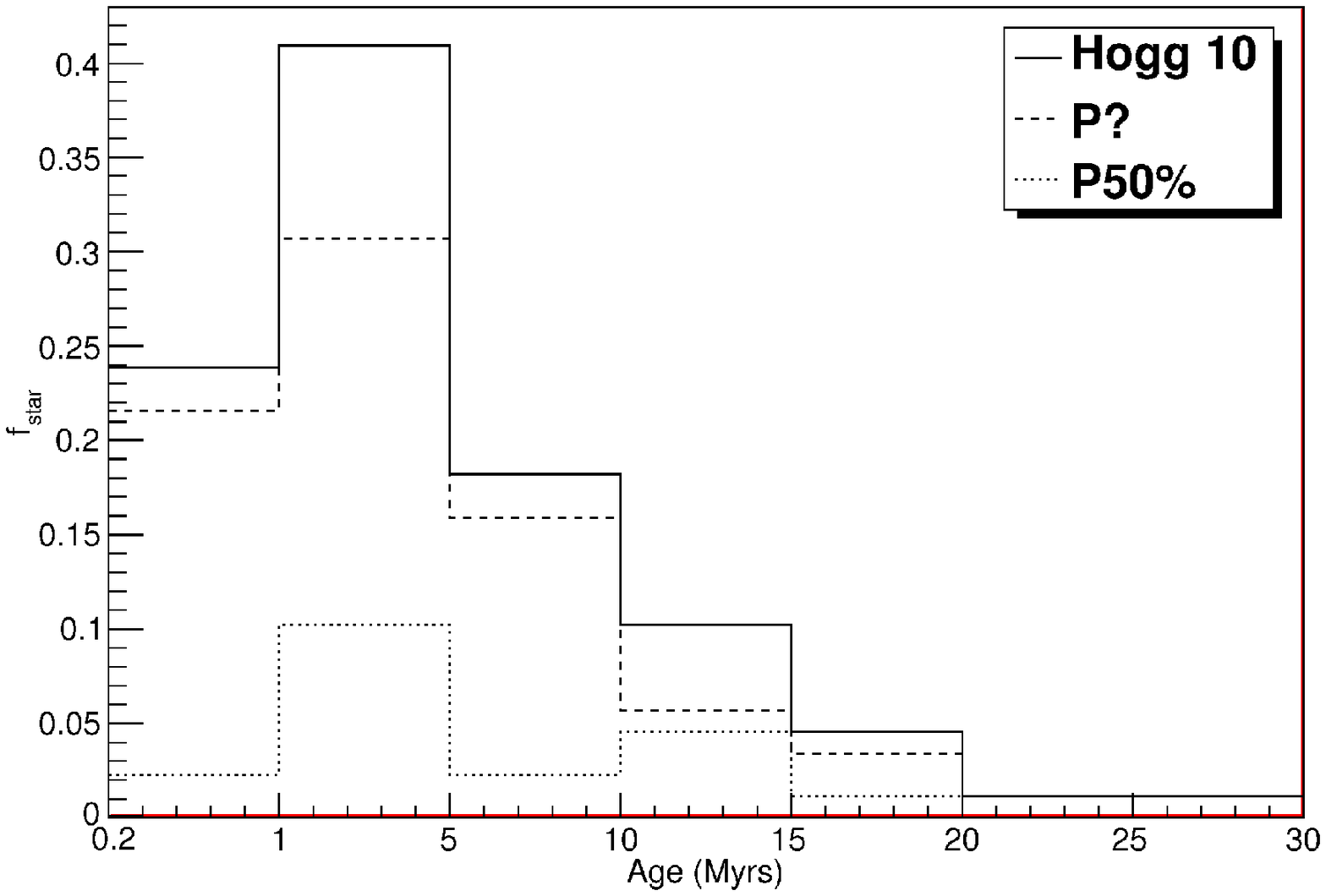}
\includegraphics[width=4.6cm]{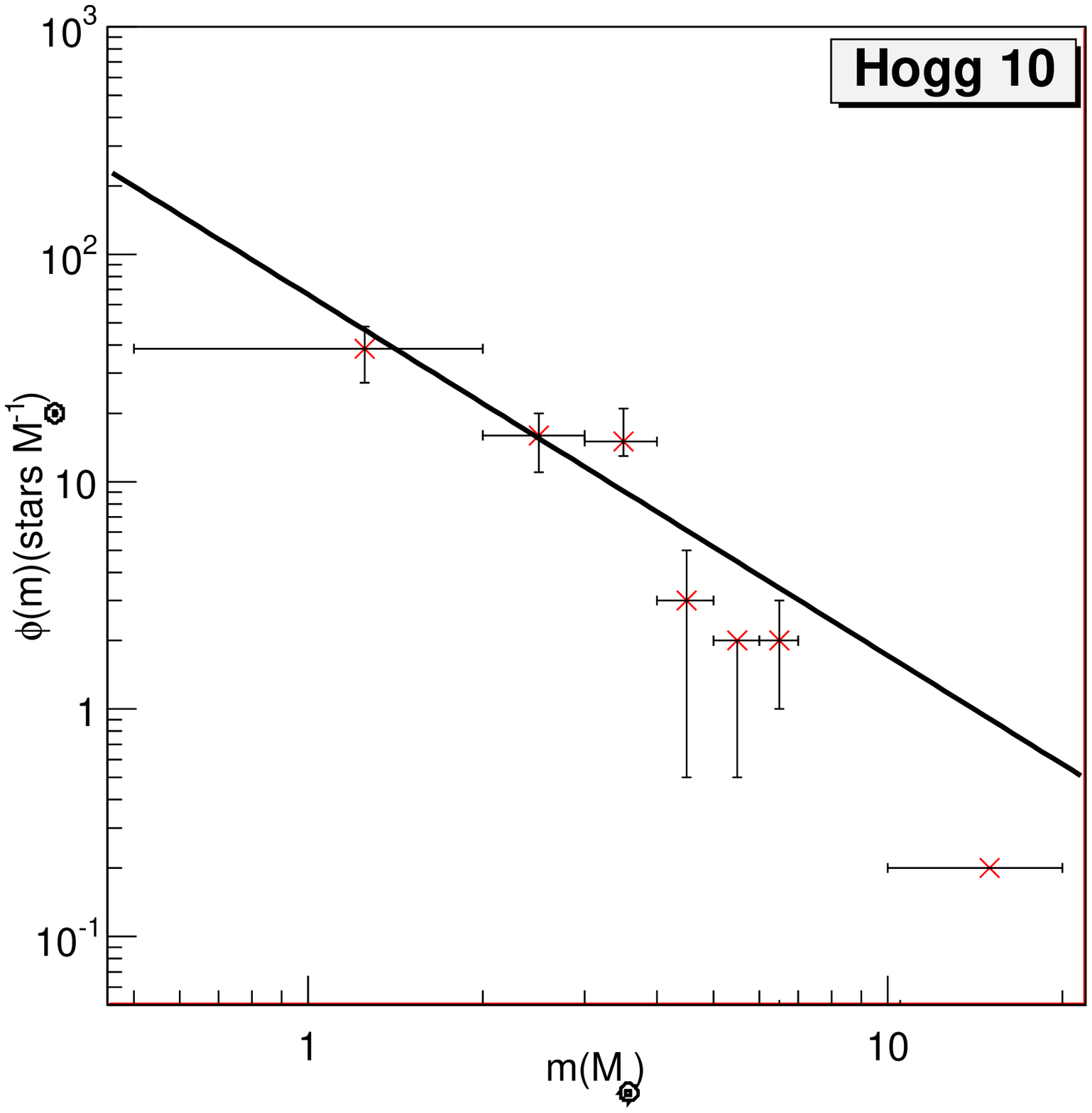}
\includegraphics[width=6.3cm]{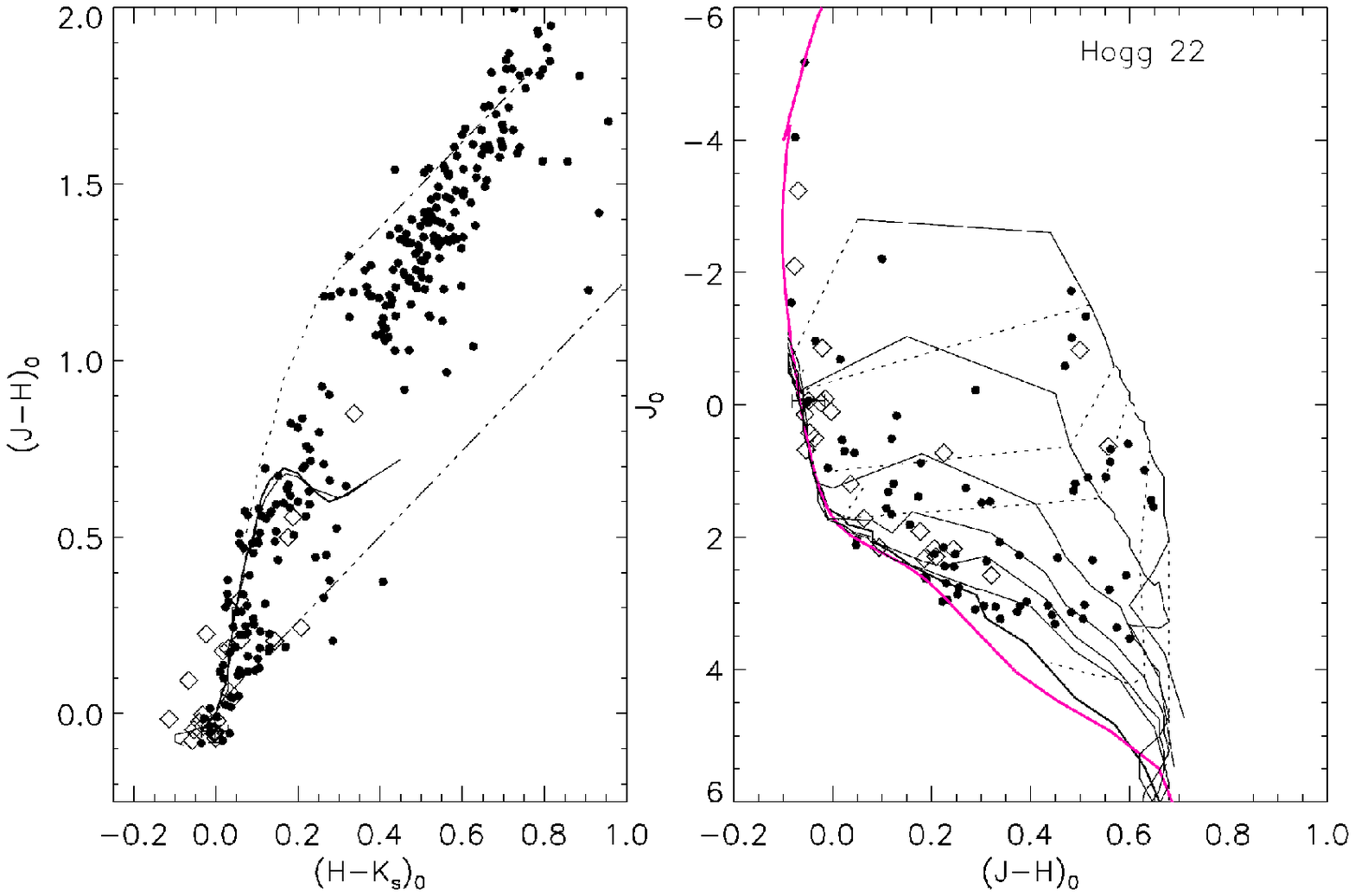}
\includegraphics[width=6.8cm]{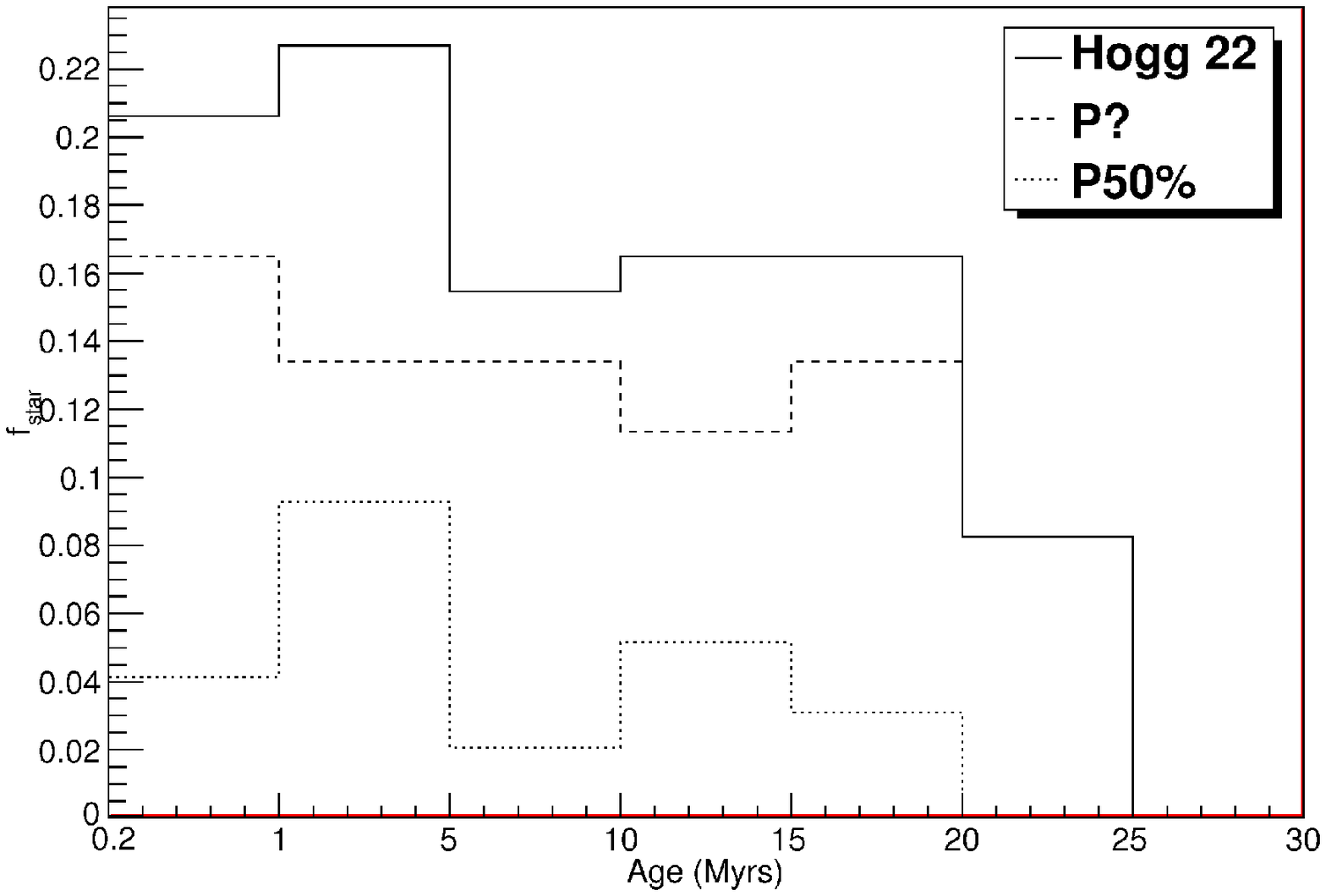}
\includegraphics[width=4.6cm]{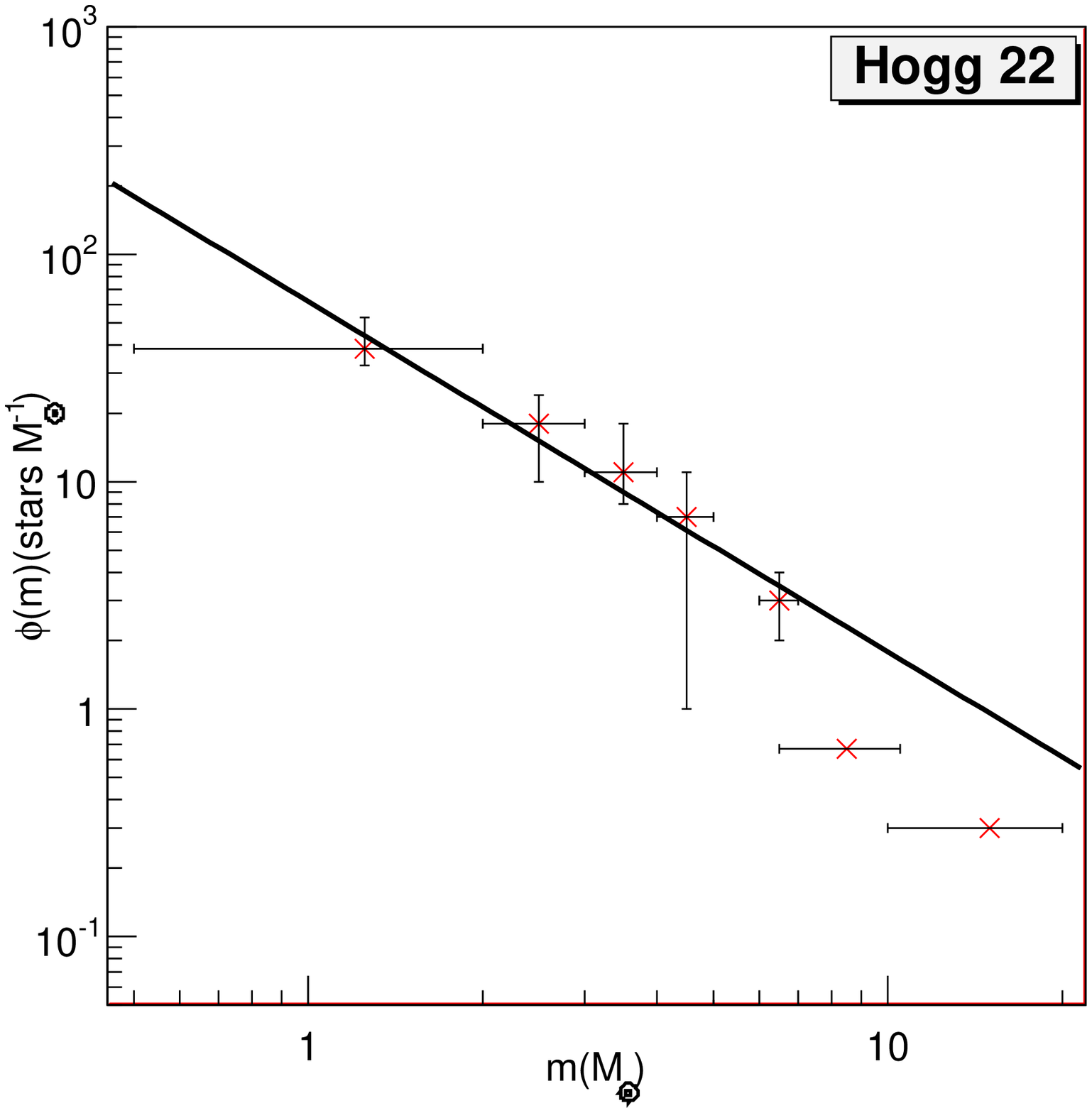}
\includegraphics[width=6.3cm]{diag_ly14.eps}
\includegraphics[width=6.8cm]{histo_id_ly14.eps}
\includegraphics[width=4.6cm]{histo_massa_ly14_X.eps}
\includegraphics[width=6.3cm]{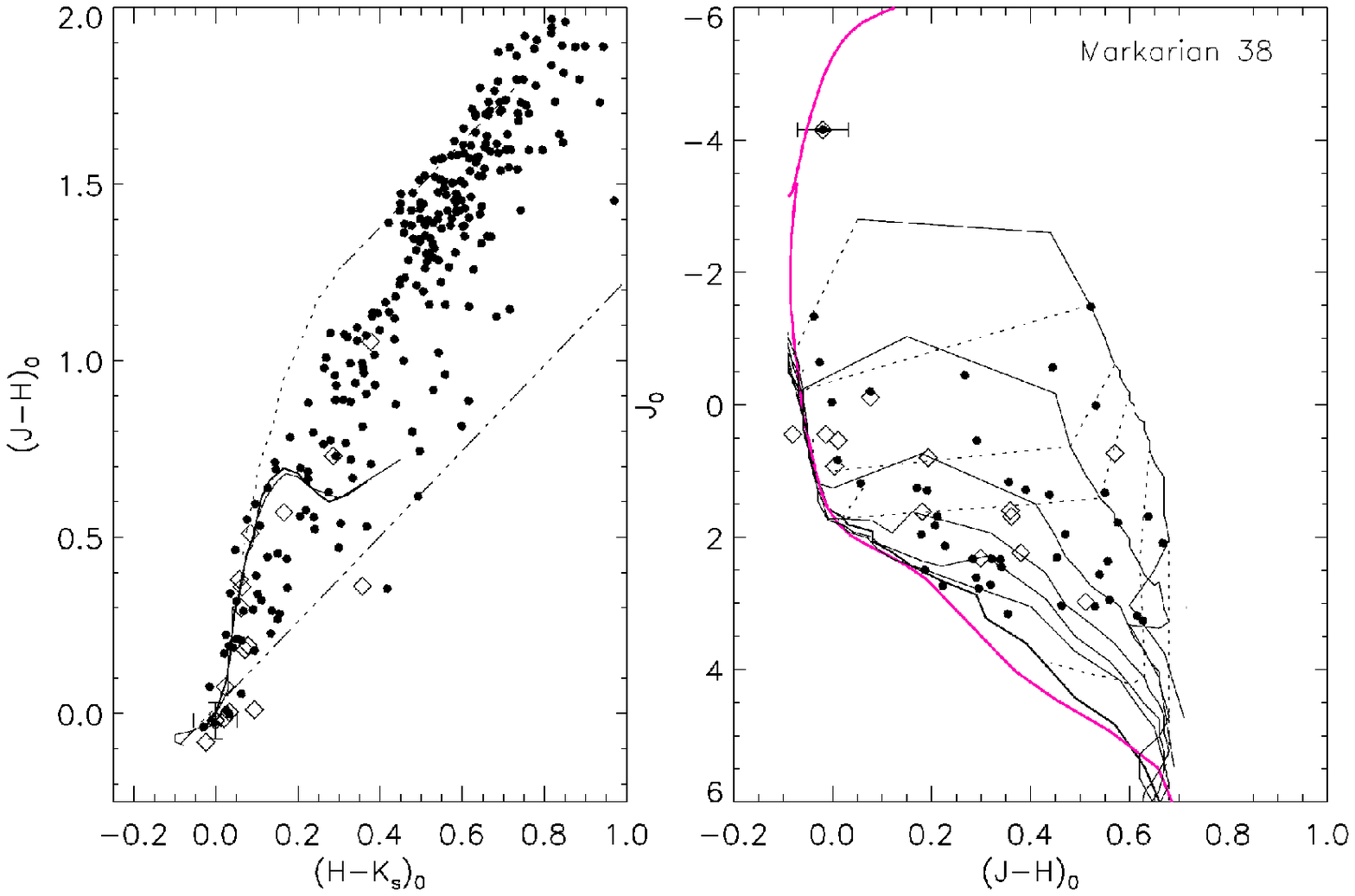}
\includegraphics[width=6.8cm]{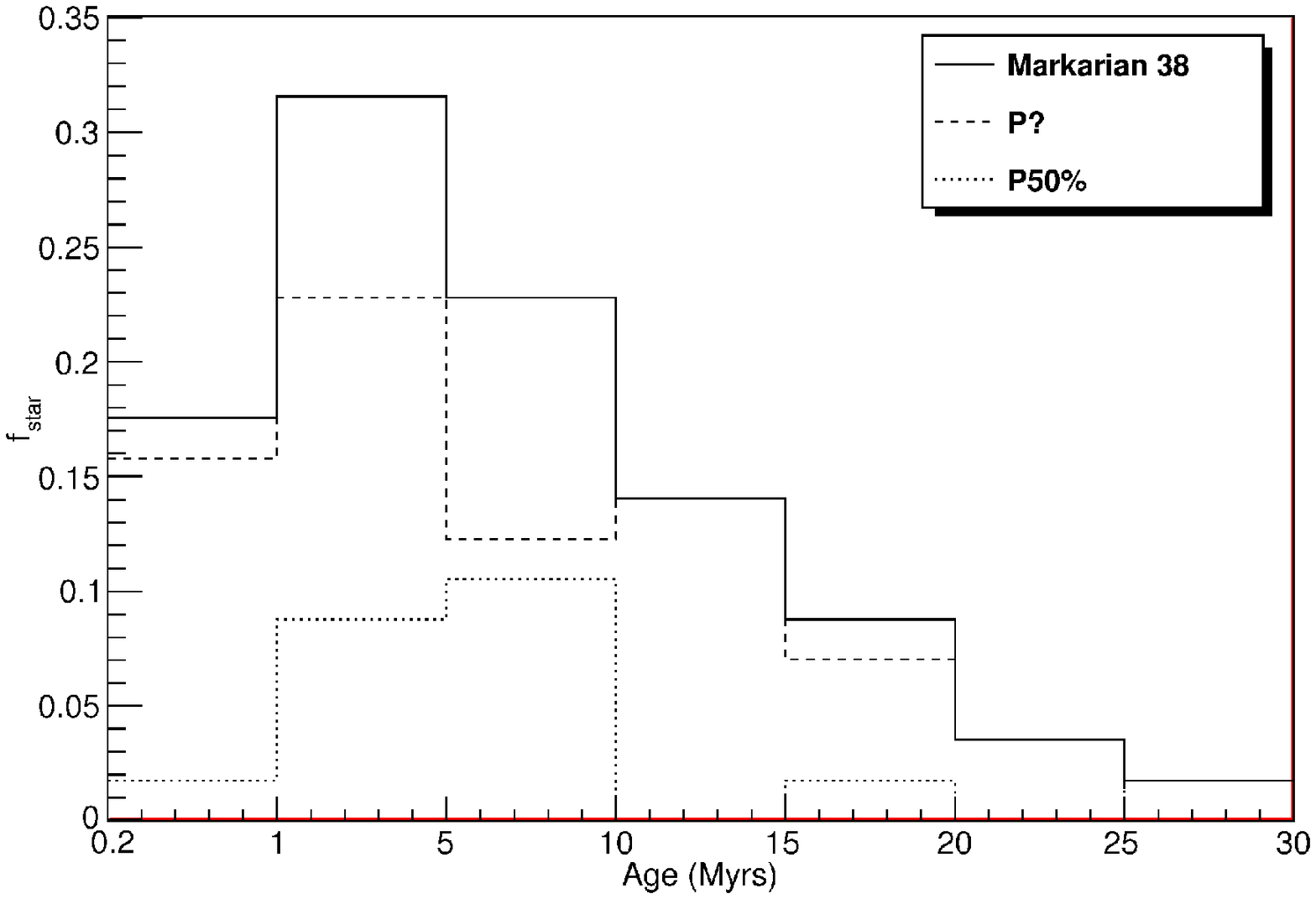}
\includegraphics[width=4.6cm]{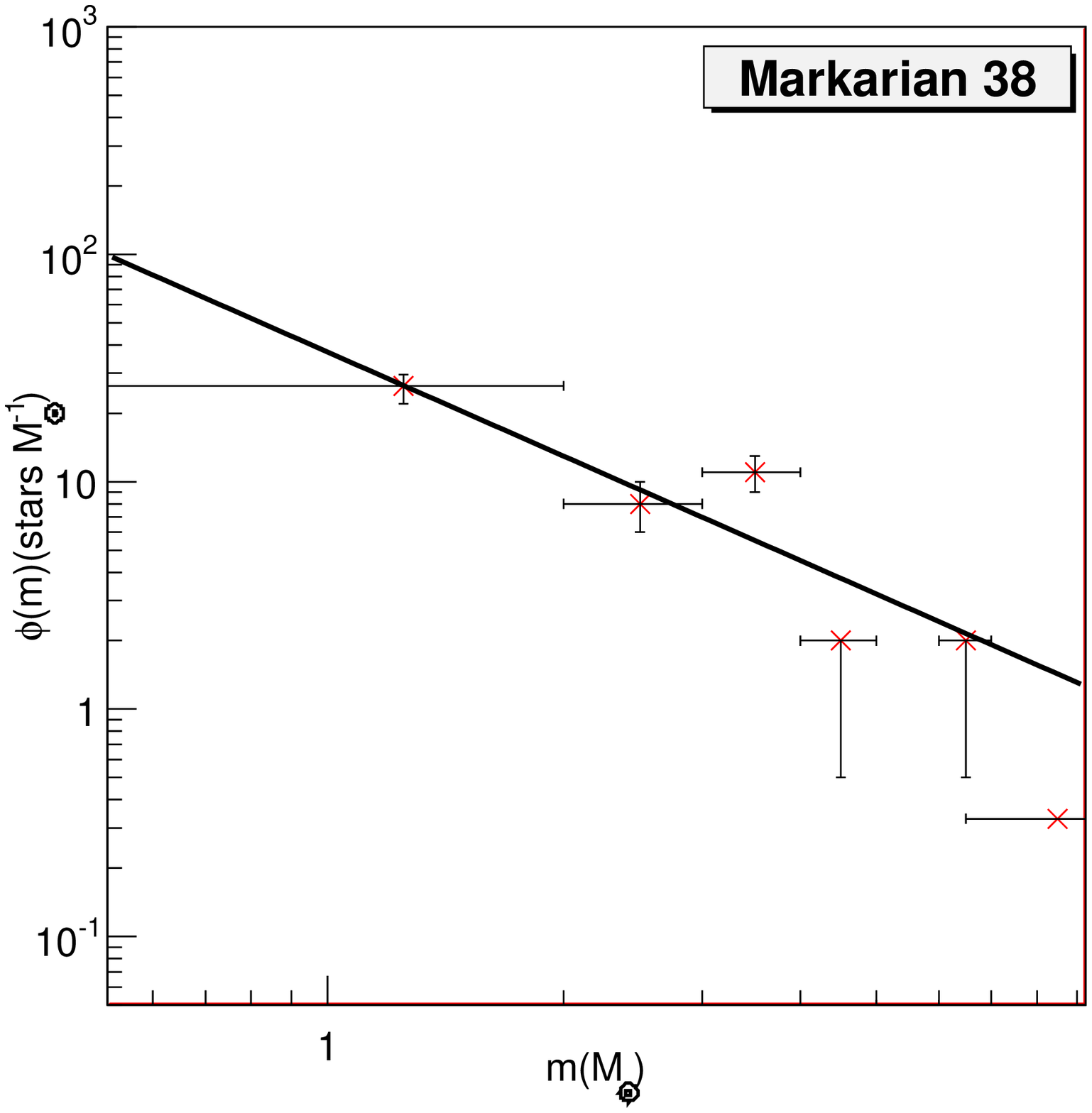}
\includegraphics[width=6.3cm]{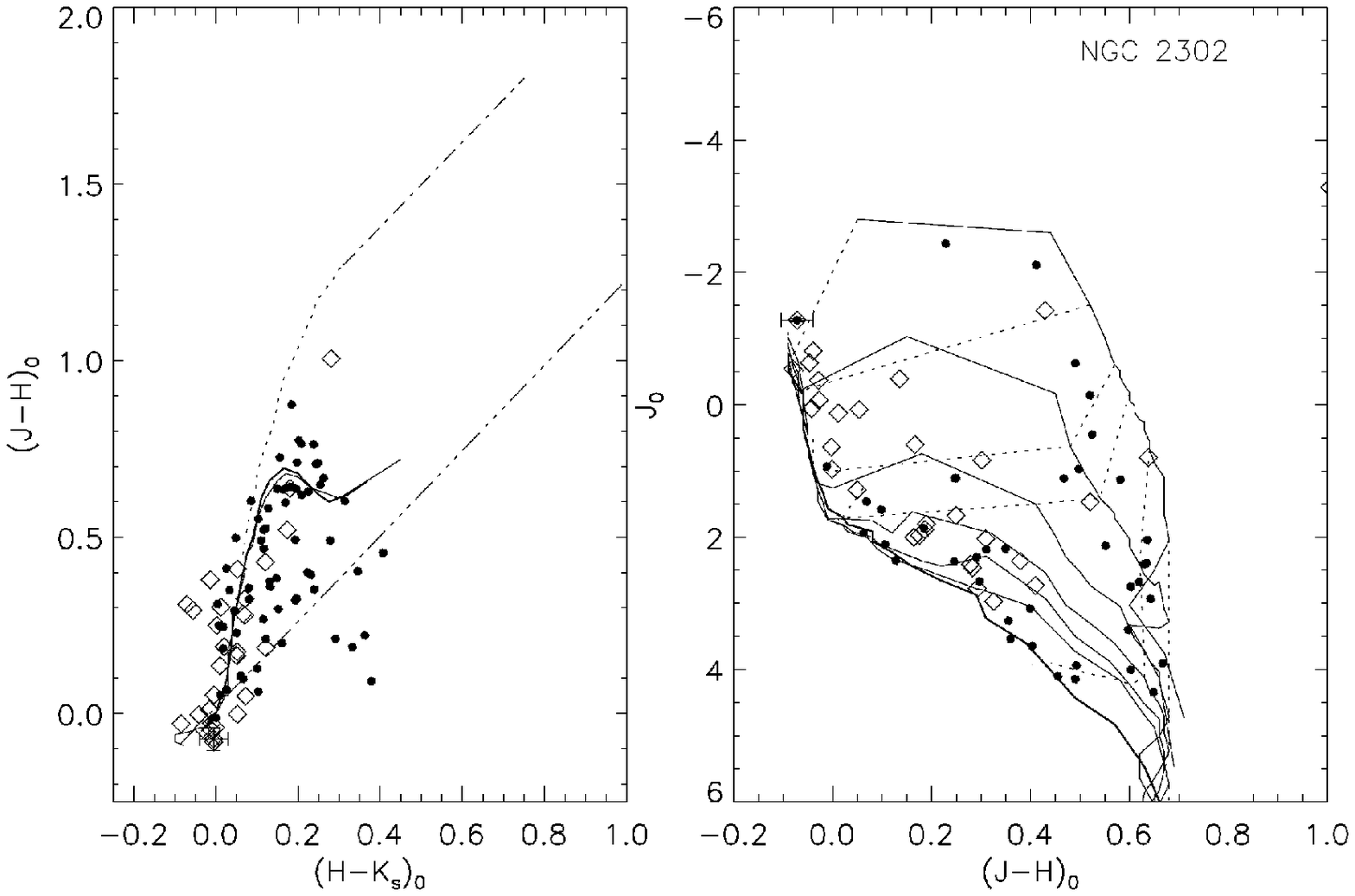}
\includegraphics[width=6.8cm]{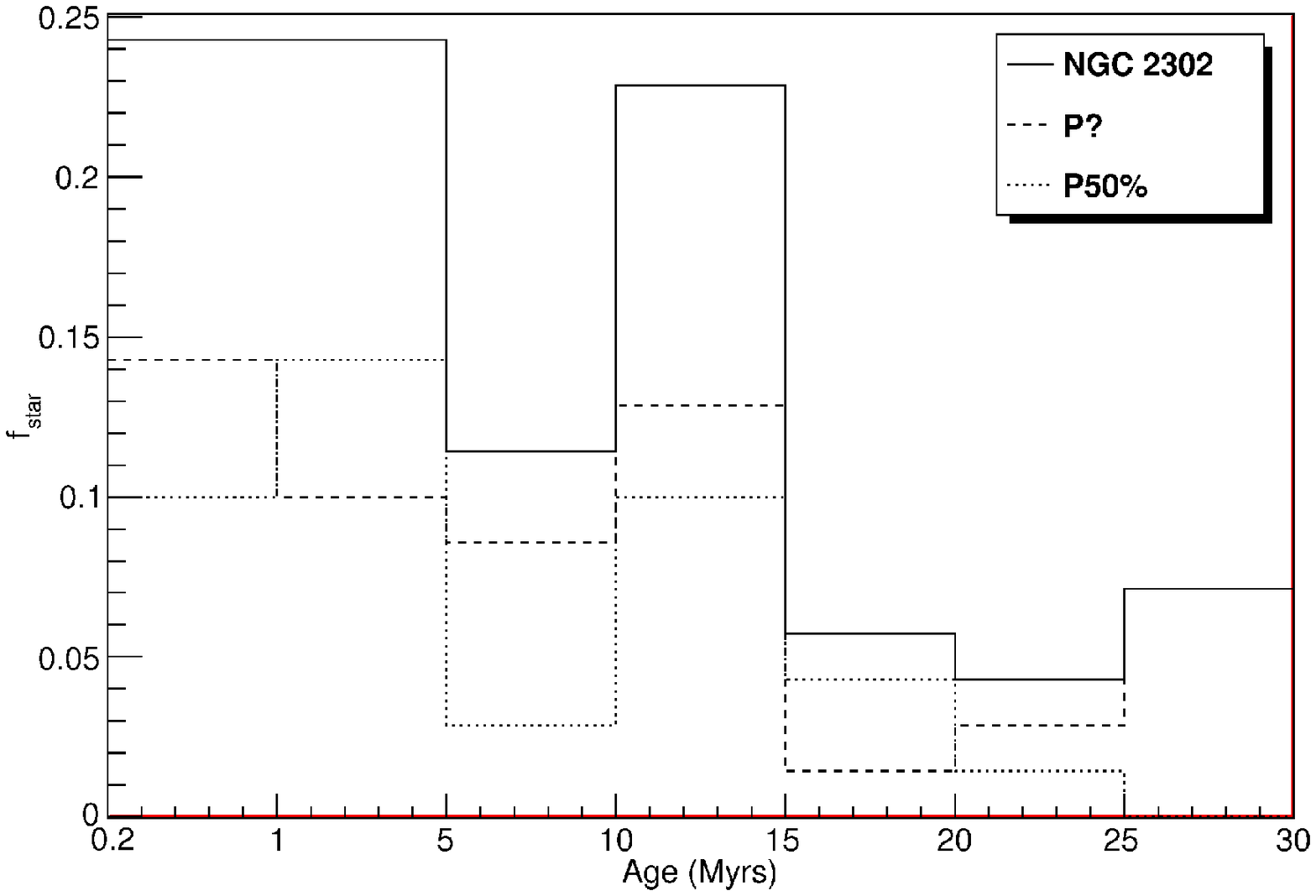}
\includegraphics[width=4.6cm]{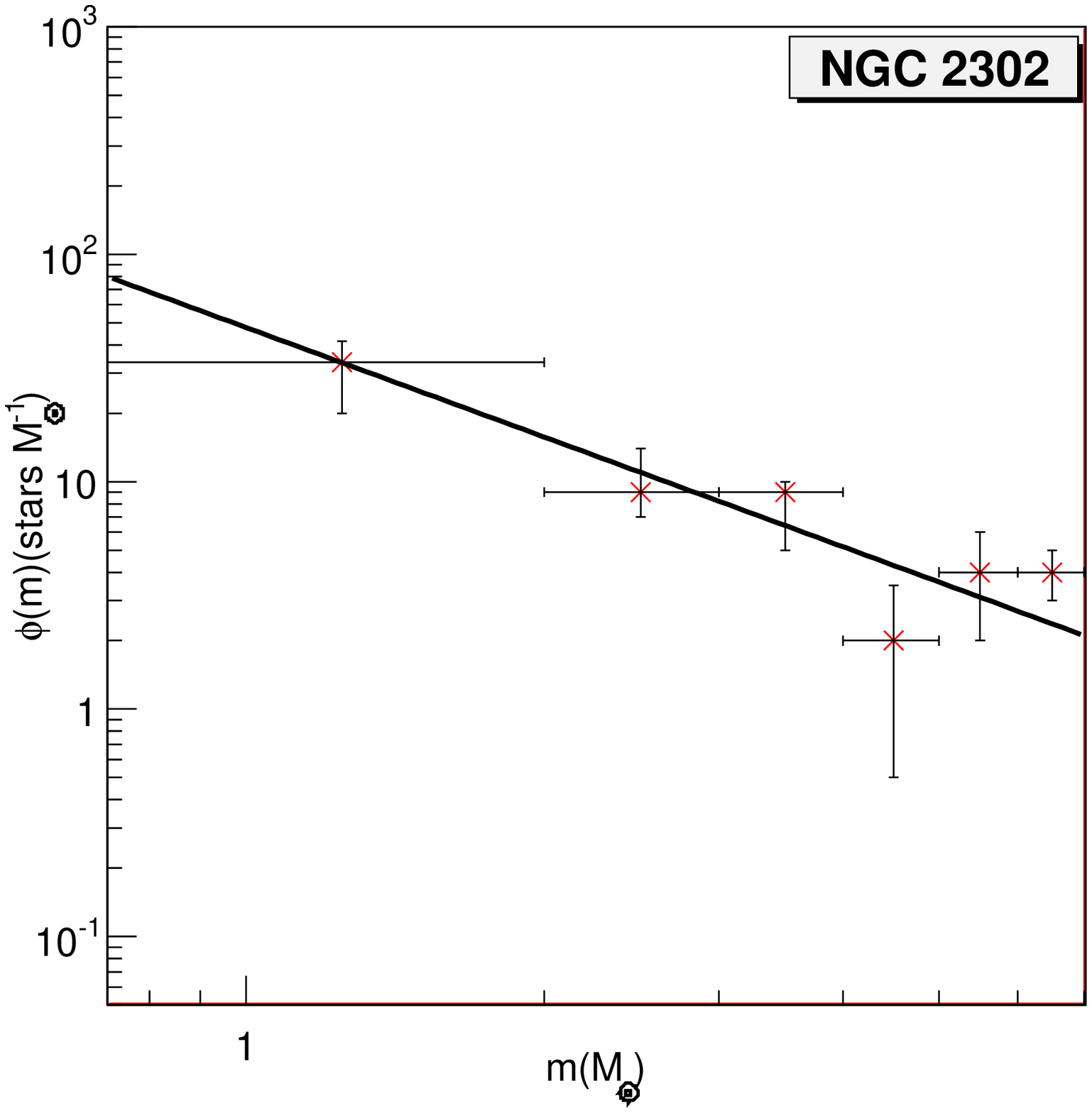}
\caption{The same as Fig A.3.}  
\label{hmass}
\end{center}
\end{figure*}

%%%%----------------------Fig. A5

\begin{figure*}[]
\begin{center}
\includegraphics[width=6.3cm]{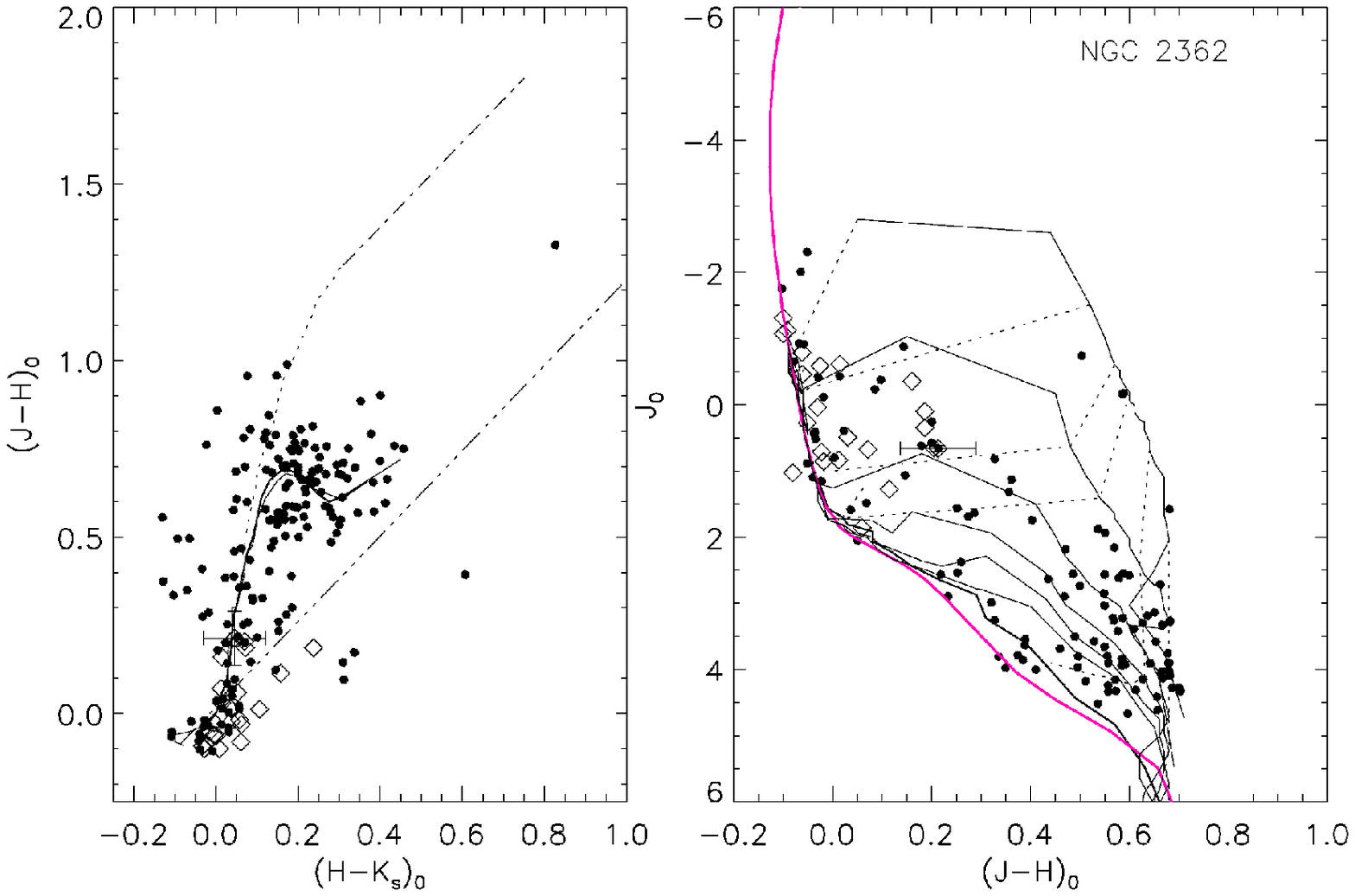}
\includegraphics[width=6.8cm]{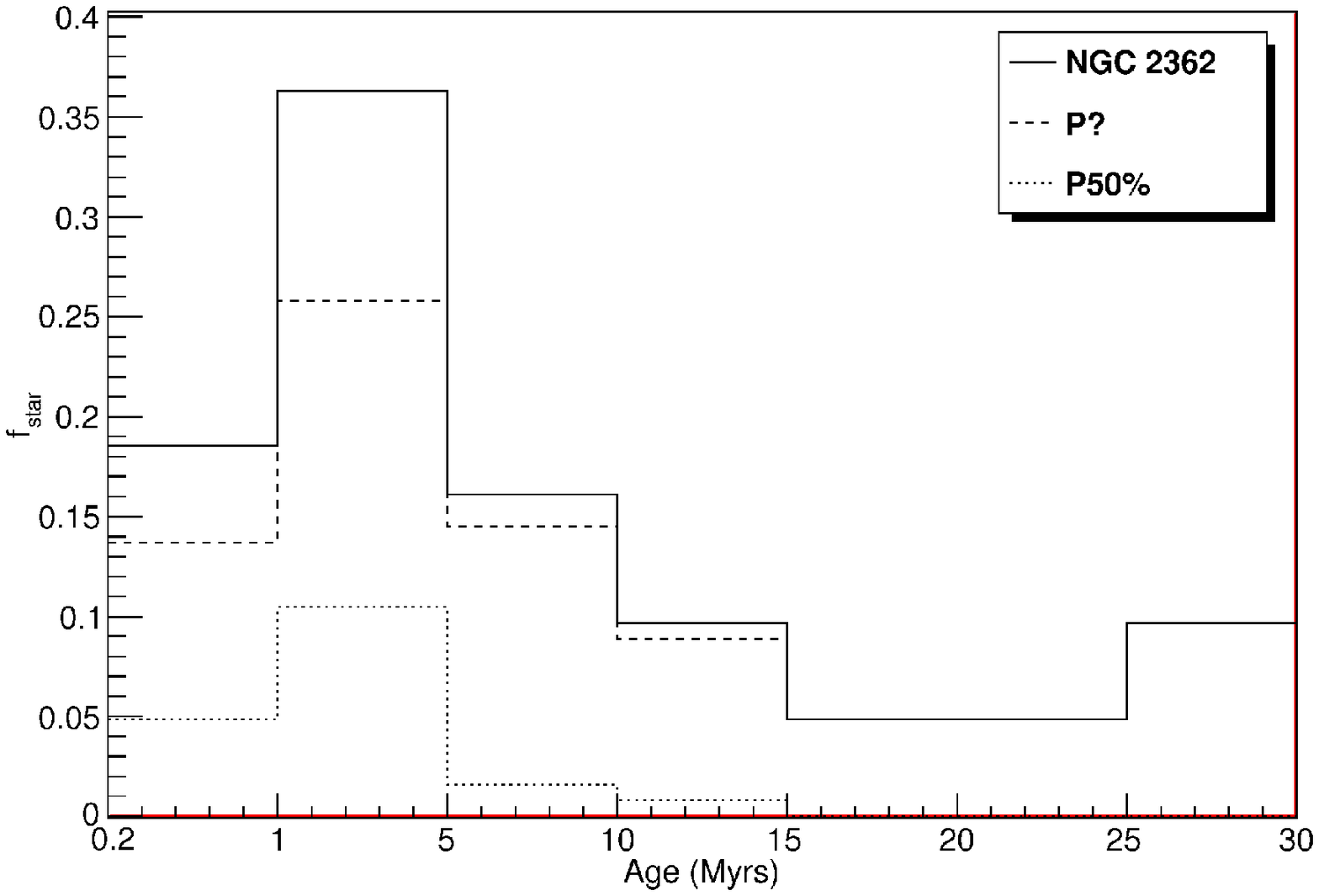}
\includegraphics[width=4.6cm]{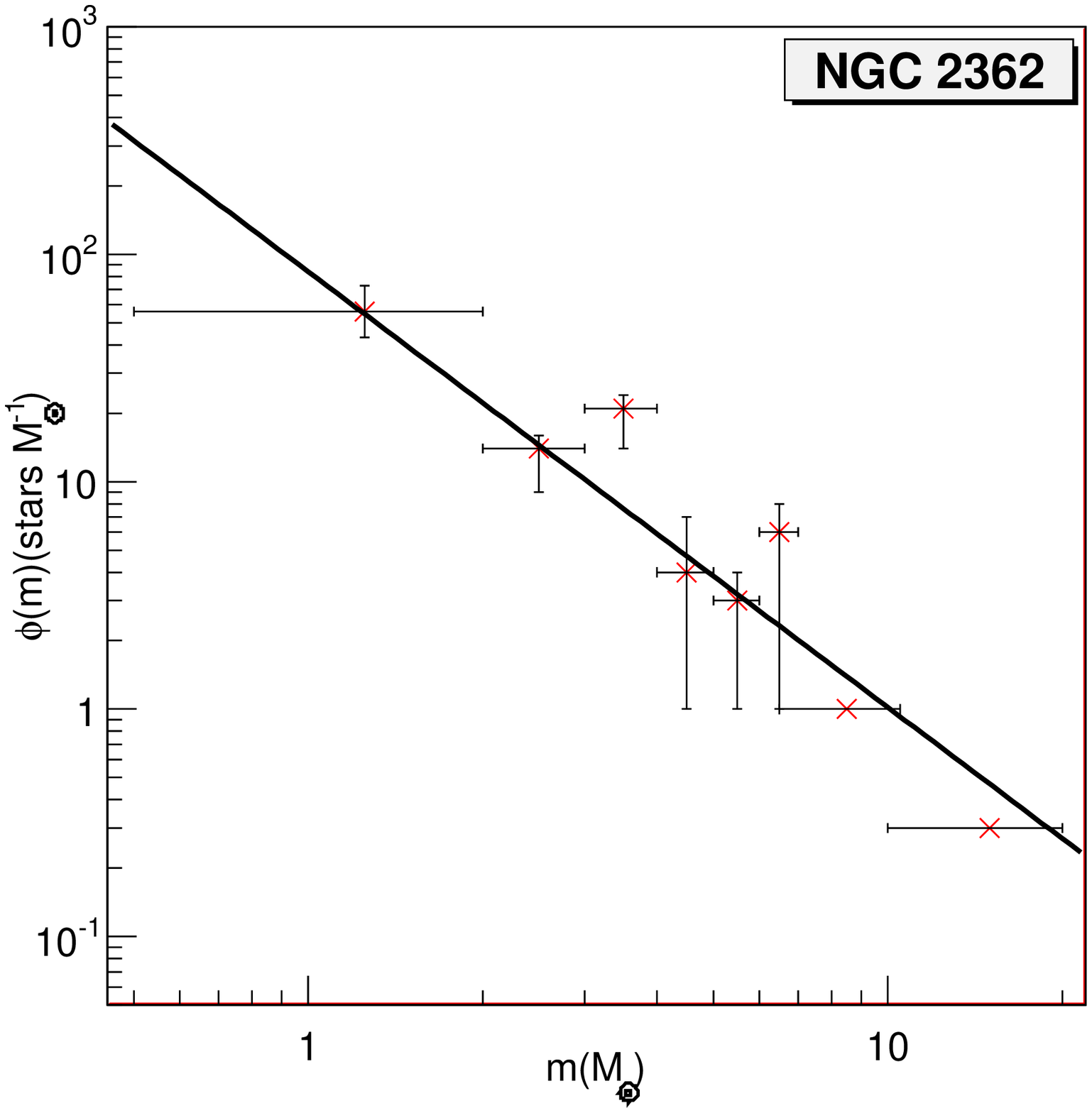}
\includegraphics[width=6.3cm]{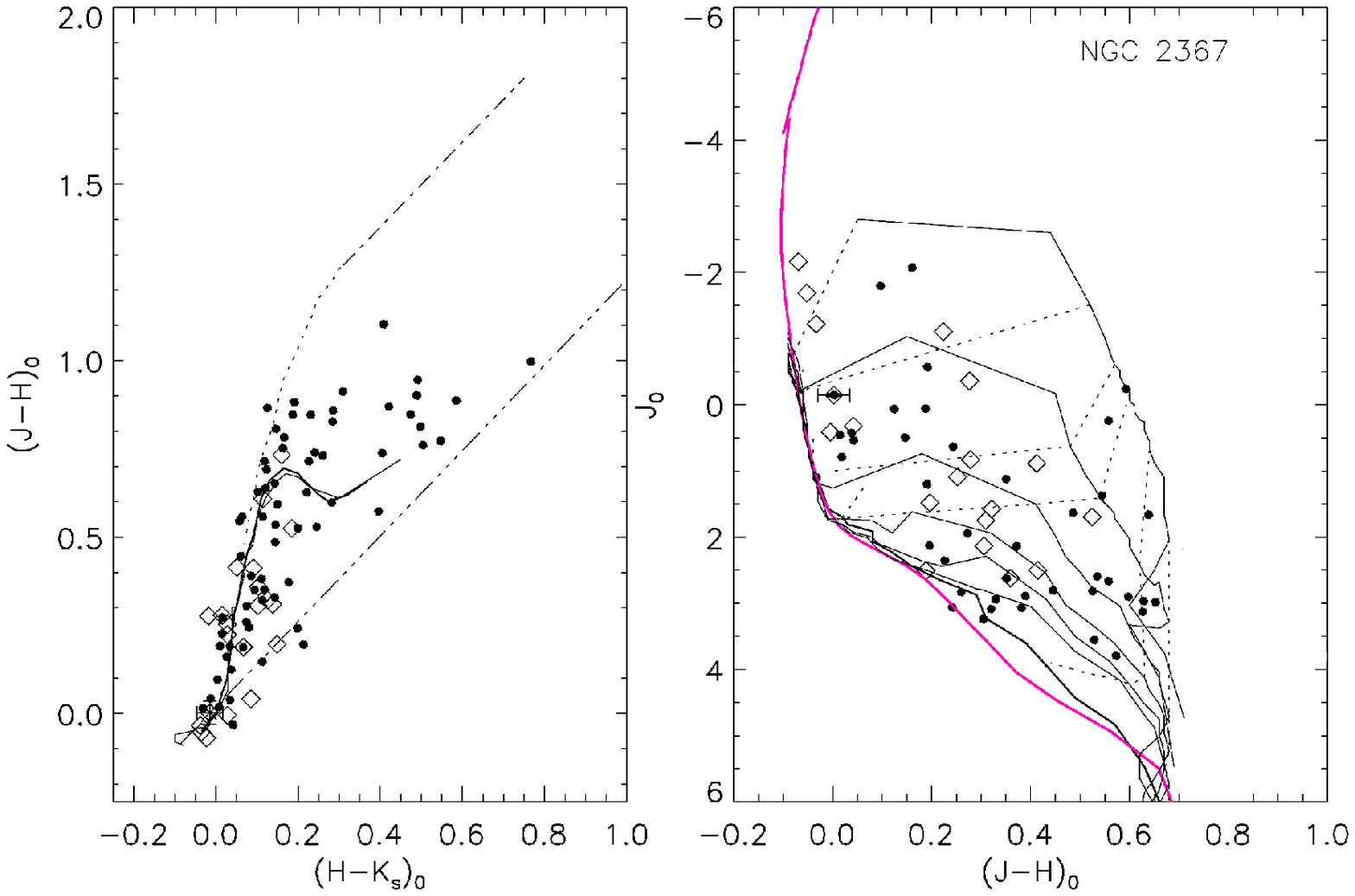}
\includegraphics[width=6.8cm]{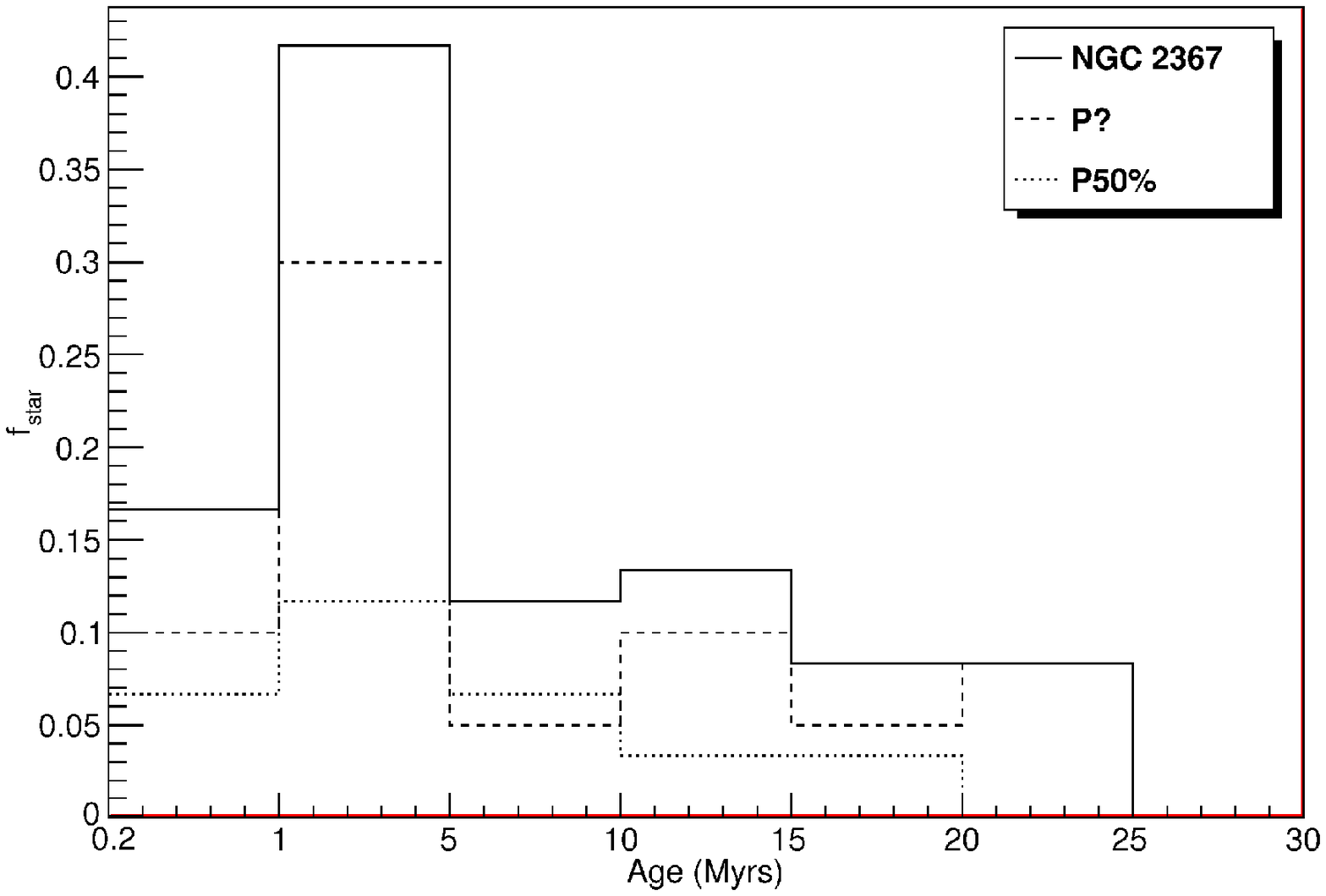}
\includegraphics[width=4.6cm]{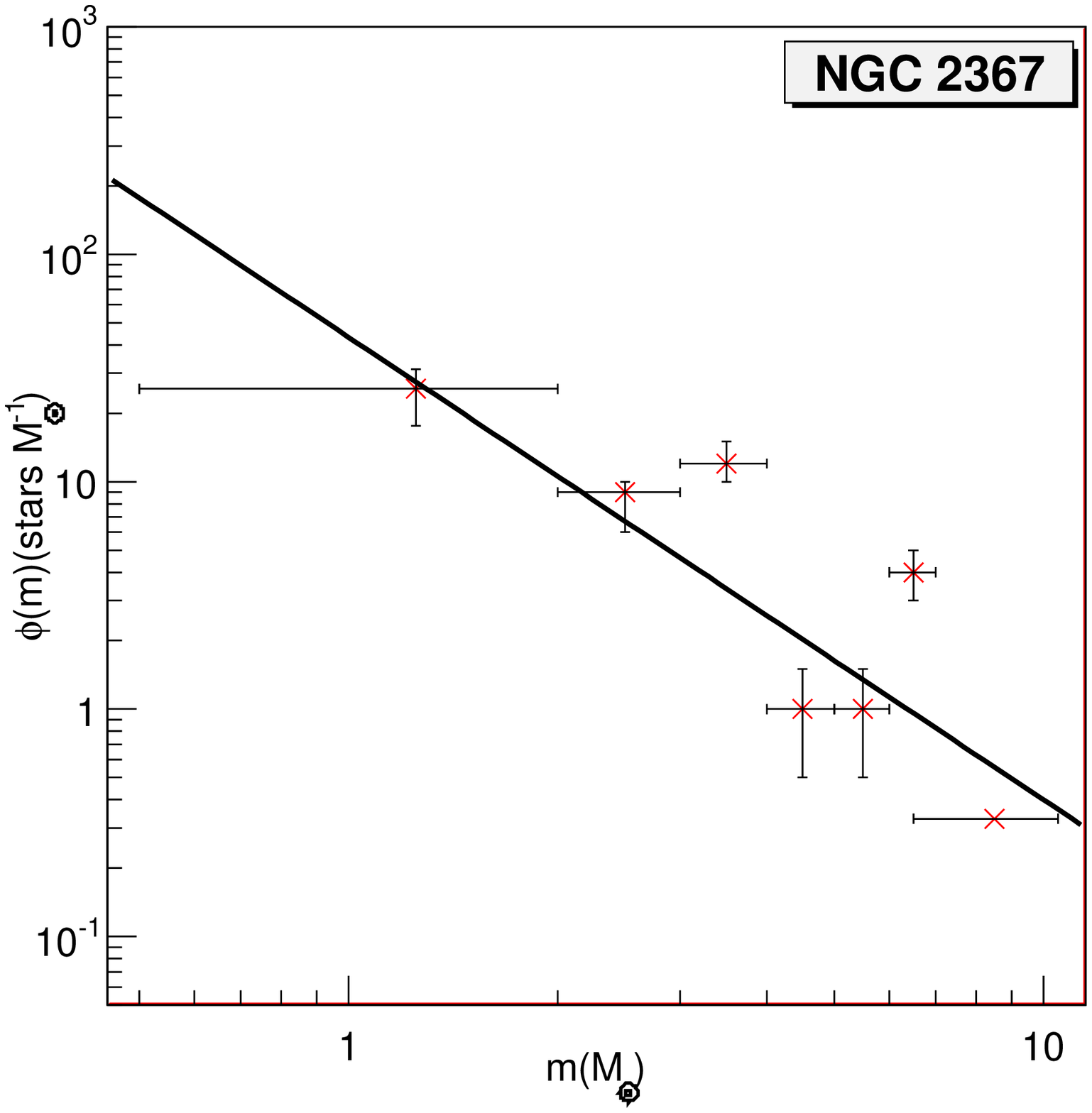}
\includegraphics[width=6.3cm]{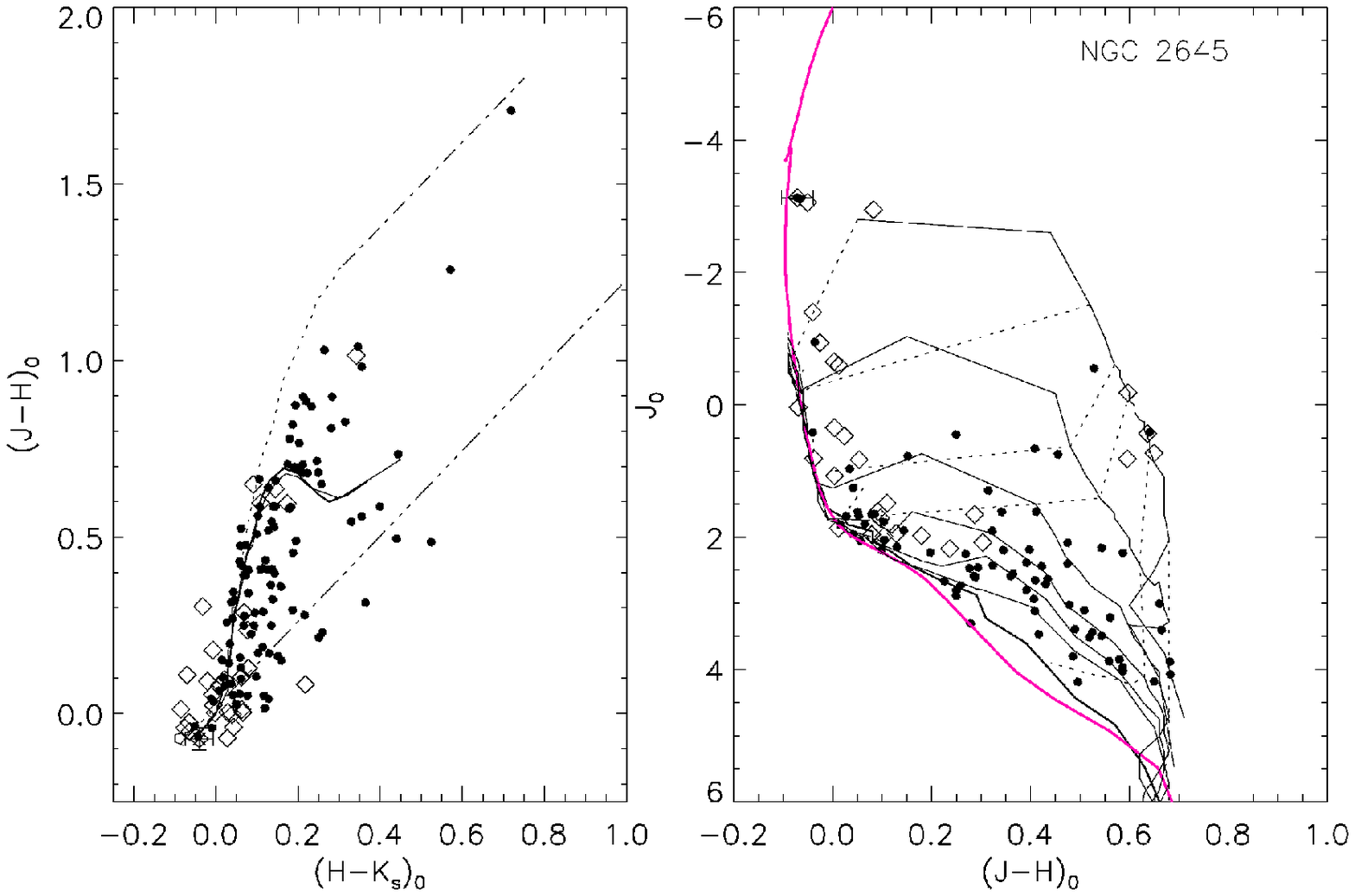}
\includegraphics[width=6.8cm]{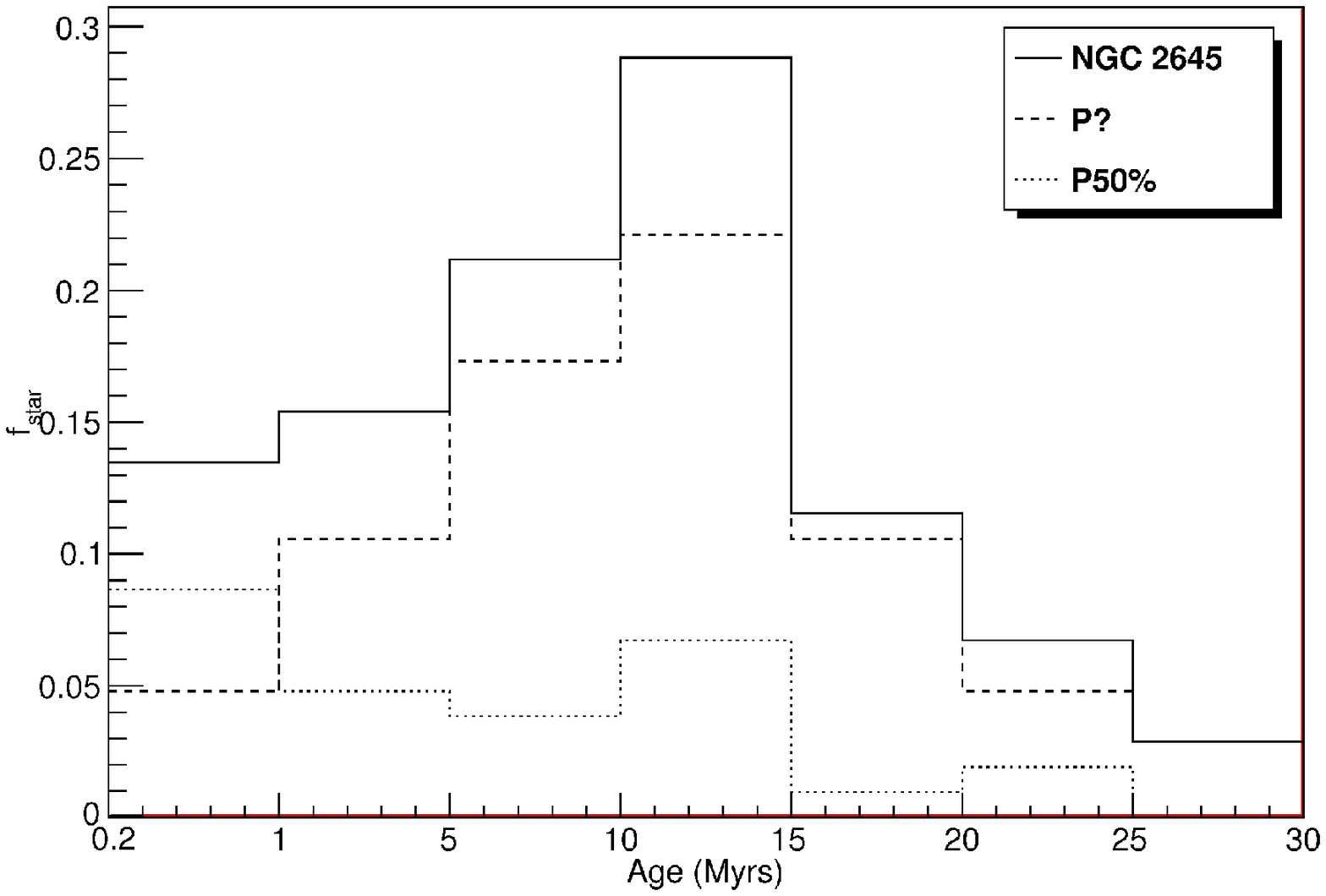}
\includegraphics[width=4.6cm]{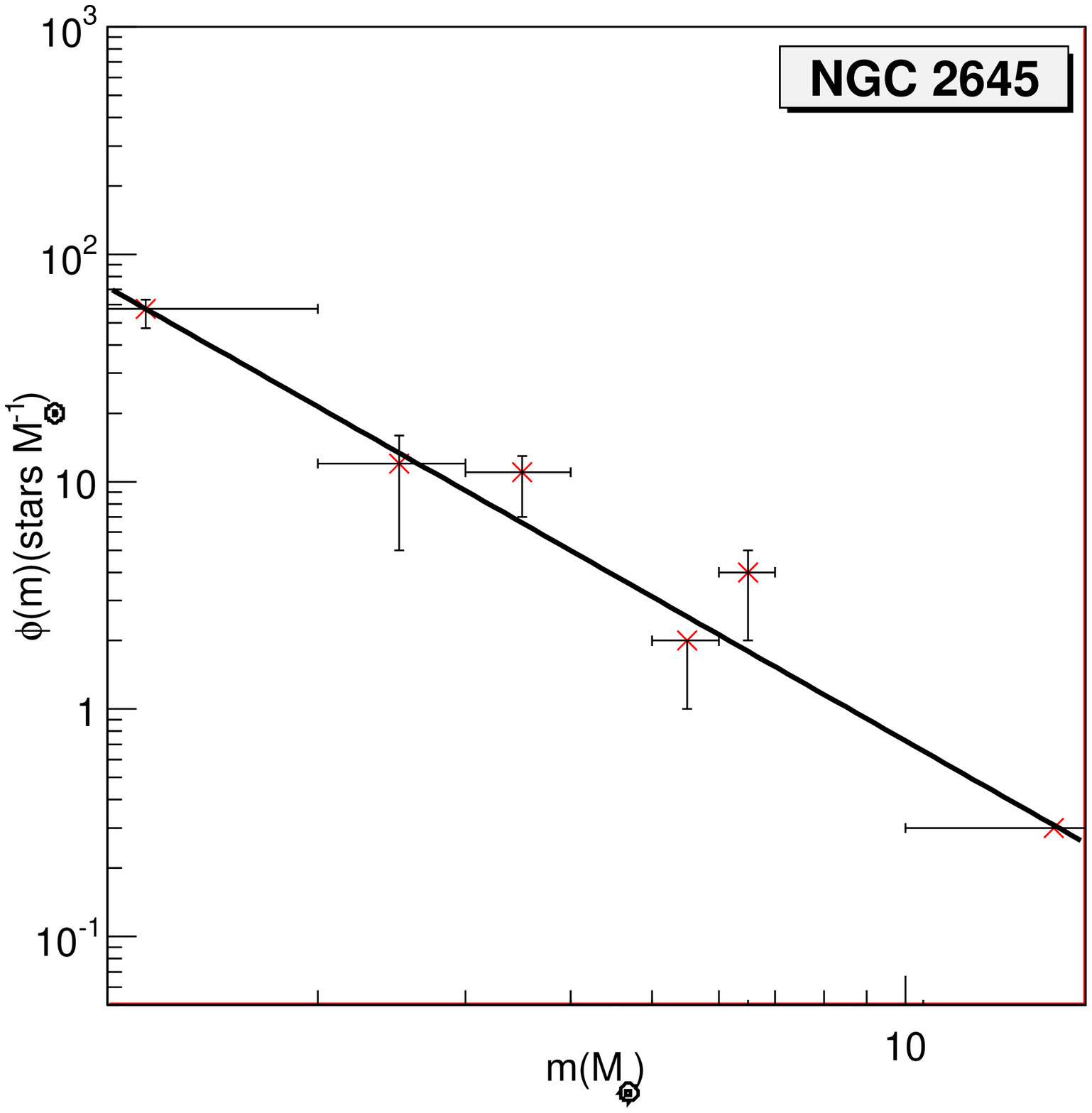}
\includegraphics[width=6.3cm]{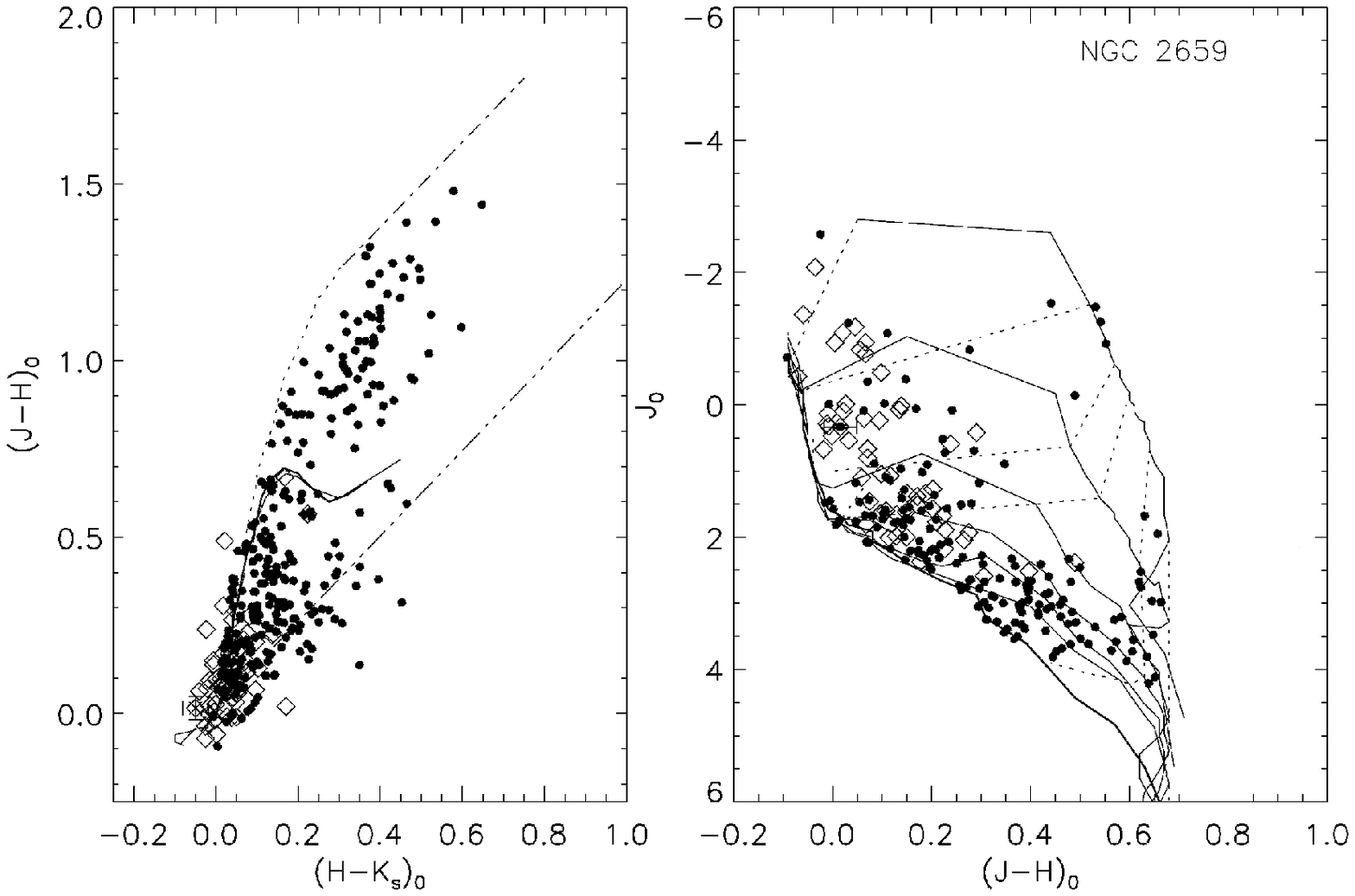}
\includegraphics[width=6.8cm]{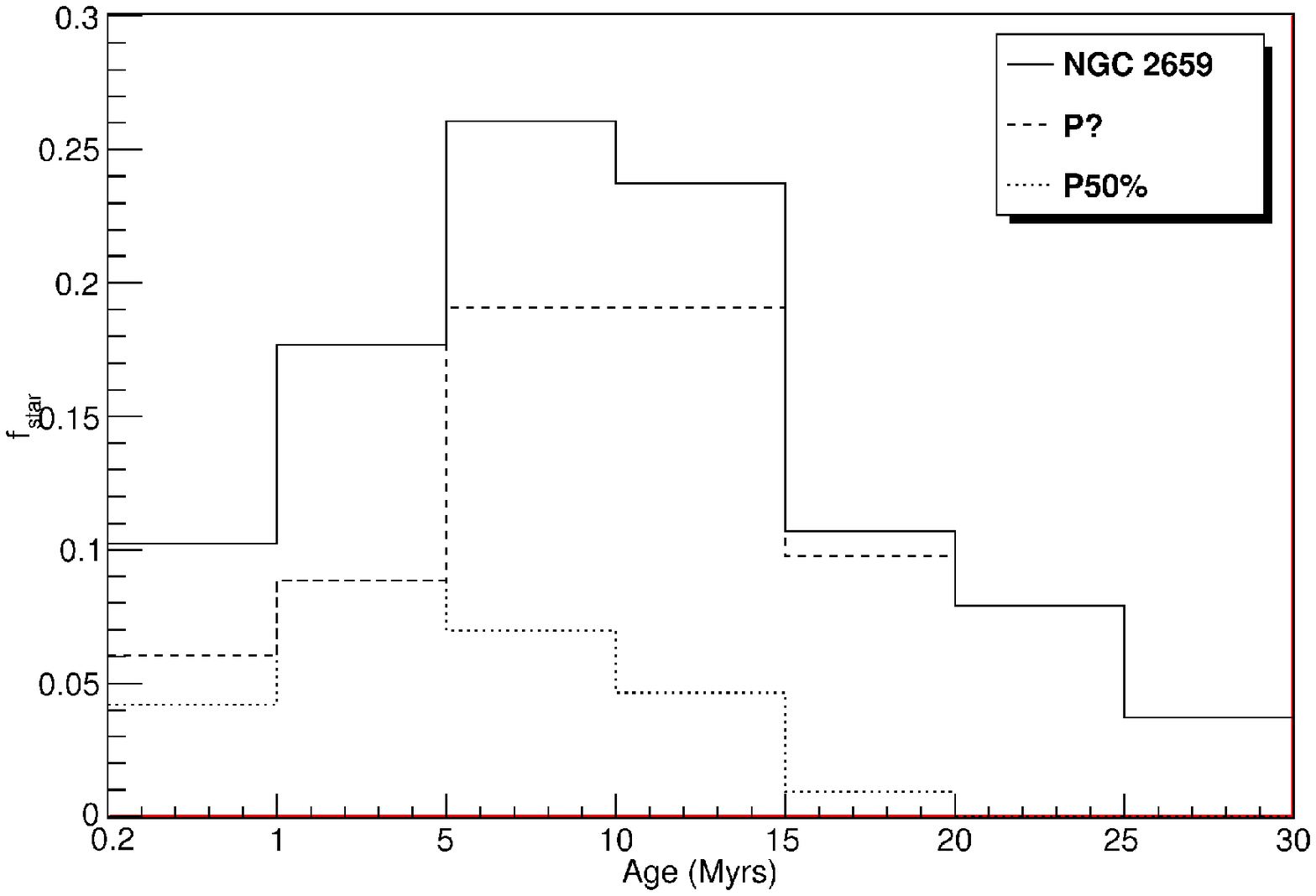}
\includegraphics[width=4.6cm]{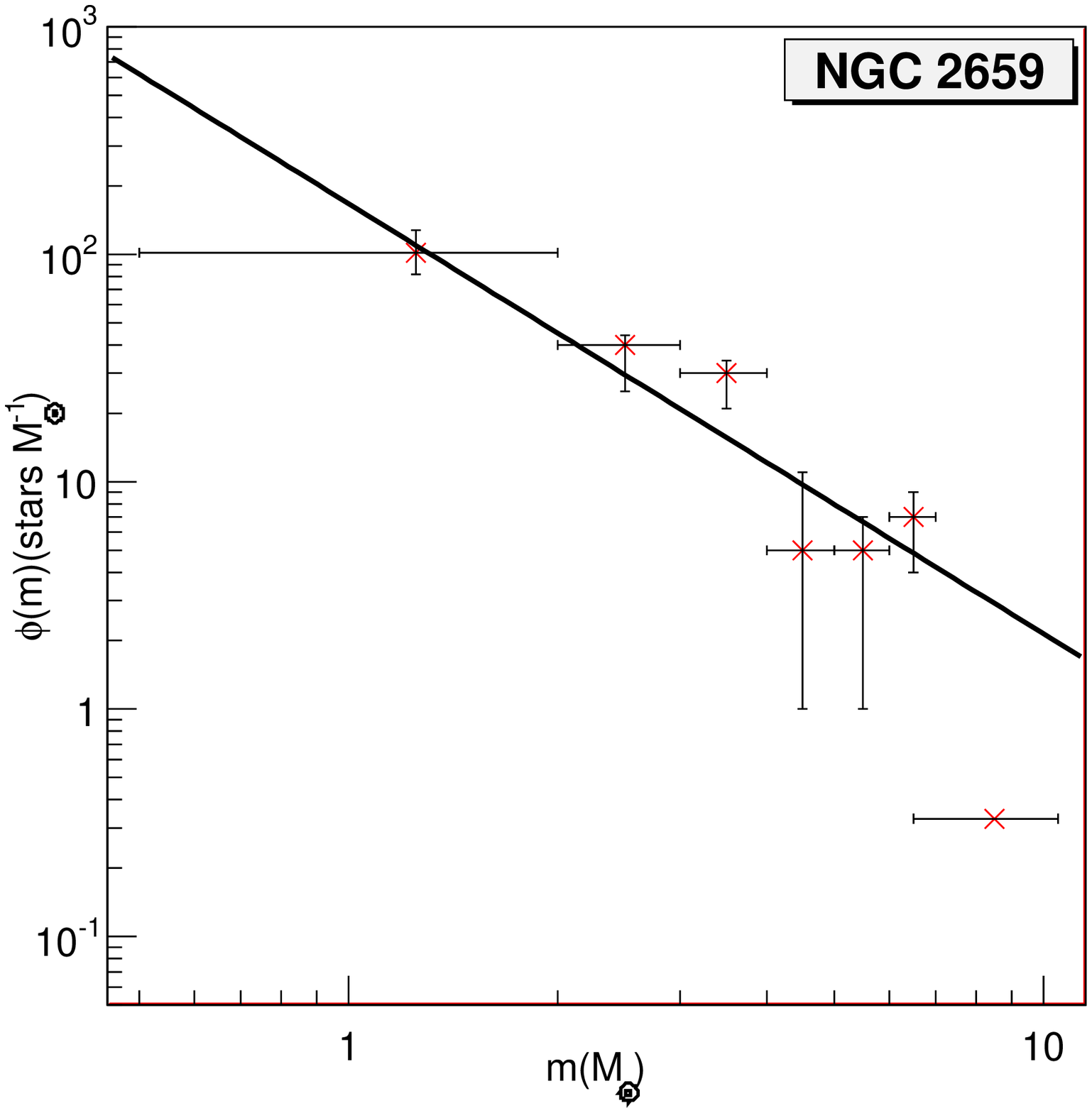}
\includegraphics[width=6.3cm]{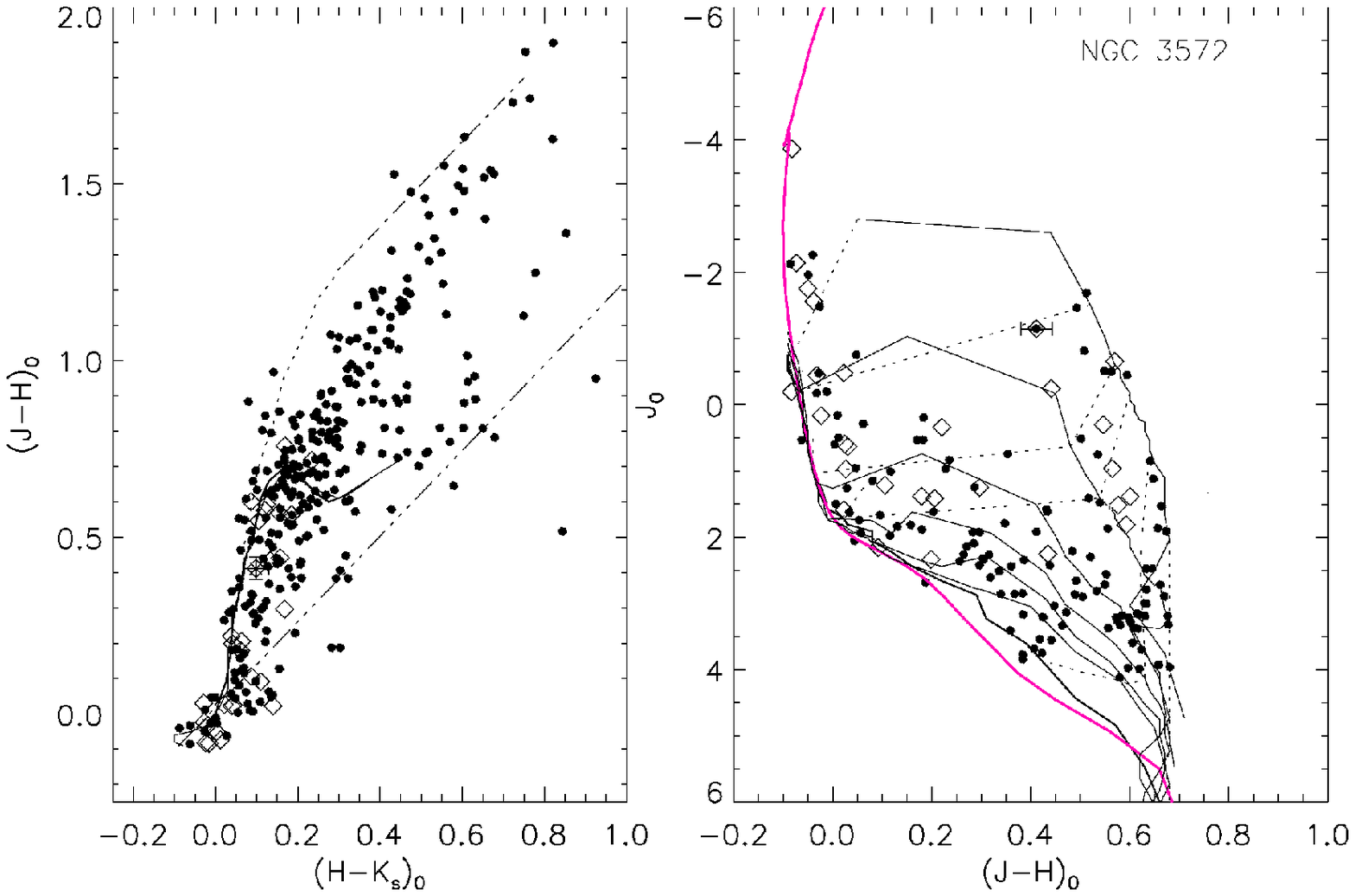}
\includegraphics[width=6.8cm]{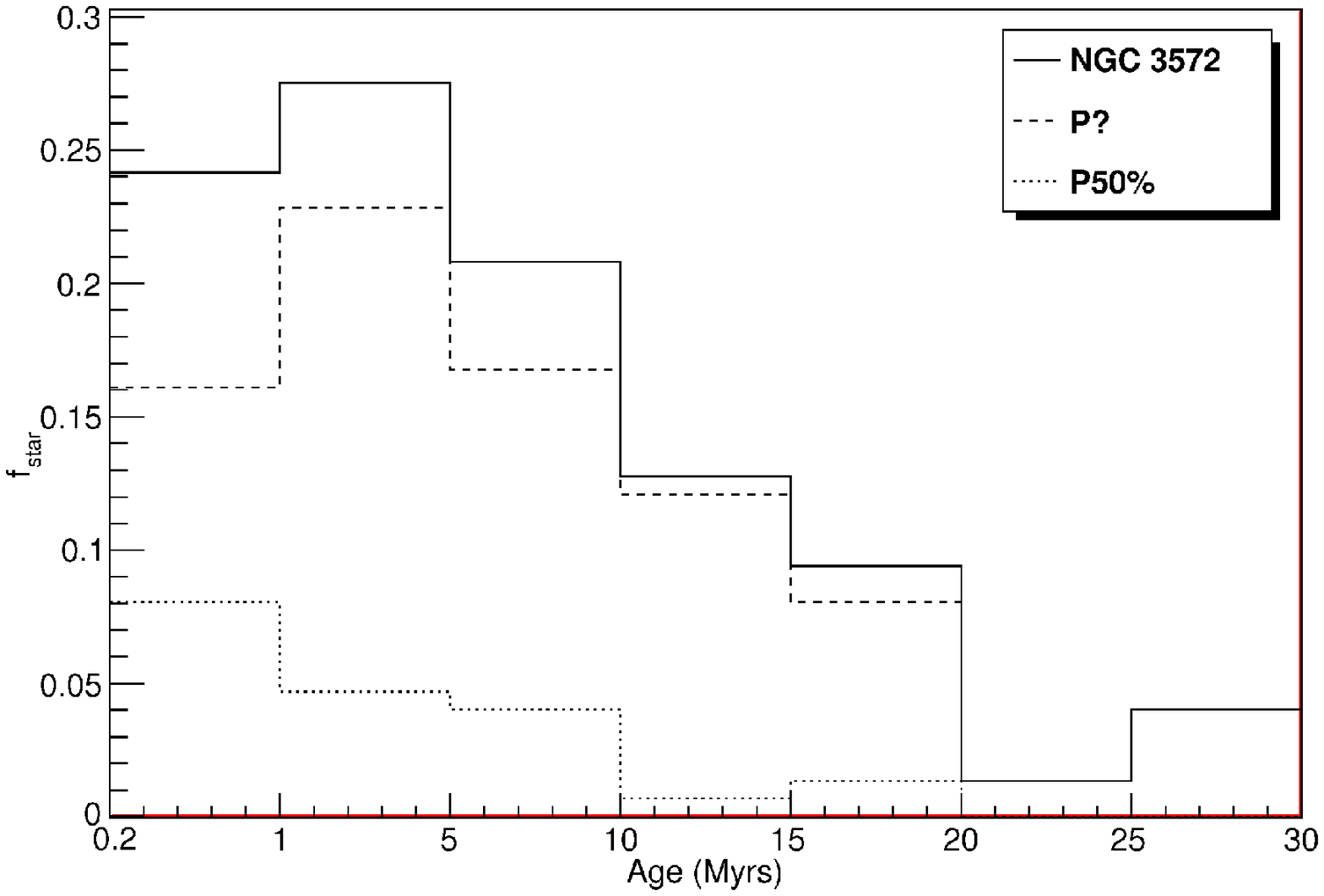}
\includegraphics[width=4.6cm]{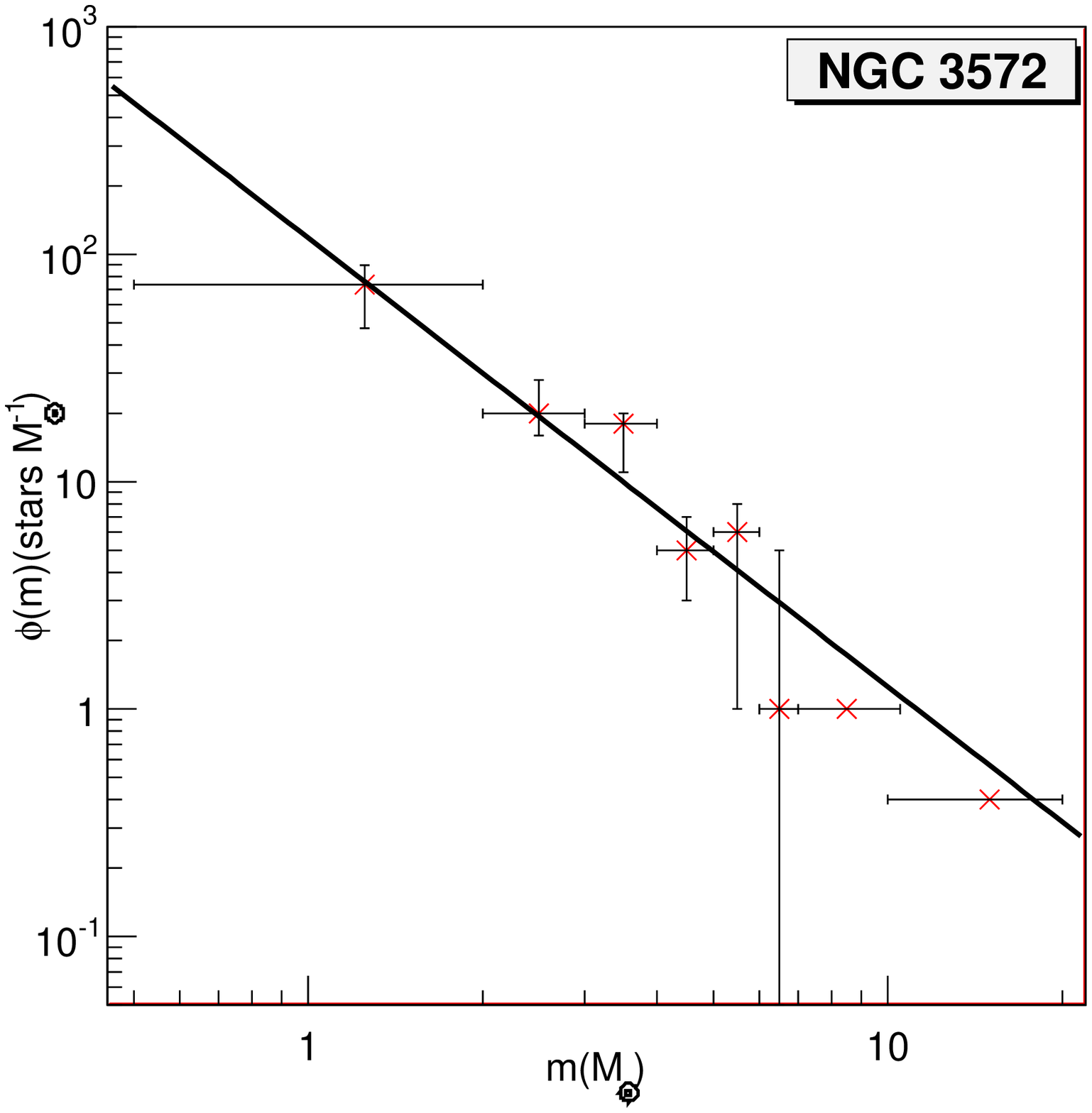}
\caption{The same as Fig A.3.}  
\label{hmass}
\end{center}
\end{figure*}

%%%%----------------------Fig. A6
\begin{figure*}[]
\begin{center}
\includegraphics[width=6.3cm]{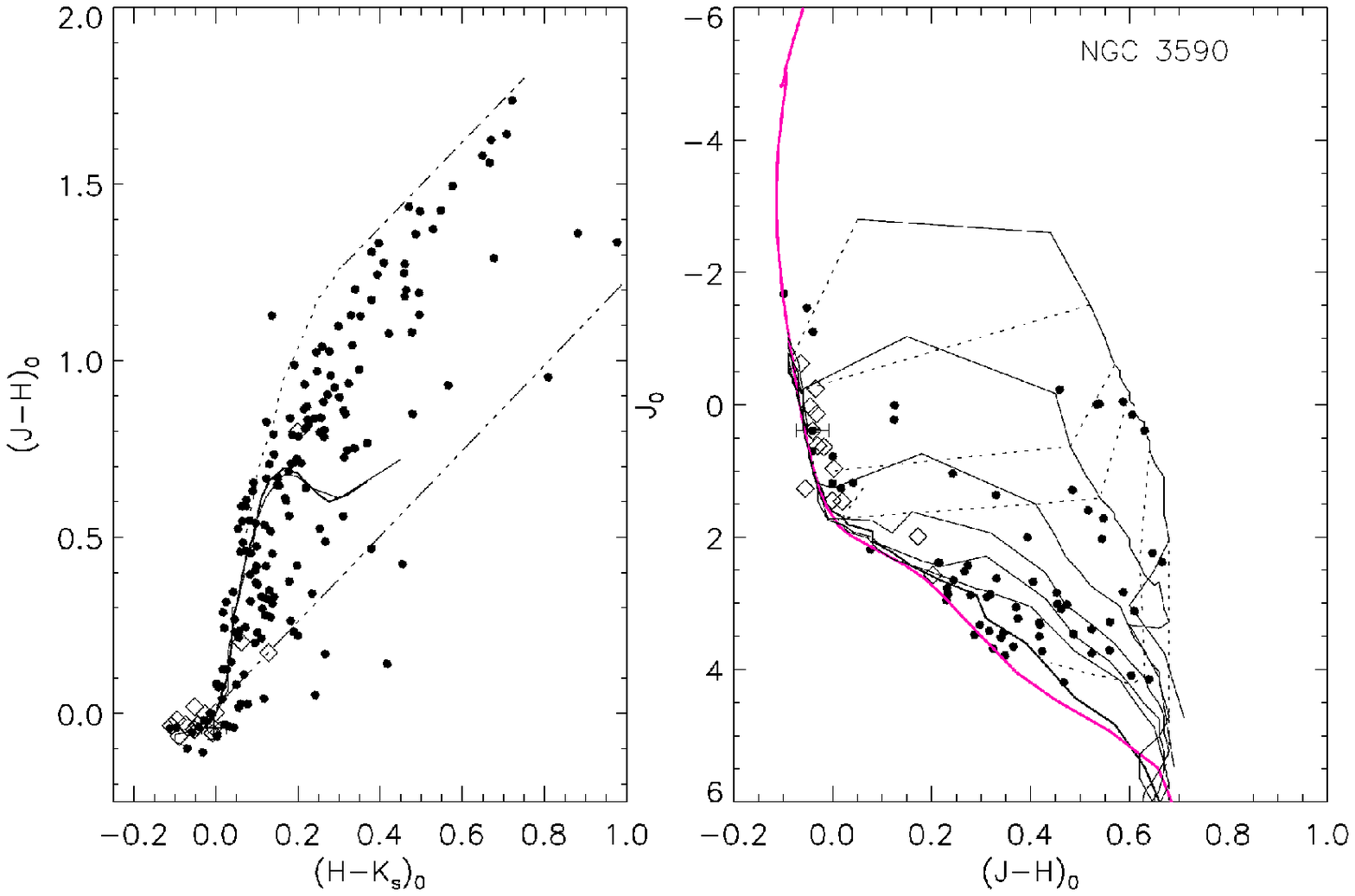}
\includegraphics[width=6.8cm]{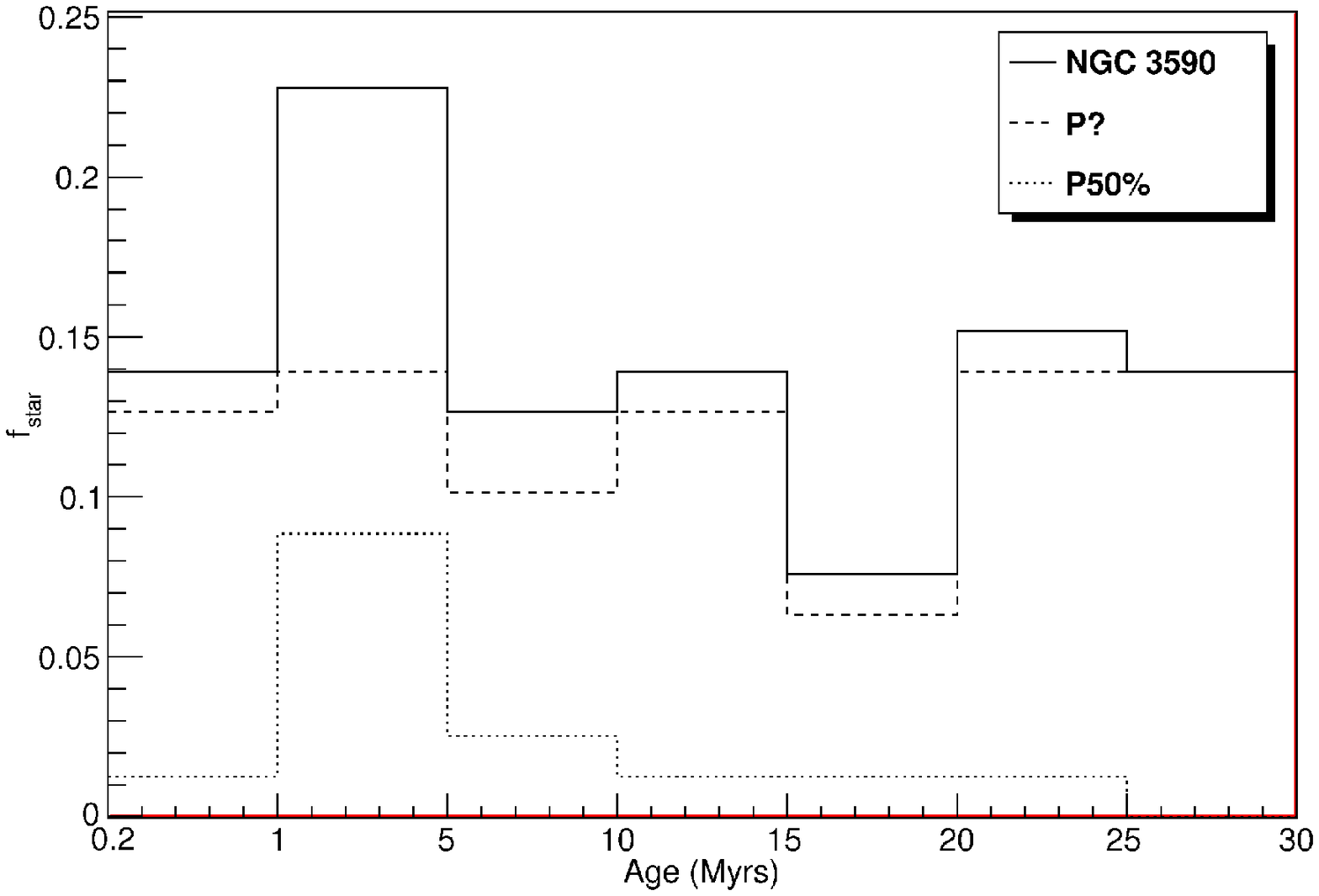}
\includegraphics[width=4.6cm]{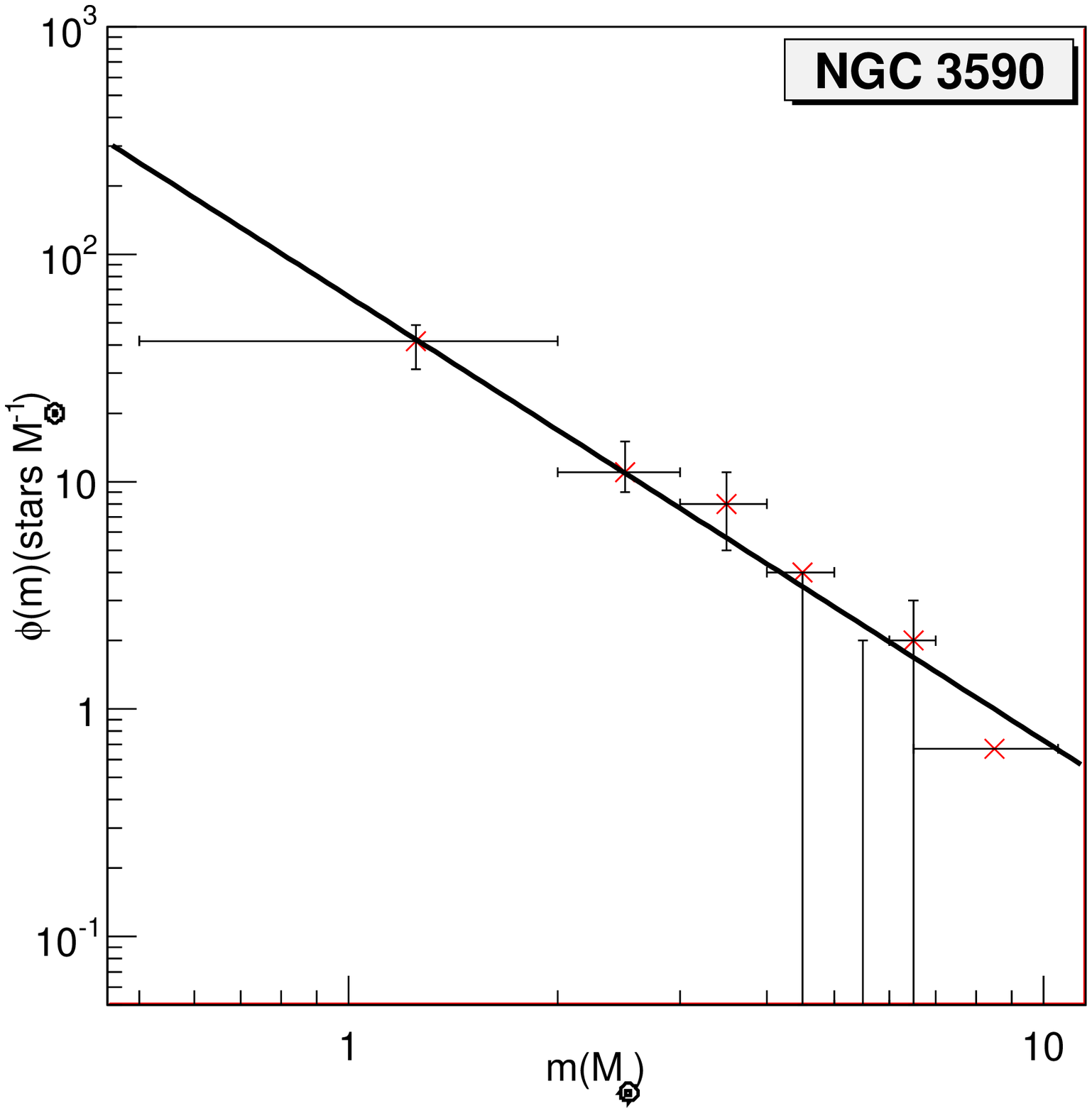}
\includegraphics[width=6.3cm]{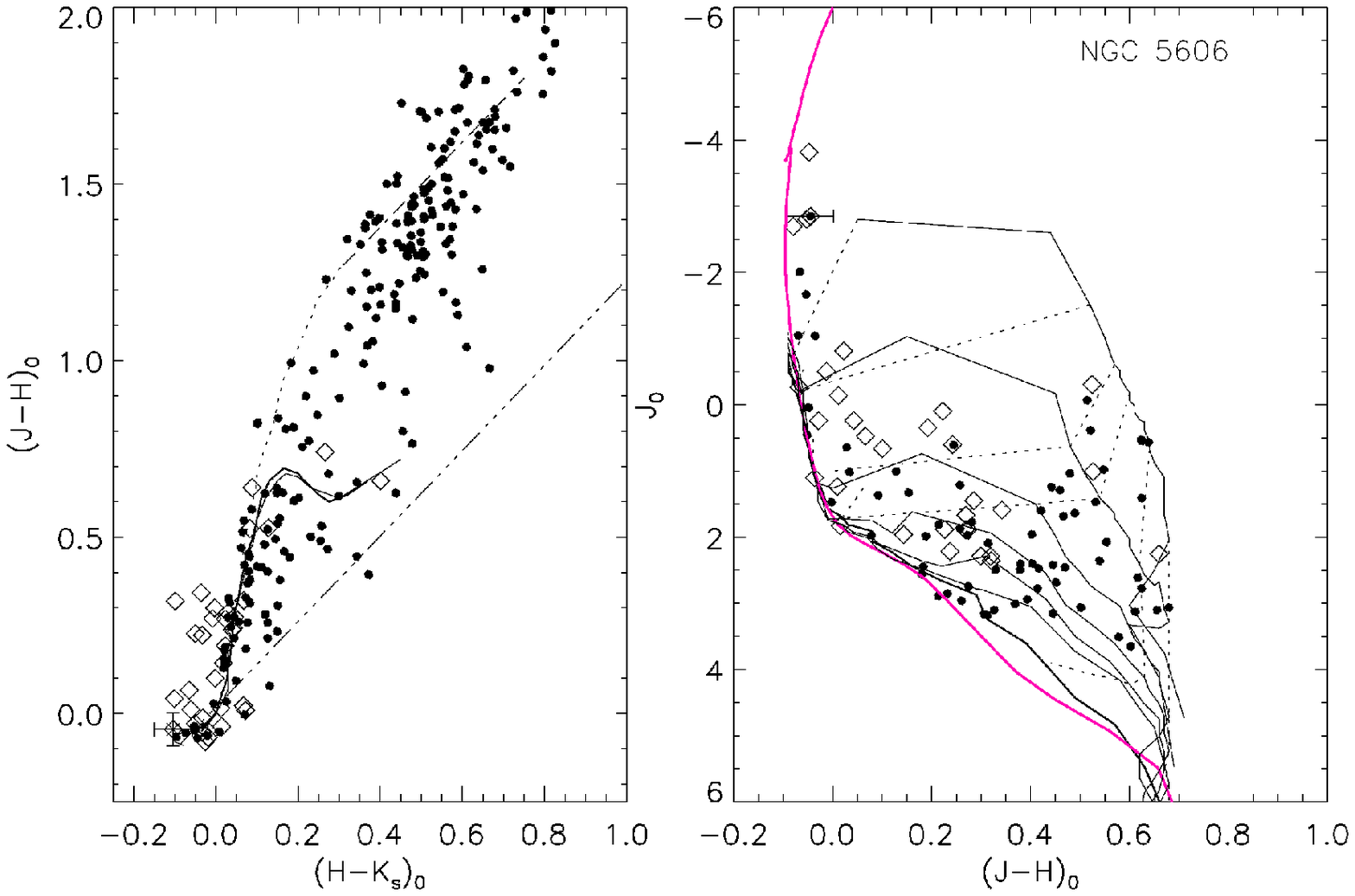}
\includegraphics[width=6.8cm]{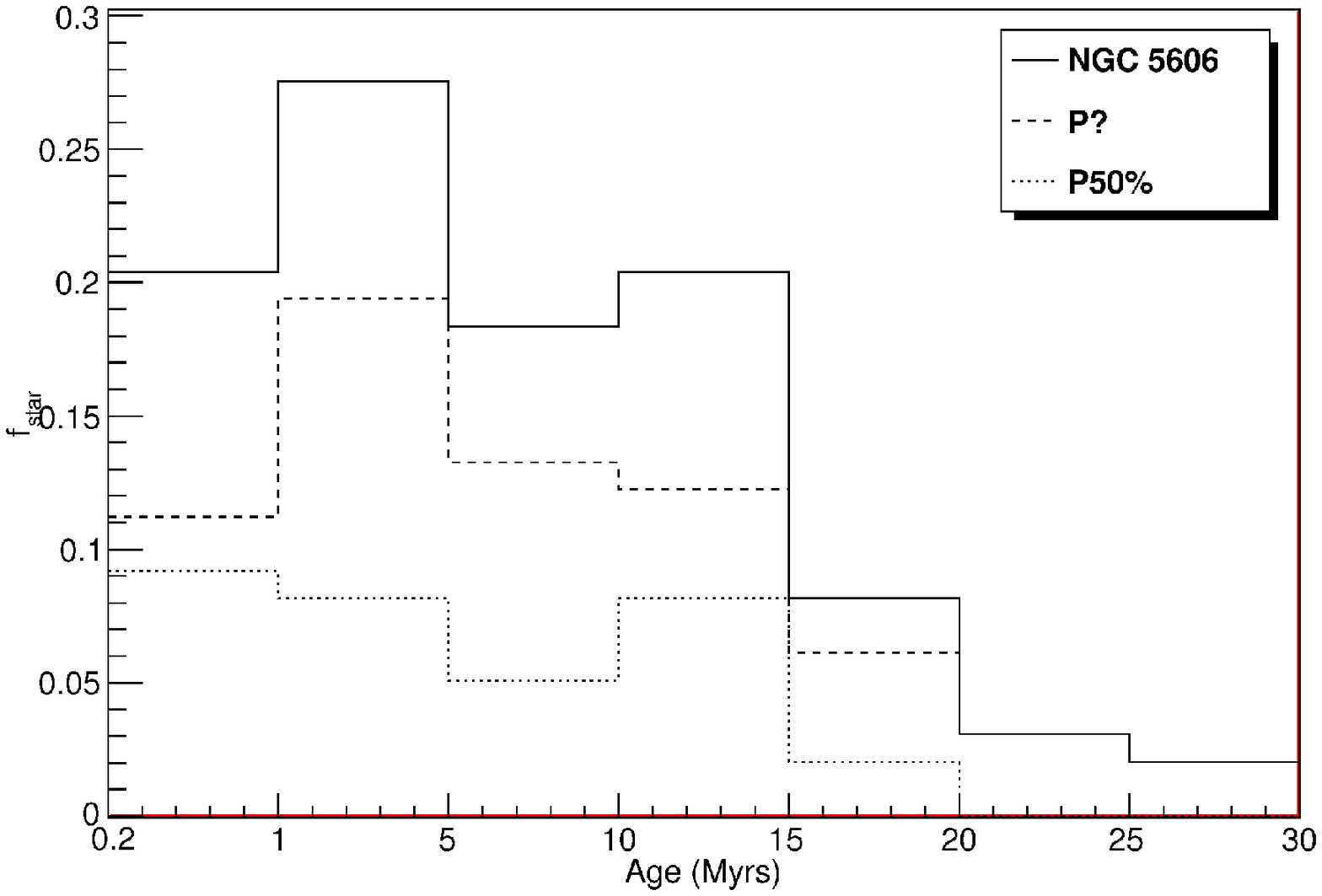}
\includegraphics[width=4.6cm]{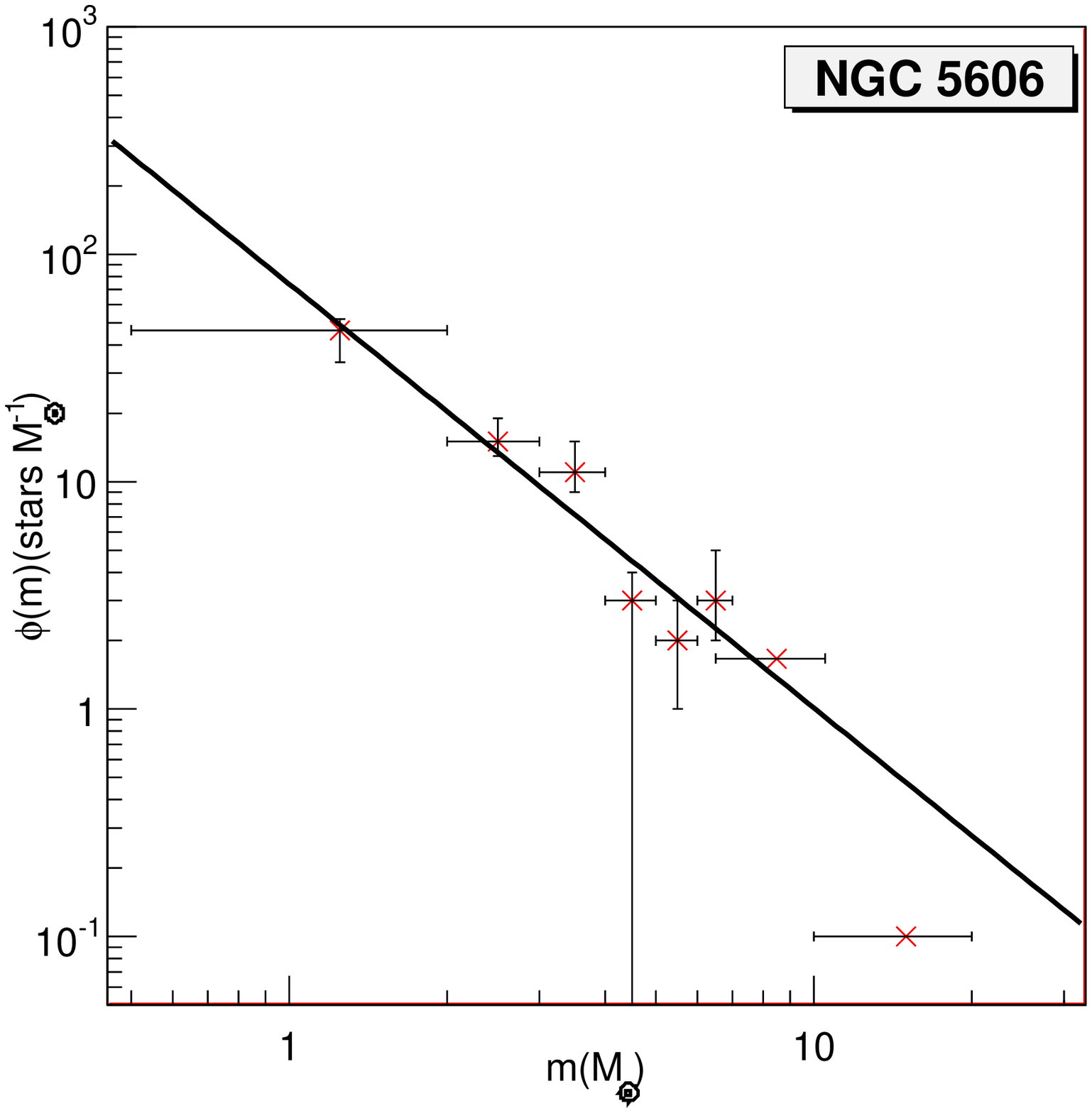}
\includegraphics[width=6.3cm]{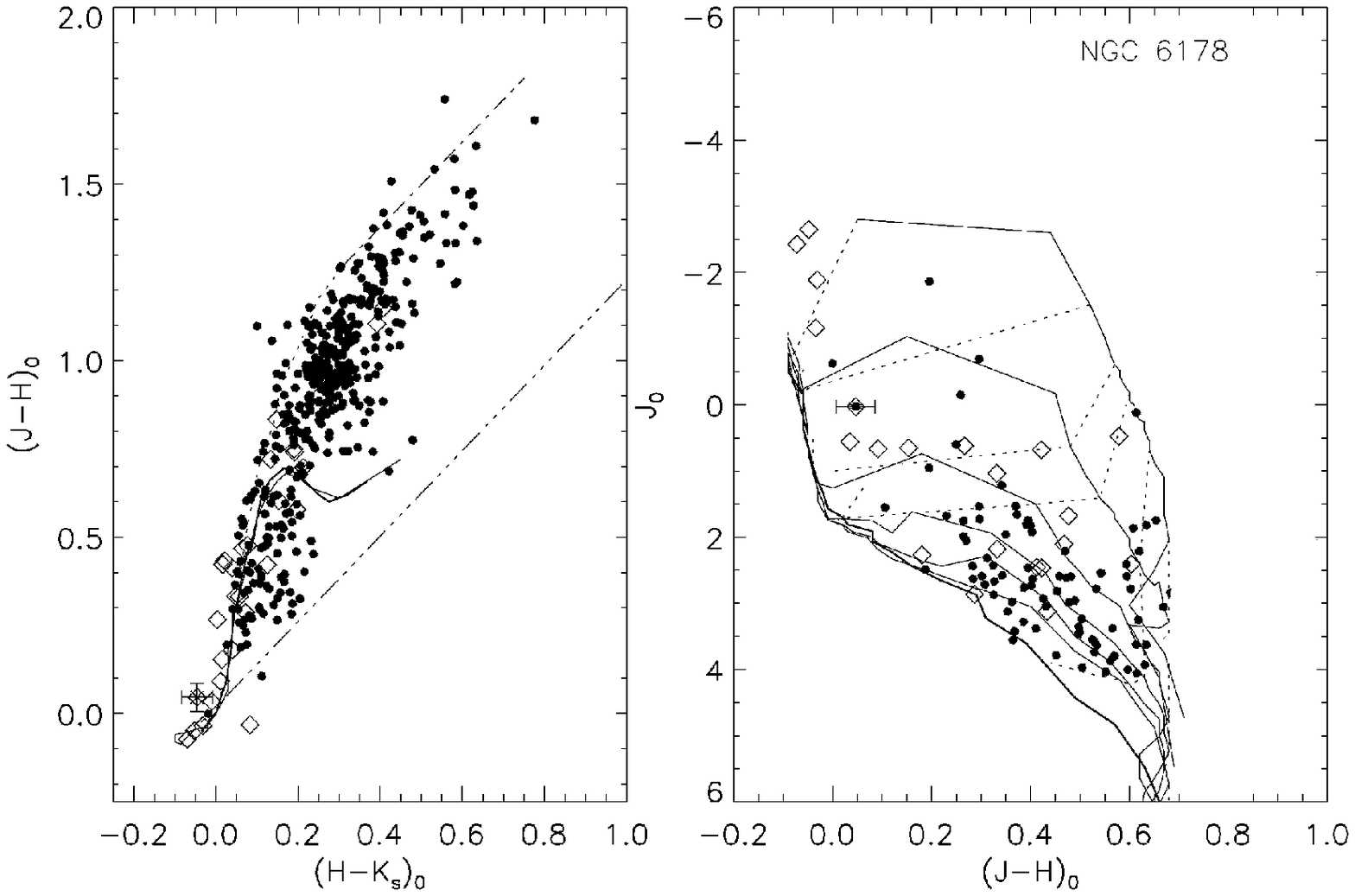}
\includegraphics[width=6.8cm]{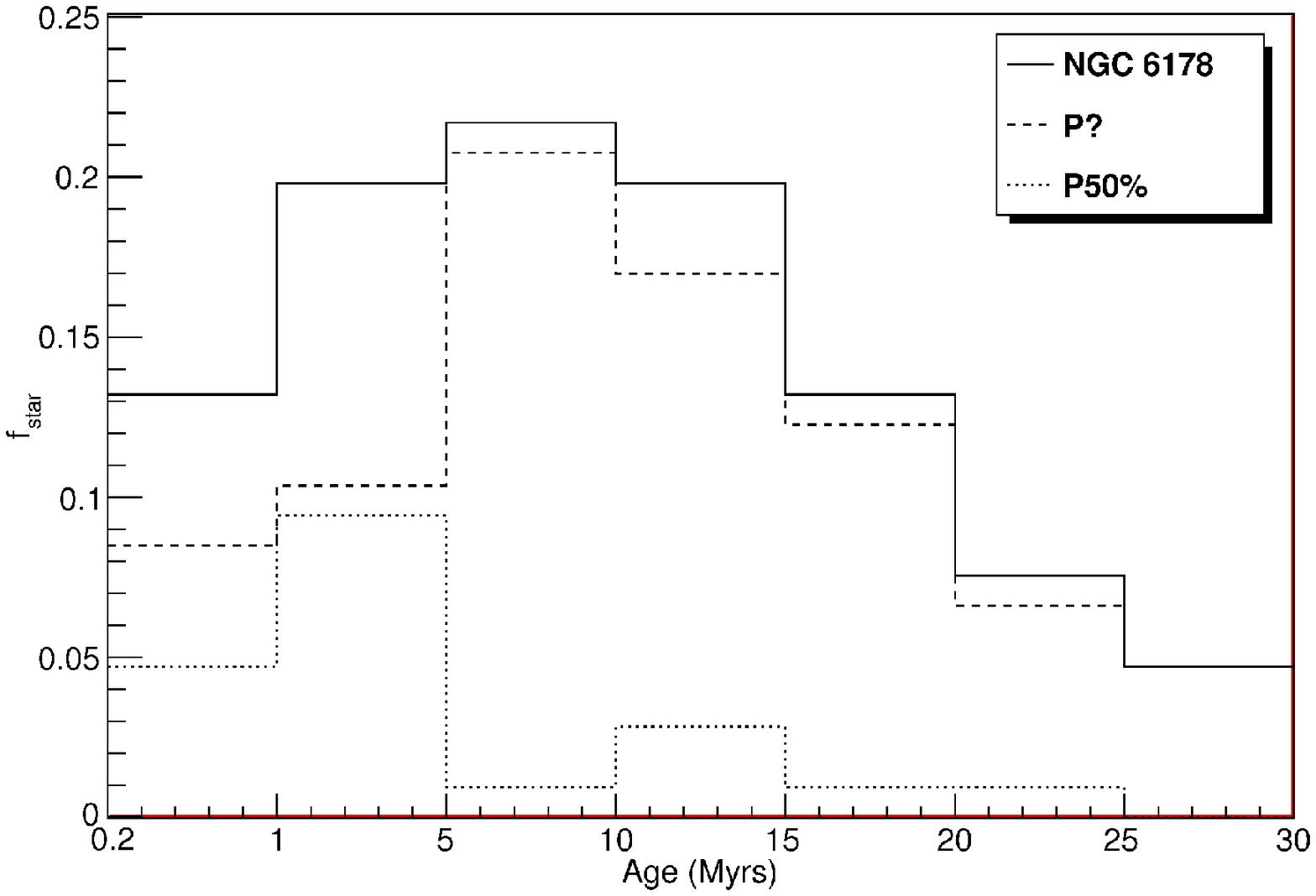}
\includegraphics[width=4.6cm]{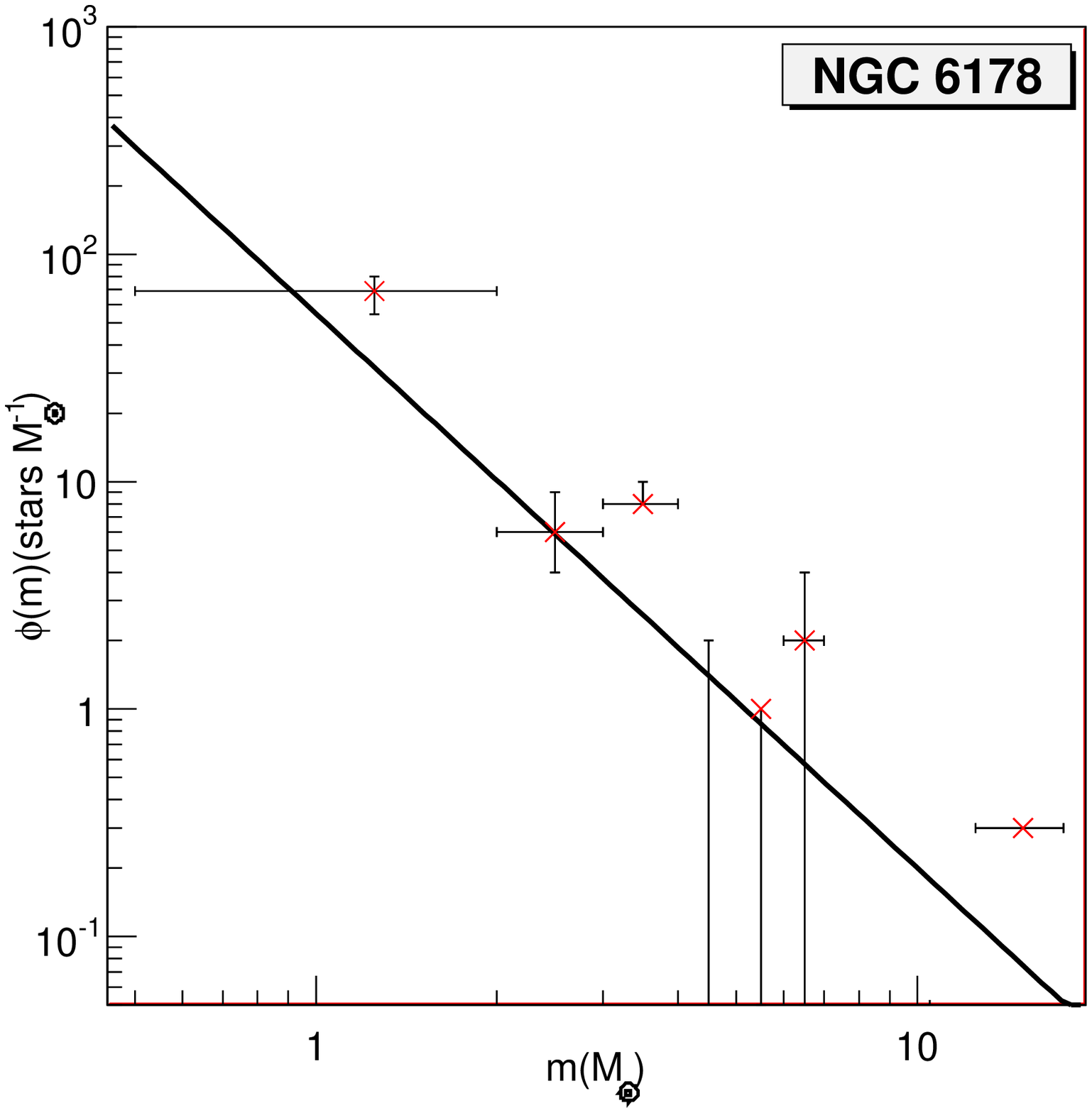}
\includegraphics[width=6.3cm]{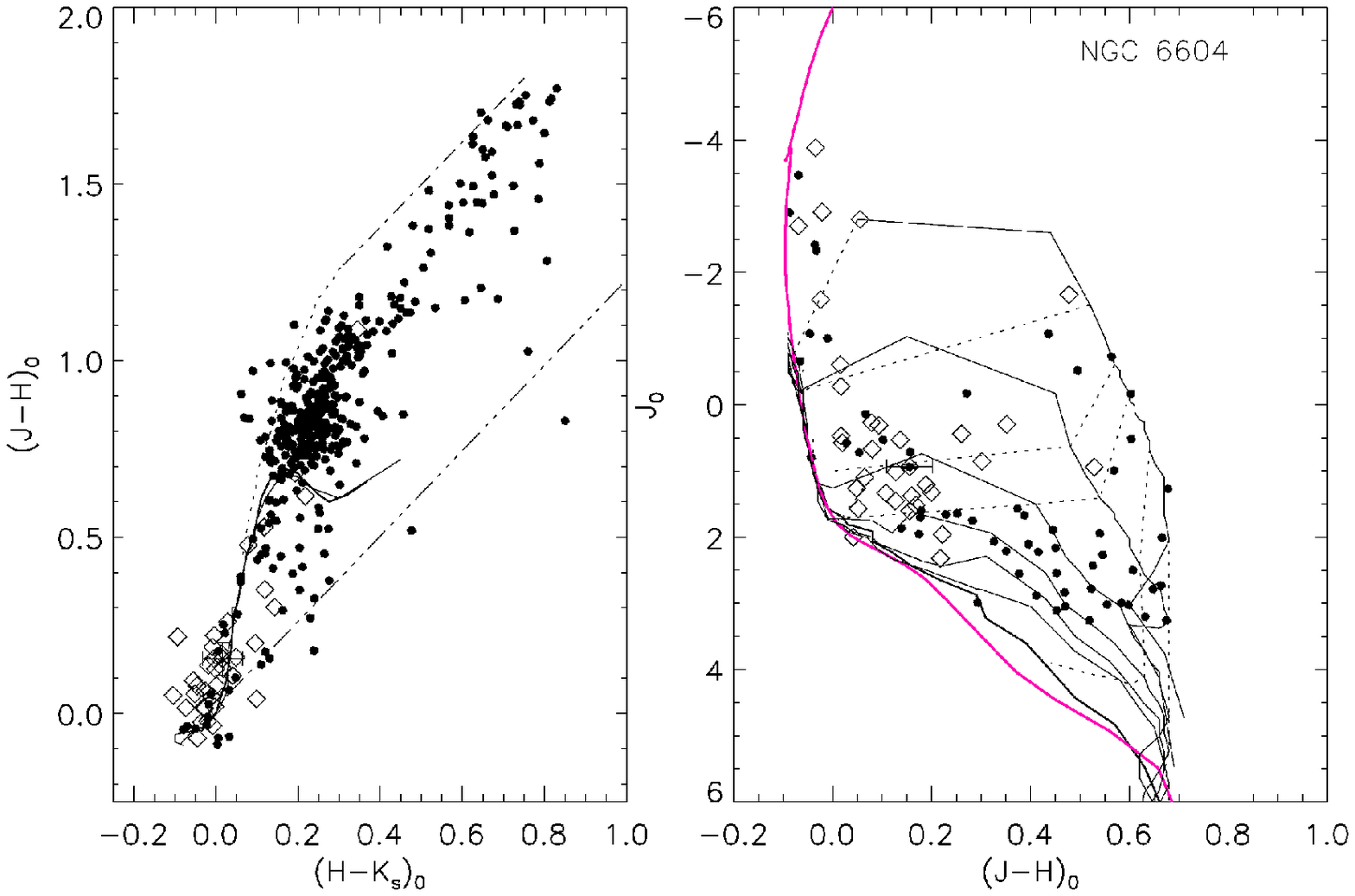}
\includegraphics[width=6.8cm]{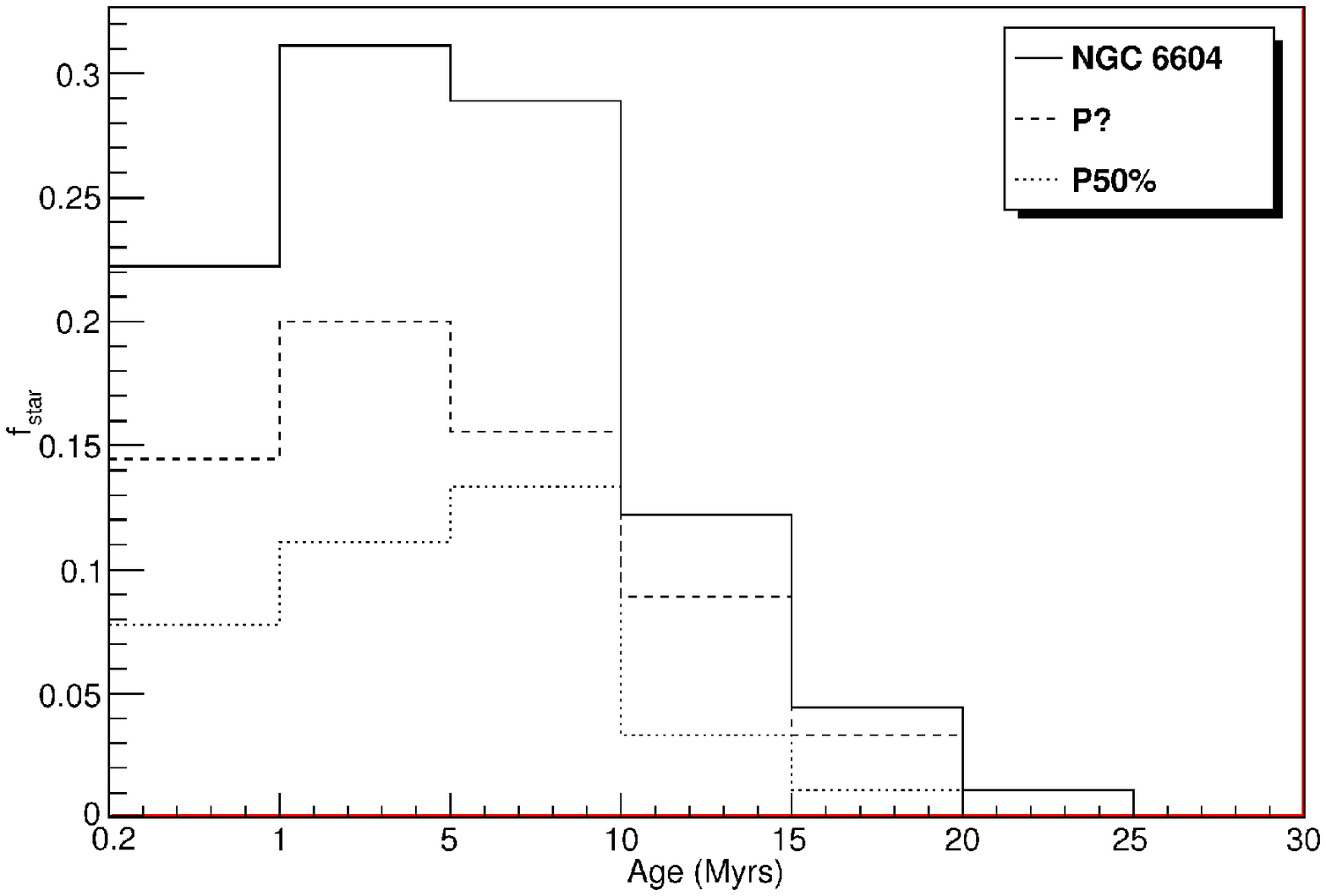}
\includegraphics[width=4.6cm]{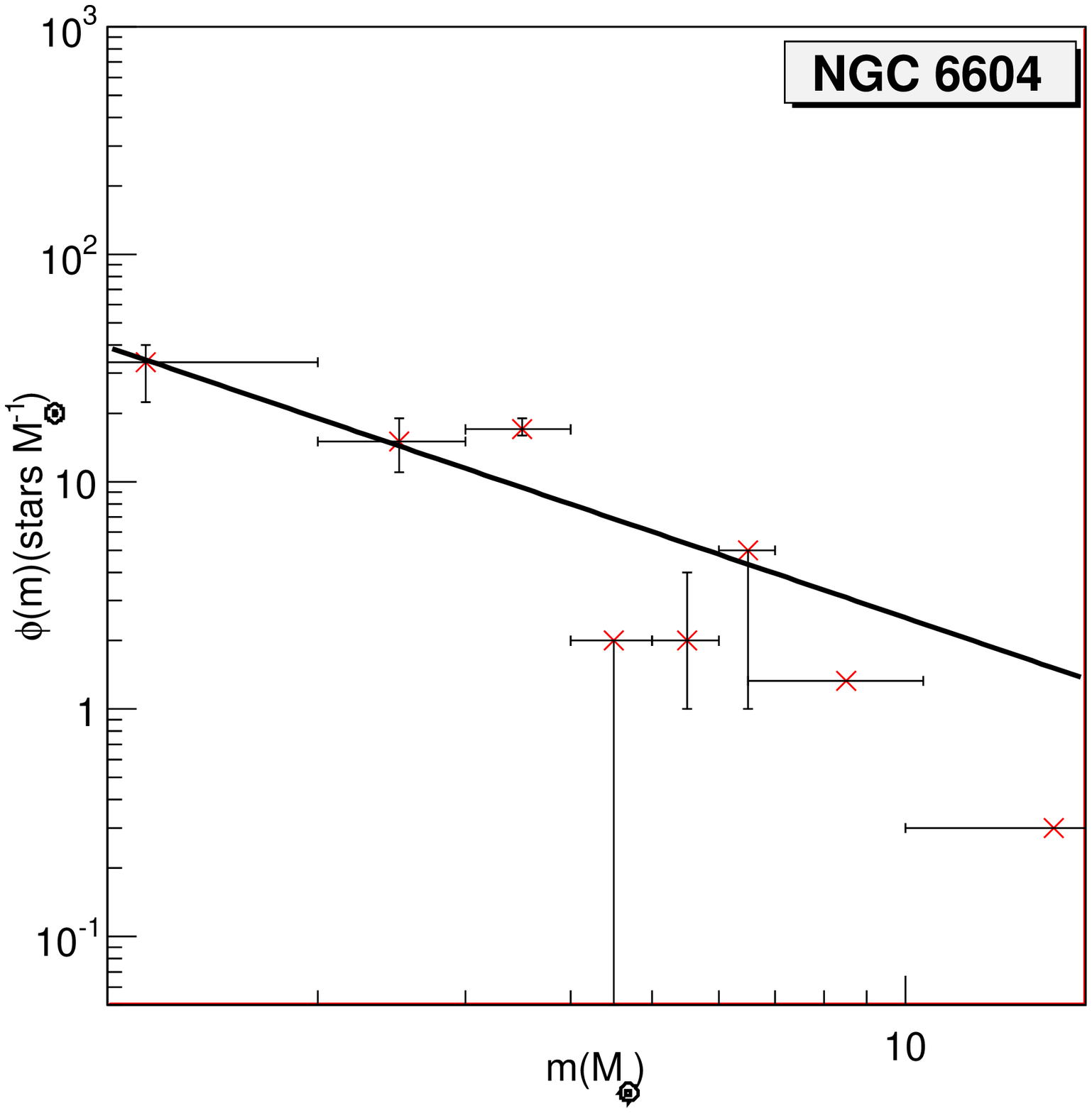}
\includegraphics[width=6.3cm]{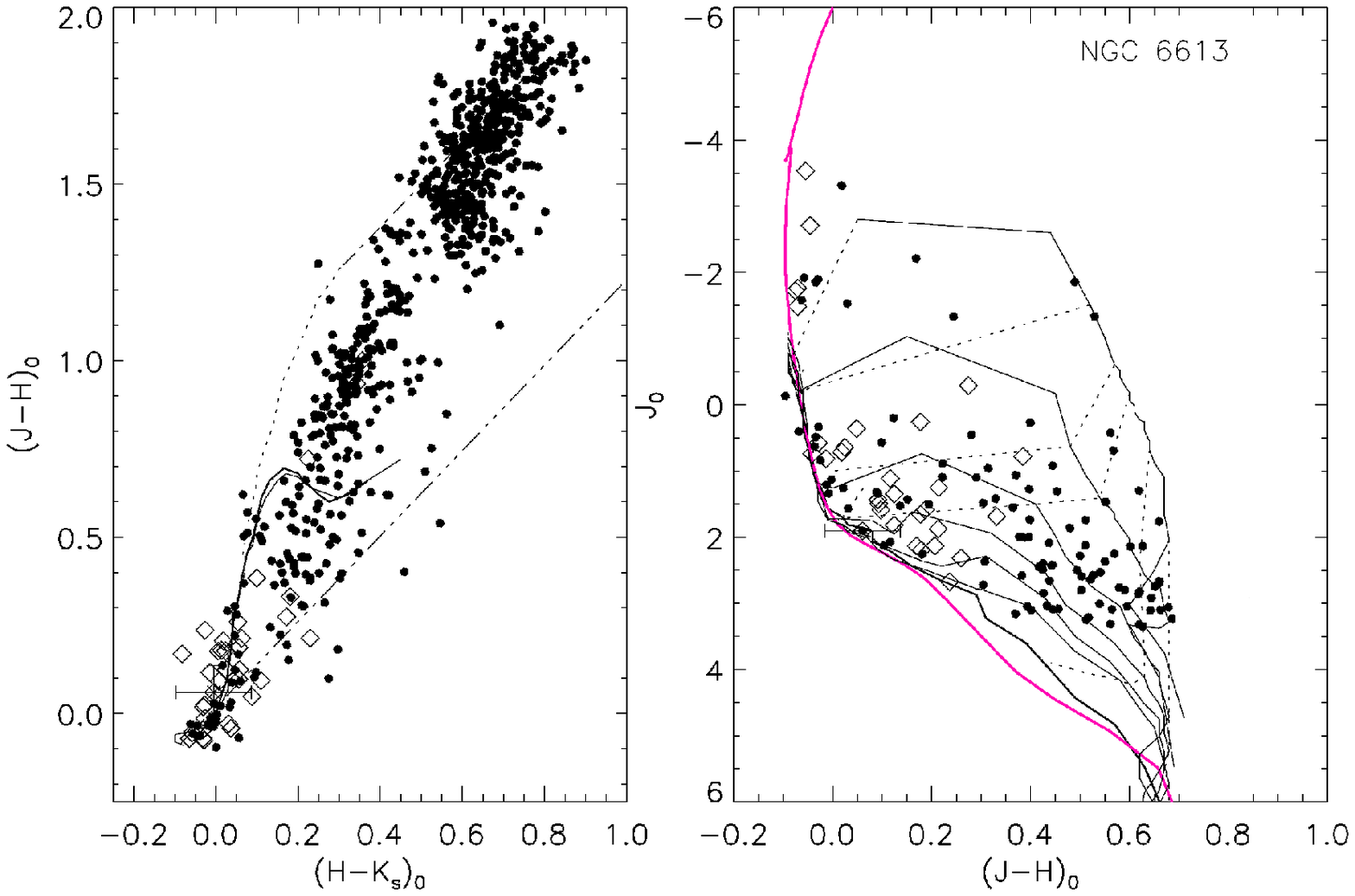}
\includegraphics[width=6.8cm]{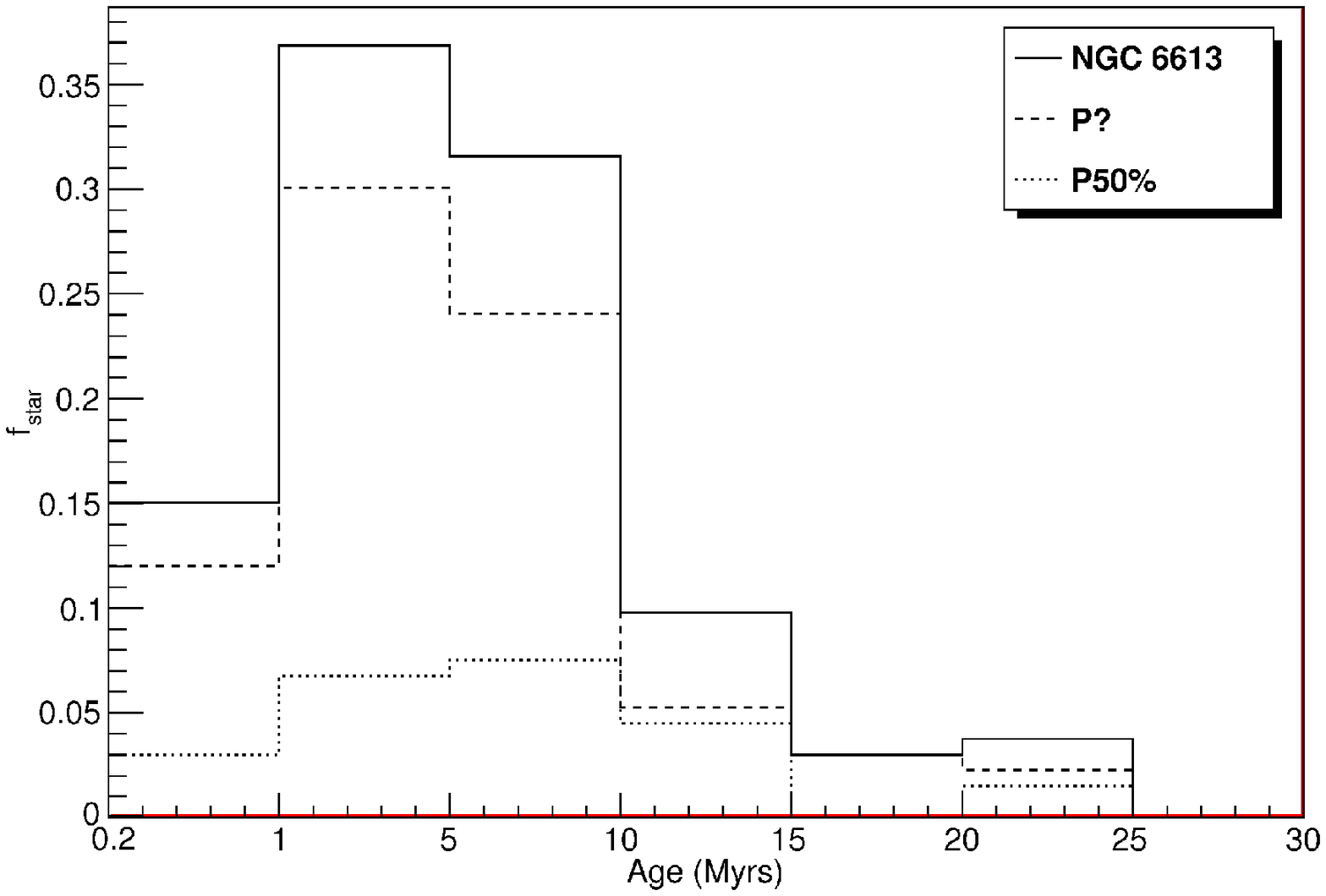}
\includegraphics[width=4.6cm]{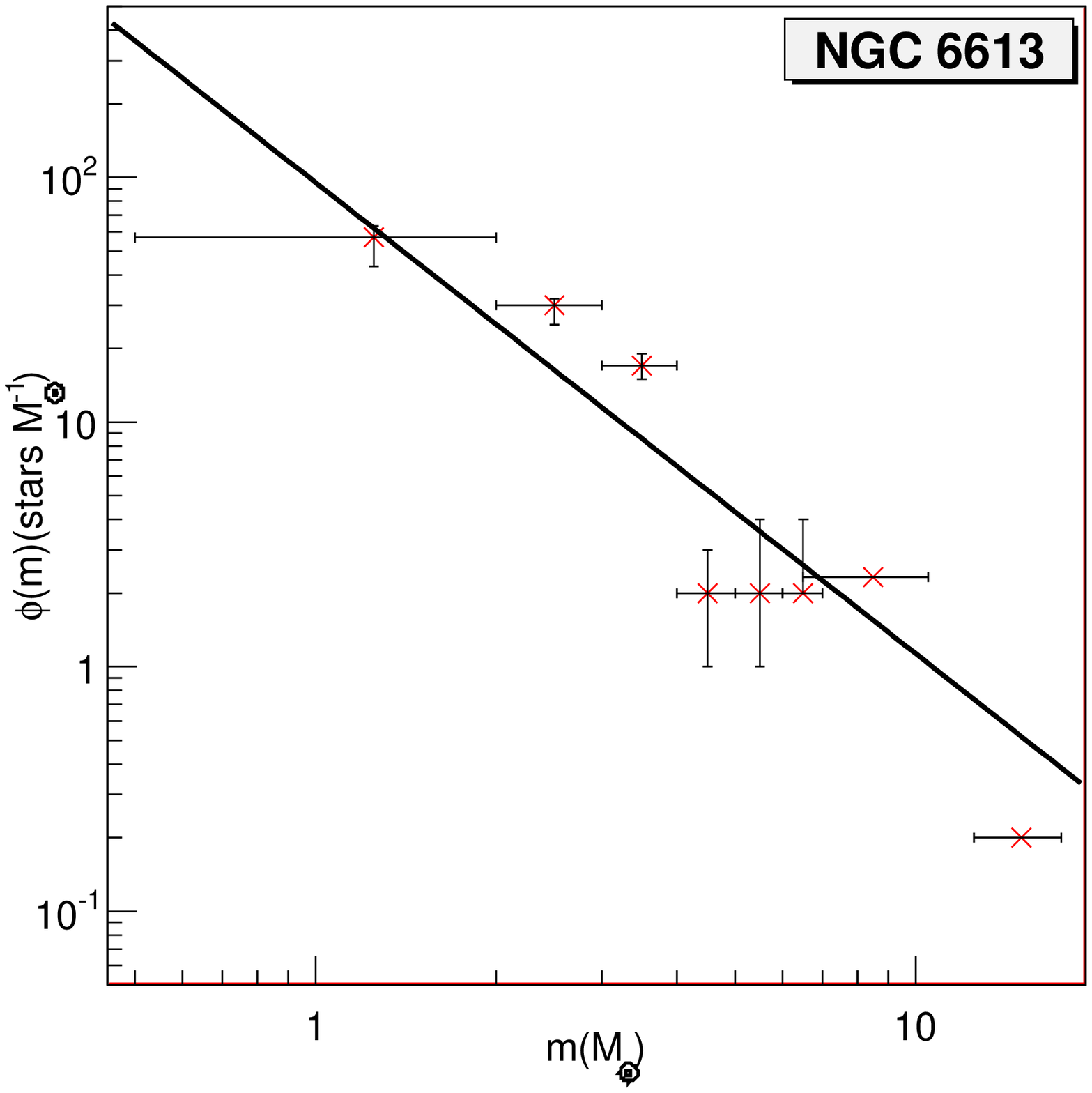}
\caption{The same as Fig A.3.}
\label{hmass}
\end{center}
\end{figure*}

%%%%----------------------Fig. A6
\begin{figure*}[]
\begin{center}
\includegraphics[width=6.3cm]{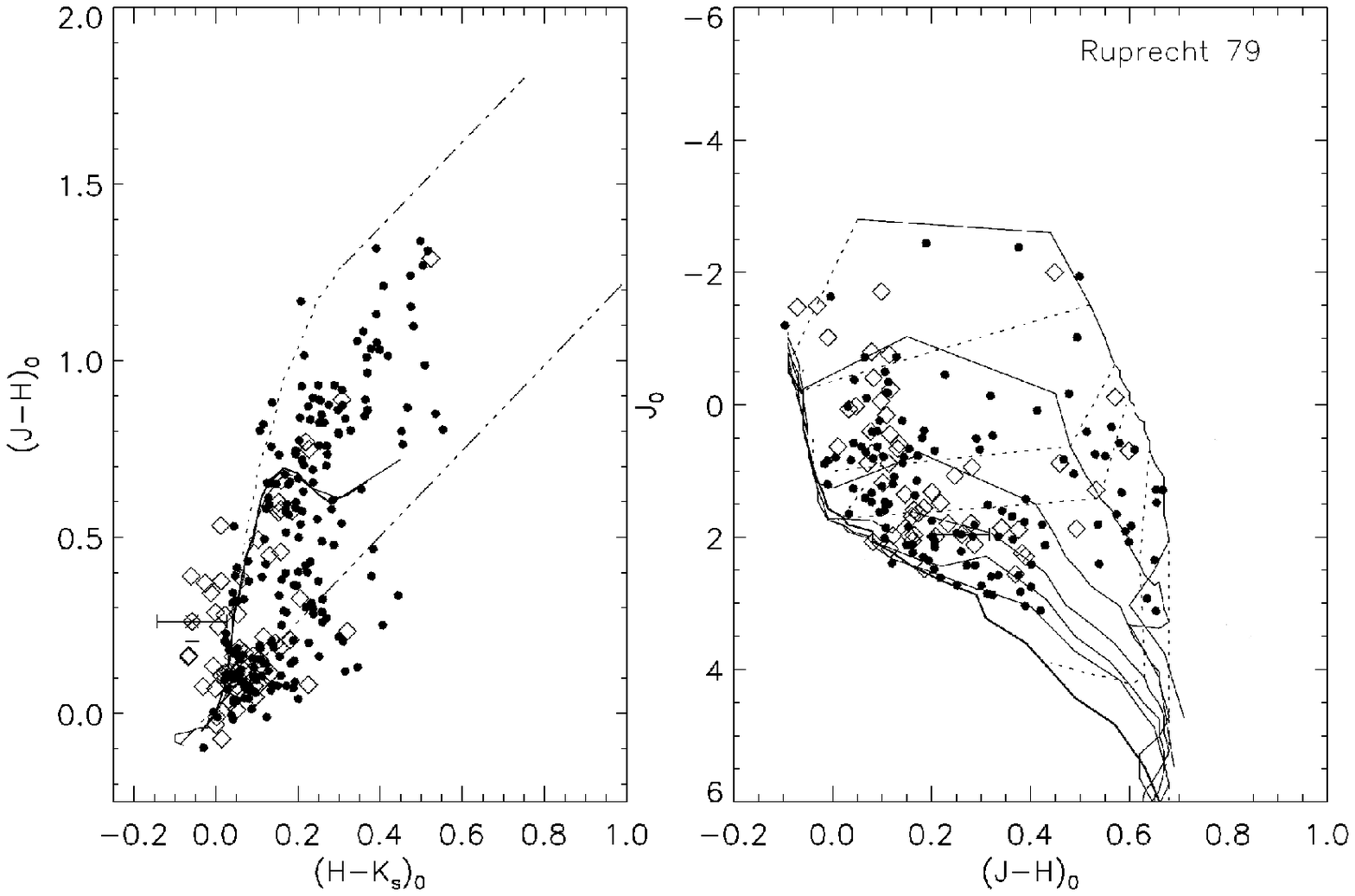}
\includegraphics[width=6.8cm]{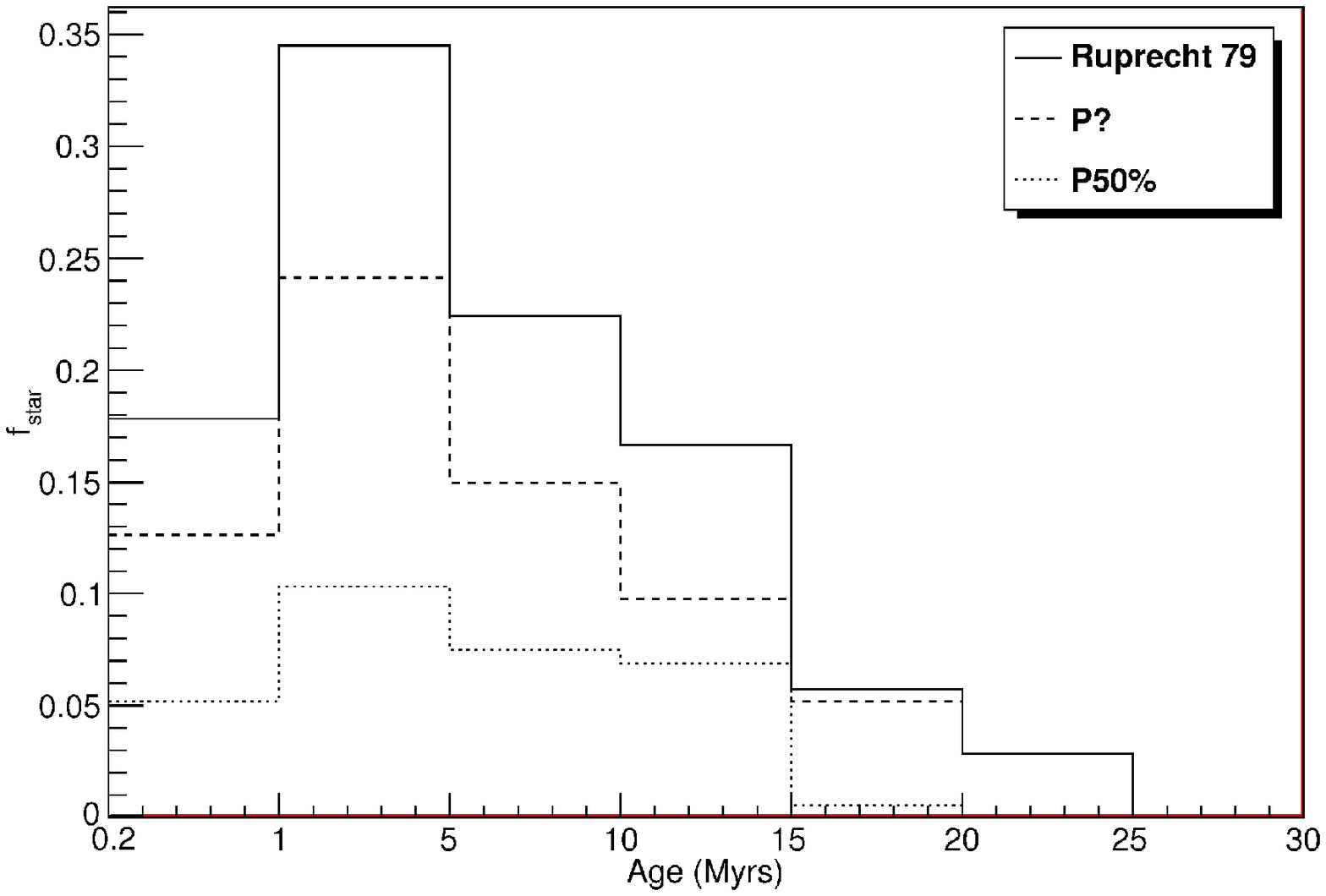}
\includegraphics[width=4.6cm]{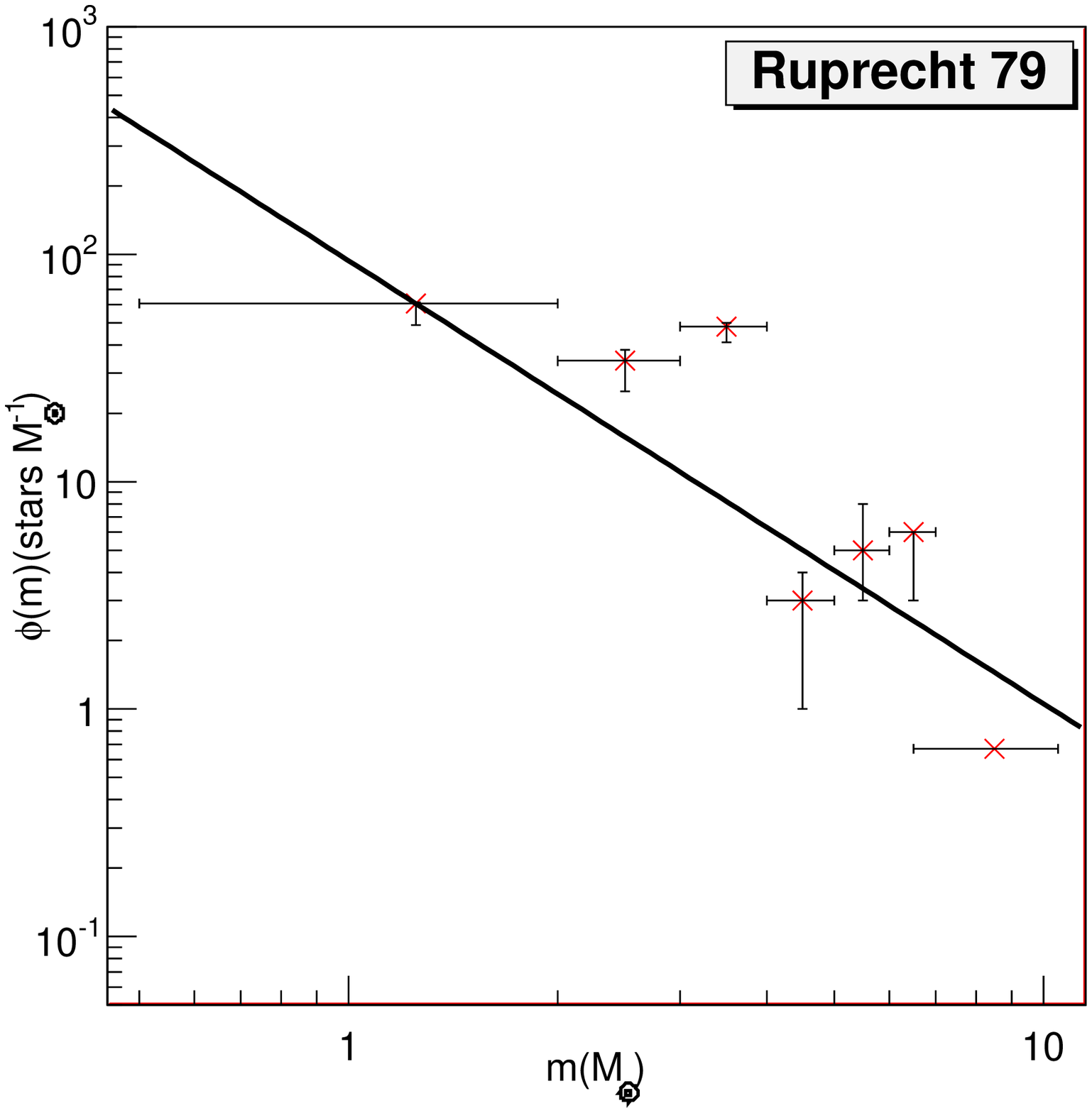}
\includegraphics[width=6.3cm]{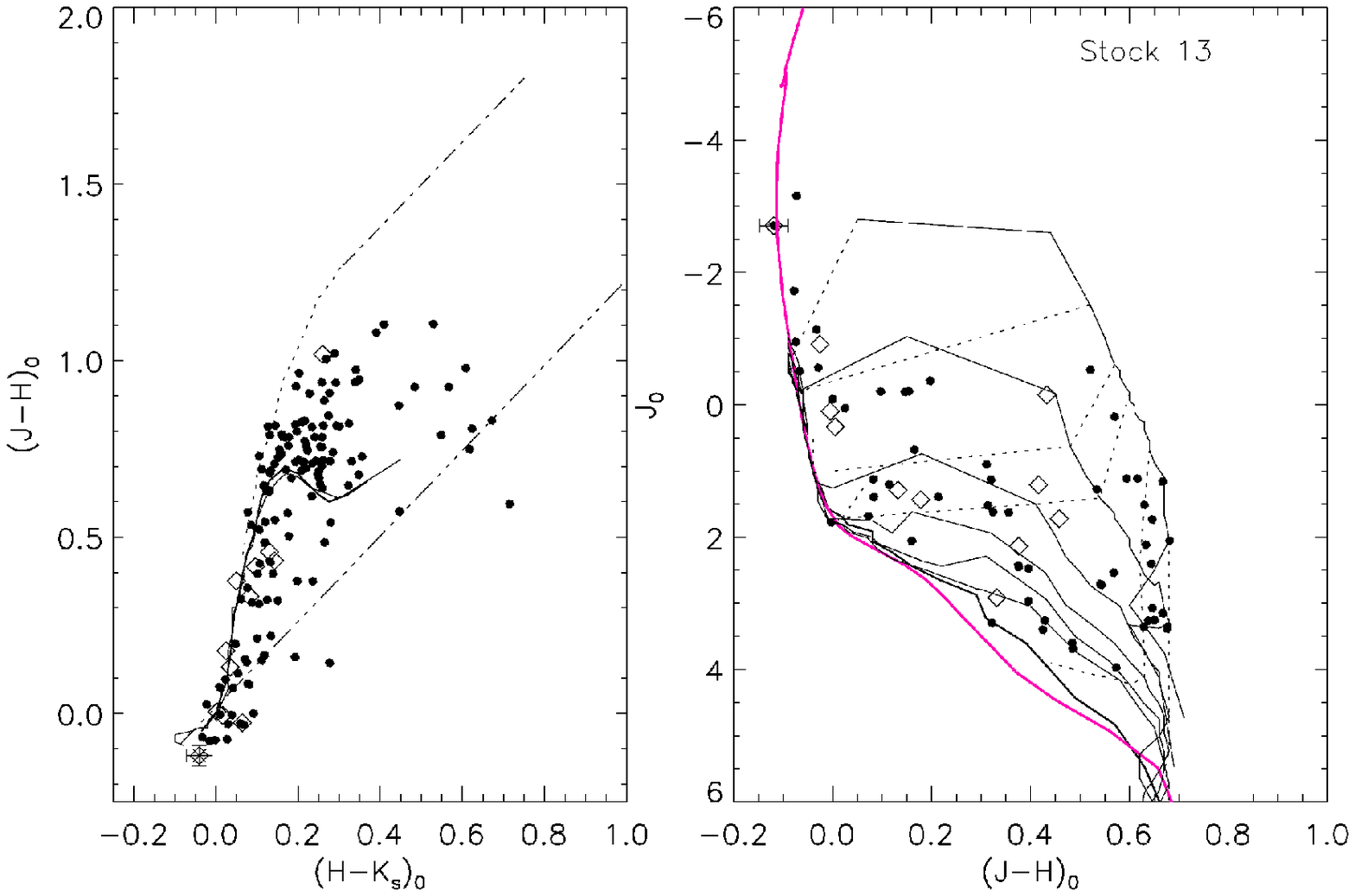}
\includegraphics[width=6.8cm]{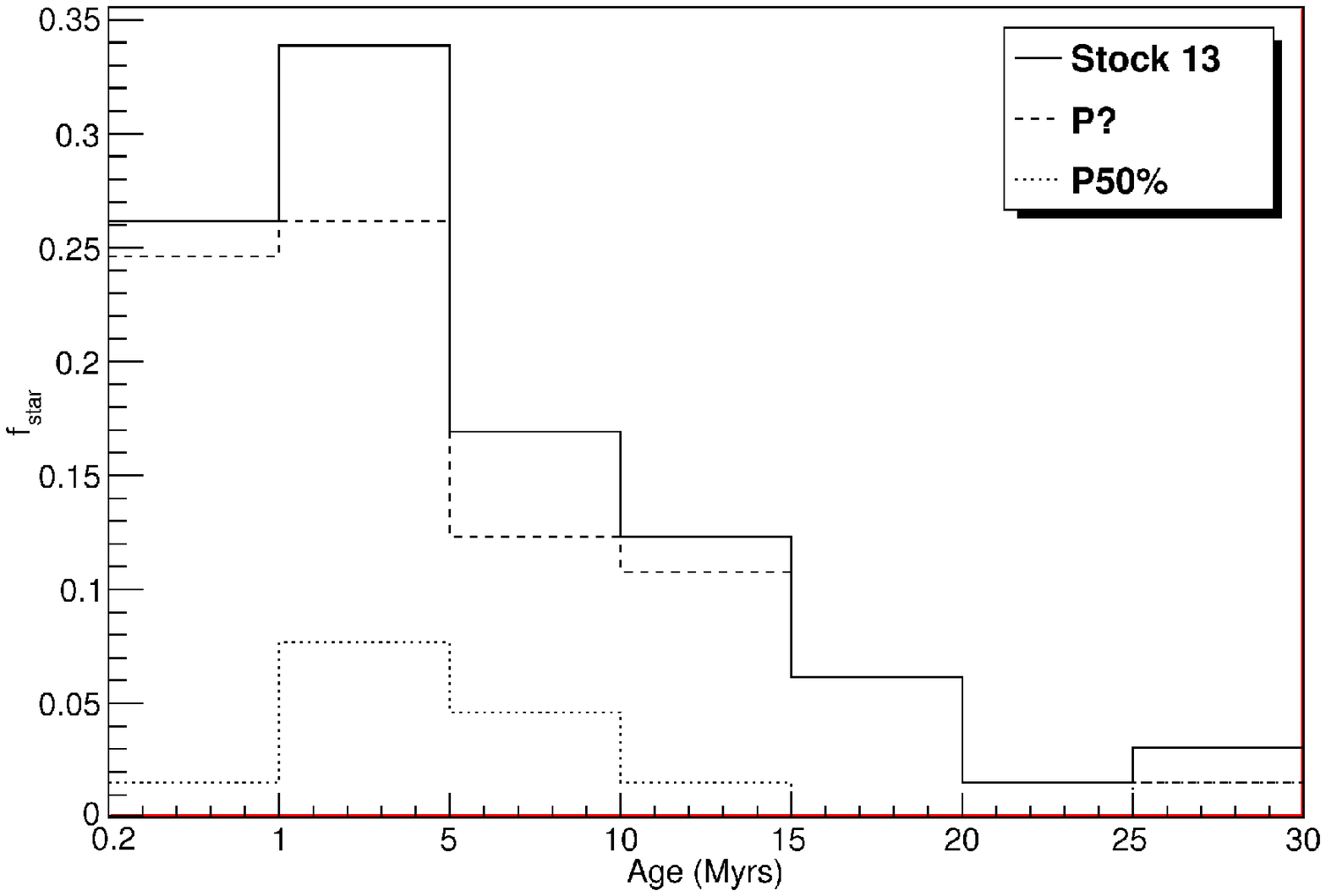}
\includegraphics[width=4.6cm]{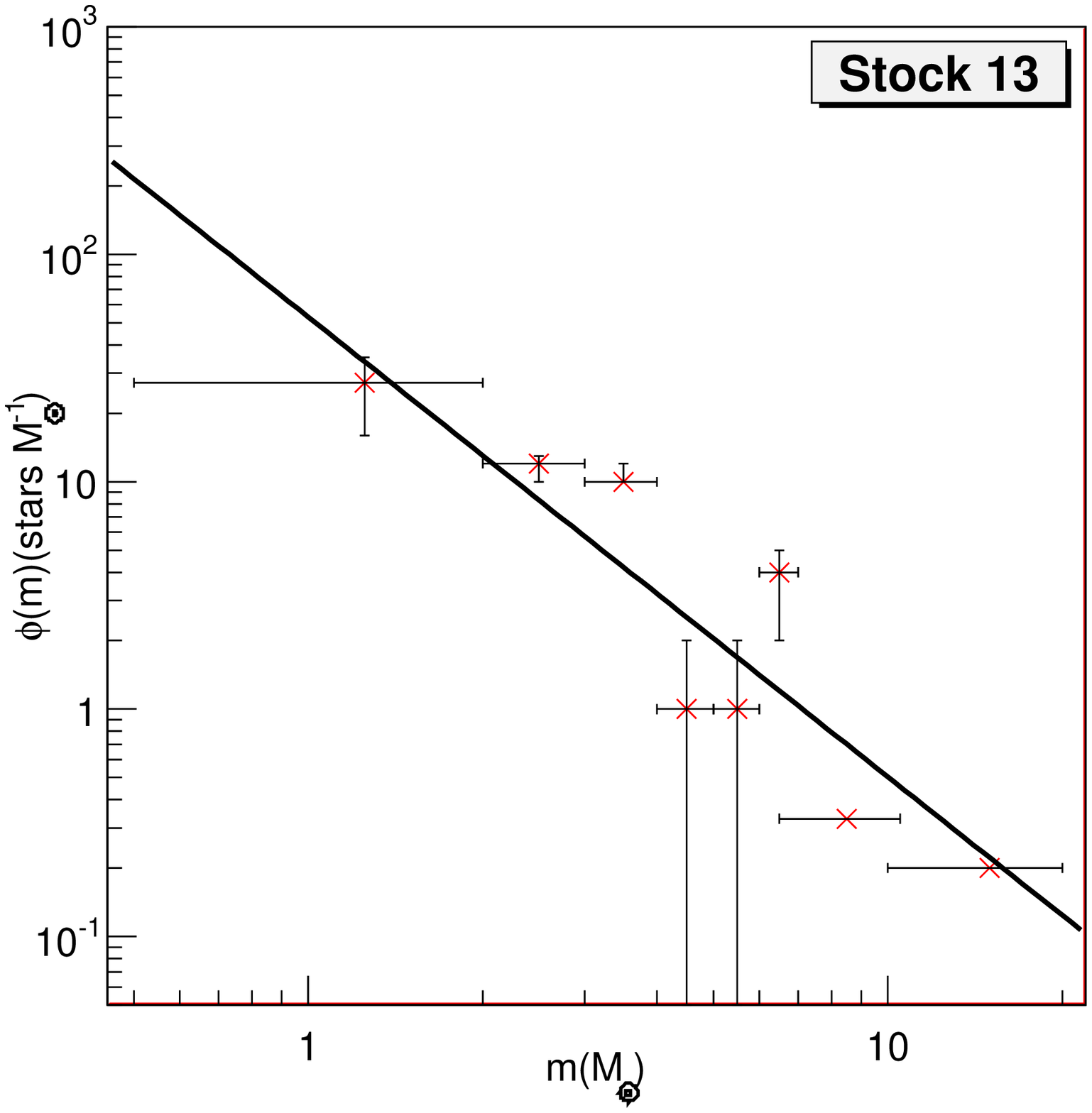}
\includegraphics[width=6.3cm]{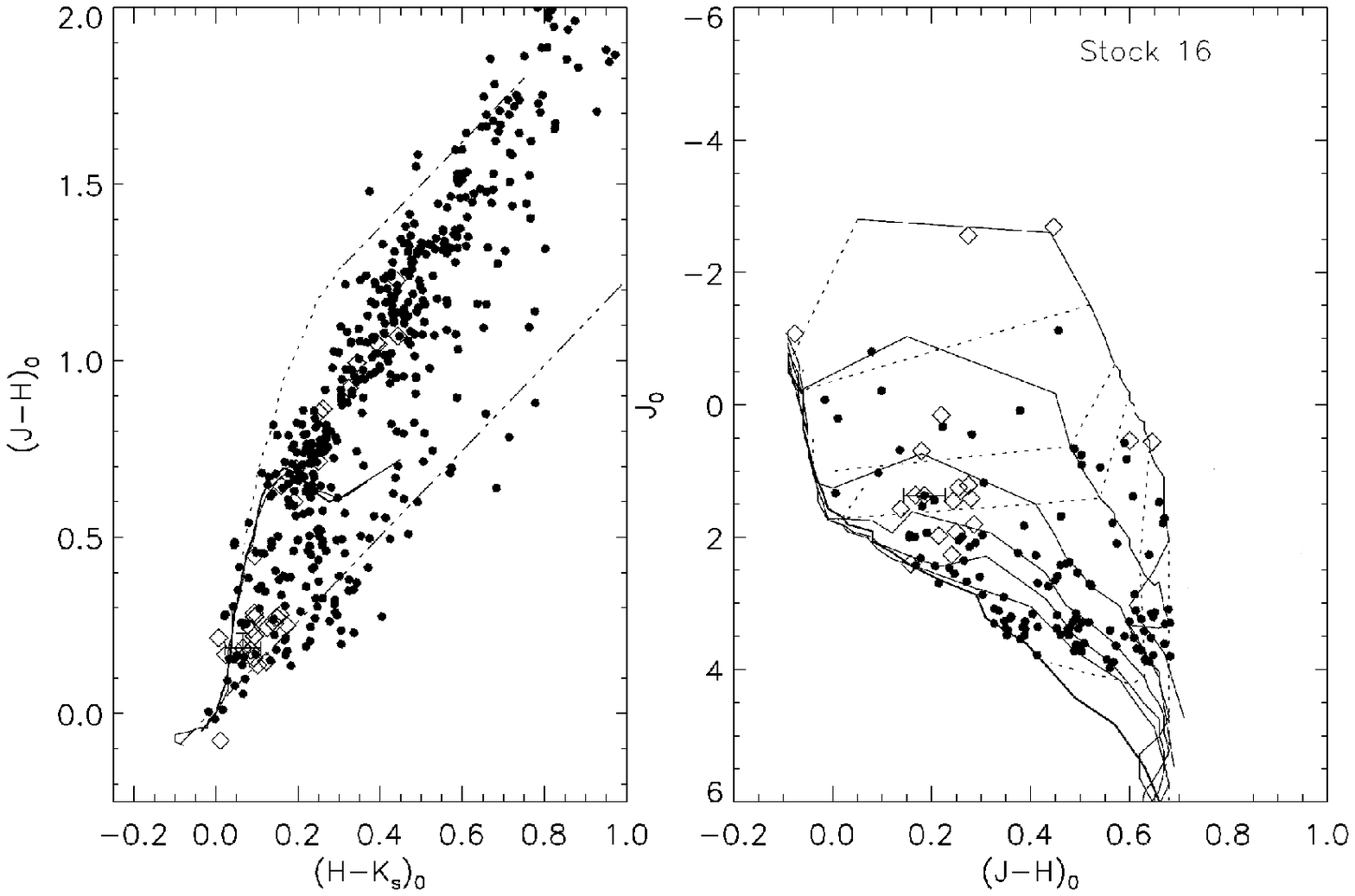}
\includegraphics[width=6.8cm]{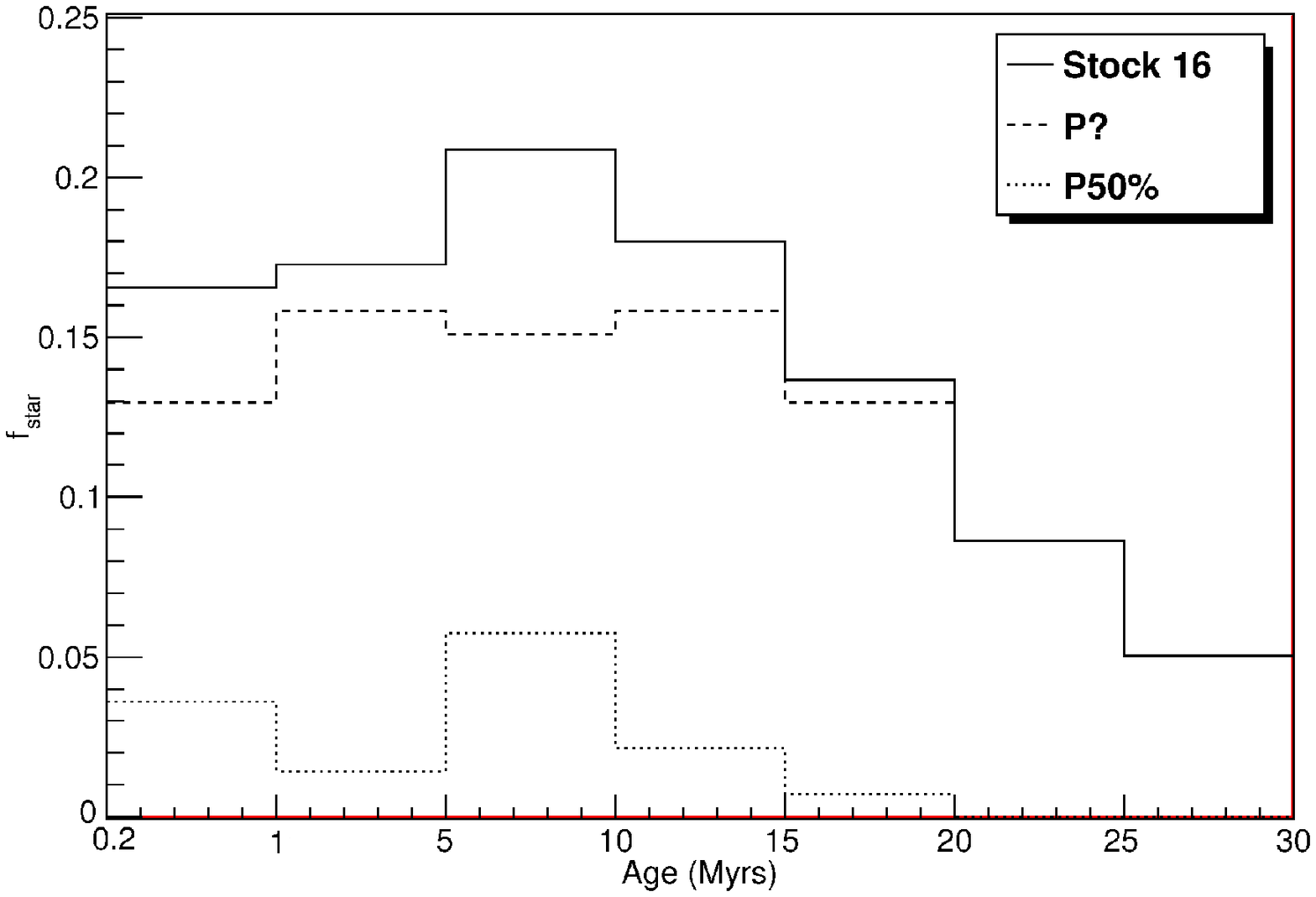}
\includegraphics[width=4.6cm]{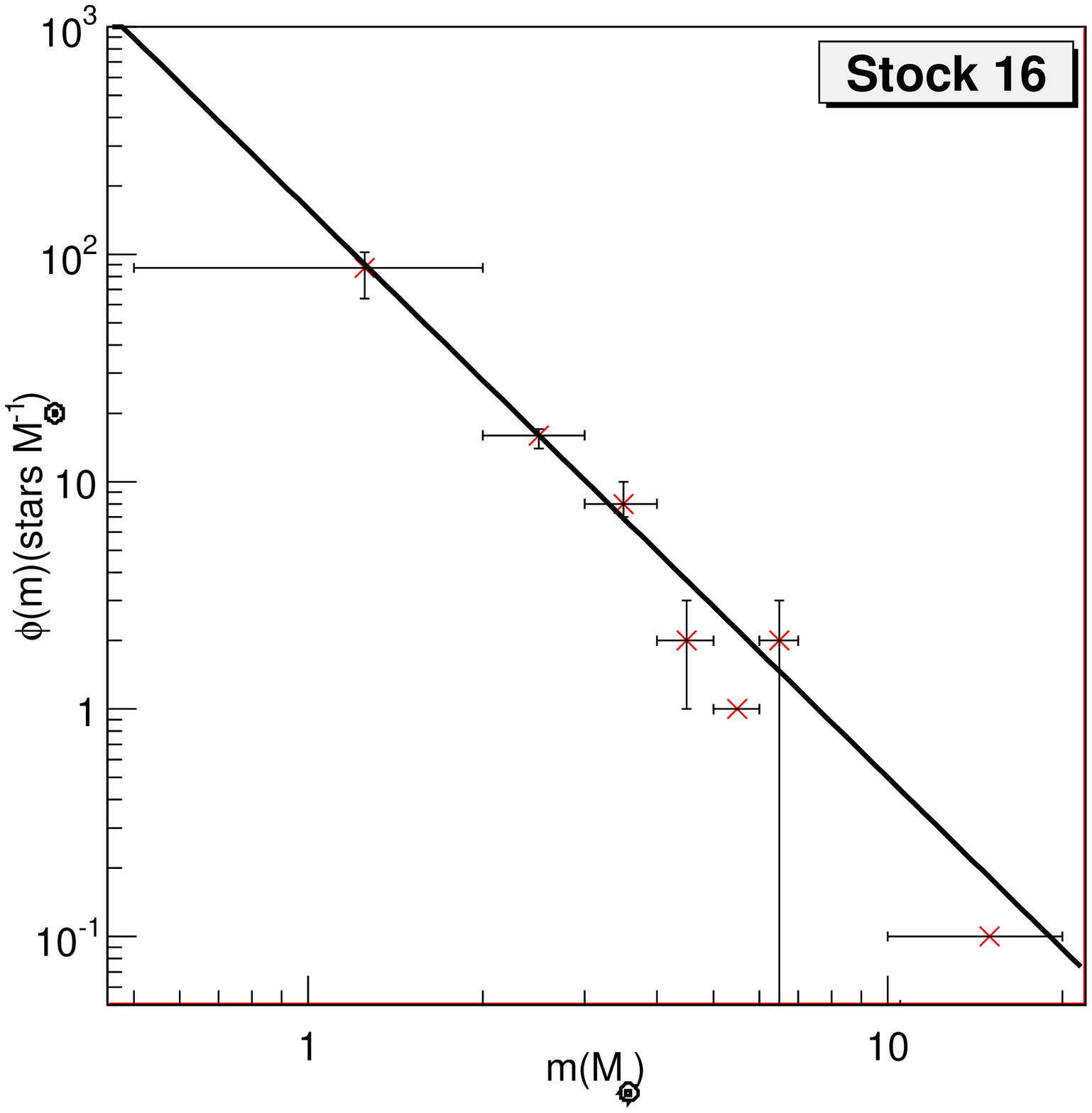}
\includegraphics[width=6.3cm]{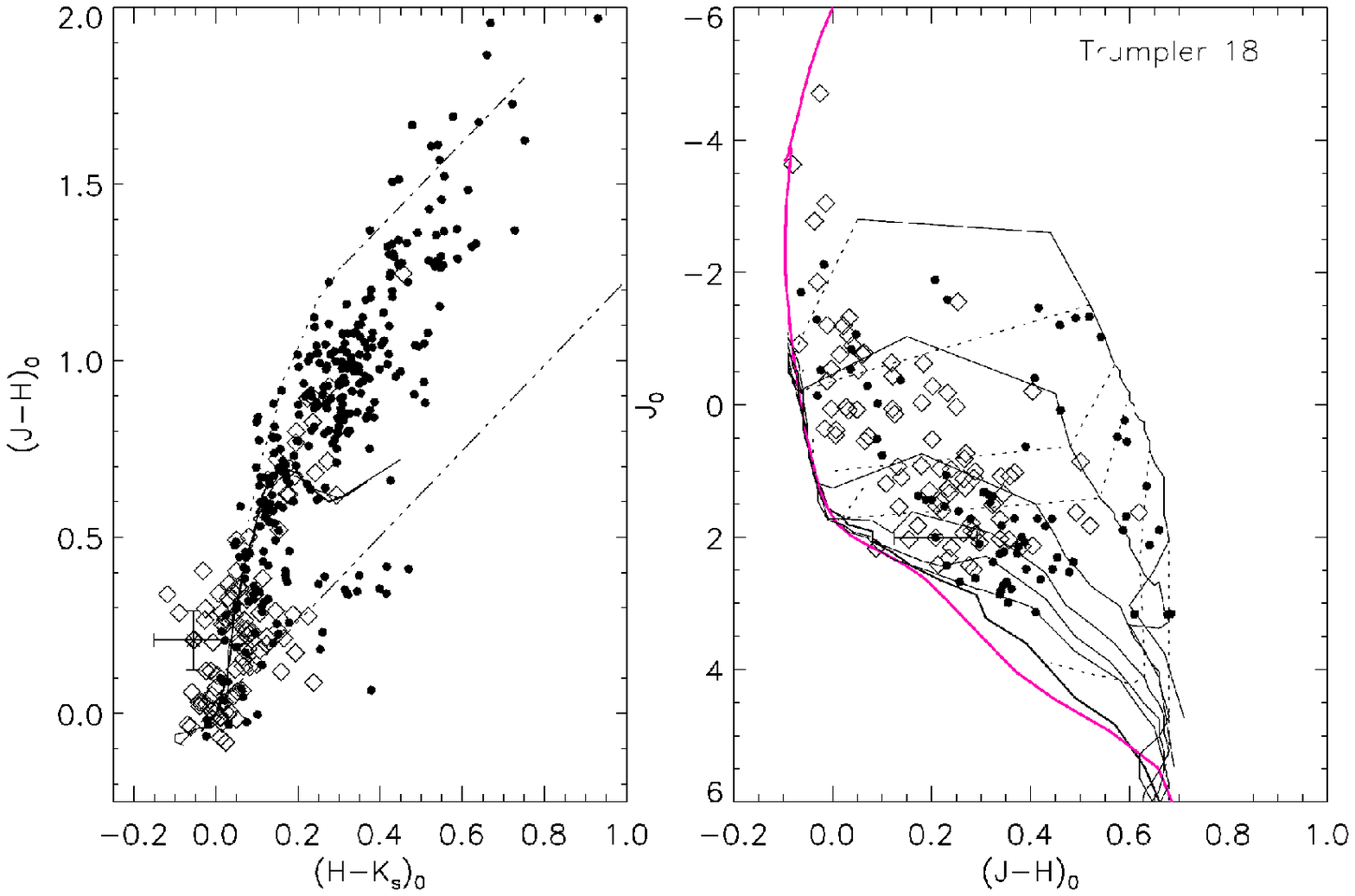}
\includegraphics[width=6.8cm]{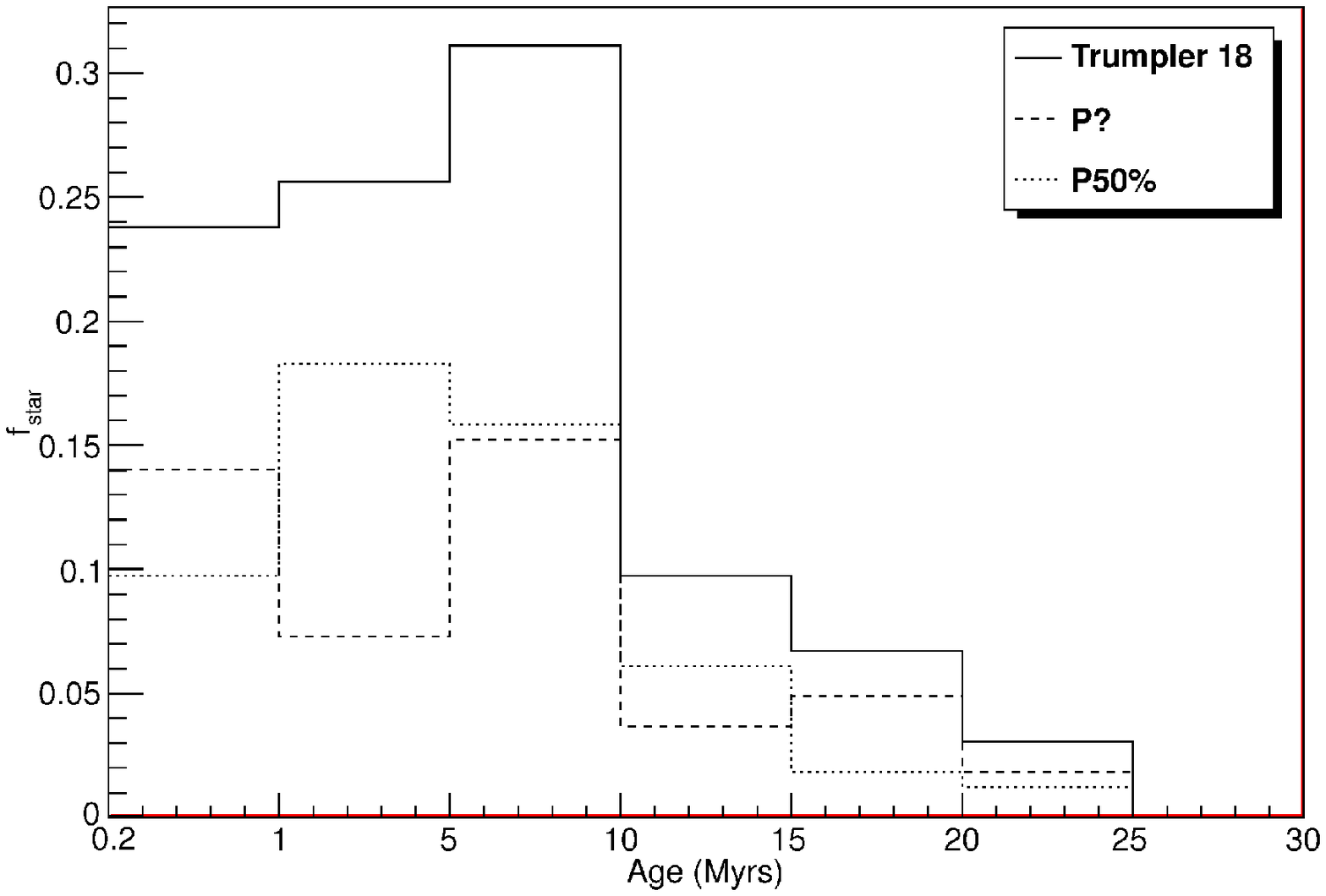}
\includegraphics[width=4.6cm]{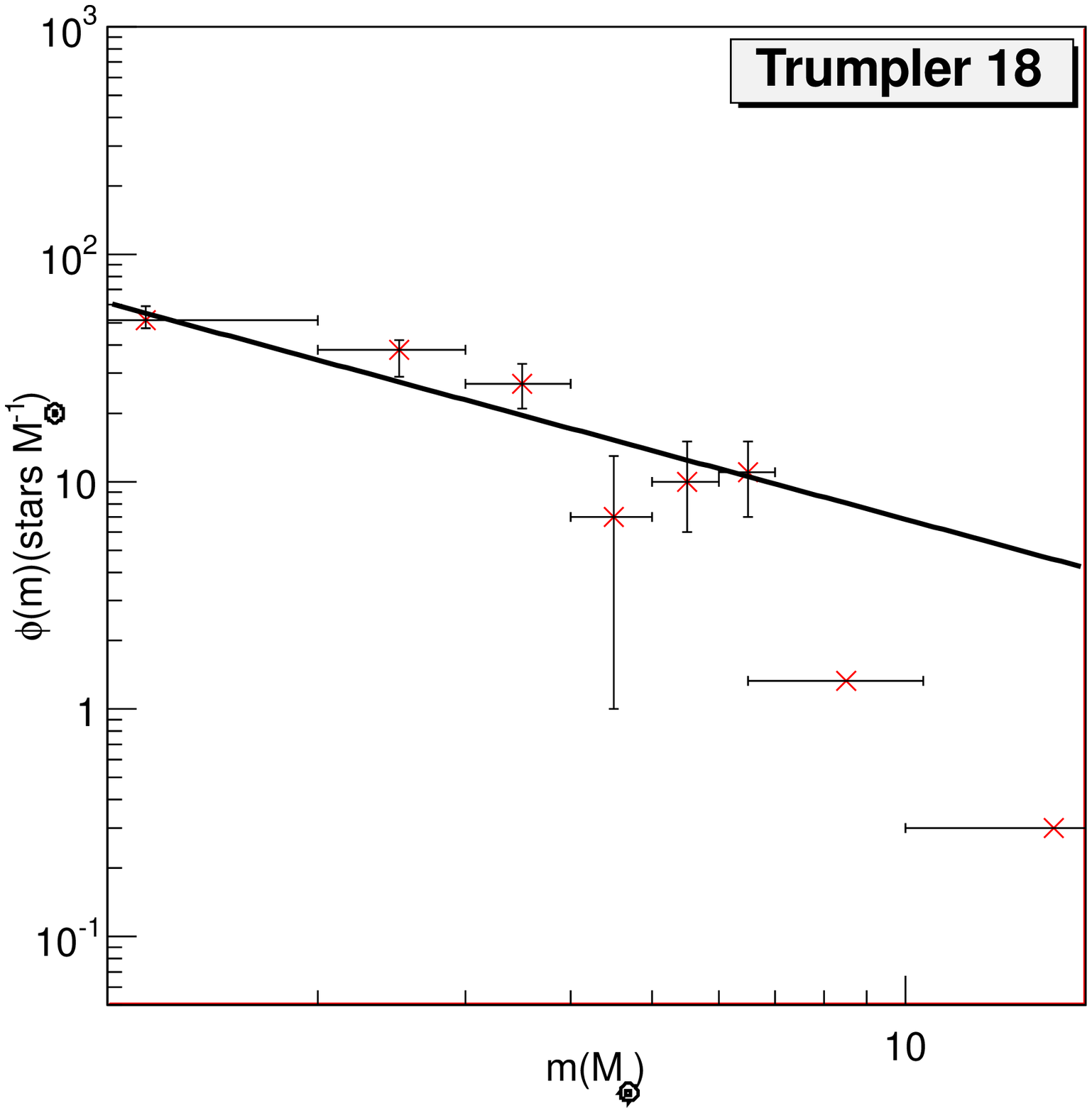}
\includegraphics[width=6.3cm]{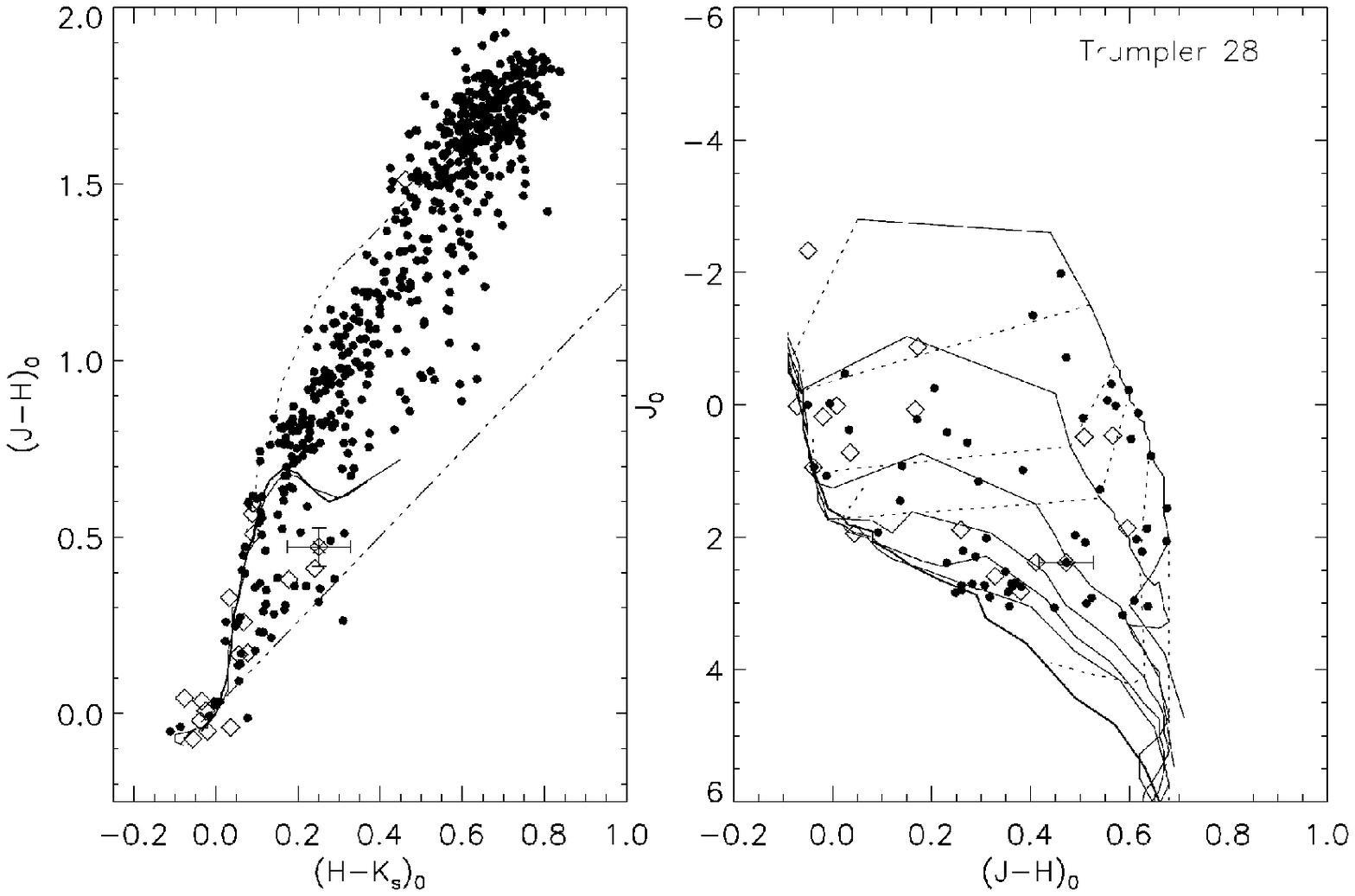}
\includegraphics[width=6.8cm]{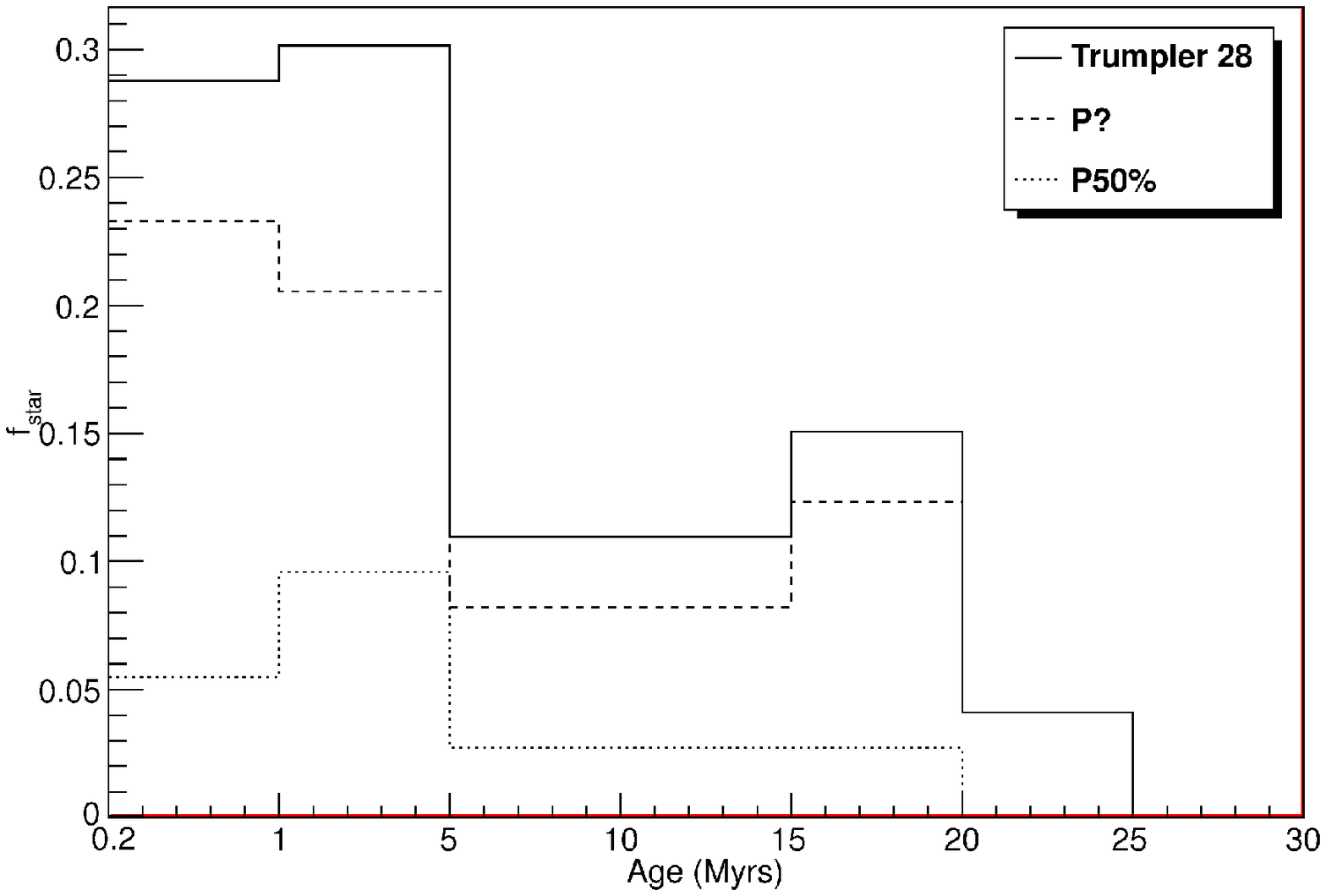}
\includegraphics[width=4.6cm]{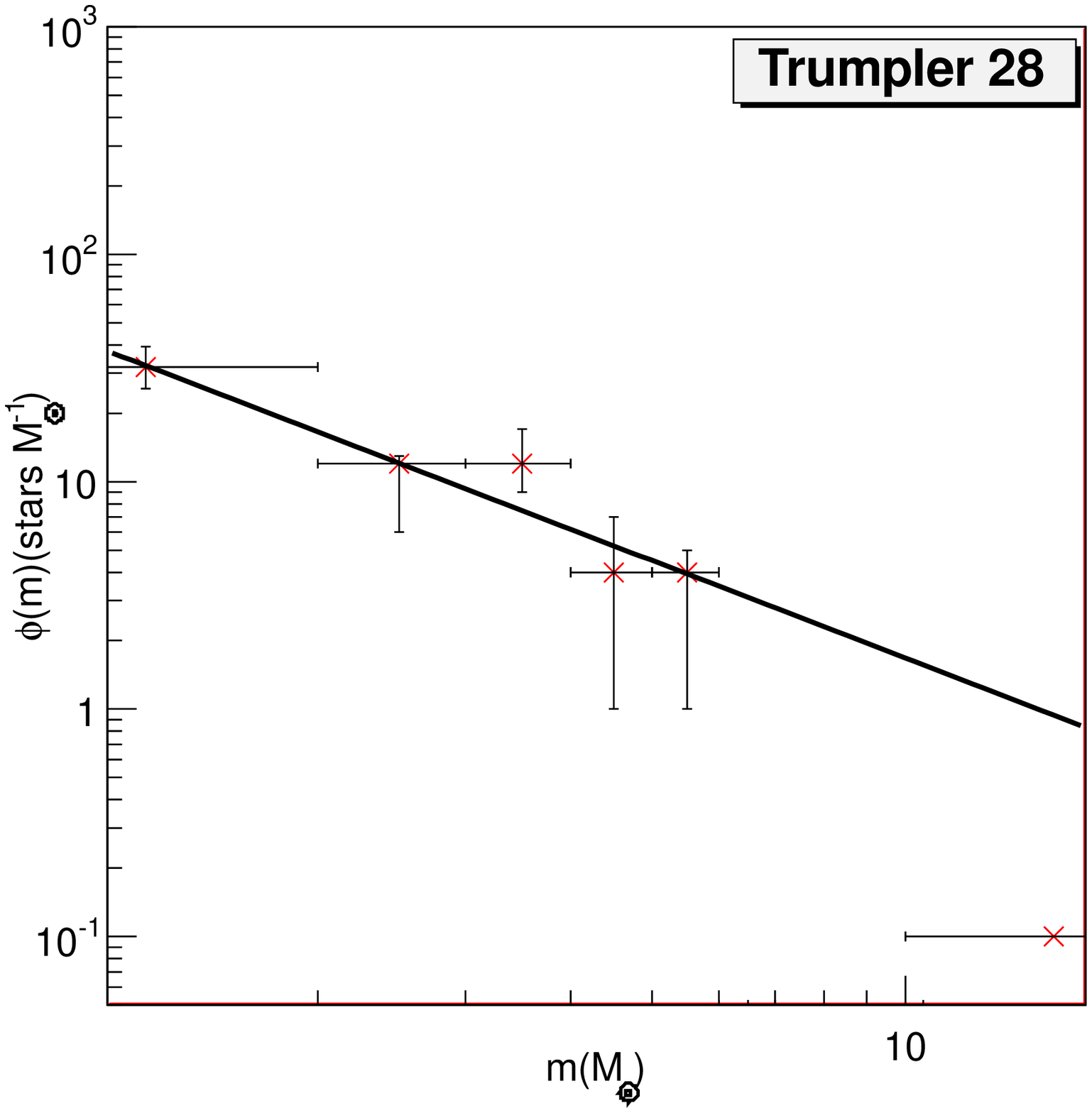}
\caption{The same as Fig A.3.}
\label{hmass}
\end{center}
\end{figure*} 

\end{appendix}

\end{document}